# Compositional and Interface Engineering of Hybrid Metal Halide Perovskite Thin Films for Solar Cells

Submitted in partial fulfillment of the requirements

of the degree of

**Doctor of Philosophy**

*by*

**Kashimul Hossain**

**(Roll No. 174120006)**

Supervisor

**Prof. Dinesh Kabra**

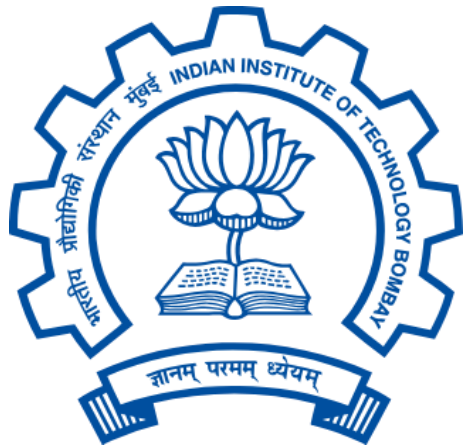

## Department of Physics

## INDIAN INSTITUTE OF TECHNOLOGY BOMBAY

**(Year 2024)**

# Approval Sheet

This thesis/dissertation/report entitled **"Compositional and Interface Engineering of Hybrid Metal Halide Perovskite Thin Films for Solar Cells"** by Kashimul Hossain (174120006) is approved for the degree of Doctor of Philosophy.

Examiners

_______________________

_______________________

_______________________

Supervisor (s)

_______________________

Chairman

_______________________

**Date:** 17$^{th}$ May 2024

**Place:** IIT Bombay, Powai-400076, Mumbai, Maharashtra, India.

# Declaration

I declare that this written submission represents my ideas in my own words and where others' ideas or words have been included, I have adequately cited and referenced the original sources. I also declare that I have adhered to all principles of academic honesty and integrity and have not misrepresented or fabricated or falsified any idea/data/fact/source in my submission. I understand that any violation of the above will be cause for disciplinary action by the Institute and can also evoke penal action from the sources which have thus not been properly cited or from whom proper permission has not been taken when needed.

_____________________________
(Signature)

Kashimul Hossain

(Roll No. 174120006)

Date: 17th May 2024

# Abstract

Perovskite solar cells (PSCs) are the fastest-growing photovoltaic (PV) technology in the solar cell community and have reached an efficiency close to that of commercial silicon (Si) solar cells. The organic-inorganic halide perovskite solar cell is an emerging PV technology and grabbed much attention due to its low cost, high efficiency, and ease of fabrication at lower temperatures 100°-200° C by solution-processed spin coating or thermal evaporation techniques. However, the photophysics and device engineering of PSCs have become a hot topic due to their exotic optoelectronic properties. This thesis demonstrates the compositional and interface engineering of hybrid metal halide perovskite thin film solar cells. Perovskite is a class of materials with a crystal structure $ABX_3$. However, the stability of the perovskite crystalline material depends on its composition. The different composition based perovskite crystal structures can be obtained by choosing multi-cations A, B, and multi-halides X. Therefore, we studied the role of monovalent cations (A) in the hybrid organic-inorganic metal halide perovskite solar cells in terms of the dielectric relaxation process. The multi-monovalent cation-based perovskite has lower defect states, confirmed via various optoelectronic measurements. Further, the correlation between defects and stability is established using the concept of the first principle calculation of the perovskite crystal structure. The increased hydrogen bonding of the FA-containing triple-cation perovskite system shows faster dielectric relaxation, which is the signature of lower defects and correlates with the stability of the perovskite solar cells. Further, we introduce a self-assembled monolayer (SAM) based hole transport layer (HTL) in the *p-i-n* device architecture PSC. In this work, we used the mixing engineering strategy of SAM with a conjugated polyelectrolyte. We dealt with the hydrophobicity and tailored the work function of the mixed SAM based HTL. Therefore, the HTL/perovskite interface is engineered, and associated device physics is discussed. However, we observed that the specific mixing ratio of SAM and polymer i.e. 9:1, shows excellent device performance (reproducible efficiency and stability), the lowest dark current, and the ideality factor close to the unity. Furthermore, the unity ideality factor is a primary signature of solar cells working near the radiative limit. Hence, we studied the dominant recombination mechanism on a complete solar cell using steady-state and transient measurements. We developed a consistent characterization scheme for studying solar cells operating under dominant bimolecular recombination. Our characterization schemes are validated using multiple experiments, such as predicting the reverse saturation current density both from the dark current and pseudo $J - V$, which are comparable to the experimental results. After that, we replace the opaque (Ag) electrode with a semi-transparent electrode (IZO) of the best performing opaque PSC to fabricate the semi-transparent solar cells. We demonstrate an efficient four-terminal tandem solar cell architecture of PSCs with silicon (Si) and cadmium telluride (CdTe) solar cells. In addition, we observed the lowest dark current for specific mixed (9:1) HTL, which is a prerequisite in photodetector applications. Therefore, a detailed photodetection analysis is discussed to study the scalable photodetector device. This thesis thoroughly studies perovskite compositional and interface engineering via various optoelectronic measurements. An in-depth device physics is discussed to study the interfacial defects between the charge transport and the perovskite layers. This thesis will be helpful in exploring a new class of perovskite materials and interface modification engineering for fabricating reproducible, stable, and highly efficient hybrid organic-inorganic metal halide perovskite solar cells.

# Publications included in this thesis:

1. **K. Hossain**, S. Singh, and D. Kabra "Role of Monovalent Cations in the Dielectric Relaxation Processes in Hybrid Metal Halide Perovskite Solar Cells" ACS Appl. Energy Mater. **2022**, 5 (3), 3689–3697.

2. **K. Hossain**, A. Kulkarni, U. Bothra, B. Klingebiel, T. Kirchartz, M. Saliba, and D. Kabra "Resolving the Hydrophobicity of the Me-4PACz Hole Transport Layer for Inverted Perovskite Solar Cells with Efficiency >20%" ACS Energy Lett. **2023**, 8 (9), 3860–3867.

3. **K. Hossain**, D. Sivadas, D. Kabra, and P. R. Nair "Perovskite Solar Cells Dominated by Bimolecular Recombination - How Far Is the Radiative Limit?" ACS Energy Lett. **2024**, 9, 2310–23127.

4. A. Paul[†], A. Singha[†], **K. Hossain**[†], S. Gupta, M. Misra, S. Mallick, K. R. Balasubramaniam, A. Munshi, and D. Kabra "4-T CdTe/Perovskite Thin Film Tandem Solar Cells with Power Conversion Efficiency >24%" ACS Energy Lett. **2024**, 9, 3019-3026. († indicates equal contribution to the work).

5. **K. Hossain**, B. Bhardwaj, and D. Kabra "Low Dark Current with High Speed Detection in a Scalable Perovskite Photodetector" Device. **2024**, 2, 100513.

6. **K. Hossain**, S. Nayak, and D. Kabra "Challenges and Opportunities in High Efficiency Scalable and Stable Perovskite Solar Cells" Appl. Phys. Lett. **2024**, 125, 170501-170510.

7. **K. Hossain**, A. Paul, S. Gupta, S. Koul, V. Chityala, and D. Kabra "Highly Efficient Four Terminal Tandem Solar Cells and Optical Loss Analysis" (Due for submission).

**Miscellaneous Publications:**

1. B. Sharma, S. Singh, **K. Hossain,** Laxmi, S. Mallick, P. Bhargava, and Kabra, D. Additive Engineering of 4, 4′-Bis (N-Carbazolyl)-1, 1′-Biphenyl (CBP) Molecules for Defects Passivation and Moisture Stability of Hybrid Perovskite Layer. Sol. Energy **2020**, 211, 1084–1091.

2. T. S. Chandra, A. Singareddy, **K. Hossain,** D. Sivadas, S. Bhatia, S. Singh, D. Kabra, and P. R. Nair "Ion Mobility Independent Large Signal Switching of Perovskite Devices" Appl. Phys. Lett. **2021**, 119 (2).

3. N. Gaur, M. Misra, **K. Hossain,** D. Kabra "Improved Thermally Activated Delayed Fluorescence based Electroluminescent Devices using Vacuum Processed Carbazol based Self-Assembled Monolayer" ACS Energy Lett. **2024**, 9, 1056-1062.

## Patents:

1. D. Kabra, **K. Hossain,** and N. Gaur "Tailored Work Function Based Hole Transport Layer in Optoelectronic Devices". (Application Number: 202321054565).

2. A. Paul, A. Singha, **K. Hossain,** S. Gupta, and D. Kabra "Perovskite/CdTe Thin Film Solar Cells Fabrication Methods and Thereof". (IITB Ref. No. PAT/PH/I12115-2/23-24)

## Presentations made at the Conference and symposium:

1. Resolving the Hydrophobicity of the Me-4PACz Hole Transport Layer for Inverted Perovskite Solar Cells with Efficiency >20%. **SymPhy**, Organised by IIT Bombay, India, **2024**. (Oral presentation).

2. Role of the Tailored Hole Transport Layer in the Inverted Architecture Based Perovskite Solar Cells. **NaMoSBio**. Organized by the IISER Kolkata, **India**. **2024**. (Poster presentation- **won the best poster award**)

3. Role of Monovalent Cation in the Dielectric Relaxation Process and Correlation of Defects with the Thermal Stability of Hybrid Metal Halide Perovskite Solar Cells. 2nd International Conference on Materials for Humanity (MH 22), Organized by **MRS**, **Singapore**, **2022** (Poster presentation).

4. Mixing Engineering of SAM, Me-4PACz with Ionic Polymer to Eliminate the Hydrophobic Issue of Me-4PACz/Perovskite Interface for Highly Efficient Reproducible PSCs. **IUMRS – ICA 2022**. Organized by the IIT-Jodhpur, **India**. **2022**, (Poster presentation).

5. Perovskite –Organic Monolithic Tandem Solar Cells Optimization and Progress. **SUNRISE Symposium**, **2021**. (Online-Oral presentation).

# Acknowledgments

I am profoundly grateful to my supervisor, Prof. Dinesh Kabra, whose unwavering support and guidance have been pivotal to the successful completion of my PhD. His constant motivation and encouragement for research in the area of perovskite photovoltaics always helped me in the ups and downs of my research work. I feel incredibly fortunate to have had such a dedicated mentor during my PhD career, and I sincerely appreciate his constant support and mentorship during those years. He always gave me the freedom to choose the research problem and encouraged independent thinking. His broad perspective on research problems during our discussions continually enhanced my understanding of advanced optoelectronic measurements. His extensive knowledge of optoelectronic devices greatly enriched my expertise during our research conversations. His spirit and passion for research have always motivated me and will carry with me in my future life regardless of the field I pursue. Besides the research guide, he is a lovely person and a source of positive energy with huge motivation. I am always grateful to him for giving me confidence and motivating me to make the correct decisions.

I am thankful to Prof. Pradeep R. Nair from the Department of Electrical Engineering, IIT Bombay, for the fruitful discussions, collaborative work, and scientific advice during my PhD candidature. I am thankful to Prof. K. L. Narasimhan from the Department of Electrical Engineering, IIT Bombay, for his constant helpful research feedback and scientific advice during my PhD candidature. I am thankful to my research progress committee members, Prof. Aftab Alam and Prof. Md. Aslam, for their valuable feedback and suggestions during my annual progress seminar. I am thankful to Prof. Aldrin Antony from Cochin University of Science and Technology for the fruitful discussions and suggestions during his stay at the Department of Electrical Engineering, IIT Bombay. I thank Prof. Michael Saliba, Prof. Thomas Kirtchaz, and Dr. Ashish Kulkarni for the collaborative work.

I would like to pay my deep gratitude to the IIT Bombay, National Centre for Photovoltaic Research and Education (NCPRE), Industrial Research and Consultancy Centre (IRCC), Sophisticated Analytical Instrument Facility (SAIF), UKRI Global Challenge Research Fund project, SUNRISE (EP/P032591/1), Indo-Swedish joint project funded by DST-India (DST/INT/SWD/VR/P-20/2019).

I am thankful to Dr. Shivam Singh for his immense help in the lab work, discussions, and collaboration. I am thankful to Dr. Amrita Dey for helping me learn spectroscopic measurements. I thank Dr. Ashish Kulkarni for discussing the research problems and collaborative work on the self-assembled monolayer-based hole transport layer. I thank Dr.

Dhanashree Moghe for the fruitful discussions on spectroscopic measurements. I thank Dr. Bosky Sharma for the collaborative work and for discussing research problems. I thank Mr. Saurabh Gupta and Mr. Ananta Paul for the tin oxide layer deposition using the ALD process and transparent electrode IZO deposition using the sputtering technique, respectively, for the semi-transparent perovskite solar cell fabrication. I thank Dr. Sakshi Koul for the optical loss analysis calculation in the tandem solar cells architecture. I thank Prof. Amit H. Munshi from Colorado State University for providing the CdTe solar cells to study the four terminal CdTe – Perovskite tandem solar cells.

I am grateful to have some fabulous colleagues, juniors, and friends in and out of the lab: Laxmi, Venkatesh, Shiwani, Gopa, Urvashi, Gangadhar, Kalyani, Sumukh, Neha, Phalguna, Surya, Chinmay, Dr. Sakshi, Dr. Mohit, Nrita, Manas, Saurabh, Preetam, Ananta, Abhijit, Shreyasi, Bhupesh, Vijay, Dr. Deeksha, Johns, Subir, and Karthik for research work discussions and wonderful memorable moments. Apart from the lab, life would have been incoherent in the IIT Bombay campus without my exuberant friends Dibarkar, Monish, Bikash, Dr. Poulab, Manisha, Sucharita, Pritam, Jadupati, Gurudayal, and Abhimanyu. I am also thankful to my close friends Nur, Oyaresh, Javed, Baban, Abhisek, Akashneel, Subhojit, Sushmita, Sujit, Moon, Isha, Ansar, Muktar, Selim, Samrat, Bapi, Nadia, Praveen, Kamlesh, Bhanu, Ankit, Manendra and Pintu.

I am deeply grateful to my favourite teacher, Tr. Paritosh Chandra Roy, who first ignited my passion for mathematics and consistently motivated me to pursue higher education. I extend my heartfelt thanks to my favourite dedicated mentors, Tr. Koushik Chakraborty, Prof. Ashok Ghosh, Prof. Chinmoy Basu, and Prof. Rabiul Islam for their exceptional teaching, immense support, and unwavering encouragement.

I am eternally grateful to my grandmother, Fatima Bewa, for her unwavering support in every circumstance, which has been truly invaluable. I am profoundly grateful to my beloved wife, Tania, for her steadfast support and for standing by me through every situation. I am deeply grateful to my parents (Mr. Jiaur Rahaman & Mrs. Kashima Bibi), Cousins (Sonali, Sanjib, Monisha, Samrat, Neha, Martuj, Nilufa, Nisha, Josim), relatives (Mr. Ajijul Ali, Mr. Aynat Ali) and in-laws (Mr. Tafiqueuddin Ahmed, Mrs. Maiful Khatun, Mrs. Mim, Mr. Alam, Mihan, Arman, Jewel) for their unconditional love, immense support and care without which my life would not have been flourished.

*I dedicate this thesis to my late grandparents.*

# Table of Contents:

# List of Figures:

right panel J − V characteristics shows the performance comparison of the conventional and new electrode designed PSCs. (adapted from ref.[130]).                                                           47 measure the self-absorption of the scattered light. (c) Thin film placed on the axis of the laser beam to measure the total PL.

**Chapter 4: Role of Monovalent Cation on the Dielectric Relaxation Processes in Hybrid Metal Halide Perovskite Solar Cells and Correlation with Thermal Stability** 100

**Figure 1: (a)** Schematic diagram of $ABX_3$ perovskite crystal structure. **(b)** Energy level diagram of each layer used in the p-i-n architecture based PSC, values are taken from literature.[37,38] **(c)** Typical frequency dependent photocurrent measurement ($I_{ph}(f)$) set-up used for this study and an inset image of actual device picture with two dots (blue on device & red off device) to show sites at which light is irradiated for $I_{ph}(f)$ measurements. **(d)** Quadrature part of the frequency dependent normalized photocurrent for $(FA_{1-X}MA_X)_{0.95}Cs_{0.05}PbI_3$, where X= 0.00, 0.25, 0.50, 0.75, and 1.00. **(e)** The frequency dependent device capacitance was measured for $(FA_{1-X}MA_X)_{0.95}Cs_{0.05}PbI_3$ perovskite absorbers. **(f)** The variation in Pz:PFN (9:1) PSC including the reflection (R), transmission (T), and integrated current density (Int. JSC). **(c)** Boxplot of short circuit current density (JSC), open circuit voltage (VOC), fill factor (FF), and power conversion efficiency (PCE%) over 30 devices. 

**Figure 11: (a)** $J - V$ characteristics of the photovoltaic devices under 1-Sun (100 mW/cm$^2$) condition in the forward and reverse scan direction for 0.805 cm$^2$ active area (11.5 mm x 7 mm) PSC device. The inset figure represents the PSC of active area 0.805 cm$^2$. **(b)** $J - V$ characteristic curves of photovoltaic devices device under AM1.5G illumination conditions employing different mixing ratios of Pz:PFN HTL, under forward and reverse scan for 0.175 cm$^2$ active area (2.5 mm x 7 mm) PSC device. **(c)** The power output of the PSCs taken under maximum power point tracking for Pz:PFN (6:4) with a stabilized efficiency of 17.32% (at 920 mV). The inset figure represents the PSC of active area 0.175 cm$^2$. 

**Figure 12: (a)** EQE/IQE & reflectance (R) spectra and **(b)** EQE & integrated JSC of PSC devices employing different mixing ratios of Pz:PFN HTLs. 

**Figure 13:** The performance of the solar cells corresponds to the active area of the device and has a bandgap of the perovskite layer of $E_g \sim (1.60 \pm 0.04)$ eV. 

**Figure 14: (a)** Structure of PFN-Br showing non-polar groups responsible for hydrophobic nature and **(b)** water contact angle showing hydrophobicity of PFN-Br. 

**Figure 15:** Solubility of PFN-Br in DMF and DMSO solvents. 

**Figure 16: (a)** XRD pattern of normalized (110) peak of the CsFAMA perovskite film deposited on bilayer Pz/PFN and Pz:PFN(9:1) coated substrates. The grain size was calculated from the FESEM images of the perovskite films (using Image-J software) deposited on bilayer **(b)** Pz/PFN and **(c)** Pz:PFN(9:1) coated substrates. 

**Figure 17:** Photographic image of the MAPbI$_3$ perovskite film deposited on the **(a)** MeO-2PACz and **(b)** Pz:PFN (9:1) HTLs coated glass substrates. FESEM images of the MAPbI$_3$ perovskite film deposited on the **(c)** MeO-2PACz and **(d)** Pz:PFN (9:1) HTLs coated glass substrates. 

**Figure 18: (a)** XRD pattern and **(b)** PL spectra (self-absorption corrected) of the MAPbI$_3$ perovskite thin film ( thickness ~ 250 nm ) deposited on the MeO-2PACz and Pz:PFN(9:1) coated glass substrates.[44] **(c)** $^1$H NMR of the CsFAMA, CsFAMA +PFN-Br and PFN-Br solution prepared in DMSO d6 solvent. **(d)** $^1$H NMR of the MAI, MAI +PFN-Br and PFN-Br solution prepared in DMSO d6 solvent. **(e)** $^1$H NMR of the FAI, FAI +PFN-Br, and PFN-Br solution prepared in DMSO d6 solvent. **(f)** The work function of the Pz:PFN HTLs was measured using KPFM study. 

**Figure 19: (a)** The photoluminescence (PL) spectra and the **(b)** absorption spectra of the MAPbI$_3$ perovskite film deposited on the MeO-2PACz and Pz:PFN (9:1) HTL. **(c)** The PL spectra after self-absorption correction of the MAPbI$_3$ perovskite films deposited on the MeO-2PACz and Pz:PFN (9:1) HTLs. 

**Figure 20: (a)** The KPFM scanning image of a standard gold (Au) sample deposited on ITO substrate. **(b)** KPFM scanning image of perovskite film deposited on the ITO substrate. The work function of the Pz:PFN HTL at two different locations and the average of them are listed in Table S5. All the Pz:PFN HTL samples were deposited on the ITO substrates. **(c)**

and **(d)** are the KPFM scanning image at two different points for the Pz:PFN (6:4). **(e)** and **(f)** are the KPFM scanning image at two different points for the Pz:PFN (8:2). **(g)** and **(h)** are the KPFM scanning image at two different points for the Pz:PFN (9:1). **(i)** and **(j)** are the KPFM scanning images at two different points for the Pz:PFN (9.5:0.5). 163

**Figure 21: (a)** Work function of Me-4PACz without and with PFN-Br interlayer and with different Pz:PFN mixing ratios, perovskite, PFN-Br, ITO, and Au metal electrode measured using KPFM. **(b)** The valence band maximum and the Fermi-level of the Me-4PACz and PFN-Br mixed HTLs are measured with respect to the vacuum using ultraviolet photoelectron spectroscopy (UPS). 165

**Figure 22:** The built-in potential (Vbi) is estimated from the dark current-voltage characteristics for the diode of ITO/Pz:PFN(X:Y)/Perovskite/PCBM/BCP/Ag. 166

**Figure 23: (a)** The dark current density plotted in semi-log scale as a function of applied bias. **(b)** The log (JSC/J0) vs. mixing ratio Pz:PFN (X:Y), where X and Y represent the volume ratio for Me-4PACz and PFN-Br respectively. 167

**Figure 24: (a)** The $V_{OC}$ of the representative devices as a function of short circuit (JSC) and reverse saturation current density (J0). **(b)** The $V_{OC}$ of the representative devices as a function of the perturbed charge carrier lifetime measured from TPV measurement at 1-Sun illumination condition. **(c)** Electroluminescence (EL) quantum efficiency plotted as a function of JSC and J0. 168

**Figure 25: (a)** Normalized transient photovoltage (TPV) decay profile of the Pz:PFN based PSCs under 1-Sun light bias. **(b)** Intensity dependent $V_{OC}$ (please see Figure 26) of Pz:PFN HTL based PSCs, solid line is fit to the Shockley diode equation.[55] **(c)** The voltage shift ($\Delta V$) (please see Figure 27) of the illuminated current and dark shifted current. **(d)** The series resistance calculated from the voltage shift by $R_S = \Delta V JSC$ . [56,57] 169

**Figure 26:** Raw data plot of Figure 25b. Intensity dependent $J - V$ characteristics of Pz:PFN HTL based PSCs of **(a)** Pz:PFN (6:4), **(b)** Pz:PFN (7:3), **(c)** Pz:PFN (8:2), **(d)** Pz:PFN (9:1), and **(e)** Pz:PFN (9.5:0.5). The ND 0.0 indicates 1-Sun light intensity and the rest are indicating 1-Sun light is passing through ND filters of optical density 0.1, 0.2,0.3, etc. 171

**Figure 27:** Raw data plot of Figure 25c. The scattered point with the line indicates the illuminated current, and the solid line plot indicates the dark current shifted by JSC and merged with the illuminated JSC for the Pz:PFN HTL based PSCs of **(a)** Pz:PFN (6:4), **(b)** Pz:PFN (7:3), **(c)** Pz:PFN (8:2), **(d)** Pz:PFN (9:1), and **(e)** Pz:PFN (9.5:0.5). The difference between the voltage for the same current is represented by $\Delta V$ (see Figure 25c) and used to calculate the series resistance (see Figure 25d) from the voltage shift by $R_S = \Delta V JSC$ . [56,57] 172

**Figure 28:** The plots represent the pseudo $J - V$ (solid point) measured from Suns-VOC and real $J - V$ (half solid points) at the 1-Sun condition for the Pz:PFN HTL based PSCs of **(a)** Pz:PFN (6:4), **(b)** Pz:PFN (7:3), **(c)** Pz:PFN (8:2), **(d)** Pz:PFN (9:1), and **(e)** Pz:PFN (9.5:0.5).[62] 173

**Figure 29:** The reverse saturation current density is calculated from the dark current (Jd) i.e. J0, d and from pseudo $J - V$ i.e. J0, svwhich is plotted in a semi-log scale along with the dark current. The pseudo $J - V$ is measured from the Suns-VOC measurement for the Pz:PFN HTL

based PSCs of **(a)** Pz:PFN (6:4), **(b)** Pz:PFN (7:3), **(c)** Pz:PFN (8:2), **(d)** Pz:PFN (9:1), and **(e)** Pz:PFN (9.5:0.5). The J0, d and J0,sv are tabulated in Table 8. [62–64] 

**Figure 30: (a)** The electroluminescence quantum efficiency (QE$_{EL}$) is a function of injected current density and the inset represents the digital image of the device during a 5 mA injected current. **(b)** The QE$_{EL}$ at the injected JSC of the representative PSCs. 

**Figure 31:** The J − V scans were taken periodically of the unencapsulated devices, measured in RH ∼ 40% for >3000 hours. The normalized J − V parameters of the **(a)** Pz:PFN (6:4) and **(b)** Pz:PFN (9:1) based PSCs over 3000 hours were plotted. The individual parameters are compared in **(c)** JSC **(d)** VOC **(e)** FF **(f)** PCE. 

**Figure 32:** Top surface SEM images of perovskite thin films deposited on Pz:PFN(9:1) coated ITO substrates. The perovskite films were annealed at 100º C for 30 minutes and cooled down to room temperature. Further, the perovskite films were annealed at 85º C for **(a)** 0.0 hours, **(b)** 1.5 hours, **(c)** 3.0 hours **(d)** 4.5 hours **(e)** 6.0 hours, and **(f)** 7.5 hours respectively. 

**Figure 33: (a)** Absorption spectra and **(b)** crystallographic XRD plot of the 85º C annealed perovskite films at 0.0 hour, 1.5 hours, 3.0 hours, 4.5 hours, 6.0 hours, and 7.5 hours respectively. 

**Figure 34: (a)** The operational stability of the un-encapsulated device was measured under N$_2$ environment and kept under constant 1-Sun illumination. **(b)** For thermal stability, we kept the device at 85º C and measured the photovoltaic performance. We kept the device at 85º C for approximately 8 hours and then kept it at room temp. Again, on the next day, we kept the device at 85º C and measured the PV device performance and so on. 

**Chapter 6: Perovskite solar cells dominated by bimolecular recombination- how far is the radiative limit?** 

**Figure 1**: Current density vs. Voltage (J-V) characteristics of the photovoltaic devices under 1-sun (100 mW/cm2) illumination conditions in the forward and reverse scan direction. The active area of the devices used are 0.175 cm$^2$ and 0.805 cm$^2$. The inset figure represents the cross-sectional view of the device (glass/ITO/Pz:PFN/Perovskite/PCBM/BCP/Ag). (b) EQE, IQE spectrum with reflection spectra (R), and the integrated current densities (Int. J$_{SC}$) over the AM 1.5G spectra. (c) Comparison of our devices against the state-of-the-art from literature. The open-circuit voltage (VOC) denoted by the open symbols and the efficiency (solid symbols) are plotted against the active area of the inverted architecture-based PSCs having a bandgap of Eg = 1.6 ± 0.02 eV ( please see Table 2 for references, etc.). Colour symbols:: ◯ -black circle: ref.1, △-red up-pointing triangle: ref.2, ▽- violet down-pointing triangle: ref.3, ◇-blue diamond: ref.4, ◁ -cyan left-pointing triangle: ref.5, ▷- magenta right-pointing triangle: ref.6, ⬡-dark yellow hexagon: ref.7, ☆- olive star: this work on the active area of 0.175 cm$^2$, ☐- wine square: this work on the active area of 0.805 cm$^2$. 

**Figure 2: (a)** The steady state efficiency at maximum power point tracking under 1-Sun condition for 0.175 cm$^2$ active area PSC. **(b)** EQE, IQE spectrum with reflection spectra (R), and the integrated current densities (Int. J$_{SC}$) over the AM 1.5G for 0.805 cm$^2$ PSC. 

**Figure 3**: Boxplot histogram of short circuit current density ($J_{SC}$), open circuit voltage ($V_{OC}$), fill factor (FF), and power conversion efficiency (PCE%) over 30 devices of device architecture glass/ITO/Pz:PFN(9:1)/Perovskite/PCBM/BCP/Ag. 203

**Figure 4: (a)** Schematic of the experimental set-up of steady-state Current vs. Voltage characteristics measurement at different intensities. **(b)** Current density vs. Voltage ($J - V$) characteristics of the photovoltaic devices under varied illumination conditions. The intensity of the 1-sun light changed using a set of neutral density (ND) filters. The legends in the figure indicate that the DC light is passing through different ND filters of optical density 0.1, 0.2, 0.3, and so on. For example, ND 0.1 means the DC background 1-sun light passing through ND filter of optical density 0.1 and the resultant intensity to the device is ($100 10 O. D. mW/cm^2$) 79.43 mW/cm$^2$. **(c)** The variation in JSC with respect to different illumination intensities. 205

**Figure 5**:**(a)** Schematic of the experimental set-up of transient photovoltage (TPV) measurement. Decay profile of the transient photovoltage at different DC background intensities while the perturb excitation laser pulse (ON time 500 ns) intensity. **(b)** The TPV signal amplitude ($\Delta VOC, max$) decay profile of a perovskite solar cell with respect to different DC background illumination. The legends indicate that the DC light is passing through different ND filters of optical density 0.1, 0.2, 0.3, and so on. ND 0.1 means the DC background 1-sun light passing through ND filter of optical density 0.1 and the resultant intensity to the device is ($100 10 O. D. mW/cm^2$) 79.43 mW/cm$^2$ . **(c)** The normalized TPV signal at different background DC light intensities. 207

**Figure 6**: Comparison of the perturbation light and the background light. During the measurement, the laser intensity was kept fixed at 10 mW with pulsed with 500 ns and repetition rate 1 kHz whereas the background intensity changed from 100 mw/cm$^2$ to 5.01 mw/cm$^2$. 208

**Figure 7**: The effect of capacitance on the charge carrier recombination lifetime from transient photovoltage measurement (TPV). The red plot represents the measured data, the black plot represents the time due to the capacitance effect, and the blue plot represents the corrected charge carrier recombination lifetime. 210

**Figure 8**: Optoelectronic characterizations to explore radiative detailed balance limits of perovskite solar cells. The top panel shows the schematic of the experimental setup (not drawn to scale). The solar cell is subjected to a constant illumination which is modulated by laser pulses, and the resultant open circuit voltage transients are measured. The bottom panel shows the time dependence of illumination intensity (I), corresponding photo-generation rate (G) and open circuit voltage (VOC). 212

**Figure 9**: Intensity dependence of VOC and dark $J - V$ of perovskite solar cells. **(a)** Variation of the open-circuit voltage ($VOC, 0$) with steady state illumination intensity. The ideality factor obtained[35] from $VOC, 0$ vs. ln(I0) is 1.05, which indicates that the device could be operating close to the radiative detailed balance limit. **(b)** The dark current characteristics measured from $J - V$ scans and Suns VOC measurements.[36–38] The $Jdark, J - V$ on the left y-axis indicates measurement from dark $J - V$ scans whereas the $Jdark, SV$ on the right y-axis indicates the dark current estimated using Suns VOC method.[39,40] Solid lines indicate numerical fits to obtain parameters like reverse saturation current density and ideality factor. 218

**Figure 10**: Comparison of experimental results against theoretical benchmark criteria (eq. 6.5) for perovskite solar cells. **(a)** $\Delta VOC$, max and $\tau - 1$ variation with I0 are in accordance with eq. 6.5b,c **(b)** Direct extraction of k1 and k2 using $\tau - 1$ vs. I0 plot, from the intercept and the slope respectively (i.e., using eq. 6.5c). **(c)** $\ln(\tau)$ varies linearly with $VOC, 0$ with slope close to $-q/2kT$, as anticipated by eq. 6.5d. These trends indicate that the device is limited by bimolecular recombination.                                                  220

**Figure 11**: **(a)** The transient photovoltage (TPV) signals a rising profile for different background intensities. **(b)** The estimation of k2 from eq. 6.5b for different background intensities.                                                                                        222

**Figure 12**: **(a)** The dark current measured (experimental) from the dark $J - V$ characteristic and from the Suns VOC measurement. The Suns VOC plot is obtained from the intensity dependent $J - V$ characteristics.[37,38] The dark current from the Suns VOCis obtained by shifting the pseudo $J - V$ by JSC towards the dark current.[39,40] **(b)** The dark current is estimated from the achievable limit calculation by using $Jdark, SRH = qk1niWeqV2kT$ and $Jdark, BB = qk2ni2WeqVkT$.                                                             227

**Figure 13**: **(a)** AM 1.5G spectrum is taken from the PV lighthouse, which is used to calculate the integrated current density over the EQE spectrum of the PSCs.[65] **(b)** The ideal EQE spectrum (no loss) i.e. 100% EQE is considered, and the corresponding integrated JSC is calculated for 1.6 eV (300 to 776 nm) solar cells. **(c)** the achievable limit of the PV parameters calculated using integrated $JSC, SQ \approx 26$ mA/cm$^2$ (calculated from **Figure 13b**) and $k1 = 104$ s$^{-1}$ $k2 = 10 - 10cm^3s^{-1}$, $k3 = 10 - 28 cm^6s^{-1}$, and $ni = 0.5 \times 106$ cm$^{-3}$ in equation (6.6). **(d)** The achievable limit of the PV parameters calculated using integrated $JSC, SQ \approx 26$ mA/cm$^2$ (calculated from **Figure 13b**) and revised values of $k1 = 0$, $k2 = 10 - 10 cm^3s^{-1}$, $k3 = 0$, and and $ni = 0.5 \times 106$ cm$^{-3}$ in equation (6.6).                       230

**Figure 14**: **(a)** Transient photocurrent (TPC) using a 490 nm modulated laser of pulse duration 5 μs, and repetition rate 10 KHz with duty cycle 5%. **(b)** Trap density of states (DoS) as a function of demarcation energy calculated from transient photocurrent (TPC).  232

**Figure 15**: Simulated dark $J - V$ characteristic considering the **(a)** shallow traps **(b)** mid-level traps and **(c)** both shallow and mid-level traps. The shallow traps have $ET - EI = 0.6$ eV, where ET is the effective trap level, Eg ~ 1.6eV. The simulation results were obtained through numerical solutions of non-linear coupled Poisson and drift-diffusion simulations. Methodology and calibration of simulation scheme are discussed in our prior literature.[72–75,77–80]                                                                               233

**Figure 16**: The solar cell is used as a light-emitting diode (LED). **(a)** Electroluminescence external quantum efficiency of the solar cell device as a function of injected current density. **(b)** The photographic image of the device during LED operation.                         235

**Figure 17**: **(a)** The achievable limit of the PV parameters calculated using integrated $JSC, SQ \approx 26$ mA/cm$^2$ (calculated from **Figure 13b**), $k1 = 0$, $k2, rad = 10 - 11 cm^3s^{-1}$, $k3 = 0$, and $ni = 0.5 \times 106$ cm$^{-3}$. The PV parameters are listed in table 7.                         235

**Figure 18**: **(a)** Current density vs. Voltage (J-V) characteristics under $1 - sun$ Illumination condition and pseudo $J - V$ measured from Suns VOC measurement. **(b)** EQE & IQE spectrum of the PSC including the reflection (R) spectra measured from the glass/ITO side. EQE represents the external quantum efficiency, R represents the reflection and IQE

represents the internal quantum efficiency spectrum. The integrated current densities are estimated by integrating the EQE spectrums over AM 1.5G spectrum. 

**Figure 1:** Schematic representation of the working principle of tandem solar cells. **(a)** Schematic representation of below bandgap loss and thermalization loss. **(b)** Silicon solar cells have lower below bandgap loss and higher thermalization loss. **(c)** Perovskite solar cells have higher below bandgap loss and lower thermalization loss. **(d)** Schematic representation of tandem solar cells using large and smaller bandgap absorber material. **(e)** Working of tandem solar cells using two semiconductors of different bandgap. **(f)** Minimizing the below bandgap loss and thermalization loss in the tandem solar cells. 

**Figure 2: (a)** Schematic of semi-transparent PSC. Glass/ITO is the transparent conductive electrode, Pz:PFN (9:1) is the hole transport layer, perovskite is the active material, PCBM/BCP acts as the electron transport layer, ALD $SnO_2$ acts as an electrons transport as well as sputtering damage protecting layer and IZO is the transparent back electrode. **(b)** The cross sectional view of the semi-transparent perovskite solar cells. 

**Figure 3: (a)** J-V characteristic curves of the semi-transparent (ST)-PSC and opaque PSC device of active area 0.175 $cm^2$ and 0.805 $cm^2$ under AM1.5G illumination conditions. **(b)** Photographic image of the ST-PSC of active area 0.175 $cm^2$ and **(c)** 0.805 $cm^2$ respectively. 

**Figure 4: (a)** J-V characteristic curves of the semi-transparent (ST)-PSCs device of active area 0.175 $cm^2$, 0.805 $cm^2$, and PERC Si solar cell under AM1.5G illumination conditions. **(b)** IPCE spectrum, including EQE, transmission, and integrated JSC of Monocrystalline PERC Si and perovskite solar cells. 

**Figure 5: (a)** J-V characteristic curves Silicon Hetero Junction (SHJ) solar cell by LONGi[28] under AM1.5G illumination conditions. **(b)** IPCE spectrum including EQE, transmission, and integrated JSC of SHJ Si and perovskite solar cells. 

**Figure 6:** Loss analysis by optical simulations: Loss analysis of the current density of **(a)** Monocrystalline PERC Si and **(b)** Silicon Hetero Junction (SHJ) solar cell by LONGi.[28] 

**Figure 7: (a)** J-V characteristic curve of CdTe solar cell under AM1.5G illumination conditions. **(b)** IPCE spectrum including EQE, transmission, and integrated JSC of CdTe and perovskite solar cells. 

**Figure 1: (a)** Work function of the mixed Pz:PFN HTLs measured from KPFM study (UPS results discussed in Chapter 5). **(b)** Current density vs. voltage (J − V) characteristics of different bandgap perovskite solar cells (PSCs) with different Pz:PFN mixed HTLs. **(c)** The external quantum efficiency of the different bandgap PSCs. 

# List of Tables:

# CHAPTER 1

# Introduction to the perovskite solar cells

# CHAPTER 1

## Introduction to the perovskite solar cells

## 1.1 Background

Energy plays a crucial role in sustaining life and supporting various activities in our daily lives. Energy in the form of heat, light, chemical, mechanical, electrical, etc., is essential to meet our diverse demands in daily life needs. Still, we depend on non-renewable energy sources such as coal, fossil fuels, oil and natural gas, etc. But these are limited and will end very soon. Besides the limited resources, consuming these energies leads the world to dangerous situations like climate change and global warming, which are further uninhabitable for humans and other creatures.[1] Keeping this in mind, using renewable energy sources is being started with wind, biomass, geothermal, solar light energy, etc. The worldwide energy generation in 2020 data indicates that non-renewable energy sources are still dominant over renewable sources **Figure 1a**.[2] Among all the 11.73% renewable energy applications, the worldwide 36% contribution comes from the solar PV industry in 2020 **Figure 1b**.[2] Recently, the International Energy Agency published data on the share of power capacity from 2010 to 2027, and it is shown that solar power will lead to energy production in the near future because of the continued presence of sunlight and the growing PV industry **Figure 1c**.[3] The dependency on energy will increase daily as the population is increasing linearly with time.[4] Hence, the security of the energy requirement is essential to live a hassle-free, sustainable life.

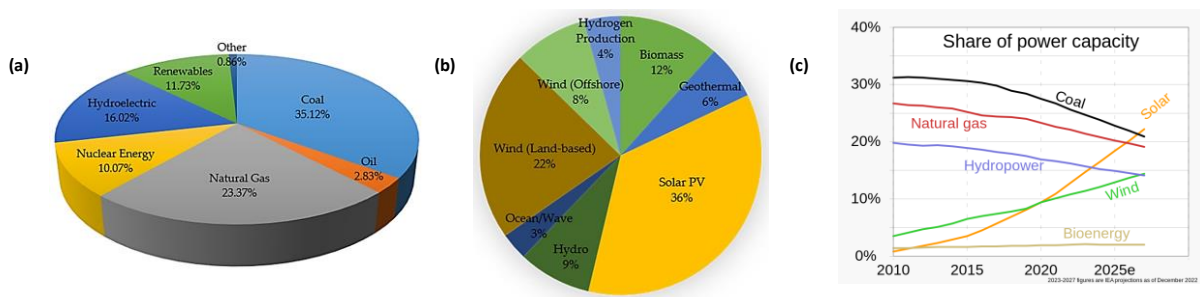

***Figure 1:*** *(**a**) Pie chart of all mixed energy generation worldwide.[2] (**b**) Pie chart of likely worldwide renewable energy generation.[2] (**c**) Share of power capacity in the upcoming future.[3]*

As per the report proposed by the *'Ministry of Power, Government of India'*, in February 2019, the projected gross electricity generation during the year 2029-2030 will be 23% from solar photovoltaic renewable energy **Figure 2a**.[5] Long-term studies observed that coal-based plants capacity expansion is insignificant. In the near future, renewable energy production will lead to green energy production **Figure 2b**.[5] India projects to reduce the use of coal by 50% in 2030 by promoting renewable green energy production. As per the recent report on 31.10.2023 by the Ministry of New and Renewable Energy, Government of India, the total solar power installed in India is 72.02 GW.[6] The annual solar power installation is increasing rapidly in India to avoid an energy crisis from the limited conventional resources. Currently, the Indian government is highly enthused by the green energy revolution, especially in the solar cell energy sector **Figure 2b**.

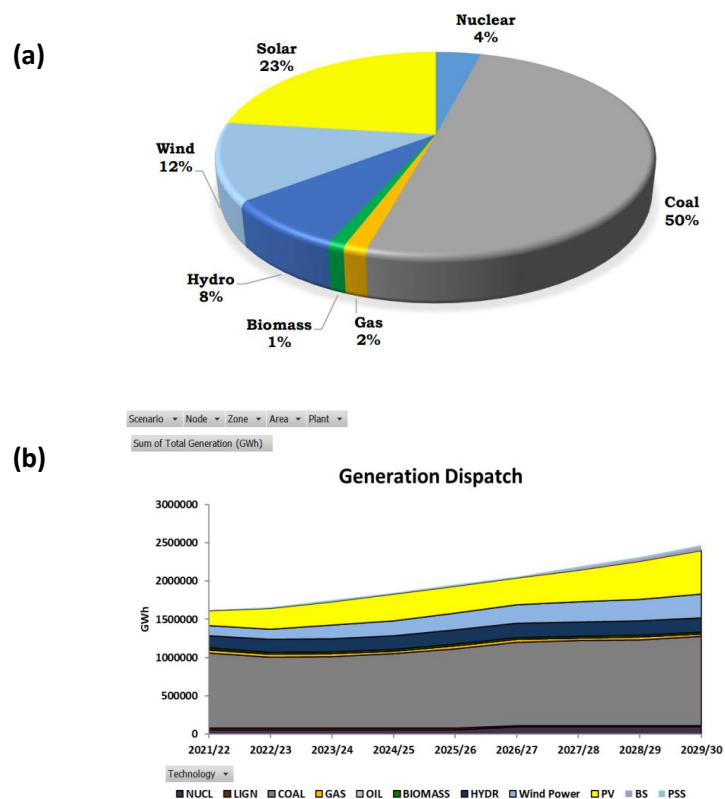

***Figure 2:*** *(**a**) Pie chart of likely gross generation (831 GW) capacity in 2029-2030 (Source: Central Electricity Authority, Government of India).[3] (**b**) Long term generation planning of the mixed energy sources.*

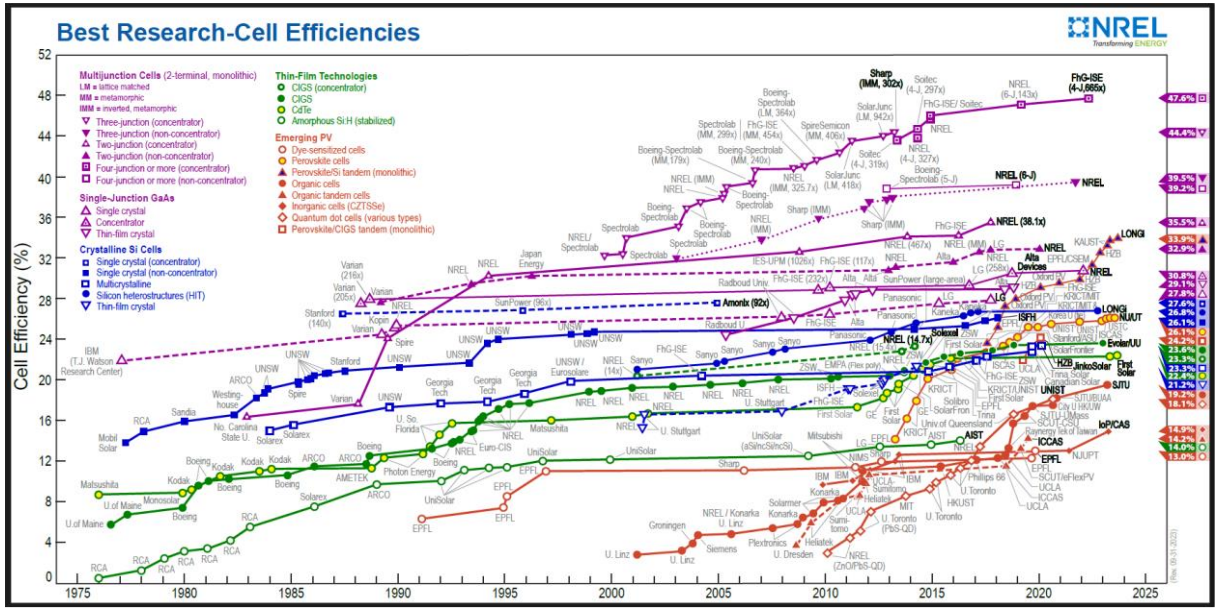

*Figure 3: Best research solar cell efficiency chart from NREL.[7]*

This thesis will focus on solar energy production considering the geographical locations. There are diverse ways to harvest solar energy, such as solar water heater[8], solar cooker[9], concentrator solar power, i.e., steam engine[10], photovoltaic (PV) or solar cell[11]. Among all, solar cell technology is efficient and popular in renewable green energy production. Even in solar cell technology, numerous photovoltaic technologies evolved, such as crystalline silicon (Si) solar cells[12], single junction GaAs[13], multi-junction cells[14], CIGS[15], CdTe[16], Amorphous Si:H[17], thin film solar cells[18], etc. Si-based solar cells are commercialized for terrestrial applications, and continuous efforts have been made to reduce manufacturing costs. However, there is not enough space to reduce the manufacturing cost, and the market is almost saturated.[19] Therefore, a new photovoltaic technology must be adopted to harvest cost-effective solar energy where the material and manufacturing costs are cheaper and easy to fabricate. The emerging photovoltaic technology stands as one of the promising candidates to harvest cost-effective solar energy. Currently, the emerging PV technologies are dye-sensitized solar cells[20], organic cells[21], quantum dot cells[22], inorganic (CZTSSe) cells[23], perovskite cell[24], etc. Among all emerging thin-film PV, the perovskite solar cells (PSCs) are the fastest-growing and show a huge jump in efficiency from 3.8% to ≥ 25% in just one decade, which is close to the market-leading single junction crystalline Si solar cell **Figure 3**.[25–27] This motivates people to engage more in the perovskite photovoltaic devices. Recently, perovskite solar cells become very popular due to their application in tandem solar cell architecture. A tandem solar cell, also known as a multi-junction solar cell, consists of multiple semiconductor layers with different bandgaps stacked on top of each other. Each layer is designed to absorb a specific portion of the solar spectrum, allowing the cell to capture a broader range of sunlight and convert it into electricity more efficiently.[28] Currently, the Si perovskite tandem solar cell has reached an efficiency of >33%, which promises the future of more efficient light-to-electrical energy conversion.[29–31]

## 1.2 Levelized cost of energy (LCOE)

Apart from the efficiency, it is essential for the investor/industry to understand the lifespan and production cost of the cells before starting the business. Levelized cost of energy (LCOE) is a techno-economical term that is being used to evaluate the cost of energy production rate considering the material cost, fabrication cost, efficiency of the cell, lifetime of the cell, etc. The levelized cost of energy is defined as

$$LCOE = \frac{sum\ of\ costs\ over\ lifetime}{sum\ of\ the\ energy\ produced\ over\ lifetime}$$

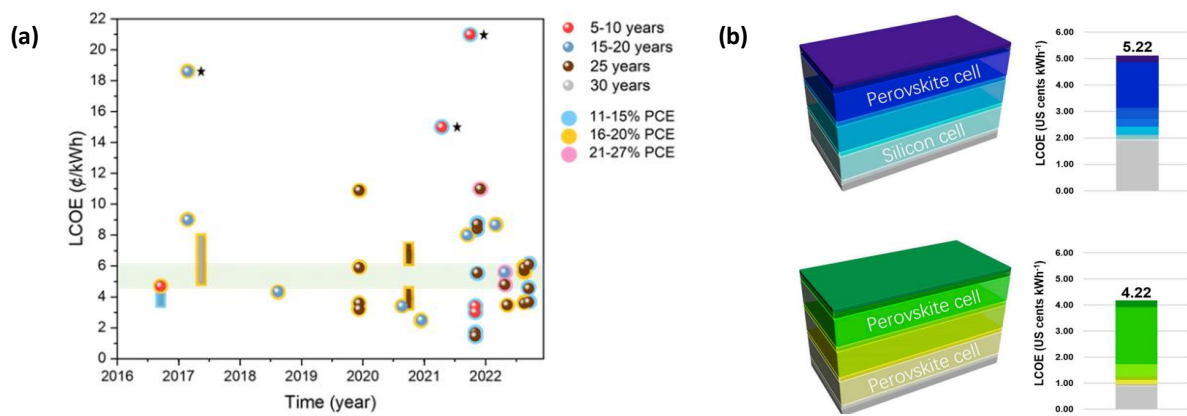

***Figure 4:*** *(a) LCOE prediction of the perovskite photovoltaics over the last seven years. (b) Cost analysis of the perovskite-silicon and perovskite-perovskite tandem solar cells.*

Recently, it is shown that the LCOE of the perovskite solar cells varies from 3 to 6 US cents per kWh⁻¹ **Figure 4a**.[32] Whereas, for the Si solar cell, the LCOE decreased from $76 to 30 US cents in 2015 and is almost saturated.[33] Another report by Zongqi *et. al.* in 2018 shows that the LCOE of the Si –perovskite tandem solar cell comes down to 5.22 US cents per kWh⁻¹ and further decreased to 4.22 US cents kWh⁻¹ for the perovskite –perovskite tandem cell **Figure**

**4b**.[34] Interestingly, the LCOE of perovskite is much lower than that of commercialized Si solar cells and will be further decreased by increasing stability, i.e., the lifespan of perovskite solar cells. The cost-effective production of energy for the perovskite solar cells grabbed much attention in the solar cells community, and companies like Microquanta Semiconductor (China), RenShine Solar (China), Kunshan GCL (China), Hanwha Q CELLS (South Korea), Oxford PV (UK & Germany), and so on involved into R&D for aiming commercialization of perovskite solar cells in single junction and tandem device architecture. Microquanta demonstrated 20.2% efficiency on 20 cm$^2$ solar cells area and 14.24% efficiency for large area (200 cm × 800 cm) solar modules.[35] RenShine Solar achieved a certified conversion efficiency of 19.42% on a 30 cm × 40 cm perovskite solar module and passed IEC61215 reliability tests.[36] Kunshan GCL showed an efficiency of 19.04% on the single junction perovskite module of area 100 cm × 200 cm, and aiming for the world's first large-scale 2GW perovskite production line.[37] However, these recipes are not available in the public domain. Evidently, further progress is required to minimize the LCOE for commercialization of perovskite solar cells.[34]

**Figure 5a** shows the perovskite panel of commercial components $\alpha$, size 1245 × 635 × 6.4 mm by Microquanta, with a 12-year linear power output warranty along with a 25-year product material and workmanship warranty.[38] As per release time (03/12/2023) by Microquanta, they demonstrated the 1 MW perovskite ground–mounted photovoltaic power station.[39] They showed the world's largest perovskite grid-connected ground power station, truly realizing the large-scale manufacturing and application of photovoltaic technology **Figure 5b**.[39]

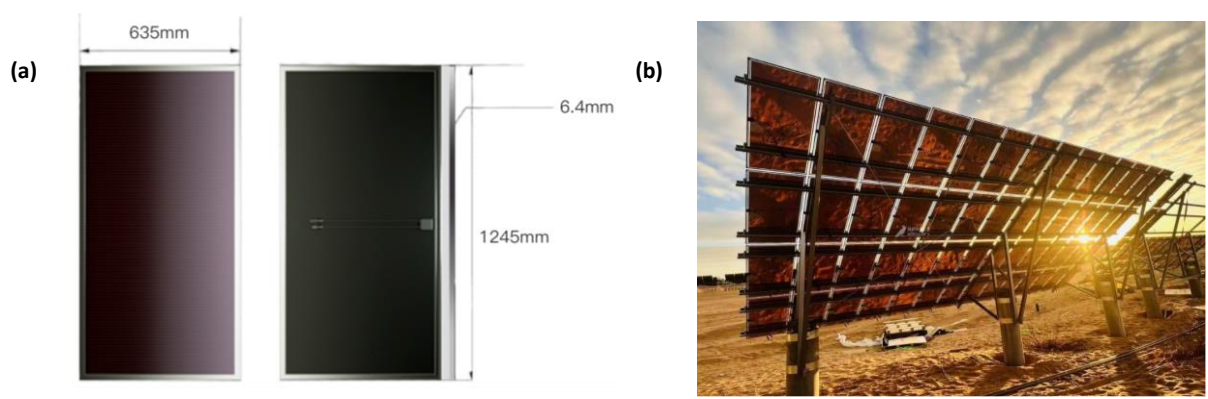

***Figure 5:*** *(a) Microquanta Semiconductor  demonstrated perovskite commercial components α, size 1245 × 635 × 6.4 mm: 25-year product material and workmanship warranty, 12-year linear power output warranty.[38] (b) World's largest perovskite grid connected ground power station on 03/12/2023, demonstrated by Microquanta Semiconductor.[39]*

## 1.3 Importance of perovskite solar cells

Perovskite is a class of semiconducting material with the crystal structure of $ABX_3$, similar to the calcium titanium oxide ($CaTiO_3$) mineral. The mineral was discovered by Prof. Gustav Rose in 1839 and is named after the Russian mineralogist Lev Perovski. The crystal structure was discovered by Victor Goldschmidt in 1926 in his tolerance factor work[40] and later published in 1945 from X-ray diffraction data on barium titanate ($BaTiO_3$) by crystallographer Helen Dick Megaw.[41] In the $ABX_3$ crystal structure, A and B are the positively charged atoms/molecules of different sizes, and X is the halide. **Figure 6** is the schematic of the typical crystal structure where the black ball at the center of the crystal represents the monovalent A cations ($Cs^+$, $FA^+$, $MA^+$), the blue balls at the center of each octahedral represent the divalent B cations ($Pb^{2+}$, $Sn^{2+}$) and the red balls at the corner of the octahedral represents X halides ($I^-$, $Br^-$, $Cl^-$).[42]

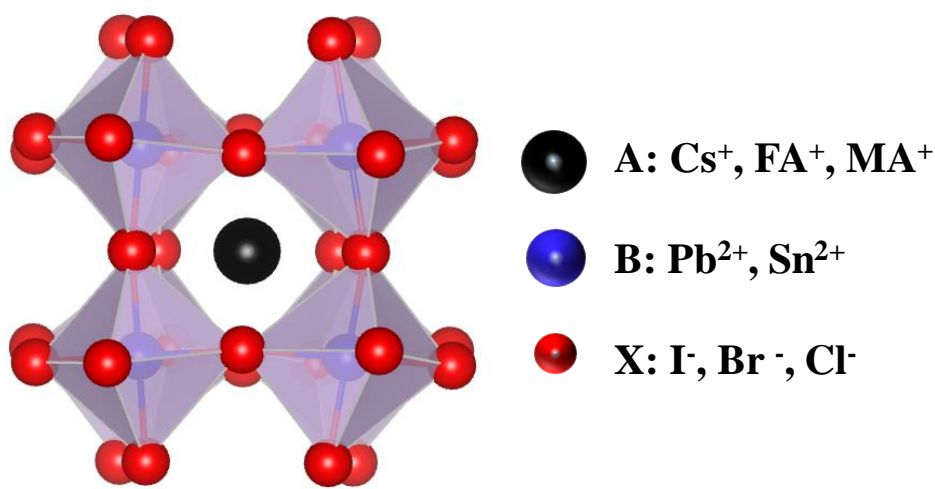

**Figure 6**: *Schematic of the $ABX_3$ perovskite crystal structure where the black ball at the center of the crystal represents the monovalent A cations ($Cs^+$, $FA^+$, $MA^+$), the blue balls at the center of each octahedral represent the divalent B cations ($Pb^{2+}$, $Sn^{2+}$), and the red balls at the corner of the octahedral represents X halides ($I^-$, $Br^-$, $Cl^-$).*

A class of perovskite materials can be developed by choosing various combinations of the different compositions of A, B cations, and X halides. To get a stable perovskite crystal structure, the choice of the A, B, and X is such that the crystal structure must follow the Goldschmidt tolerance factor ($t$) and octahedral factor ($\mu$) rules.[43,44]

The tolerance factor ($t$) limit

$$0.81 < t = \frac{R_A + R_X}{\sqrt{2}\,(R_B + R_X)} < 1.11$$

Octahedral factor limit

$$0.44 < \mu = \frac{R_B}{R_X} < 0.9$$

Where $R_A$, $R_B$ and $R_X$ are the radii of the corresponding ions.

In an ideal case, the cubic perovskite crystal structure can be obtained when the tolerance factor $t \approx 1$.[45] In this case, the size of the A cation must be larger than the B cation. Generally, the B site is occupied with large Pb or Sn atoms for the halide perovskite, so A cation must be much larger than B.[46] However, Cs is almost the largest group-I element in the periodic table but still not large enough to hold the cubic perovskite crystal structure, so it must be replaced with a larger atom/molecule. However, various additive engineering chemicals are developed to maintain the stable cubic crystal structure of the $CsPbI_3$ perovskite.[47] Therefore, the A cation is chosen as an organic molecule, such as $CH_6N^+$, $CH_5N_2^+$, $CH_6N_3^+$, etc., to obtain a stable organic-inorganic halide perovskite crystal structure without additives.

Perovskite is a direct bandgap semiconductor with a high optical absorption coefficient, making it superior to the other commercialized silicon (Si) and gallium arsenide (GaAs) solar cells. The absorption coefficient of the perovskite material is $10^4 - 10^6$ cm$^{-1}$, which is higher than the commercialized Si, Ge, etc., semiconductor.[48] This results in a 300-600 nm thick perovskite layer that is sufficient to absorb almost all the 1- Sun light; hence, it becomes a potential candidate for solar cell application.[49,50] Whereas the thickness of the first generation (silicon) solar cell absorber layer is 100-300 μm[51–53] and the thickness of the second generation (GaAs) solar cell is 1-3 μm[54–56]. The spin-orbit coupling plays an essential role in the band structure, hence the optoelectronic properties.

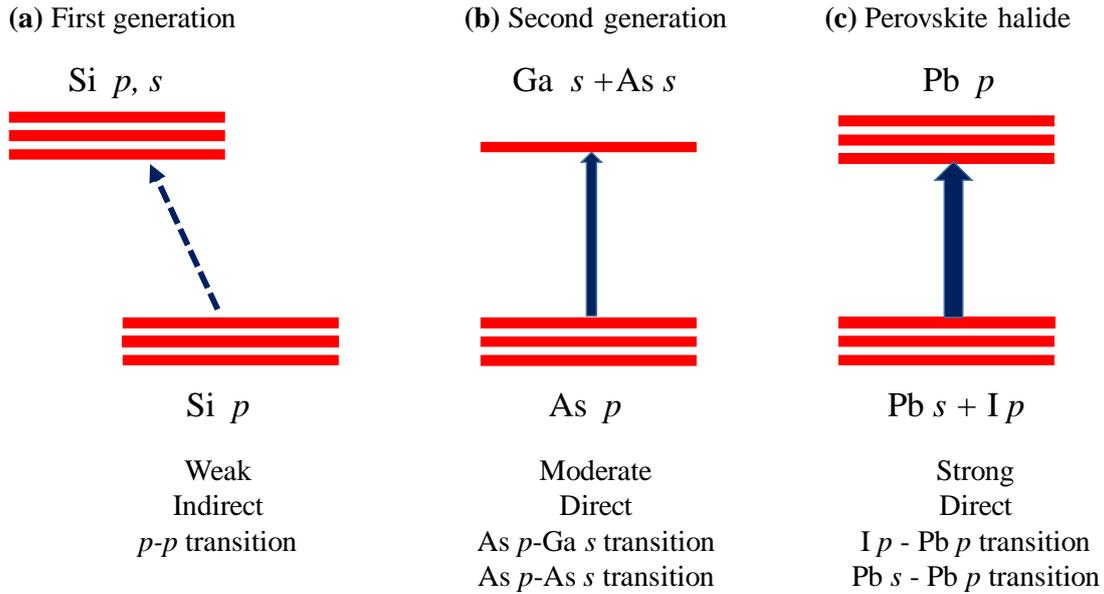

*Figure 7: The schematic of the optical absorption of the (a) first generation, (b) second generation, and (c) halide perovskite solar cell absorber.[46] The GaAs absorber has been chosen as a second generation, and MAPbI₃ as a halide perovskite solar cell absorber.*

The optical absorption process for the first generation, second generation, and the perovskite solar cell absorber is shown in **Figure 7**. **Figure 7a** For the first generation (Si) solar cells, the optical absorption takes place between the Si p orbital and the Si p and s orbital. However, silicon is an indirect bandgap semiconductor whose transition probability between the band edges is two orders of magnitude lower than that of the direct bandgap semiconductor. This results in the absorber layer being two orders of magnitude thicker and increasing the cost of solar cell fabrication. Despite being the indirect band gap of the Si absorber, the most commercialized solar cell in the PV community is the Si solar cell. The reason is that the Si absorber is fabricated from the sands by Czochralski (CZ) crystal growth or float zone(FZ) growth at temperature ~1400° C.[57,58] At such a high temperature, the Si absorber crystal growth is so pure that the charge carrier lifetime can be high as ~ 1 ms, diffusion length 100- 300 µm, results efficient reproducible Si solar cells fabrication.[59] However, all these fabrication processes are expensive and require alternative approaches such as low cost and easy fabrication of perovskite solar cells. The bandgaps of the second generation (GaAs) and perovskite absorber are direct and hence possess stronger optical absorption than that of Si solar cells. The absorption coefficient of the perovskite absorber is one order of magnitude higher in the visible range than GaAs due to strong spin-orbit coupling.[60] The electronic structure of GaAs and perovskite absorbers are different. The valence band of GaAs is made up of p orbitals of As, and the conduction band is composed of the s orbital of Ga and the s orbital of As **Figure 7b**. However, the valence band of the perovskite is formed of the s orbital of Pb and p orbital of halides (Cl[-], Br[-], I[-]), whereas the conduction band is made up of p orbital of Pb **Figure 7c**. The conduction band of GaAs is primarily made up of dispersive s orbital, whereas for the perovskite, it is made up of degenerate p orbitals. The atomic p orbitals are more localized than the s orbitals. As a result, the density of states (DoS) at the bottom of the conduction band is significantly higher for the perovskite than for the GaAs absorber.[46] The intra-atomic transition occurs between the valence band (Pb s) and conduction band (Pb p), facilitating a higher transition probability in the perovskite absorber. Hence, the optical absorption strength is higher in the perovskite than in GaAs. Apart from the high absorption coefficient, the other optoelectronic properties such as easy bandgap tunability (1.2 eV-3.2 eV), low binding energy ($\sim k_B T$), long diffusion length ($\sim \mu m$), higher charge carrier mobility ($10^1 - 10^2$ cm²V⁻¹s⁻¹), low temperature (100-200°C) solution and thermal evaporation processable technique promise the future of perovskite solar cells in the PV community.[61,62]

## 1.4 Motivation

To commercialize any PV technology, the essential requirements are (i) reproducible high efficiency, (ii) long-term stability, and (iii) scalability. The perovskite solar cells have already passed the efficiency number compared to the commercialized silicon solar cells, but long-term stability and scalability are still significant challenges. Most of the reported perovskite solar cells have an active area of $\sim 0.1$ cm²· and the anti-solvent approach is applied to control the crystallization to get the pinhole free and smooth perovskite film by solution processable technique. However, for scalable perovskite solar cells, the pinhole free uniform deposition of the perovskite layer over a large area is challenging using anti-solvent treatment. Hence, additive engineering, spray coating, thermal evaporation, screen printing, doctor blades, and carbon-based devices can be alternative approaches. However, the thermal evaporation-based perovskite layer deposition technique is promising for efficient and reproducible device fabrication.[63] Apart from efficiency, stability is essential to commercializing solar cells. Various engineering techniques are adopted to increase the solar cell devices moisture/thermal stability, such as additive engineering using ionic liquids or small molecules, interface passivation using 2D materials, encapsulation of the device using epoxy, etc.[64,65] There are various ways to measure the stability of the solar cells, such as maximum power point tracking (MPPT) under continuous illumination, moisture stability at relative humidity (RH) of 50% or 85%, thermal stability at 65˚C or 85˚ C and these are classified using standard ISOS protocols.[66] The interface between the perovskite and charge transfer layers plays a crucial role in the devices performance and stability operation. The charge transfer layers should be chosen to match the energy band well with the perovskite layer. However, during the stability operation study under continuous illumination, the halide ions come at the interface, which impedes the charge collection and deteriorates the photovoltaic performance. So, proper selection of the charge transfer layers and additive engineering solve such problems. Furthermore, the ITO/FTO substrate sheet resistance is also problematic in large-area devices, and hence, a mini-module kind of architecture is adopted in large-area devices.[67,68] However, apart from the device engineering, the photophysics of the perovskite materials surprises every day. A few of them are that the perovskite semiconductor has a positive temperature coefficient[69,70], double emission of PL peak at low temperature,[71] contrast behaviour of enhanced PL intensity *vs.* decreased charge carrier lifetime at low temperature with respect to room temperature,[72] photon recycling *vs.* defect-mediated recombination,[73] the origin of non-radiative recombination[74], the role of ions vs. ferroelectricity,[75] etc. Perovskite semiconductors are full of defects, unlike silicon, and possess disorders (static or dynamic), dislocation, distortion, ion migration, and phase segregation (mixed halide perovskites), resulting in the unique physics of this novel class of semiconductor.[76,77] Hence, it is necessary to understand these imperfections (defects/disorders) by various optoelectronic measurements to improve the quality of the perovskite material further and understand the device physics.

## 1.5 Outline of the thesis

This thesis focuses on the compositional and interface engineering of hybrid metal halide perovskite thin film solar cells. Chapter **2** demonstrates the fundamentals of solar cells with a literature survey. This chapter covers the challenges and future perspectives of fabricating scalable, efficient perovskite solar cells. Chapter **3** describes the perovskite thin film/solar cell fabrication procedure and the characterization techniques to support our experimental results and outcome. Chapter **4** describes the role of monovalent (A) cation in the $ABX_3$ perovskite crystal structure in terms of the dielectric relaxation process and correlates with the thermal stability of the perovskite solar cells. This chapter will introduce the study of dielectric relaxation and defects using frequency-dependent photocurrent measurement of a complete solar cell device and eventually correlate with the thermal stability of perovskite solar cells. Chapter **5** deals with the hydrophobic self-assembled monolayer (SAM) hole transport layer in the inverted (*p-i-n*) architecture-based device. In this chapter, we studied the mixing engineering strategy of SAM with polyelectrolyte polymer and tailored the work function of the mixed HTL for 1.6 eV perovskite solar cells. The interface of the device is modified for different mixing ratios of HTL, and associated device physics is discussed. However, the device shows excellent efficiency performance, the lowest dark current, and the ideality factor close to the unity. The lowest dark current is a prerequisite for a photodetector application. Hence, we carried out a photodetector study of the best-performing HTL-based device in Appendix A. However, the unity ideality factor is a primary signature of solar cells working near the radiative limit. Hence, Chapter **6** deals with the dominant recombination mechanism on a complete solar cell using steady-state and transient measurements. We developed a consistent characterization scheme for studying a solar cell operating near the radiative limit. Chapter **7** describes the application of perovskite solar cells in the four-terminal tandem solar cell structure. In this case, we replaced the opaque (Ag) electrode with a semi-transparent electrode (IZO) to fabricate semi-transparent solar cells and use it in the four-terminal tandem architecture with Si and CdTe solar cells. Finally, chapter **8** concludes all the research work in the thesis and demonstrates the future outlooks related to this thesis.

# CHAPTER 2

# Fundamentals of solar cells and literature survey

# CHAPTER 2

# Fundamentals of solar cells and literature survey

## 2.1 Introduction

Chapter **1** describes the motivation and outlines of the thesis to understand the compositional and interface engineering of hybrid organic-inorganic metal halide perovskite thin film solar cells. The fundamentals of solar cells and the literature survey will be discussed in Chapter 2. This chapter will cover the literature review on the fundamental aspects of perovskite solar cells and their progress in terms of efficiency, stability, and scalability. This chapter also justifies the need for work on the research problems addressed in this thesis.

## 2.2 Working principle of solar cells

A solar cell is an optoelectronic device that transduces light energy to electrical energy through the photovoltaic effect. The photovoltaic (PV) effect was first discovered by the French physicist Alexandre Edmond Becquerel in 1839 while investigating the light interaction with electrochemical cells.[1] After that, it took 115 years to demonstrate the 6% efficient PV solar cell using silicon (Si) absorber by C.S. Fuller and G. Pearson at Bell Laboratories in 1954.[2] After that, tremendous progress can be seen in the PV solar cell field as a consequence of the first-generation, second generation and today emerging PV technology came in the solar cell community.[3] The performance of the solar cells depends on the semiconductor used in the device and the corresponding fundamental mechanisms involved in it. The fundamentals of the solar cell are discussed below.

When photons are absorbed by a semiconductor absorber of energy equal to or more than the energy bandgap ($E_g$), the electron and hole pairs are generated at the conduction and valence bands respectively. The incident sunlight consists of photons with energies ranging from infrared to visible to ultra-violet region **Figure 1a**.[4] If the energy of the incident photon is less than the bandgap of the absorber, the photon will not be absorbed, leading to below bandgap Loss **Figure 1b**. If the energy of the incident photon is more than or equal to the bandgap of the semiconductor, it will be absorbed and create electron and hole pairs. For the photons of energy more than the bandgap, the excited electrons will go to the higher energy states of the conduction band, undergo various non-radiative fast relaxation processes, and eventually arrive at the bottom of the conduction band. These fast relaxation processes happen via phonon interaction and release heat to the semiconductor, referred to as thermalization loss.[5] However, the free electrons at the conduction band and the holes at the valence band take part in charge transport, i.e., current flow in the device. The interaction of light with the photovoltaic semiconductor absorber leads to (i) absorption/generation and (ii) recombination process.

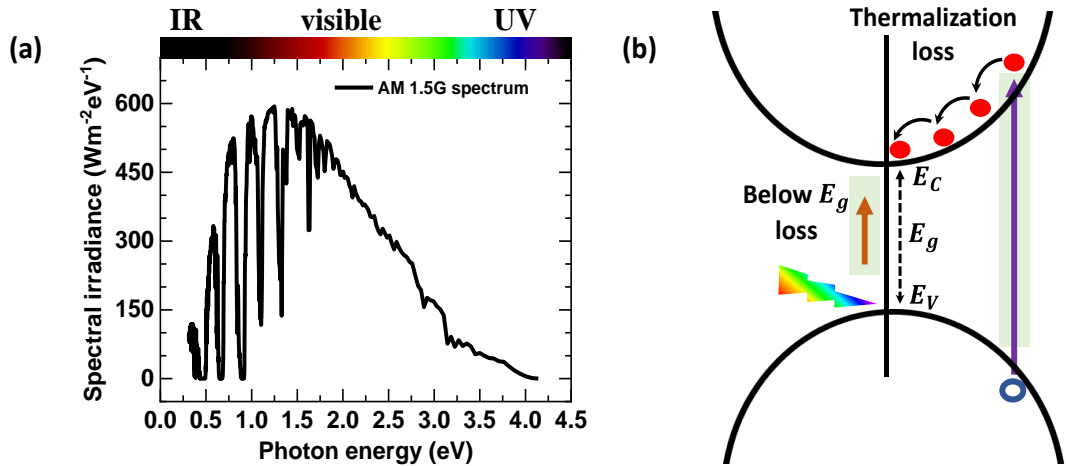

***Figure 1:*** *(a) 1-Sun solar spectrum (AM 1.5G).[4] (b) Schematic of absorption and generation of the charge carriers by optical excitation.*

### 2.2.1 Absorption and Generation

For an efficient solar cell, the rate of generation of the electrons and holes should be as high as possible. The generation of electrons and holes in the solar cell occurs when the incident sunlight has energy higher than the bandgap of the absorber. However, the generation of the charge carriers in the semiconductor happens via direct band-to-band transition or through sub-gap states.[6] If the sub-gap states are close to the conduction or valence band, they are referred to as shallow trap states or tail states. However, if the sub-gap states lie in the middle of the energy band, called the mid-gap states.[7] The optical absorption process through direct band-to-band and sub-bandgap states is shown in **Figure 2**. The optical transitions through the direct band are called intrinsic transitions, whereas the transitions through the sub-gap states (due to impurity/defects) are referred to as extrinsic optical transitions.[8] The extrinsic optical transitions are not desirable for efficient solar cell application. Apart from the extrinsic transitions, the semiconductor materials can also be direct and indirect bandgap. In the direct bandgap semiconductor, the minimum of the conduction band and the maximum of the valence band are in the same crystal momentum space (**k**). In contrast, in the indirect bandgap, the momentum space is different.[9] The absorption coefficient α(λ) of the direct bandgap semiconductor is sharp at the band edge (Perovskite, CdTe, CuInSe2)[10–12] whereas for the indirect bandgap, it slowly increases (Si)[13] **Figure 3**.

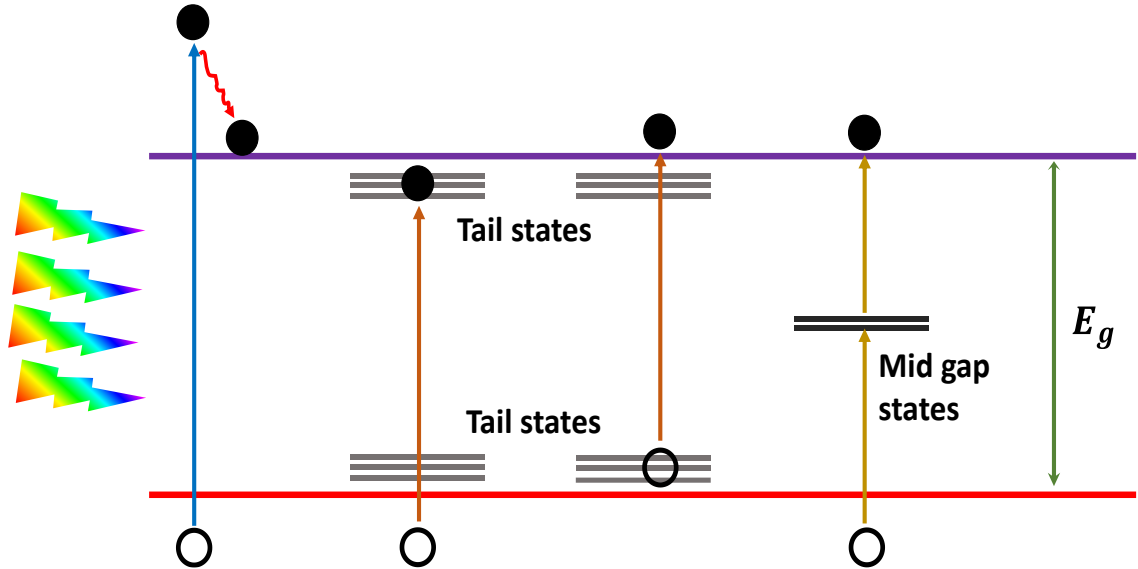

***Figure 2:*** *Optical absorption process through direct band-to-band and sub-gap states transition.*[7]

The bandgap of a semiconductor as a function of the absorption coefficient is defined by the Tauc method as[14]

$$(\alpha h v)^{1/\gamma} = B(h v - E_g) \qquad (2.1)$$

Where $h$ is the Planck constant, $v$ is the frequency of the photon, $E_g$ is the energy bandgap of the semiconductor, and B is a constant. The factor γ depends on the nature of electron transfer and is equal to 2 for indirect and 1/2 for direct transition bandgaps respectively. The bandgap of a semiconductor is calculated using the above eq. 2.1.[9,15]

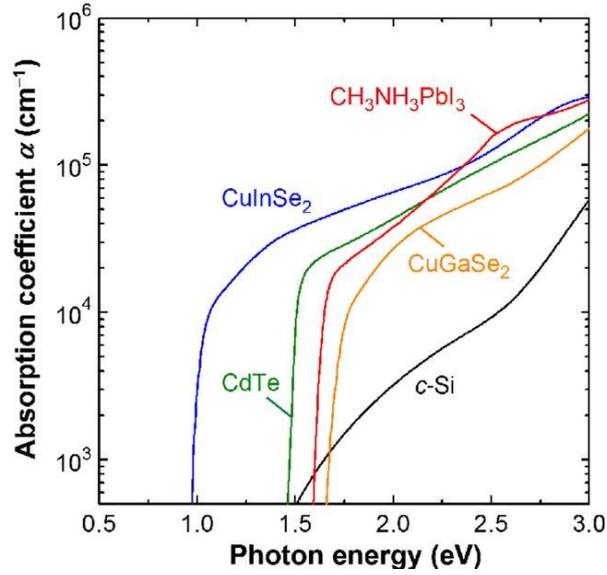

***Figure 3**: Absorption coefficient of direct and indirect bandgap semiconductors.[13]*

The sub-bandgap states do not contribute to the absorption coefficient significantly, but they provide information about the distribution of the band structure.

The Urbach energy ($E_u$) is the measure of disorder in the bulk of the perovskite semiconductor and can be estimated from the absorption spectra using the following equation,[16,17]

$$\alpha(E) = \alpha_0 \exp\left(\frac{E - E_g}{E_u}\right) \qquad (2.2)$$

Where E denotes the photon energy, and $\alpha_0$ is the material-dependent absorption coefficient.

### 2.2.2 Recombination

Upon optical excitation of a semiconductor, electrons and holes are generated in the bulk of the semiconductor. However, the generated electrons and holes recombine through various relaxation channels and decrease the charge carrier density. At steady state condition, the carrier dynamics under illumination) for an undoped semiconductor is,

$$\frac{\partial n}{\partial t} = G - k_1 n - k_2 n^2 - k_3 n^3 \qquad (2.3)$$

where $\frac{\partial n}{\partial t}$ is the rate of change of charge carrier density at steady state condition G is the generation of the carriers upon optical excitation, and the parameters $k_1$, $k_2$, and $k_3$ denote the monomolecular, bi-molecular recombination, and Auger recombination rates, respectively. The recombination parameter $k_1$ is contributed by trap-assisted recombination centers, also called Shockley Read Hall (SRH) recombination, $k_2$ is corresponds to the bimolecular recombination and $k_3$ is due to the Auger recombination process.

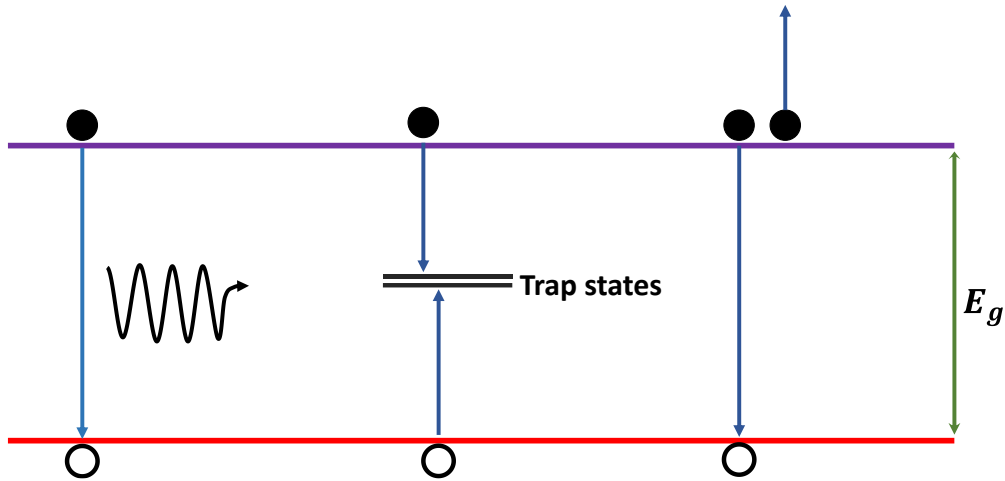

*Figure 4*: *Schematic representation of band-to-band, trap-assisted, and Auger recombination process.*

The band-to-band recombination process merely involves the direct annihilation of the conduction band electron to the valence band hole and is significant in the direct bandgap semiconductor only. The electron jumps from the allowed conduction band minimum to the vacate valence band maximum, and the excess energy is released in the form of light of energy equal to the bandgap of the semiconductor. This recombination process is also referred to as bi-molecular recombination.[18] The trap-assisted recombination, also called the Shockley Read Hall (SRH) recombination, is a non-radiative process through the localized trap states. Since the traps can absorb the difference in momentum between the carriers, SRH is dominant in Si or other indirect bandgap semiconductor. However, such SRH can also be dominant in direct bandgap semiconductors under very low carrier densities (low-level injection) or materials with high trap density, such as perovskites. The Auger recombination is a three-particle interaction. The energy released by an excited state electron during band-to-band transition or trap-assisted transition can transfer to an electron in the conduction band and move to a higher energy state. However, this process is significant only when the charge carrier density is very high.

## 2.3 Shockley – Quisser (SQ) limit

In 1961, William Shockley and Hans-Joachim Queisser estimated the efficiency of 30% for a single junction solar cell of absorber material bandgap 1.1 eV using a 6000K black-body spectrum as an approximation of the solar spectrum.[19] Further, the theoretical maximum efficiency of 33.7% was achieved at a bandgap of 1.34 eV considering the AM 1.5G (1000 mW/cm$^2$) solar spectrum.[20] The model is done considering the cell temperature at 300K and the sun temperature at 6000K. The device performance of the solar cell is modeled as a function of the bandgap of the absorber, as shown in **Figure 5.** According to the estimates, if the bandgap of the absorber is below 1.34 eV, then the energy loss is due to the thermalization relaxation process, and for a bandgap higher than 1.34 eV, the energy loss is due to the absorption loss or below bandgap loss **Figure 5a**. Each photon in the solar spectrum produces electron – hole pairs and results current flow in the solar cell. At higher bandgap, there are fewer photons in the solar spectrum, results in lower current which is attributed to spectrum loss **Figure 5b**. **Figure 5c** the red dotted line represents the $V_{OC}$ of the solar cell and is equal to the bandgap of the absorber when there is no recombination, but due to radiative recombination it is lower represented by the black solid line.

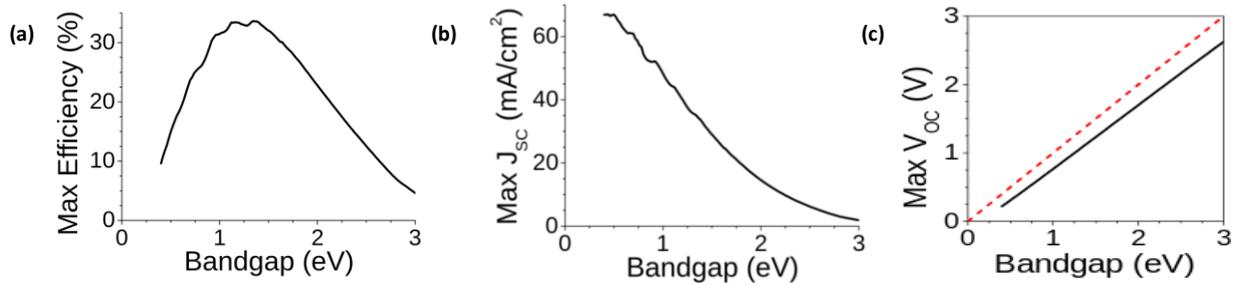

***Figure 5***: *The plot between the (**a**) SQ limit efficiency, (**b**) SQ limit short circuit current density, and (**c**) SQ limit open circuit voltage vs. bandgap of the semiconductor.*[19]

The SQ theoretical efficiency limit is based on the five assumptions as discussed below,

### 2.3.1 SQ limit assumptions:

    (i)    If the energy of the incident photon, $h\nu > E_g$, the photon will be absorbed, whereas for photons of energy $h\nu < E_g$, do not interact with the cell, i.e., the absorptivity

A($h\nu$) of the cell is a step function, 0 for h$\nu$ < $E_g$, and 1 for $h\nu > E_g$. Where, $E_g$ represents the bandgap of the absorber.

(ii)     The absorption of one photon generates exactly one electron-hole pair that contributes to the photocurrent at short-circuit conditions ($J_{SC}$).

(iii)    For the high-energy photons, the excess energy $h\nu - E_g$, releases in the form of heat in femtosecond (fs), and the cell is always in equilibrium with the environment at 300K.

(iv)    At this point, the generated electron–hole pair either get collected at their respective electrodes or recombine within the cell. The only allowed recombination in the cell is the radiative with emission of photon. The emitted photon further reabsorbs, creating a new electron-hole pair (photon recycling).[21] The relation between absorption and emission of photons in a semiconductor results from the principle of detailed balance limit. This can be described as every microscopic process must have the same rate as its inverse process in thermal equilibrium, or else thermal equilibrium could not be reached.[22]

(v)     Each contact that exchanges only one carrier type (electrons or holes) with the absorber layer is ideal and has negligible resistance.

## 2.3.2 Violation of SQ limits in practical solar cells

In reality, no practical solar cell holds the assumption 1, i.e., the step function of the absorptivity, because, in semiconductor absorbers, there are always sub-gap states resulting from the impurity/defects. Further, the parasitic absorption of photons in contact layers or by free carriers (electrons) in the bulk of the absorber reduces the average number of photogenerated electron-hole pairs and violates assumption 2. In assumption 3, the thermalization losses release heat from the cell but remain in equilibrium with the surroundings, which is not possible in real life. In assumption 4, the allowed recombination is radiative, but the violation of assumptions 1 and 2 indicates the presence of sub-gap states and leads to non-radiative recombination.

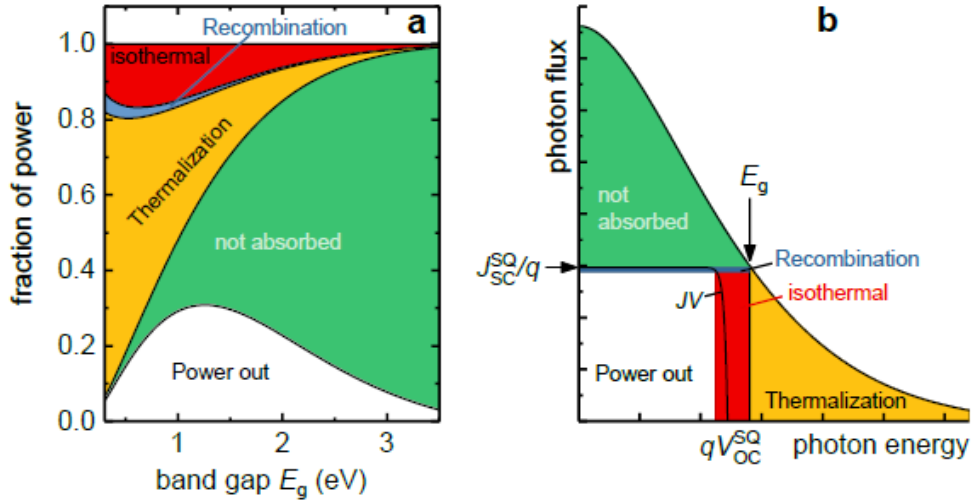

***Figure 6****: (a) Illustration of SQ –model as a function of bandgap (always at maximum power point) using AM 1.5G solar spectrum. (b) Energy losses for a given bandgap energy are depicted as a function of photon flux* vs. *photon energy (adapted from ref.[23]). By dividing the flux with elementary charge q and multiplying the energy with q, the axes can be current density vs. voltage (J vs. V).*

In assumption 5, the contact electrodes are assumed to be ideal with negligible resistance, but in reality, it is not possible due to the finite mobility or carrier concentration of the selective layer.

However, in the practical device single junction solar cells, the total current under illumination is due to a short circuit current ($J_{SC}$) along with diode current ($J_d$) eq. 2.4. The diode current is dependent on the reverse saturation current density $J_0$ which is due to the recombination mechanism. Thus, the violation of assumptions 1 and 2 results in a lower short circuit current than the SQ limit. 0

$$J = J_{SC} - J_0 \left[ \exp\left(\frac{qV}{nkT}\right) - 1 \right] \qquad (2.4)$$

Violation of assumptions 2, and 3 increases cell temperature as the recombination current is thermally activated, increasing exponentially with temperature. This increase in recombination current reduces the $V_{OC}$ of the device compared to the SQ limit. [23]

$$V_{OC} = \frac{nkT}{q} \ln\left(\frac{J_{SC}}{J_0} + 1\right) \qquad (2.5)$$

Further, violation of assumption 5 increases the series resistance of the device and reduces the fill factor ($FF$). **Figure 6a** shows the fraction of power as a function of bandgap considering

SQ limit assumptions. **Figure 6b** shows the SQ limit $J - V$ plot considering SQ limit assumptions. Thus, in a single junction practical solar cell, the efficiency is always lower than the SQ limit. However, the efficiency can be improved in the practical solar cells by minimizing the defect states and choosing the right bandgap absorber. However, the model shows a multi-junction cell with an infinite number of absorber layers, the efficiency limit is 68.7% using AM 1.5 G solar spectrum and 86.8% using concentrated sunlight.[24,25] The multi-junction technique has been applied in the various tandem solar cells and surpasses the single junction SQ limit efficiency.[3,26]

## 2.4 I-V characteristics and PV parameters of solar cells

The fundamental way to test a solar cell is by measuring the current density *vs.* voltage $(J - V)$ characteristics under dark and illumination conditions with AM 1.5G spectrum. The current flow in a solar cell under illumination (considering the single-diode model) can be expressed as[27]

$$I = -I_L + I_0 \left[ \exp\left( \frac{qV}{nkT} \right) - 1 \right] \qquad (2.6)$$

Where $I_L$ is the illumination current, $I_0$ is the reverse saturation or leakage current, which flows in the opposite direction to the illumination current, and $I$ is the total current. The $n$ represents the ideality factor, which lies between 1 and 2 (depending on the bulk and interface quality)[27] $q$ is the electronic charge, and $k$ is the Boltzmann constant. The numerical factor 1 can be avoided for a forward bias of 100 mV (or higher) as the diode current increase is exponential. On the other hand, at lower applied bias or in reverse bias, the illumination current $I_L$ dominates over $I_0$. However, the device's total current depends on the active area of the solar cells. Hence, the current density $(J)$ is used to estimate the device's efficiency. Thus the above expression can be written in terms of $J$ as

$$J = -J_L + J_0 \exp\left( \frac{qV}{nkT} \right) \qquad (2.7)$$

The typical illumination $J - V$ characteristic is shown in **Figure 7a**, and the power conversion efficiency as a function of the applied bias voltage is shown in **Figure 7b.**

The power conversion efficiency (PCE) of solar cells can be expressed as

$$PCE(\%) = \frac{P_{max}}{P_{in}(100\ mW/cm^2)} \times 100\% \qquad (2.8)$$

Which can be written in terms of the PV parameters as

$$PCE(\%) = J_{SC} \times V_{OC} \times FF\ (\%) = J_{max} \times V_{max} \qquad (2.9)$$

Where, $J_{SC}$ is the short circuit photocurrent density, $V_{OC}$ is the open circuit voltage, and $FF$ is the fill factor (Figure 7a). The FF is defined by the maximum square of the $J - V$ curve, and it can be represented as,

$$FF = \frac{J_{max} \times V_{max}}{J_{SC} \times V_{OC}} \qquad (2.10)$$

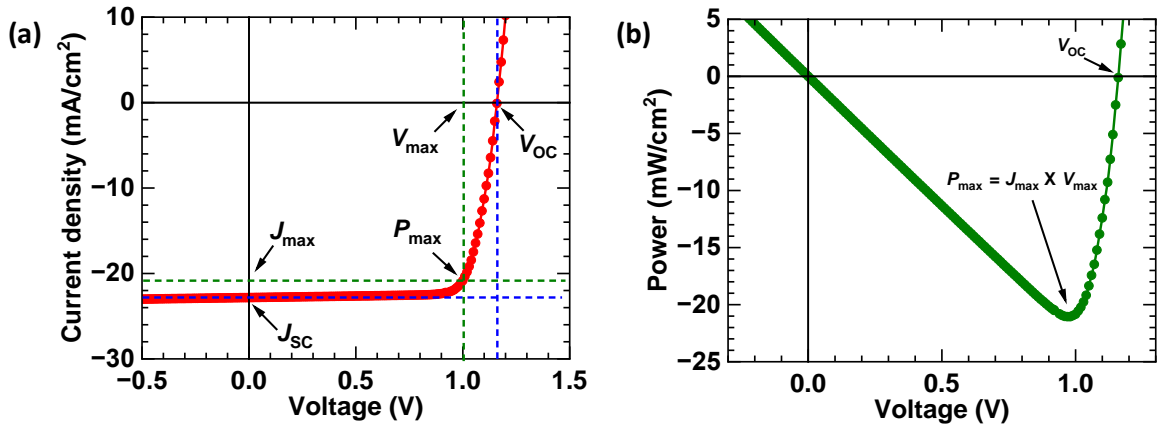

***Figure 7**: (a) Schematic representation of current density vs. voltage ($J - V$) characteristics with the PV parameters for a typical perovskite solar cell. (b) The power plot as a function of applied bias voltage.*

In practical solar cells, series ($R_s$) and shunt ($R_{sh}$) resistances are the two important parameters that affect the device's performance. The series resistance of a solar cell originates from the contact resistance between the absorber layer and the charge transport layer, along with the sheet resistance of the bottom and top contacts. However, the shunt resistance originates from the physical and electronic defects. [28,29] The specific origin of electronic defects is yet to be known.[82] However, the physical defects (point defects, grain boundaries, pinholes, dangling bonds at the surface of the film, etc.)[30,31] can be prominent considering low temperature (~100° C) solution processable techniques widely used for perovskite layer deposition and chances of non-uniformity over the whole substrate.[29,32,33]

Therefore, upon applying the 1-Sun (100 mW/cm²) light, the photogenerated charges take place, and the total current in terms of $R_s$ and $R_{sh}$ can be written as [34]

$$J = -J_L + J_0 \left[ \exp\left( \frac{V - JR_S}{\frac{\eta kT}{q}} \right) - 1 \right] + \frac{V - JR_S}{R_{Sh}} \qquad (2.11)$$

These $R_s$ and $R_{sh}$ resistances can be evaluated from the illuminated $J - V$ as follows

$$R_s = \frac{dV}{dJ} \; at \; V = V_{OC} \qquad (2.12)$$

$$R_{sh} = \frac{dV}{dJ} \; at \; V = 0 \qquad (2.13)$$

However, these $R_s$ and $R_{sh}$ parameters can be extracted more accurately from the dark current as the illuminated curve can be affected by slight fluctuations in light intensity, low diffusion length, poor collection properties, etc.

Under dark conditions, the eq. 2.11 becomes [35]

$$J_d = \left[ \exp\left( \frac{V - J_d R_s}{\frac{\eta kT}{q}} \right) - 1 \right] + \frac{V - J_d R_S}{R_{Sh}} \qquad (2.14)$$

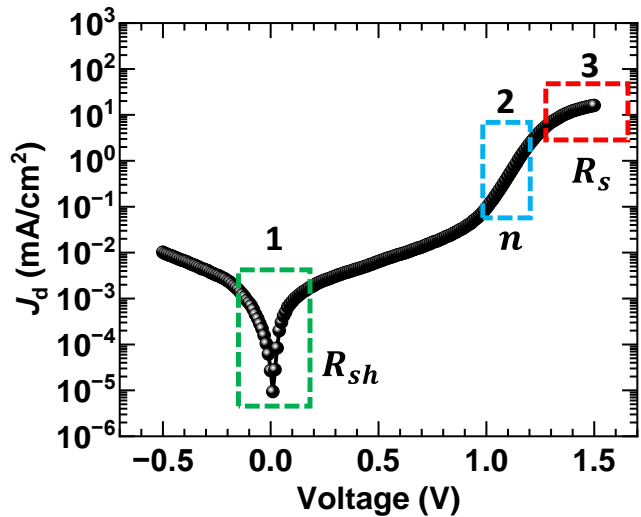

**Figure 8**: *The typical dark current of an inverted (p-i-n) architecture based perovskite solar cell of device active area 1 cm².*

At the low voltage, the dark current flows merely through the shunt path and is dominated by $R_{sh}$ (region 1 **Figure 8**). The 2nd term of eq. 2.14 represents the shunt current, and region 1 is used to measure the shunt resistance from the inverse of the slope. When the applied bias increases the dark current increases exponentially in region 2. This portion represents the recombination current, and the fitting of region 2 estimates the ideality factor ($n$). Further increase of the applied bias the current is limited by $R_s$, and it saturates at higher voltage regions, as represented by region 3. The inverse slope of region 3 estimates the value of $R_s$.

## 2.5. Suns - $V_{OC}$ characterization and pseudo $J - V$ estimation

The device performance depends on the series resistance of the solar cell device for instance larger areas of solar cells, as the $FF$ gets affected the most. The Suns-$V_{OC}$ measurement is a characterization technique of solar cells where the effect of series resistance is avoided, and the pseudo $J - V$ characteristic can be generated. The Suns-$V_{OC}$ technique is widely used for the characterization of silicon (Si) solar cells.[36] This technique more properly can be termed as illumination $V_{OC}$.[37] The basic concepts of Suns-$V_{OC}$ technique was outlined in 1963 but was not popular until Sinton et.al. in the year 2000 and the subsequent availability of the commercial tool.[38,39] Generally, in the Suns-$V_{OC}$ technique, the $V_{OC}$ of a solar cells is measured with varying illumination intensity (1-Sun to 0.01 Sun.) **Table 1**. In the solar cells device, the short circuit current density ($J_{SC}$) related to the intensity ($P$) as $J_{SC} = P^{\alpha}$, where the exponent $\alpha \approx 1$. Hence the intensity ($P$) can be replaced by $J_{SC}$ at each intensity and a pseudo $J - V$ i.e ($J_{SC} - V_{OC}$) can be constructed, where the voltage ($V$) is the $V_{OC}$ at each intensity. The detail application of the Suns- $V_{OC}$ technique will be shown in Chapter **6**. It is worth noting that, the $J - V$ plot (**Figure 9a**) is shown in the first quadrant for convenience (as Si solar cell community do) and the pseudo $J_{SC}$ estimated as per column D in **Table 1**. For better understanding we have shown the 1-Sun $J - V$ and the pseudo $J - V$ in the same figure (Figure 9a) in the first quadrant. The green plot represents the pseudo $J - V$, where the voltage axis corresponding to $V_{OC}$ and the current density axis is $J_{SC}$ at each intensity (Table 1). The reason for higher FF of the pseudo $J - V$ is due to avoiding the series resistance at $V_{OC}$ point where the current flow in the circuit is zero (**Figure 9b**).

**Table 1:** *Example of the pseudo $J-V$ estimation from the intensity dependent $J-V$ measurement.*

| Col(A) | Col(B) | Col(C) | Col(D) |
|---|---|---|---|
| Intensity | $J_{SC}$ | $V_{OC}$ | Pseudo $J_{SC} = J_{sc,0} - $Col(B) |
| $I_0$ | $J_{sc,0}$ | $V_{OC,0}$ | $J_{sc,0} - J_{sc,0}$ |
| $I_1$ | $J_{sc,1}$ | $V_{OC,1}$ | $J_{sc,0} - J_{sc,1}$ |
| $I_2$ | $J_{sc,2}$ | $V_{OC,2}$ | $J_{sc,0} - J_{sc,2}$ |
| $I_3$ | $J_{sc,3}$ | $V_{OC,3}$ | $J_{sc,0} - J_{sc,3}$ |
| $I_4$ | $J_{sc,4}$ | $V_{OC,4}$ | $J_{sc,0} - J_{sc,4}$ |
| $I_5$ | $J_{sc,5}$ | $V_{OC,5}$ | $J_{sc,0} - J_{sc,5}$ |
| $I_6$ | $J_{sc,6}$ | $V_{OC,6}$ | $J_{sc,0} - J_{sc,6}$ |
| $I_7$ | $J_{sc,7}$ | $V_{OC,7}$ | $J_{sc,0} - J_{sc,7}$ |
| $I_8$ | $J_{sc,8}$ | $V_{OC,8}$ | $J_{sc,0} - J_{sc,8}$ |
| $I_9$ | $J_{sc,9}$ | $V_{OC,9}$ | $J_{sc,0} - J_{sc,9}$ |
| $I_{10}$ | $J_{sc,10}$ | $V_{OC,10}$ | $J_{sc,0} - J_{sc,10}$ |

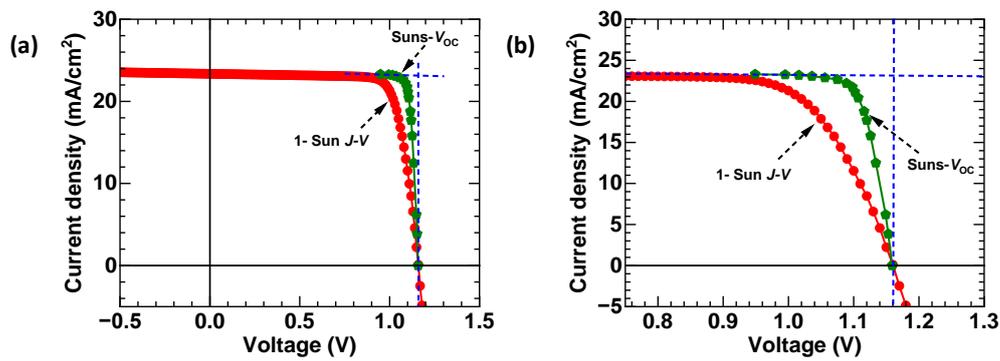

**Figure 9**: (**a**) *Comparison of the 1-Sun $J-V$ and the pseudo $J-V$ estimated from Suns$-V_{OC}$ measurement. The intensities varied for the Suns$-V_{OC}$ measurement is 1-Sun to 0.01 Sun. (**b**) Represents the zoomed X-axis for a better understanding of the FF.*

## 2.6 Introduction to perovskite solar cells

### 2.6.1 Progress in perovskite solar cells

The journey of perovskite solar cells began with the concept of dye-sensitized solar cells (DSSCs). The historical evolution of hybrid organic-inorganic perovskite solar cells, starting from the DSSC is shown in **Figure 10**.[40] In 2006 and 2009, Miyasaka *et. al.* used hybrid organometallic halide perovskite MAPbBr$_3$ and MAPbI$_3$ in the DSSCs and achieved an efficiency of 2.6% and 3.8% respectively.[41,42] In 2011, Park *et. al.* used MAPbI$_3$ of 2-3 nm sized nanocrystal and surface-treated TiO$_2$ and achieved an efficiency of 6.5%.[43] The electrolytes used in the DSSC are detrimental to the device as it has corrosion problems. In 2012, M. Gratzel and N.G. Park replaced the electrolyte with solid-state Spiro-MeOTAD and immersed perovskite into the TiO$_2$ scaffold.[44] Interestingly, the efficiency of the cell boosted to 9.7%.[44] In the same year, Snaith *et. al.* replaced the TiO$_2$ with electrically inert Al$_2$O$_3$, and the efficiency was boosted to 10.9%.[45] In the DSSC, the TiO$_2$ acts as an electron acceptor to assist electron transport. Snaith *et. al.* showed that the perovskite could transport electrons itself without scaffold TiO$_2$.[45,46] However, these findings indicate that the perovskite itself has good electrical transport properties and is promising for PV applications. Further, Snaith *et. al.* replaced the solution-processed perovskite deposition with thermal evaporation-based perovskite in the same device architecture and showed an efficiency of 15.4%. This indicates that the efficiency can be improved by improving the perovskite quality. Later, Gratzel and Kim *et. al.* used quantum dot (QD) SnO$_2$ as an electron transfer layer (ETL) and FAPbI$_3$ perovskite as an active layer, showing an efficiency 25.4%, which is comparable to the commercialized Si solar cells.[47] The above device structure originated from liquid-based DSSC, and it is a regular (*n-i-p*) device architecture. In parallel, an inverted architecture (*p-i-n*) was used by Guo *et. al.* in 2013 using PEDOT:PSS as HTL and PCBM as ETL, which showed an efficiency of 3.9%.[48] The *p-i-n* architecture is beneficial as it deals with the low-temperature processed charge transport layers (CTLs) such as poly(3,4-ethylenedioxythiophene) polystyrene sulfonate (PEDOT: PSS), poly (triaryl amine) (PTAA), self-assembled monolayers (SAM), PCBM, ICBA, fullerene (C$_{60}$), etc.[48–53] Currently, the efficiency achieved on the *p-i-n* architecture based PSC is 25%, similar to the *n-i-p* architecture.[54] In addition, the *p-i-n* architecture-based device is crucial in monolithic tandem solar cells such as perovskite-perovskite, perovskite–organic, and Si-perovskite tandem solar cells. Such a speedy journey of efficiency improvement of the perovskite solar cells (PSCs) (3.9% to >25%)[42,55,56] can be credited to the unique optoelectronic features of perovskites such as a high absorption coefficient, low exciton binding energy, and a tuneable bandgap.[57–61] Unprecedentedly, Etgar *et. al.* found that the MAPbI$_3$ perovskite could be p-type[62] whereas You *et. al.* showed n-type behavior[63]. Thus, MAPbI$_3$ is bipolarly dopable, and different processes lead to different doping behaviors (p-type or n-type). However, various perovskites can be prepared by choosing different mixing compositions of the A, B, and X ions in the ABX$_3$ perovskite. The underlying mechanism of perovskite is still unclear due to the low temperature processed deposition, which leads to point defects, making perovskite unintentionally doped. Perovskite is full of defects and disorder, but the unique properties of perovskite are that the dominating defects are not harmful as they do not create detrimental deep-level defect states within the bandgap.[64]

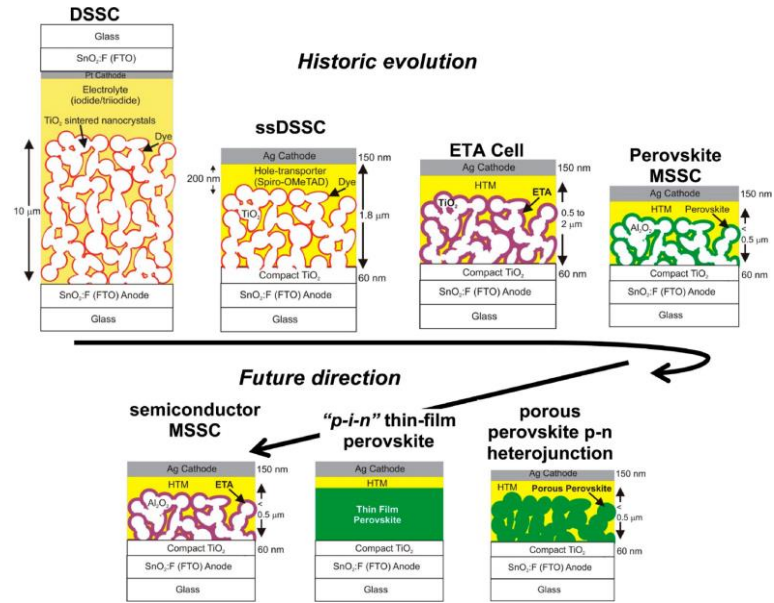

***Figure 10***: *Historical evolution of hybrid organic-inorganic perovskite solar cells, starting from the DSSC. (adapted from ref.[40] )*

## 2.6.2 Defects in perovskite semiconductor

Ideally, each atom in a crystal lattice should be located at its designated position and any deviation from this results in a defect state **Figure 11a**. The structural non-periodicity or any foreign atom in the periodic lattice introduces disorder into the crystal lattice bonding, and impacting the physical properties and density of states. Such disorders are not desirable for PV application as they produce localized sub-gap electronic states. When the charge carrier transitions happen, these disorders play a significant role as non-radiative centers **Figure 11b**. The trapping and detrapping of the charge carriers depend on the position of the trap states. If the trap states are located in the mid of the bandgap, they are called mid-gap states. If the trap states are located close to the conduction or valence band, they are called shallow traps. If the charge carriers trap at the mid-gap states, then it will be difficult to detrapped, however, if the charges trap at the shallow traps, then they can be detrapped by additional energy (activation energy), via further optical excitation or thermal energy ($kT = 25.7$ meV at room temperature) **Figure 11c**. A brief of the origin of various defects and defect passivation engineering will be discussed below.

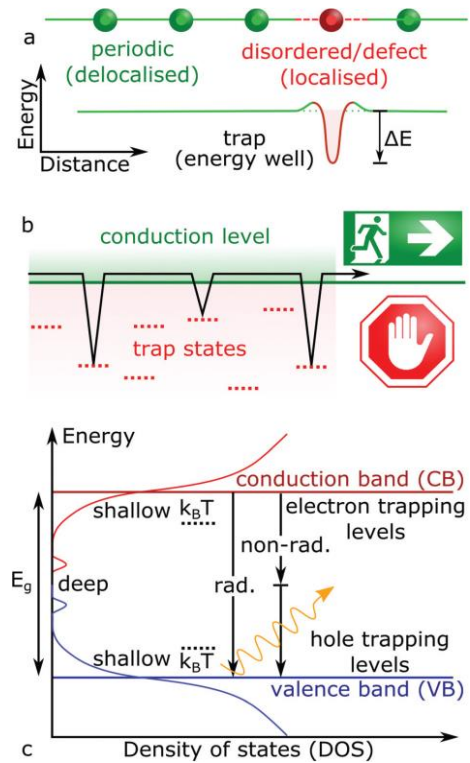

**Figure 11**: **(a)** *Schematic of a charge localization caused by non-periodic disorder and defect in the lattice.* **(b)** *Schematic of charge trapping kinetics in a semiconductor, which inhabits charge transport by trapping and detrapping events.* **(c)** *Schematic representation of the density of states (DoS) is a disordered semiconductor whereby both the radiative (band-to-band) and non-radiative (trap-assisted) recombination can happen. (adapted from ref.[65]).*

The efficiency of the PSCs can be limited due to the presence of non-radiative defect centres and uncontrolled uniform deposition of perovskite films over the large area substrates. The low temperature (~100° C) processed perovskite films are generally polycrystalline films with a thickness of ≈ 500 nm and create impurities in the bulk and surface of the film. The imperfections in the perovskite are classified into two categories: (i) intrinsic and (ii) extrinsic. The extrinsic imperfections are caused by environmental surroundings and unsaturated bonds, which result in polycrystalline networks, grain boundaries, line defects, surface defects (due to finite thickness), etc. However, the intrinsic defects are also known as point defects, which depend on the growth of the crystal structure under the chemical environment and the used composition of the perovskite system. These defects play the role of trap centers of the charge carriers, as discussed above, and result in the lower performance of the solar cell. Hence, it is essential to understand the origin of such defects and passivation mechanisms to obtain improved performance in perovskite solar cells.

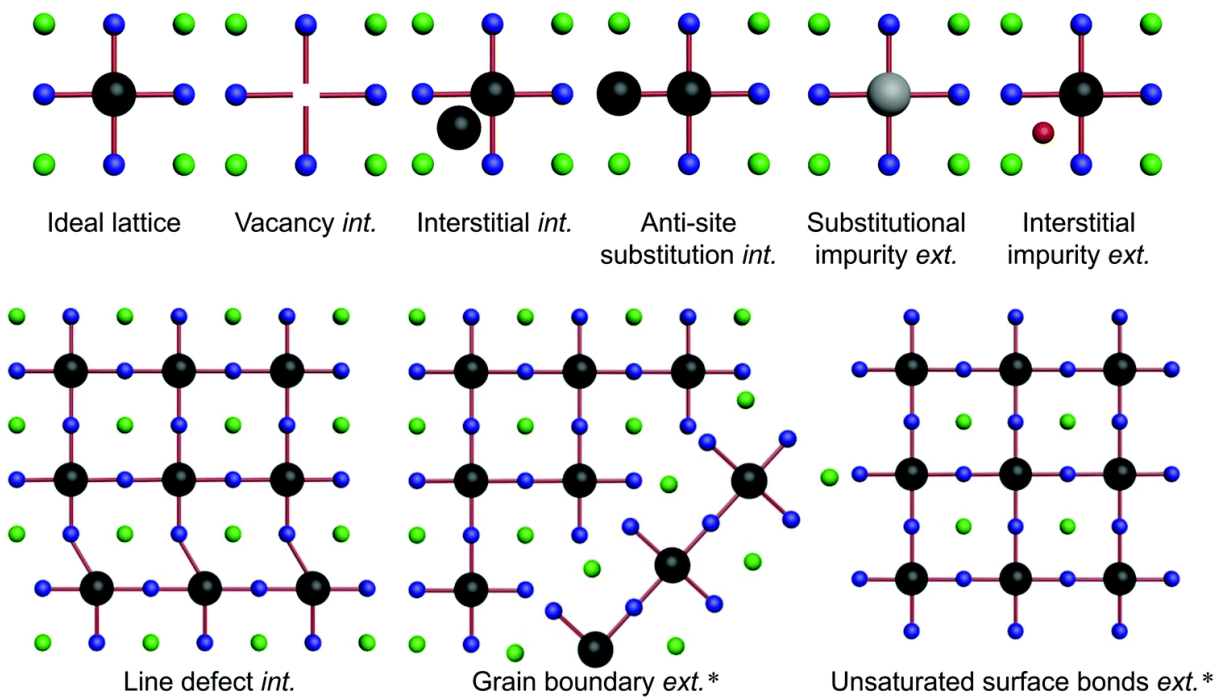

*Figure 12*: *Schematic of intrinsic (int.) and extrinsic (ext.) defects in perovskite layer compared to the ideal crystal lattice. For all the crystal structures, green, black, and blue spheres represent A, B, and X elements in the ABX₃ perovskite crystal structure, while grey and red represent different impurities. The defects which are identified experimentally are marked by '\*'.[66,67] (adapted from ref.[65]).*

**Figure 12** shows the schematic representation of intrinsic and extrinsic defects. For instance, the intrinsic defects are vacancy (missing atoms)[68], interstitial (same or different atom occupies interstitial site)[69], antisite (atoms of different type exchanges positions)[70], impurity(foreign atoms that replace some atom)[71], etc. In contrast, the extrinsic defects are line defects (whole atoms in a row arranged anomalously)[72], grain boundaries (separation line between polycrystalline grains)[73], unsaturated surface bonds (boundary dangling atoms)[74], etc. In figure 12, "int." refers to intrinsic defects, and "ext." corresponds to extrinsic defects. Some of the defects that can be directly observed experimentally are marked with '*'. However, there are various device engineering techniques by which these intrinsic and extrinsic defects can be minimized.

### 2.6.3 Defects passivation engineering of perovskites solar cells

Even though the first experimental demonstration of perovskite solar cells was done by Miyasaka *at. el.* in 2006 using MAPbBr$_3$ perovskite, but the theoretical investigation using the first principle theory studied before that. In 2004, Change *et. al.* used local density approximation (LDA) theory to study the cubic phase of inorganic perovskite CsPbI$_3$ and hybrid organic–inorganic halide perovskite MAPbX$_3$ (where X= Cl, Br, I).[75] They observed that the effective mass of holes was higher than the electron mass, which is not the case with conventional semiconductors.[75] However, various perovskites are later being used to study the intrinsic defects in the perovskite crystal structure system using the first principle calculation.[76–79] There are various kinds of point defects present in the perovskite crystal structure, and the main such point defects are (i) interstitial, (ii) vacancy, (iii) antisites or substitutional (iv) Frankel, (v) Schottky, (vi) impurity, etc. However, the MAPbI$_3$ is the widely studied perovskite, and it observed that there are mainly 12 dominating point defects present in the crystal structure e.g., three vacancies ($V_{MA}$, $V_{Pb}$, $V_I$), three interstitials ($MA_i$, $Pb_i$, $I_i$) and six antisites ($MA_{Pb}$, $Pb_{MA}$, $MA_I$, $Pb_I$, $I_{MA}$, $I_{Pb}$ ) defects, etc.[80] These defects act as non-radiative recombination centres and are always detrimental to the device performance.

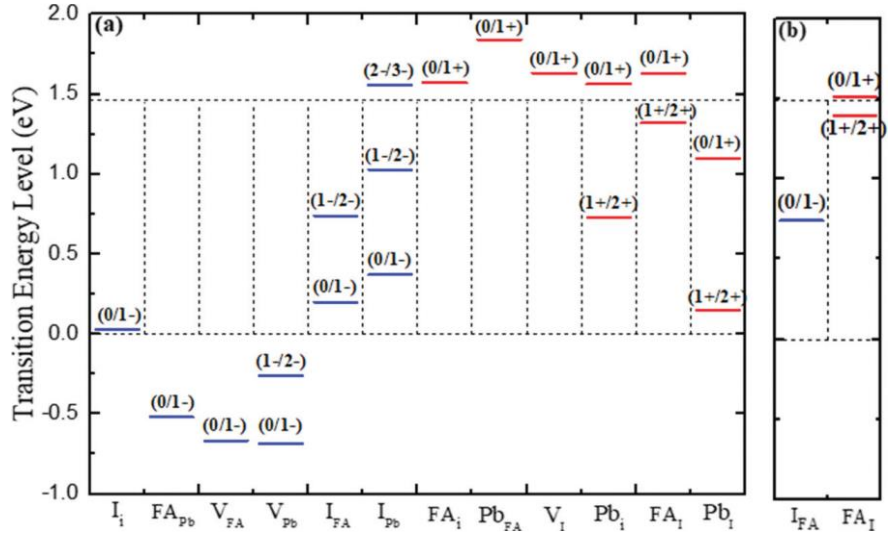

***Figure 13**: Transition energy levels of intrinsic defects (**a**) FAPbI₃ and (**b**) in the mixed organic cation FA₀.₅₂MA₀.₄₈PbI₃ perovskite. (adapted from ref.[81]).*

In 2018, Lu and Yam studied the FAPbI₃-based perovskite crystal structure under different conditions (I-rich, Pb-rich, and moderate).[81] It is observed that the $V_{FA}$, $FA_I$, and $I_{FA}$ have much lower formation energies. Hence, antisites $FA_I$ and $I_{FA}$ create deep levels in the band gap. These defects can act as non-radiative recombination centres and reduce the charge carrier lifetime, resulting in lower open circuit voltage. To avoid the formation of $FA_I$ defects, the crystal should be grown in the I-rich environment. In addition, it is observed that the mixing of organic cations increases the formation energy of deep-level defects, which is expected to reduce the non-radiative recombination **Figure 13b**. Thus, the compositional engineering of organic cations is helpful to fabricate deep-level defect free perovskite and offer improved device performance. The compositional engineering of hybrid organic-inorganic perovskite will be discussed in Chapter 4. The mixing of the cations must follow the Goldschmidt tolerance factor ($t$) and octahedral factor (μ) rules as discussed in Chapter **1**.[82,83] Apart from the organic cation mixing, in 2018, Ghosh *et. al.* showed that the addition of a small amount of Cs (≤ 25%) in the FAPbI₃ perovskite distorted the crystal structure (octahedral tilting) due to mismatch in the size of the cations.[84] The slight octahedral tilting helps to form a strong hydrogen bond (N-H….I) between the FA cation and the halide ions.[84] This eventually helps in the stabilization of crystal structure and promotes greater thermal stability with improved performance. In 2016, Saliba *et. al.* showed that the addition of 5% of CsI in the MAFAPb(IBr)₃ perovskite improves the perovskite quality and results in an improvement in stability, reproducibility, and high efficiency. [85] However, various reports show that the mix cation, i.e., compositional engineering, is a promising technique to improve the perovskite quality, which results in stable, reproducible, and efficient perovskite solar cells.[85,86]

There are various extrinsic defects present in the solution-processed perovskite film, such as grain boundaries, line defects, pinholes, unsaturated surface bonds, etc. These physical defects can be seen experimentally by scanning electron microscope (SEM) or tunneling electron microscope (TEM) images. [66,67] However, passivation of such defects is essential to obtain improved performance of perovskite solar cells. Various additive engineering and surface treatments are adopted to passivate the extrinsic defects. **Figure 14** shows the schematic of the grains and pinhole defects. For instance, in passivating defects, several large molecules in the form of Lewis acids and base are used to improve the crystal quality and improve charge carrier lifetime. [87–89] Poly(methyl methacrylate) (PMMA) is such a Lewis base and is used at the interface between HTL-perovskite[90] and ETL-Perovskite[91]. The PMMA used here by the spin coating technique suppresses the grain boundary and pinhole defects.[88]

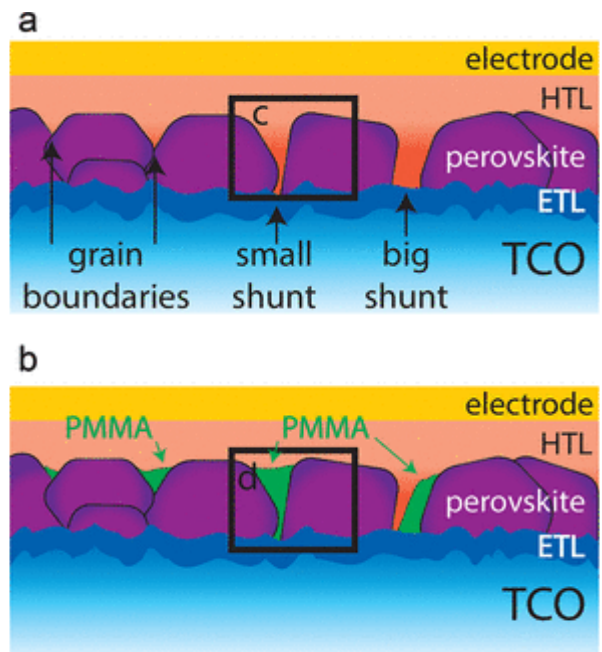

***Figure 14****: Schematic of extrinsic defects grain boundaries, pinhole. (**b**) Passivation of defects using PMAA. (adapted from ref.[88]).*

Further, additive engineering is a promising way to enhance device efficiency and lifetime. The additives applied to the perovskite bulk or at the interface interact with the perovskite absorber and the transport layers, resulting in defect passivation and ion immobilization, improving device performance and stability **Figure 15a.** The additives form strong ion bonds with uncoordinated $Pb^{2+}$ or organic cations such as MA, FA, with the halide ions ($Cl^-$, $Br^-$, $I^-$) to enhance the $Pb-I$ bonding and $MA-I$, FA-I bonding etc. Hence, the MA, FA, iodine vacancies, and other defects can be reduced. Son *et. al.* showed that adding a small amount of potassium iodide (KI) in halide perovskite is a universal approach to fabricating hysteresis-free PSCs. Wu *et al*. showed that the additive engineering of the ionic-liquid additive of methyl-ammonium acetate (MAAc) modulates the thin film's crystalline quality and overall morphology.[92]

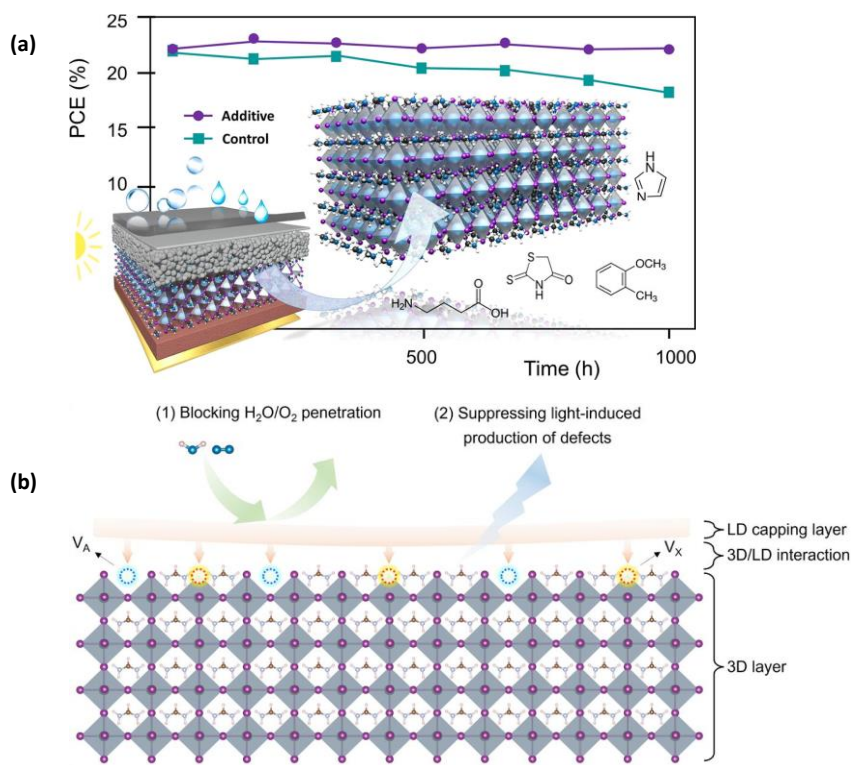

***Figure 15***: *(a) Additive engineering for the passivation of defects to enhance efficiency and stability in perovskite solar cells. (adapted from ref.[93]) (b) Schematic illustration for understanding the stability improvement of the 3D/LD structure PSCs. (adapted from ref.[94])*

The MACl additive into the perovskite solution before deposition shows an increase in grain size from 200 to over 1000 nm, where the MACl has no effect on the absorption or photoluminescence (PL) spectra of the perovskite films. [95] Li *et al.* reported a one-step solution-processing strategy using phosphonic acid ammonium additives, resulting in efficient PSCs and enhanced stability. [96]

Apart from the defects passivation, the stability of the perovskite solar cell is essential. The low dimensional (LD) capping layer plays a crucial role in surface passivation, heterojunction formation, and environmental stress resistance.[97,98] Nazeeruddin *et.al.* reported that employing a thin 2D perovskite of phenethylammonium iodide (PEAI) layer on top of a 3D perovskite film shows improved efficiency and stability. The devices retained 85% of the initial efficiency even stressed under 1-Sun illumination for 800 hours at 50 °C.[99] Liu *et. al.* introduced a 2D capping layer of $A_2PbI_4$ over the 3D perovskite layer and showed devices retained 90% of their initial efficiency for 1000 hours in humid air and 1-sun light.[100] Wang *et al.* reported the incorporating n-butyl-ammonium cations of 2D perovskite into a 3D perovskite $FA_{0.83}Cs_{0.17}Pb(I_yBr_{1-y})_3$ shows an efficiency of 19.5% which maintains the 80% of the initial efficiency of its post-burn after 1,000 hours in air, and close to 4,000 hours when encapsulated.[101]

### 2.6.4 Charge transport layers and interfacial engineering in perovskite solar cells

The journey of perovskite solar cells (PSCs) began with the concept of dye-sensitized solar cells (DSSCs) in the *n-i-p* (aka "regular") device architecture.[42] In 2009, Miyasaka *et. al.* used hybrid organometallic metal halide perovskite $MAPbI_3$ in the DSSCs and achieved an efficiency of 3.81%.[42] Currently, the reported efficiency of the perovskite solar cells is ~26% in both *n-i-p*[102] and *p-i-n* (aka "inverted")[103] device architecture, where the practical achievable limit[104] could be ~30%. This thesis deals with the *p-i-n* architecture-based solar cells. The *p-i-n* device architecture has several advantages, including low-temperature processability of the charge transport layers (CTLs) and long-term operational stability derived from non-doped hole transport layers.[105]    In 2013, the *p-i-n* architecture perovskite solar cell started with PEDOT:PSS as the hole transport layer and PCBM or ICBA as the electron transport layer.[48] However, in the *p-i-n* architecture, although C60 and PCBM are widely used as electron transport layers (ETL), numerous hole transport layers (HTLs) have evolved. The widely used hole transport layers are, inorganic ($NiO_x$, $Cu_2O$)[106,107], polymeric (PEDOT:PSS, PTAA, Poly-TPD)[108–110], small molecules[111], and self-assembled monolayer (Me-2PACz, MeO-2PACz, Me-4PACz,).[112–114] The PEDOT:PSS is hygroscopic and acidic in nature, which impedes the efficient performance and stability of perovskite solar cells.[115] The chemical stability of $NiO_x$ is challenging when connected with the perovskite, and hence, an additional interface passivation layer is used on the top of the $NiO_x$ layer.[116,117] The PTAA-based HTL shows a offset of 200-300 meV energy band alignment mismatch for perovskite of bandgap 1.6 eV.[113,118] Recently, various phosphonic acid group-anchored carbazole-based SAM HTLs have grabbed significant attention due to the high performance and excellent stability of perovskite solar cells. **Figure 16a** shows the molecular structure of 2PACz SAM deposited on ITO-coated glass substrates.[119] The phosphonic acid group strongly binds with ITO and shows a strong monolayer fingerprint.[113] In 2021, Levine *et. al* showed that, in the *p-i-n* architecture solar cells, the dominating loss mechanisms are associated with hole-selective buried interfaces. Compared to other SAMs, the Me-4PACz-based device showed suppressed non-radiative recombination at the perovskite/Me-4PACz interface, resulting in a faster hole transfer rate **Figure 16b**.[112] In 2020, Al-Ashouri *et al.*, used various phosphonic acid group SAMs such as 2PACz, MeO-2PACz,

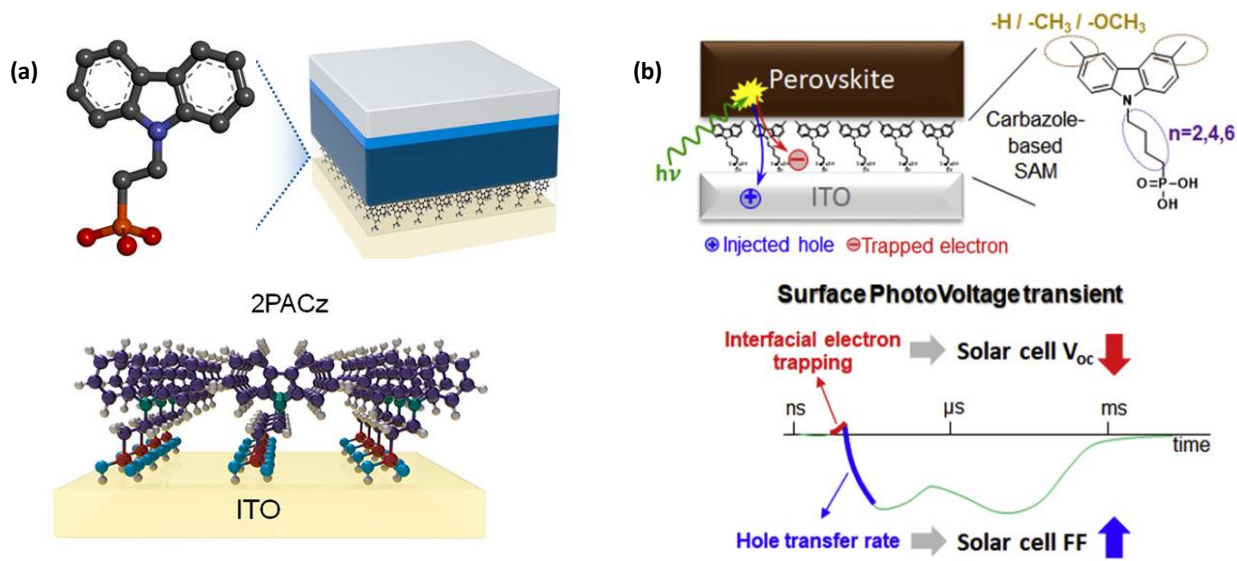

***Figure 16***: ***(a)*** *The chemical structure of 2PACz SAM, and it is deposited on the ITO-coated glass substrate. (adapted from ref.[119])* ***(b)*** *Different SAMs used at the ITO-perovskite interface and surface photovoltage transients with a minimalistic kinetic model applied to extract interfacial electron trap densities and hole transfer rates and their correlation with open-circuit voltages and fill factors. (adapted from ref.[112])*

and Me-4PACz as HTL, demonstrating superior device performance than the widely employed PTAA as HTL.[113,120,121] In particular, by using Me-4PACz SAM as HTL, Al-Ashouri *et al*. showed that, a certified PCE of ~29% in silicon-perovskite two-terminal tandem solar cells with improved stability.[121] Recently, Li *et al*., employed a mixture of MeO-2PACz and 2PACz to improve the charge extraction and reported a PCE of 25% for all-perovskite tandem solar cells.[122] Deng *et. al*. showed that co-assembled monolayers help in suppressing the non-radiative recombination and the resultant device showed a PCE of 23.59%.[123] Due to the exceptional performance of the SAM-based HTLs, they can promise to deliver commercialized *p-i-n* architecture-based PSCs.

## 2.6.5. Future perspective of perovskite solar cells

Perovskite solar cells (PSCs) are the fastest-growing PV technology and hold great promise for the photovoltaic industry due to their low-cost fabrication and excellent efficiency performance. To get it for the commercial readiness level (CRL), most important factor would be yield beyond 95% at PSC module levels. The current essential requirements for PSC are the reproducibility of high efficiency devices, scalability and stability. The reported certified high-efficiency (24-26%) papers are based on FAPbI$_3$ perovskites with bandgap of $E_g \approx 1.5$ eV and the typical device's active area of $\approx 0.1$ cm$^2$ to maximum of 1 cm$^2$. Scaling up the device's active area without compromising fill factor (FF), hence efficiency is non-trivial. Therefore, understanding the loss mechanism in large area devices is crucial. The stability analysis reported in the literature is inconsistent, preventing data comparison and identifying various degradation factors or failure mechanisms. Moreover, how the accelerated tests would be useful in predicting the real lifetime of the solar cells is yet to be developed. So, the knowledge and the technological gaps between laboratory and industry-scale production are crucial to develop. Therefore, in this section we have discussed the future perspective of perovskite solar cells in terms of the scalable and stable high efficiency PSCs.

### 2.6.5.1 Scalable perovskite solar cells and efficiency limitation

Solution-processed photovoltaic technologies grabbed much attention in the solar cell community and reported certified high-efficiency (>25%) using FAPbI$_3$ (or ~5% Cs added) perovskite absorber of bandgap ~1.5 eV over the device active area of $\leq 0.1$ cm$^2$.[56] It is evident that, in terms of lab-scale device efficiency, this value is competitive with established commercialized technologies. However, the perovskite community is still far from commercializing the perovskite photovoltaic technology as the efficiency is not similar in the slightly large area (1 cm$^2$) devices yet, regardless of the deposition techniques. In academic labs, people have demonstrated significant progress in PSCs over device active areas of up to 1 cm$^2$ and mini-modules. However, a few industries e.g., Kunshan GCL, showed efficiency >19% on the single junction perovskite module of area 2 m$^2$, but the recipe is not available in the public domain.[124] Therefore, researchers must focus on developing strategies for high performance scalable perovskite solar cells and perovskite modules fabrication. Since this is a thin-film device technology with many layers being deposited via the solution route, it seems to be one of the major challenges in scaling up. Typical thicknesses of absorber layers in other established thin-film PV technologies like; GaAS and CdTe are typical of 10-12 μm and 4-6 µm respectively, however in the case of PSCs, it is hardly 600-800 nm. Further, both established thin-film PV technologies use typically vacuum-based deposition techniques (Vacuum-transfer deposition for CdTe and MBE for GaAs PVs), whereas halide perovskite PV performance is typically processed via solution route. Vacuum-processed perovskite is lagging behind in terms of efficiency. At least in the academic laboratory, scaling up via the solution route is challenging. Recently, we have demonstrated that the fabrication of large area PSCs by spin coating is challenging as the leakage current in the device increases exponentially, which may be due to non-uniform deposition of the perovskite layer in a multi-layered stack device.[125] However, different deposition techniques have been developed, such as vacuum processing (thermal evaporation, pulsed laser deposition), solution processing (blade coating, spray coating, slot dye coating, inject printing, screen printing) apart from spin coating technique to make uniform perovskite layer over large area substrates. Even though the blade coating and evaporation based technique showed an efficiency ~25% over a small area (≤0.1 cm$^2$) devices, however, the up-scaled perovskite solar modules (PSMs) of area ~100 cm$^2$ showed better efficiency (~10%) in evaporated devices due to conformal deposition over large area substrates. However, the solution based techniques such as spray coating, slot die coating, and inject printing-based devices have lower cell efficiency in small area cells (0.1 cm$^2$) but relatively improved efficiency performance in PSMs. Therefore, contrasting the photovoltaic performance with cell and module sizes of different deposition techniques makes it difficult to compare their performance disparities fairly and consistently. Chalkias *et. al.* develop a mathematical equation to ascertain the scaling-up factor ($f_{scaling\_up}$)[126,127] as

$$f = \frac{1 - \frac{\eta_{module}}{\eta_{cell}}}{\log\left(\frac{A_{module}}{A_{cell}}\right)} \times 100\%$$

Where, $\eta$ is the PCE and A is the active area of the cell or module. The lower value of the $f$ indicates less scaling-up loss from PSCs to PSM, which is essential for commercialization. Interestingly, it is observed that $f$ = 3.48% for the inject printing process ($A_{cell}$ = 0.105 $cm^2$, $A_{module}$ = 804 $cm^2$, $\eta_{cell}$ = 20.7%, $\eta_{module}$ =17.9%) and $f$ = 8.57% for evaporation based process ($A_{cell}$ = 0.16 $cm^2$, $A_{module}$ = 228 $cm^2$, $\eta_{cell}$ = 24.8%, $\eta_{module}$ =18.1%), indicating the possible direction of scaling up the process in the future.[128] People have reported in various review papers that scaling up of the PSCs active area results in poor efficiency, but which PV parameter affects the most is not well elaborated. Additive and interface engineering were adopted to improve the efficiency over large area PSC devices but the efficiency and reproducibility both decrease on increased device's active area, which restricts the PSCs toward commercialization.[129] Even though the PSCs research started with the millimeter square device area, people have recently focused on taking it to the centimeter square device area. Deplorably, the literature study shows that there is a lack of representation of the performance of PSC in large areas. Many of the reported papers have demonstrated high efficiency PSC's performance but don't report the device area, we deliberately avoid to refer them here, however, would like to pass this is an important parameter to be checked at all levels, i.e., authors, editors, and reviewers. Even some groups report the device's active area in the SI file, which is a bad practice, and the reader needs to open another file to get the information on the device area. It would be good if the authors report the device's active area with dimension (length × width) in the main text along with the device image. Also, the sheet resistance of the transparent conductive oxide (TCO) substrate on which the device is fabricated could influence the device's performance, hence it is important to mention the sheet resistance ($\Omega$/Sq) of the substrates.[130] It is also important to include the list of materials, device characterization tools, and detailed information on characterization techniques. Saliba *et. al.* studied 16000 papers and showed that the short circuit current density ($J_{SC}$) measured from $J - V$ scans is 4-5% higher than the $J_{SC}$ measured from integrating the EQE spectrum.[131] This mismatch could be attributed to either edge effect[132] from the active area of the device or pre-bias measurement condition [133]. Some groups use a non-reflective metal mask aperture on the top of the active area during $J - V$ scans, which could avoid edge effect.[56] Since the dimension of the aperture masks influences the PV parameters, such as $V_{OC}$ and $FF$, so it should be $\geq 80\%$ of the device's active area.[134]

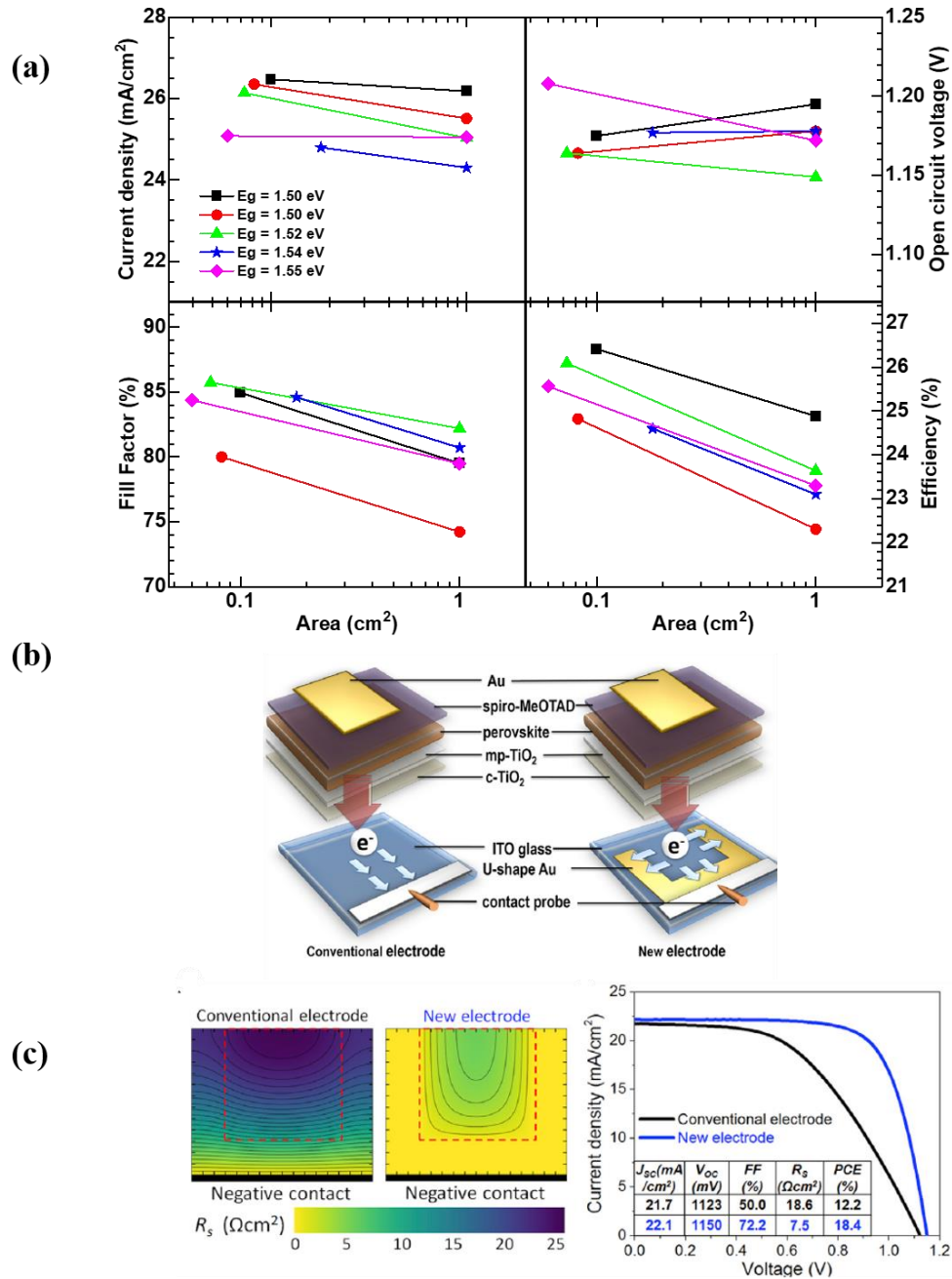

**Figure 17:** *Photovoltaic (PV) parameters as a function of device active area for the highest efficiency reported perovskite solar cells (PSCs).*[102,103,135–137] *(b) 1-cm² active area n-i-p architecture based PSC with the conventional electrode (left) and U-shaped new electrode design (right) . (adapted from ref.*[130]*). (c) Left images indicate the simulated series resistance distribution for* 10 × 10 *mm² active area device (red dashed lines define active area) and right panel J − V characteristics shows the performance comparison of the conventional and new electrode designed PSCs. (adapted from ref.*[130]*).*

Apart from that, recently, people reported excellent efficiency on the small area (<0.1 cm²) and relatively larger area (1 cm²) devices. However, they certify the efficiency on small area devices only or use an aperture mask of small area (<0.1 cm²), and that results in lack of confidence in large area PSCs.[136] **Figure. 17a** represents the reported high efficiency PV parameters of PSC ($E_g = 1.52 \pm 0.3$ eV) over device active area of ~0.1 cm² and 1 cm². [102,103,135–137] It is evident that the $J_{SC}$ and the $V_{OC}$ are not affected significantly by increasing the active area from 0.1 cm² to 1 cm², whereas the $FF$ does, hence the device efficiency. This effect could be severe in further large area devices. It is of utmost importance to understand the loss in the $FF$ due to increasing the device's active area to look forward the scalability of PSCs. Ho-Baillie and co-workers showed that the TCO substrate sheet resistance could influence the FF of large-area PSC devices apart from the non-uniform deposition of the device stack.[130] They fabricated *n-i-p* mesoscopic PSC device architecture on a conventional and U-shaped designed bottom gold electrode (150 nm thick) and showed certified efficiency of 19.63% with $FF$ of 75.8% over an active area of 1.02 cm² (see **Figure. 17b**). Depositing U-shaped electrodes on the TCO substrates enhances the electron collection efficiency in n-i-p device architecture. **Figure. 17c** shows the simulated series resistance distribution for 10 ×10 mm² active area device, and it is lower for the U-shaped modified electrode, and the corresponding fill factor ($FF$) increased from 50.0% to 72.2%. The FF of the device also depends on the device geometry, as shown in **Figure. 18a**. The series resistance is higher for the square-shaped electrode because the electrons travel a longer distance from the center, whereas it is lower for the strip-shaped electrode and shows an improved fill factor. However, the bottom electrode modification strategy is possible for the *n-i-p* device where the ETL (c-TiO₂ 50 nm/mp-TiO₂ 200 nm) thickness ~250 nm could be deposited by spray pyrolysis/spin coating technique over 100-150 nm thick bottom U-shaped electrode. But similar strategy is difficult to adopt in solution-processed *p-i-n* PSCs, as the HTL layer thickness is usually 2-3 nm of recently developed self-assembled monolayers (SAMs).[112,114,138] Such a 100-150 nm bottom U-shaped gold electrode would act as a barrier at the edges of the substrates and make it difficult to deposit uniform HTL using the solution-processed spin coating technique. Therefore, for large area (≥1 cm²) *p-i-n* device architecture, the U-shape electrode deposition is done (thermal evaporation or silver paste) after scratching the spin-coated thin film layers. **Figure. 18b, c, d** are the photographic images of large area (1 cm²) PSCs using different electrode shapes.

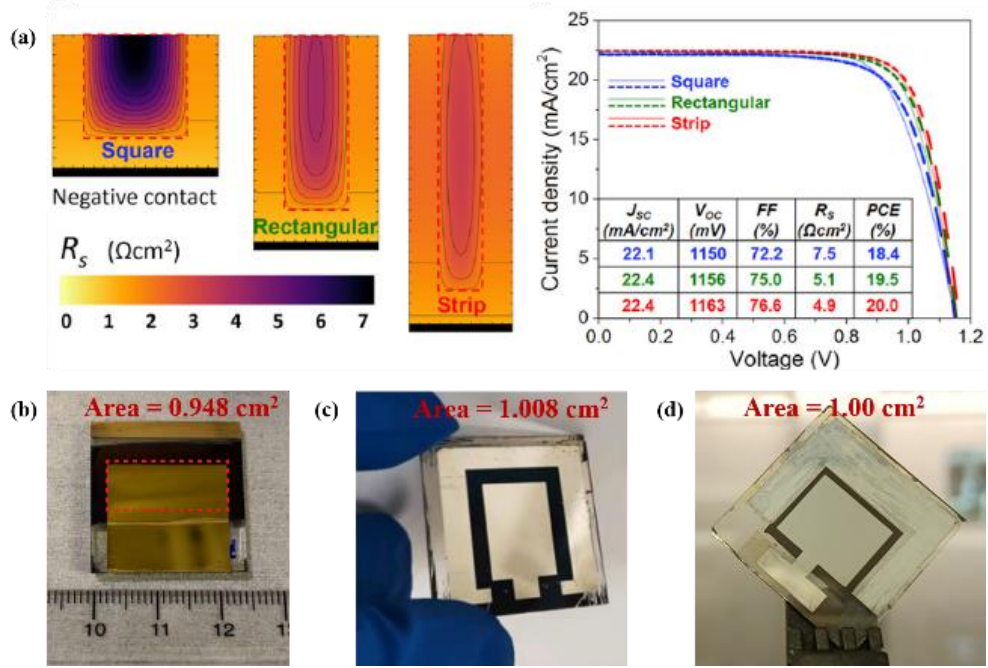

***Figure 18: (a)*** *Left panel represents the simulated series resistance distributions for* $10 \times 10$ $mm^2$, $6 \times 17\ mm^2$, *and* $4 \times 25\ mm^2$ *devices using electrode designs with different geometries. The right panel shows the simulated (solid line) and measured (dashed line)* $J - V$ *curves of devices using new electrode designs with different geometries. (adapted from ref.[130]).* ***(b)*** *Rectangular device active area of 0.948 $cm^2$ with one side TCO connection.[139]* ***(c)*** *Rectangular device active area 1.008 $cm^2$ with all side TCO connection.[140]* *and* ***(d)*** *Rectangular device active area 1.000 $cm^2$ with U-shaped (three side TCO connection) bottom electrode deposited after scribing the spin-coated layers.[141]*

Unfortunately, most of the reports do not elaborate on the electrode design for large-area devices, which is a bad practice and lacks progress toward the scalability of PSCs. We believe reporting the device engineering, including electrode design, is required for successful scalable PSC fabrication apart from uniform perovskite layer deposition. In addition, a systematic simulation or experimental study needs to be carried out to understand the series and shunt resistance losses for large-area devices to better understand the FF loss issues.[142]

### 2.6.5.2 Stability analysis of perovskite solar cells

To ensure the economic viability of perovskite technology, long-term stability is crucial apart from high efficiency and scalability. To compete with the market-leading commercialized PV technologies, the expected lifetime of the PV devices should be 20-25 years. It is impractical to use decade-long protocols for stability testing. Hence, accelerated ageing tests are required to understand the lifetime span of PSCs. The stability of the PSCs depends on extrinsic (oxygen, moisture, UV light exposure) and intrinsic (thermal and structural degradation, ion migration, electrical bias) stresses. Extrinsic stress, such as oxygen-induced degradation, moisture degradation, and UV-light degradation, can be avoided by advanced encapsulation techniques. However, the intrinsic stability of the PSCs is a crucial factor, and understanding them and corresponding solutions need to be discussed for commercializing perovskite PV technology. Considering the material stability, the mixed organic-inorganic halide perovskites represented the better choice for high efficiency and stable performance.[85,143] However, the existing accelerated aging tests described in an International Electrotechnical Commission (IEC) standards are used for testing the Si solar panel performance.[144] Unfortunately, these tests cannot be carried out in DSSC, OPV and PSCs as they are fundamentally different materials and PSCs also show ion migration, performance recovery in dark conditions, etc., which are unknown in Si photovoltaics. In 2011, a board consortium of researchers at the International Summit of Organic PV Stability developed standardized ageing tests called ISOS protocols for testing the lab-scale OPV devices to ensure the comparability of PV testing performed at different laboratories.[145] Those protocols include the following stability testing standards (i) dark storage (ISOS-D), (ii) light soaking (ISOS-L), (iii) outdoor stability (ISOS-O), (iv) thermal cycling (ISOS-T), (v) solar-thermal cycling (ISOS-LT), and each of which has three different levels determined by thermal stress, relative humidity stress, environment set-up, and light source & circuit bias (maximum power point tracking: MPPT or open circuit: OC). People reported different stability conditions and data formats in PSCs, preventing data comparison and identifying various degradation factors or failure mechanisms. The consensus statement in 2020 by Khenkin *et. al.* proposes that the ISOS stability could be carried out in PSCs along with additional ageing experiments such as (i) bias stability (ISOS-V) (ii) light cycling (ISOS-LC), and also reported a checklist that should be carried out for reporting the stability of PSCs. The reason for adding bias stability and light cycling for PSC is discussed below.

The ISOS-O-1 protocol suggests the periodic measurements of $J-V$ curves under solar simulator light illumination in MPPT or OC bias conditions, whereas ISOS-O-2 requires natural sunlight. ISOS-O-3 requires both in situ MPP tracking under natural sunlight and periodic performance measurements under a solar simulator. **Figure. 19a** shows the stability performance (ISOS-O) obtained by $J-V$ measurements and MPP tracking of PSCs do not coincide, although they generally have similar trends. Therefore, it is crucial when characterizing PSCs to describe the load and recovery time before each $J-V$ measurement. The perovskite solar cell degradation mode is known to be reversible partly or entirely in dark conditions, which is often referred to as metastability (see **Figure. 19b**). Therefore, cycling through light-dark periods to mimic day-night cycles results in a significantly different stress test than applying constant illumination (ISOS-L). The improvement in the power conversion efficiency (PCE) under illumination after storage in the dark condition could be attributed to the passivation of interfacial defects by photogenerated charge carriers or ion migration induced modification of the built-in electric field.[146] However, the PCE dynamics during a cycle depend on the present status of cell degradation condition. Therefore, the ISOS protocols were revised for PSCs and included light-dark cycling protocols to account for the recovery phenomena (ISOS-LC). In addition, the electrical bias causes PSC degradation by stimulating the ion migration or charge carrier accumulation, resulting in thermally activated trap formation.[147] Therefore, ISOS protocols revisited for PSCs to include the ISOS-V, in which the behaviour of the cell is analysed when exposed to a certain electrical forward bias (maximum power point voltage: $V_{MPP}$ or open circuit voltage: $V_{OC}$ as measured in AM 1.5G) in the dark (see **Figure. 19c**). Interestingly, people also show additional degradation studies beyond ISOS protocols, such as exposure of PSCs using 10 Suns illumination at the MPPT condition of the encapsulated devices (See **Figure. 19d**) and shows excellent thermal stability of $FA_{0.83}Cs_{0.17}PbI_{2.7}Br_{0.3}$ composition based PSC.[143] When the 3D perovskites come in contact with the environment (moisture, oxygen, UV-light), they are subjected to severe degradation and, limiting the solar cell's lifetime. Different 2D and quasi-2D perovskite materials demonstrated significant improvement in the efficiency and stability of 3D PSCs. The 2D perovskites exhibit multiple-quantum-well structures and offer tunable optoelectronic properties, defects passivation in the bulk and interface, superior thermal and light stability, hydrophobicity due to large organic cation, suppressing ion migration etc.[148] To achieve simultaneously high efficiency and high stability for 3D PSCs using 2D perovskites, numerous device engineering strategies studies have been conducted, including compositional or additive engineering, dimensional engineering, interfacial engineering, etc. Sum et. al. used different 2D (and low dimensional) perovskite materials to passivate 3D perovskite surface defects and showed an efficiency 24.1% with $T_{95}$ >1000 hours at MPPT.[94] Wolf el. al. tailored the number of octahedral inorganic sheets in 2D perovskite passivation layers for 3D PSCs, and showed efficiency 24.3% with $T_{95}$ >1000 hours under damp heat testings (85° C and RH 85%).[152] Mohite et. al. carried out dimensional tunability and grew pure 2D phase perovskites of controlled thickness on the 3D perovskite layer using solvent engineering strategy and showed efficiency 24.5% with $T_{99}$ >2000 hours under continuous 1-sun light at 55° C and RH 65% (see **Figure. 19e**).[150] However, for the practical application of perovskite solar cells, the encapsulation of the PSC device is mandatory to prevent the penetration of oxygen, moisture, and lead leakage from the lead-based PSCs.

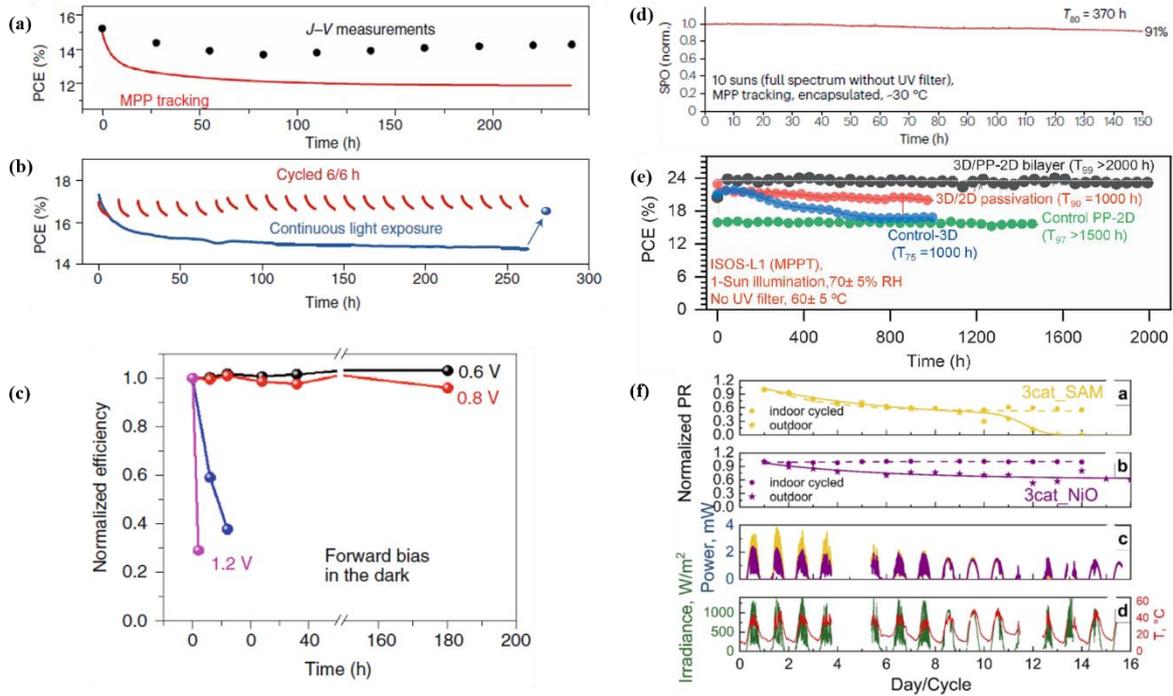

***Figure 19***: *(**a**) Power conversion efficiency (PCE) extracted from current density vs. voltage ($J-V$) scans (black dots) and continuous MPP tracking (red curve) for the same PSC. (adapted from ref.[149]). (**b**) PCE variation of PSCs upon continuous (blue curve) or cycled (6/6 h, red curves) illumination by white light emitting diodes. (adapted from ref.[149]). (**c**) Normalized PCE of PSCs changes with forward bias in the dark condition. (adapted from ref.[149]). (**d**) Stabilized power output (SPO) of the perovskite devices, continuously recorded under 10 Suns concentration illumination. (adapted from ref.[143]) (**e**) ISOS-L-1 stability measured at MPPT in ambient condition under continuous 1-sun illumination (55°C) for epoxy encapsulated control-3D and 2D passivated 3D PSC. (adapted from ref. [150]) (**f**) Indoor cycled and outdoor stability of encapsulated PSCs with triple cation perovskite and different transport layers. The outdoor power output of the representative cells is depicted at the corresponding irradiance and temperature. (adapted from ref.[151]).*

Emery *et. al.* recently performed the outdoor stability (ISOS-O3) of $FA_{0.85}Cs_{0.15}PbI_{2.55}Br_{0.45}$ composition based PSC after encapsulating using glue-based epoxy and Polyolefin Elastomer (POE) + butyl lamination in a glass−glass stack.[153] They observed that the glue-based encapsulation lasts for only 3 months whereas the POE-based encapsulation lasts for ten months. However, the POE-based encapsulation is technically challenging to implement, but it provides excellent stability even under damp heat and outdoor test conditions. The POE + butyl lamination in a glass-glass stack of PSC restricts its use as it requires an annealing temperature of 150 °C for 20 minutes for the lamination process, which could damage the methylammonium iodide (MAI) containing perovskites. Therefore, developing encapsulation technology for perovskite solar cells is of utmost importance for commercialization of it. Nevertheless, it is crucial to understand that the indoor stability test results as per ISOS protocols do not match the outdoor stability results due to the transient behaviour of PSCs not being accounted for in the standard ISOS constant illumination testing conditions. Abate *et. al.* showed that light cycling shows much better agreement with outdoor observations, at least qualitative (and even quantitative in some cases) as shown in **Figure. 19f**. Apart from all these different accelerated ageing tests, the important question is how can we predict the practical lifetime of the PSCs from these ageing tests? As far as we know, no methods have been developed yet to estimate the practical lifetime from these accelerated tests for PSCs. So, to speed up the commercialization of PSCs, it is of utmost importance to develop a simulation model to predict the real field lifetime of the PSCs using the accelerated ageing test results. However, Solar Energy Technologies Office (SETO) in DOE, USA and Perovskite PV Accelerator for Commercializing Technology (PACT), have made a joint effort to perform the field deployment testing. PACT has planned to develop a standardized testing protocols to establish outdoor testing to ensure their accelerated ageing tests accurately reproduce the real field degradation mechanisms and allow researchers to corroborate their prototypes in the real field.[154]

# Chapter 3
# Methodology

# CHAPTER 3:

# Methodology

This chapter will introduce all the experimental techniques employed to fabricate thin films and solar cell devices, as well as the characterization method used to study both the perovskite thin films and solar cell devices. To understand the photophysics and device physics of perovskite semiconductors, we have used different experimental techniques such as absorption spectra, steady-state and time-resolved photoluminescence spectra, photoluminescence quantum yield, field emission scanning electron microscopy, dektak profilometer, X-ray diffraction, atomic force microscopy, kelvin probe force microscopy, contact angle measurement, nuclear magnetic resonance, dark and illuminated current density *vs.* voltage $(J - V)$ characteristics, intensity-dependent $(J - V)$, Suns$-V_{OC}$, incident photon to current efficiency (IPCE), current density - voltage - light $(J - V - L)$ measurement, steady-state electroluminescence, transient photovoltage/photocurrent measurement, and frequency-dependent photocurrent measurement.

## 3.1 Thin films and device fabrication

### 3.1.1 Substrates preparation

The indium tin oxide glass substrates ( 10-15 $\Omega/sq$) of dimension 60 mm × 60 mm are purchased from Luminescence Technology Corp. The dimensions of the patterned ITO substrates used for device fabrication are 15 mm × 15 mm. The following steps are involved in preparing the substrates (see **Figure 1**)

- (i)    Cut the substrates in 60 mm × 15 mm (i.e. in four strips) and then place a 12 mm width thermal tape on one side of the ITO substrate.

- (ii)    In the open area where the ITO etching will be done, deposit zinc power paste (in water) and dry it for 5 minutes.

- (iii)    Slowly drop 37% HCl (purchased from Sigma Aldrich) solvent on the zinc-coated area to etch the ITO.

- (iv)    Dip the substrates in DI water to remove HCl completely.

- (v)    Remove the thermal tape, wipe it with isopropanol, and dry it in the air for 5 minutes.

(vi)     Cut the patterned 60 mm × 15 mm substrates into four pieces i.e. 15 mm × 15 mm.

(vii)    For a 1 cm² device, the substrate's size is 20 mm × 20 mm. In this case, the 60 mm × 60 mm substrates are cut in 60 mm × 20 mm i.e. in three pieces, and use a thermal tap of 15 mm width. Repeat the above ITO etching process with zinc powder.

(viii)   The substrates were sequentially cleaned with soap solution, deionized (DI) water, acetone, and isopropanol for 10 minutes each in an ultra-sonicator.

(ix)     The cleaned substrates are dried with a nitrogen gun and kept on the hot plate at 100°C for 10 minutes to remove the residual solvents.

(x)      Take the substrates inside the oxygen plasma ashing chamber for 15 minutes. The plasma ashing is done at an RF power of 18 watts.

(xi)     After plasma ashing, take the substrates immediately inside the $N_2$ environment glove box ($O_2$<0.1 ppm, $H_2O$<0.1 ppm) for the spin coating process.

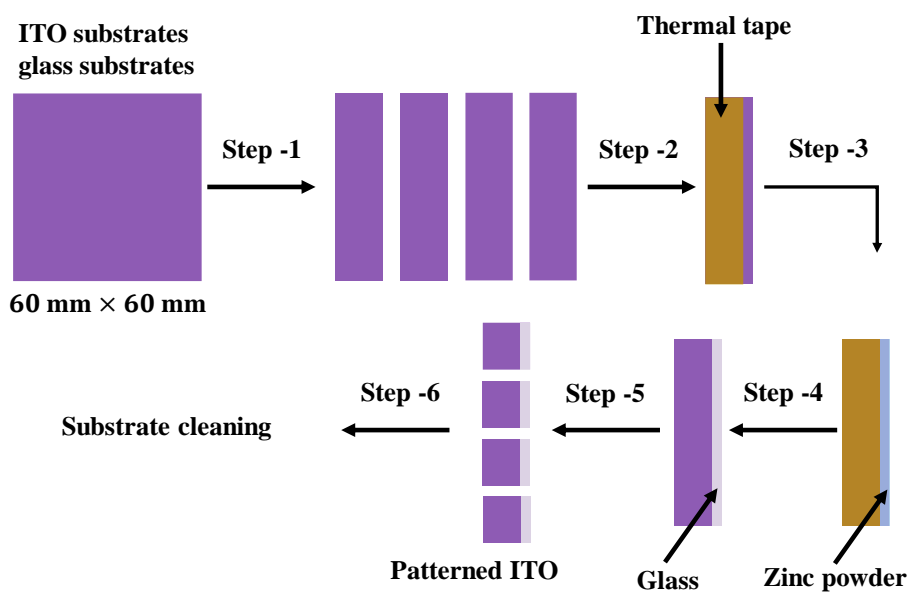

***Figure 1:*** *Steps involved in pattering the ITO substrates. The 60 mm × 60 mm ITO-coated glass substrate was cut in 60 mm × 15 mm i.e. 4 pieces (step -1). Then, place a 12 mm width thermal tap on one side (step-2). After that, put zinc paste on the open area (step-3). Drop HCl to remove the ITO (step-4). Wipe the substrate with IPA (step -5). Cut the substrate into four pieces i.e. 15 mm × 15 mm (step-5). Finally, clean the substrates to fabricate solar cell devices.*

### 3.1.2 Perovskite solar cells fabrication

In this thesis, we have used the *p-i-n* (inverted) architecture-based perovskite solar cell (PCS) devices. The schematic diagram of the device architecture is shown in **Figure 2**. The typical device structure used in this thesis is indium tin oxide (ITO)/HTL /perovskite (CsFAMA) /phenyl-C$_{61}$-butyric acid methyl ester (PC$_{61}$BM)/ bathocuproine(BCP)/ silver (Ag). We have used two hole transport layers (HTL): PTAA or Me-4PACz. In the device structure (Figure 2), the ITO acts as a transparent electrode, PTAA or Me-4PACz acts as HTL, perovskite acts as an absorber layer, PCBM and BCP act as an electron transport layer (ETL) and buffer layer respectively, and finally silver acts as a back metal electrode. The hole transporting layer PTAA or Me-4PACz is spin-coated at 4000 rpm for 30 seconds over the plasmas-ashed ITO substrates and annealed at 100$^{\text{o}}$ C for 10 minutes. The perovskite layer was deposited on the hole-transporting layer at 5000 rpm for 30 seconds, and in the last 7 seconds, anti-solvent chlorobenzene dropped for faster crystallization. The perovskite films were immediately annealed at 100$^{\text{o}}$ C for 30 minutes. After cooling down the perovskite layer for 5 minutes at room temperature, the PC$_{61}$BM layer spin coated at 2000 rpm for 30 seconds and air dried for 5 minutes. After that, the BCP layer spin coated at 5000 rpm for 20 seconds. Finally, the substrates are taken in an N$_2$-filled thermal evaporator for opaque back electrode deposition. 150 nm silver (Ag) was deposited using a metal shadow mask under a vacuum of $2 \times 10^{-6}$ mbar pressure at a rate of 0.1 Å/second for the fast 10 nm and 1 Å/second for the rest 140 nm. The specific details of the solution preparation and spin coating are discussed in each chapter.

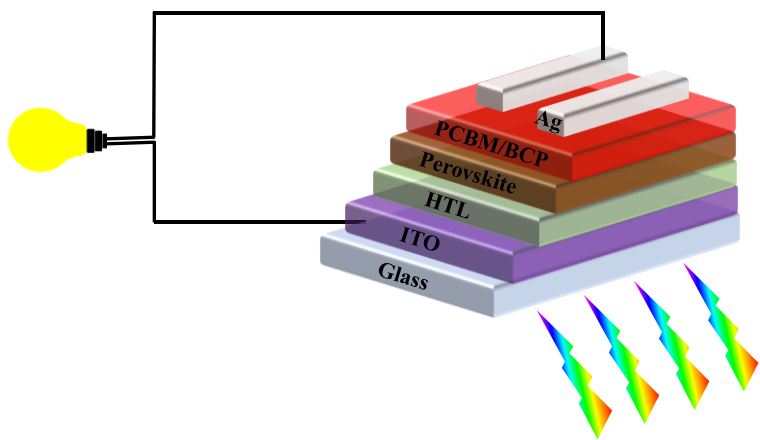

***Figure 2:*** *The schematic diagram of the p-i-n (inverted) architecture-based perovskite solar cells.*

### 3.1.3 Device contact with metal pins

After depositing the back metal electrode, we fix the customized metal pins to the device using a contact pin fitter zig. The metal pins are made of alloy Copper Tin (Phosphor Bronze) and purchased from Batten & Allen Ltd. The different active area-based solar cells are made using different metal masks but in all the devices we used the same kind of metal pin as shown in **Figure 3**.

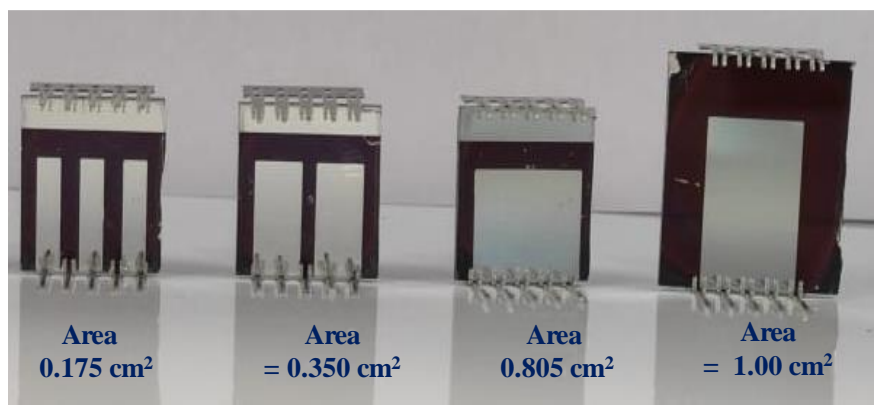

***Figure 3:*** *Digital image of the perovskite solar cells having different active areas. The devices are connected with metal pins to contact the ITO and Ag electrode.*

## 3.2. Thin film characterizations

### 3.2.1 Steady-state absorption spectroscopy

The steady-state absorption spectra of all the solution-processed thin films have been taken using the Perkin Elmer Lambda 950 spectrometer with the integrated sphere. The Perkin Elmer Lambda 950 spectrometer consists of two light sources (i) deuterium and (ii) tungsten halogen lamp. These two lamps provide a full range of UV-vis-NIR spectrum (175 nm – 3300 nm). The samples are illuminated from the front side for the absorption/transmission spectra and placed at position P1. In this case, the reflection port at place B is closed by the reflectance standard plate (coated with Barium Sulfate). For the reflection measurement, the samples are placed at position P2. The samples are illuminated from the glass side for absorption/transmission measurement of the thin films deposited on the glass substrates. For accurate optical density (O.D.) measurement of the thin films deposited on the glass substrates, first, the calibration is done using a glass substrate on which the films are deposited. The absorption measurement is carried out using glass as a baseline correction sample. The photodetectors used in the Lambda 950 spectrometer are (D1) PbS & (D2) InGaAs and are placed at the bottom of the integrating sphere as shown in **Figure 4.**

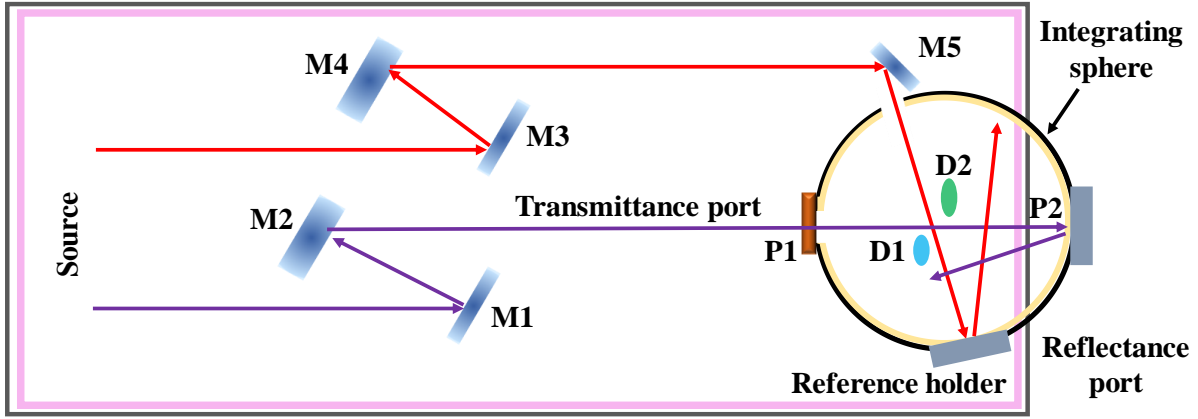

**Figure 4:** *Schematic diagram of the absorption spectroscopy measurement using Perkin Elmer Lambda 950 spectrometer.*

The absorbance of the thin film can be expressed by Beer Lambert's law as,

$$A(\lambda) = -\log\left(\frac{I}{I_0}\right) \qquad (3.1)$$

Where $I_0$ and $I$ indicate the intensities of the incident and the transmitted light respectively. The absorption coefficient can be expressed a s

$$\alpha(\lambda) = -\frac{2.303 \times A(\lambda)}{t} \qquad (3.2)$$

Where $t$ is the thickness of the perovskite absorber layer.

The bandgap of a semiconductor as a function of the absorption coefficient is defined by the Tauc method as[1]

$$(\alpha h\nu)^{1/\gamma} = B(h\nu - E_g) \qquad (3.3)$$

Where $h$ is the Planck constant, $\nu$ is the frequency of the photon, $E_g$ is the energy bandgap of the semiconductor and B is a constant. The factor $\gamma$ depends on the nature of electron transfer and is equal to 2 for indirect and $1/2$ for direct transition bandgaps respectively. The bandgap of a semiconductor is calculated using the above relation.[2,3]

The Urbach energy ($E_u$) is the measure of disorder in the bulk of the perovskite semiconductor and can be estimated from the absorption spectra using the following equation,[4,5]

$$\alpha(E) = \alpha_0 \exp\left(\frac{E - E_g}{E_u}\right) \qquad (3.4)$$

Where $E$ denotes the photon energy, and $\alpha_0$ is the material-dependent absorption coefficient.

### 3.2.2 Steady state and time resolved photoluminescence spectroscopy

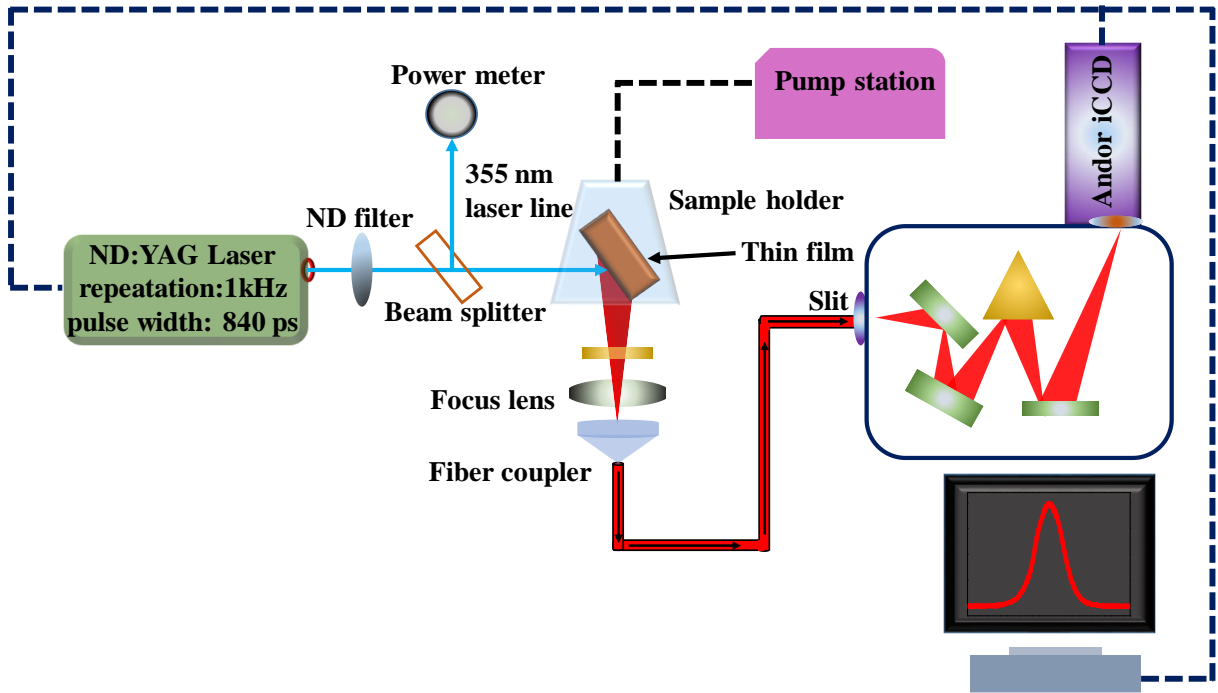

**Figure 5:** *Schematic diagram of the steady state and time resolved photoluminescence spectroscopy measurement using a pulsed laser and gated i-CCD (Andor iStar) coupled with spectrometer.*

When photons incident on a semiconductor of energy equal to or more than the bandgap, they will be absorbed and create electron and hole pairs at the conduction and valence bands respectively.[6] The absorption of the photons by the semiconductor takes place on the time scale of femtosecond (fs).[7] After the absorption process, the excited electrons undergo various fast transitions within the conduction band and relax to the minimum of the conduction band, called thermal relaxation. Further, the electrons spontaneously come down from the conduction band minimum to the valence band maxima and recombine with the holes, resulting in the re-emission of photons.[6,8] This process is called photoluminescence (PL), and the emitted photons have energy equal to the bandgap of the semiconductor material. The time scale of the photoluminescence process happens in the perovskite material in the order of a few ns to $\mu$s.[9,10] The steady state photoluminescence (SSPL) measurement of the thin films performed in the custom made chamber under a vacuum of $10^{-5}$mbar. The steady state PL spectra give information about the PL peak position and peak intensity. To get information about the charge carrier lifetime, time resolved (TR) PL measurement is useful.[11] For the TRPL measurement, we used a highly sensitive gated intensified charge coupled device (i-CCD, Andor iStar). The perovskite film is kept inside the sample holder at a pressure of $10^{-5}$mbar using a vacuum pump station as shown in **Figure 5**. To excite the film, we used a 3[rd] harmonies at 355 nm of Nd:YAD solid state laser (from Innolas) with fundamental wavelength 1064 nm, operating at 1 kHz with pulse width 840 ps. The pulse width of the laser must be lower than the expected lifetime of the charge carriers. The number of the generated electron-hole pairs depends on the power used to excite the sample. The power calibration was done using a Si-based power meter. The maximum energy limit of the Nd:YAG laser is 22 $\mu$J and it can be varied at the sample by using a varying neutral density filter wheel having an optical density of 0 to 4. The excited electrons from the conduction band decay and recombine with the holes via a radiative or non-radiate recombination process. The decay process of the charge carriers via radiative recombination can be characterized by the lifetime of the charge carriers. The lifetime of the charge carrier is defined as the time taken by the PL intensity decay by $1/e$ times of initial intensity. To get the PL decay time, the gate of the i-CCD triggered with the external circuit of the laser so that excitation and decay time difference can be accurately estimated. There is a time delay of ~75 ns between the gate opening and the optical signal reaching the gate. Therefore, we put a delay of ~75 ns to consider it as the zero time. The PL emission dispersed through a grating spectrometer with 500 nm blaze and 600 lines/mm ruling (placed inside of the Shamrock 303i spectrometer). The spectrometer is integrated with the gated i-CCD (Andor iStar). The minimum optical gating time of the i-CCD is ~780 ps with a minimum resolution of 10 ps. To increase the signal to noise (SNR) ratio, 70 accumulations per acquisition have been taken.

Generally, for the perovskite thin films, the PL decay contribution comes from fast and slow decay.[11–13] So, we fit the PL decay with a bi-exponential decay function to estimate the average lifetime of the charge carriers. The average lifetime ($\tau_{avg}$) is calculated using the following equation.

$$\tau_{avg} = \left( \frac{A_1 t_1^2 + A_2 t_2^2}{A_1 t_1 + A_2 t_2} \right) \qquad (3.5)$$

Where $A_1, A_2$ are the weightage of the fast and slow decay time $t_1$ and $t_2$ respectively.

### 3.2.3 Photoluminescence quantum yield

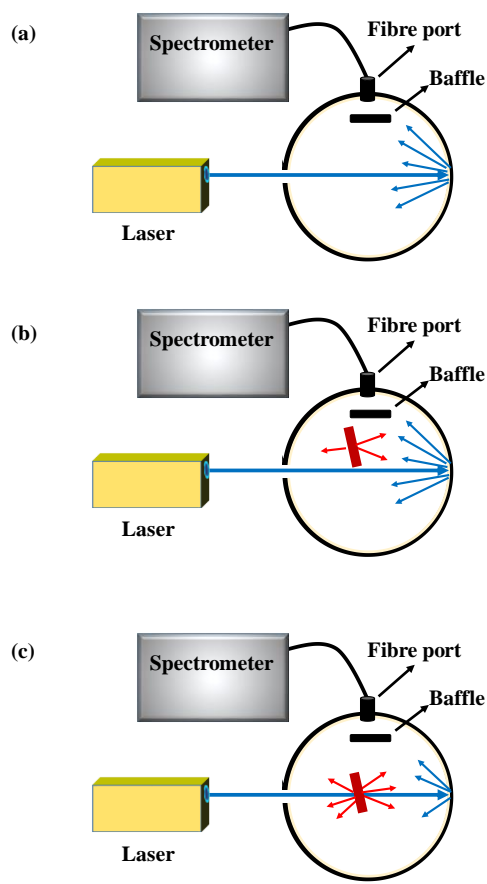

**Figure 6:** *Schematic of the absolute photoluminescence quantum yield (PLQY) measurement using the integrating sphere in the following steps, **(a)** No sample in the integrating sphere, i.e. measuring the laser intensity. **(b)** Thin film placed off the axis of the laser beam to measure the self-absorption of the scattered light. (c) Thin film placed on the axis of the laser beam to measure the total PL.*

In the steady state photoluminescence (SSPL) measurement, we just know the PL peak position and intensity of the PL spectra (depends on the calibration) but absolute PL intensity can't be measured because of not considering all emitted photons. To do so, we need an integrating sphere where the inner surface is highly reflective (~100%, Barium Sulfate coated) and results in isotropic redistribution of the light.[14] The photoluminescence quantum efficiency indicates the radiative efficiency of any sample and is defined as [15]

$$PLQY = \frac{number\ of\ emitted\ photons}{number\ of\ absorbed\ photons}$$

Higher PLQY is required for better performance of solar cells. The higher PLQY indicates the suppression of non-radiative channels. The PLQY can be increased by improving the material quality.[16] The absolute PL measurement can be done in three steps as shown in **Figure 6**. In the first step, there is no sample in the integrating sphere and only the laser intensity is measured **Figure 6a**. In the second step, a thin film is placed off the axis of the laser beam such that the laser beam directly doesn't hit the sample. This process accounts for the emission due to the reabsorption of the scattered light by the thin film **Figure 6b**. In the final step, the sample is placed on the axis of the laser beam. This process accounts for the emission due to the direct absorption of the laser beam along with the reabsorption of the scattered light **Figure 6c**.

**PLQY Calculation:**

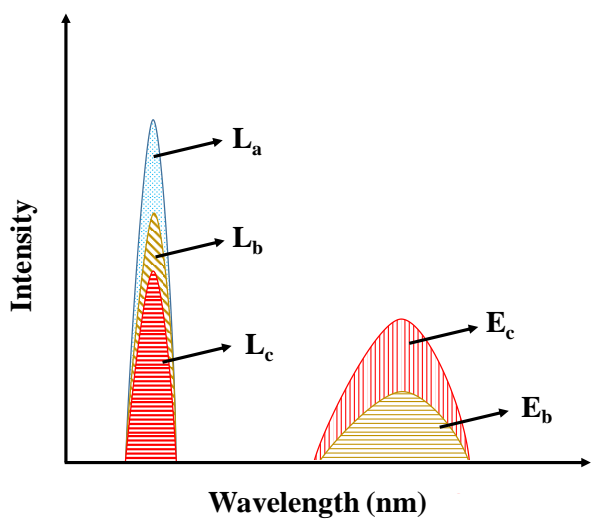

*Figure 7: Schematic of the PLQY measurement in three steps. The term 'L' represents laser spectra and 'E' for emission spectra.*

**Figure 7** represents the schematic spectral profile of the excitation laser beam and the emission spectra inside the integrating sphere, as discussed above. $L_a$, $L_b$, and $L_c$ are the areas under the curves of the laser beam spectra, and $E_b$ and $E_c$ represent the areas under the curves of the emission spectra in the three steps, as depicted in **Figure 6**. Hence, we can write,

$$L_b = L_a \times (1 - \mu) \qquad (3.6)$$

$$L_c = \{L_a \times (1 - A)\} \times (1 - \mu) \qquad (3.7)$$

Where A and $\mu$ represent the fraction of the direction absorption of the laser beam and reabsorption of the scattered light.

Simplifying the above equations, the absorption coefficient of the thin film can be expressed as,

$$A = \left(1 - \frac{L_c}{L_b}\right) \qquad (3.8)$$

The component of the scattered light which contributes to the total emission is

$$(1 - A) \times (E_b + L_b)$$

The total emission output in the 3$^{rd}$ step is

$$L_c + E_c$$

The emission due to the laser beam can be written as

$$\eta_{PL} \times A \times L_a$$

Where $\eta_{PL}$ represents the PL quantum yield of the thin film.

Thus, considering no loss of light inside the integrating sphere, we can write

$$L_c + E_c = (1 - A) \times (L_b + E_b) + \eta_{PL} \times A \times L_a \qquad (3.9)$$

Simplifying the above equation, the PLQY is given by

$$\eta_{PL} = \frac{E_c - E_b(1 - A)}{L_a \times A} \qquad (3.10)$$

### 3.2.4 Field emission scanning electron microscopy

Field emission scanning electron microscopy (FESEM) is a highly advanced technology used to study the topography of nano/micro-structured samples.[17] The FESEM can be used to study the various topographic images such as surface morphology (crystalline size, grain boundaries), the cross sectional view (thickness of the individual layer of a multi-layered thin film device). Using the energy dispersive X-ray spectroscopy (EDS) facility in the FESEM system we can study the composition of the material i.e. each constituting atom in percentage. We have used Carl Zeiss Ultra 55 FESEM instrument to study the thickness, cross section, and compositional analysis of atoms/molecules of the perovskite films. The FESEM imaging is superior over the optical microscope because of the small wavelength of the electron in the range of 0.859 Å to 0.037 Å. It facilitates the high resolution images of magnification up to 300,000×.[18] The high energy electron beam(1-30 keV) generated from the electron gun in ultrahigh vacuum ($10^{-9}$ mbar) and used as a source. The primary electrons are accelerated using the anode electrode and then directed using the electromagnetic lens (condenser lens, magnetic lens, and scanning coils) **Figure 8**. The scanning coil deflects the electron beam over the sample according to the zig-zag pattern. Due to the high-energy electrons hitting the sample's surface, the secondary electrons are emitted from the sample and collected by the secondary electron detector. These secondary electrons create the electrical signal, which is further converted to the image/videos in the monitor. For good quality images, the samples should be conductive to reduce the charging effect.[17] Therefore, we prepared the perovskite films on the ITO-coated glass substrates and before measurement, the films were grounded on the sample holder using silver paste.

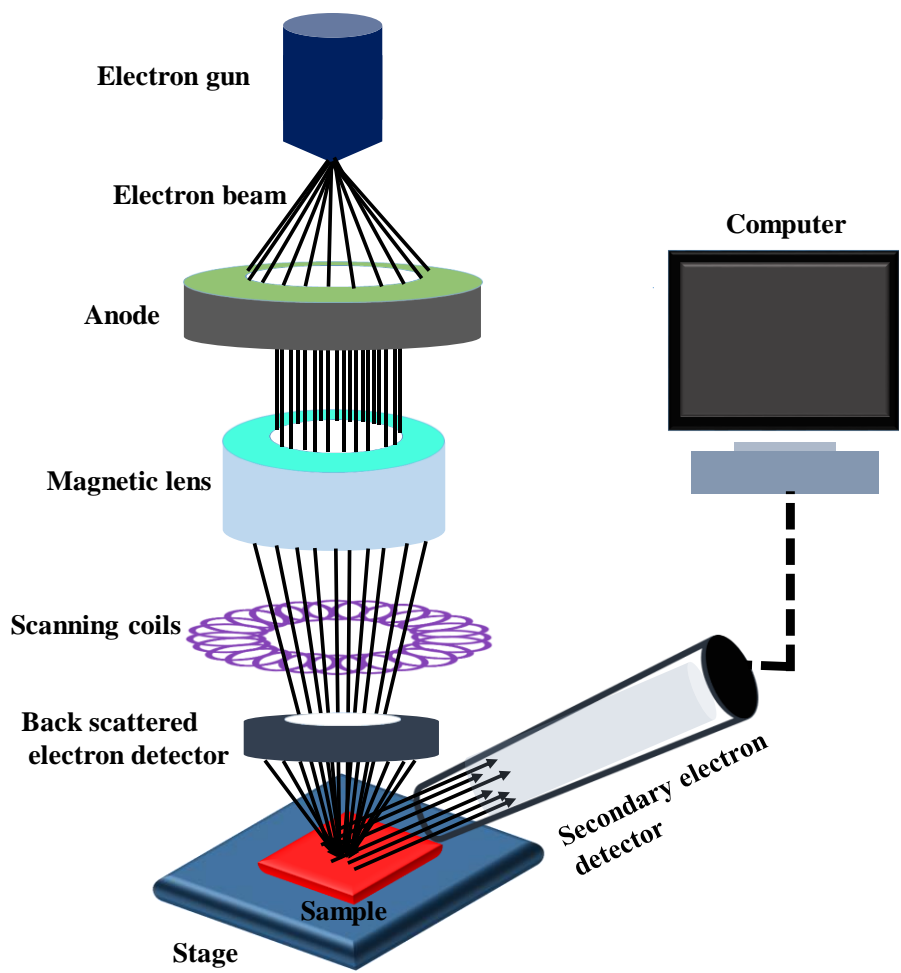

***Figure 8:*** *Schematic diagram of field emission scanning electron microscopy*

### 3.2.5 Dektak profilometer

The thickness of the solution processed perovskite thin film was measured using a Bruker dektak profilometer (DektakXT). The system takes measurements over the surface electrochemically by moving a diamond-tipped stylus. In addition to the thickness measurement, the system measures the surface roughness in the nanometre range. The scan range of the system is 6.5 $\mu$m to 524 $\mu$m with resolution 0.1 nm to 8 nm respectively. To measure the thickness, the applied stylus force can be 1 mg to 15 mg but we used 3 mg only so that the film does not get scratched during measurement which can result in inaccurate thickness. To measure the thickness of the perovskite film deposited on the glass substrates, we scratched the film at different places using a sharp needle. The difference between the top surface and scratched bottom surface (i.e. glass) gives the value of the thickness **Figure 9**.

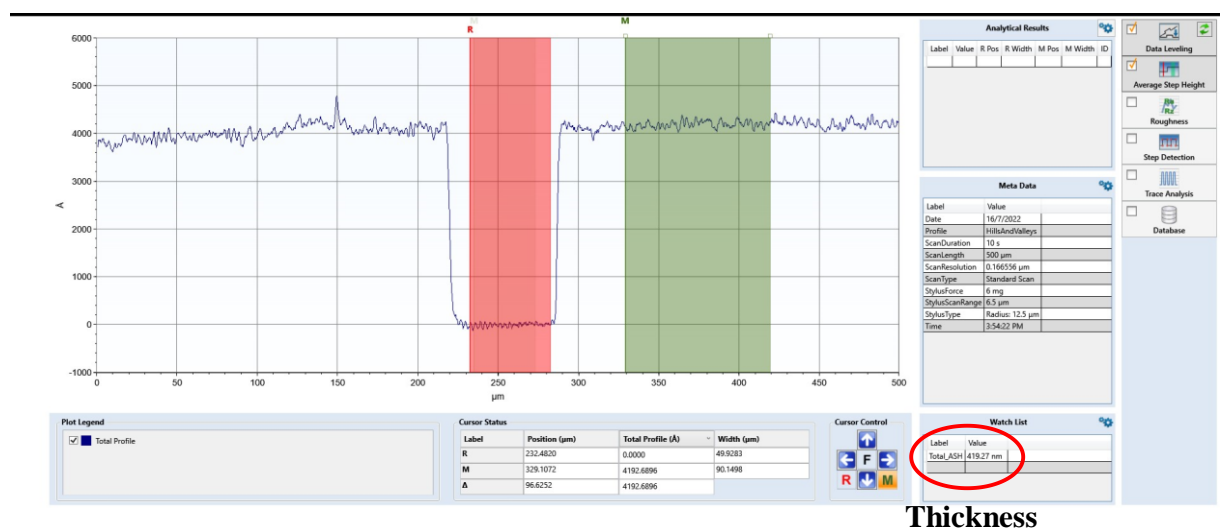

**Thickness**

***Figure 9:*** *Thickness measurement of the perovskite thin film using Dektak profilometer. The thickness of the perovskite thin film ~ 420 nm marked with a red closed loop.*

### 3.2.6 X-ray diffraction

X-ray diffraction is a non-destructive phenomenon where the high energy X-ray interacts with a crystalline material elastically and provides information about the crystallographic structure, chemical composition, physical properties, etc. of the material. A crystalline material is a regular periodic arrangement of atoms, ions, or molecules in a solid material. The X-ray light diffracted from the periodic lattice can interfere constructively or destructively. However, in the X-ray diffraction system, the detector detects only constructive interference. The schematic of the X-ray diffraction analysis is shown in **Figure 10.**

The constructive interference in the X-ray diffraction can be explained by Bragg's law[19]

$$2d\ sin\theta = n\lambda \qquad (3.11)$$

Where $d$ is the spacing between the diffracting plans, $\theta$ is the angle of incidence, $n$ is the integer number, and $\lambda$ is the wavelength of the X-ray.

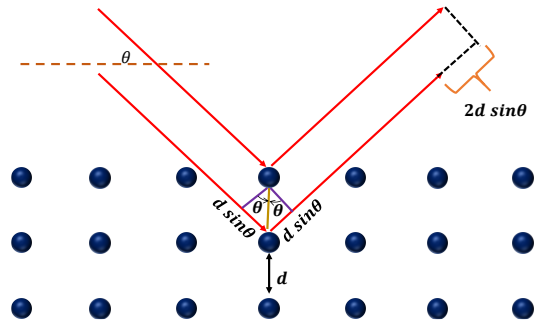

***Figure 10:*** *Schematic representation of the X-ray diffraction by the atoms of a periodic crystal lattice.*

The perovskite films used for making solar cell devices are polycrystalline. The polycrystalline perovskite film's grain size was observed from FESEM imaging with an average grain size of $\sim 200 - 300$ nm. We used a Smartlab Rigaku high-resolution X-ray diffractometer (HR-XRD) to measure the X-ray diffractograms for the perovskite thin films. The wavelength of the diffractometer is Cu K$\alpha$ radiation $\lambda = 1.54$ Å.

The XRD pattern used to calculate the crystalline size (D) of the spherical domains of the perovskite films using Debye Scherrer formula,

$$D = \frac{k\lambda}{\beta cos\theta} \qquad (3.12)$$

Where $k$ is the shape factor (~0.9)[20], $\lambda$ is the wavelength of the X-ray, $\beta$ is the full width at half maximum (FWHM), and $\theta$ is the angle of incidence.

### 3.2.7 Atomic force microscopy

An atomic force microscopy (AFM) instrument images the surface topography of a sample surface by scanning it using a highly sensitive cantilever tip.[21] The AFM provides information regarding the particle size, morphology, surface roughness, grain size, etc.[22] in the nanometre scale. The sharp tip scanning probe is attached to a flexible cantilever and moves along the surface for a very short distance, 0.2 nm to 10 nm. The AFM measures the intermolecular force between the sharp tip and the atom/molecule on the surface of the sample and results in deflection. The laser/photodiode system measures this deflection, capturing the sample's topography. We used the Bruker Multimode Nanoscope IV instrument to capture the surface image in tapping mode. The cantilever tip has a tip radius of 8-10 nm (RTESP-300) and can measure an average roughness of ~1 nm. The minimum scanning size is 0.4 $\mu$m $\times$ 0.4 $\mu$m and the maximum size is 125 $\mu$m $\times$ 125 $\mu$m. Higher scanning size has lower resolution hence we used in general 5 $\mu$m $\times$ 5 $\mu$m or is 10 $\mu$m $\times$ 10 $\mu$m at the different locations of the surface.

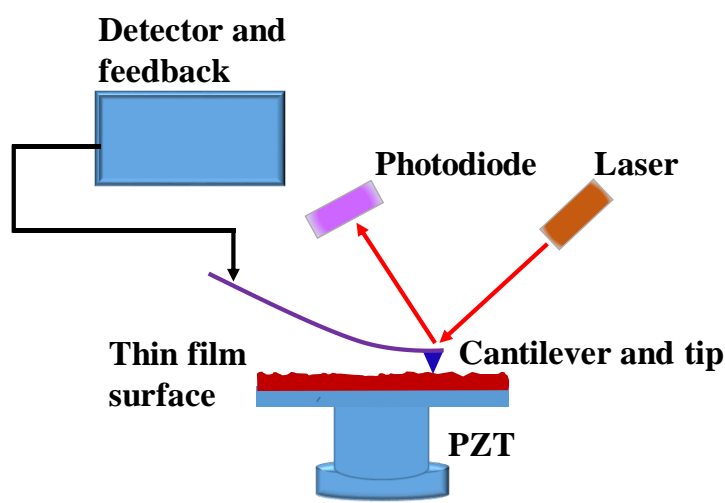

**Figure 11:** *Schematic diagram of the atomic force microscopy.*

### 3.2.8 Kelvin probe force microscopy

The kelvin probe force microscopy (KPFM) is an electrostatic force microscopy technique that is extensively used to study the surface potential of various conductive or semi-conductive samples.[23] The KPFM allows us to measure the quantitative local surface potential distribution and estimate the work function. The KPFM working principle is based on the Fermi-level alignment of the tip and the sample when they are in contact during measurement. In this case, the electrons flow from the tip to the sample or vice versa. This Fermi level alignment results in an offset in the vacuum level which in turn introduces the contact potential difference ($V_{CPD}$) between the tip and the sample.

The work function of the sample is calculated using the following formula

$$V_{CPD} = \frac{WF_{tip} - WF_{sample}}{-e} \qquad (3.13)$$

Where $WF_{tip}$ is the work function of the tip and $WF_{sample}$ is the work function of the sample.[24,25] To accurately measure the work function of the perovskite layer and the charge transfer layer, the work function of the known sample (gold W.F.= 5.1 eV ) is used to calibrate the tip work function.[25] Later, the estimated tip work function and measured $V_{CPD}$ used to calculate the sample work function.

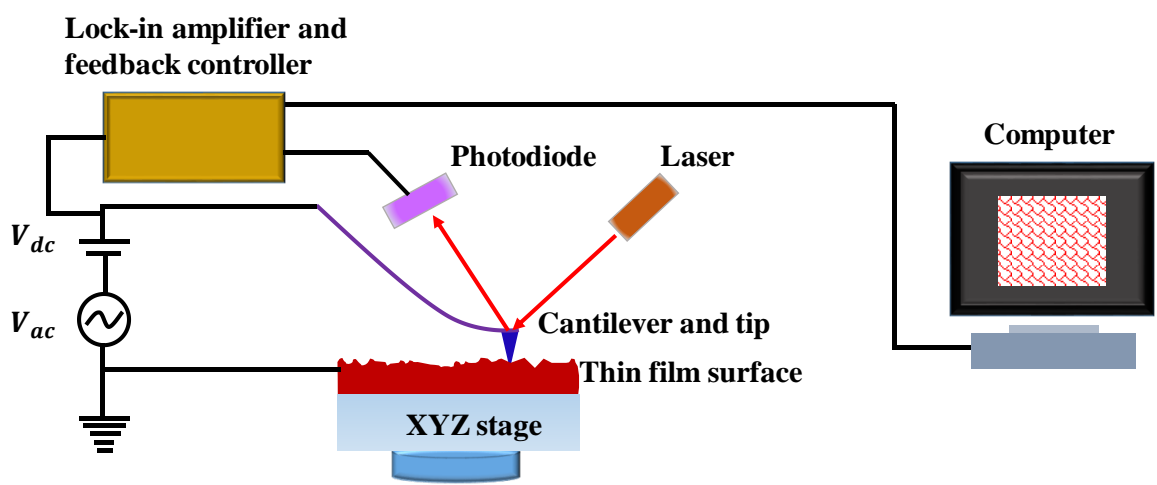

**Figure 12:** *Schematic representation of the Kelvin probe force microscopy.*

### 3.2.9 Contact angle measurement.

The contact angle is the measure of hydrophilicity/hydrophobicity of any surface. The contact angle is defined as the angle between the solid surface and the liquid where they meet.[26] A higher contact angle (i.e. hydrophobicity) is desirable where degradation due to moisture is prone to happen. However, the lower contact angle (i.e. hydrophilicity) is necessary where a layer is depositing on a surface.[27] A surface will be hydrophobic if it contains methyl group (-$CH_3$) or long alkyl chains which will be discussed in more detail in chapter 5. We have used the GBX Digidrop instrument to measure the contact angle, which also measures the surface energy, surface tension, absorption, etc. The schematic diagram of contact angle measurement is shown in **Figure 13**. A water droplet of 1-2 $\mu$L is dropped from the syringe to the surface of the sample. The volume of the water droplet is controlled by the software Windrop in the GBX Digidrop system. Once the water drop touches the substrate surface, it slowly spreads with time and after a few seconds, it becomes stable. The water drop spreading is captured by the camera. The angle between the surface and the liquid at stable conditions results in the contact angle.

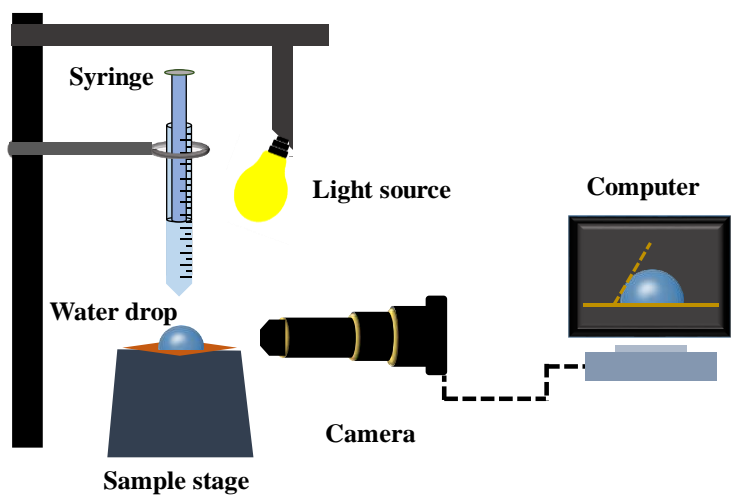

***Figure 13:*** *The schematic diagram of the water contact angle measurement between the surface and the water droplet.*

**3.2.10 Nuclear magnetic resonance spectroscopy**

Nuclear magnetic resonance (NMR) spectroscopy is a highly advanced technology based on the orientation of the atomic nuclei with non-zero spins placed in a strong external magnetic field.[28] A spinning charge generates a magnetic field and that results in a magnetic moment of strength proportional to the spin of the charge. Thus, in the presence of the external magnetic field, there are two spin states exist i.e. one in the direction of the magnetic field and another opposite to the direction of the field. The reorientation of the magnetic moments occurs with the absorption of electromagnetic radiation in the RF frequency range 4 – 900 MHz, and changes with the external magnetic field (6 - 24 T) and results in chemical shift **Figure 14**.[29] The NMR spectroscopy is widely used to study the molecular dynamics, week intermolecular interactions, ionization energy, chemical analysis, etc. There are various NMR active nuclei used to study the NMR spectroscopy such as [1]H, [2]D, [13]C, [15]N, [19]F, etc. We used JEOL, ECZR series 600 MHz NMR spectrometer for analyzing the solution state proton NMR, which will be discussed in Chapter 5.

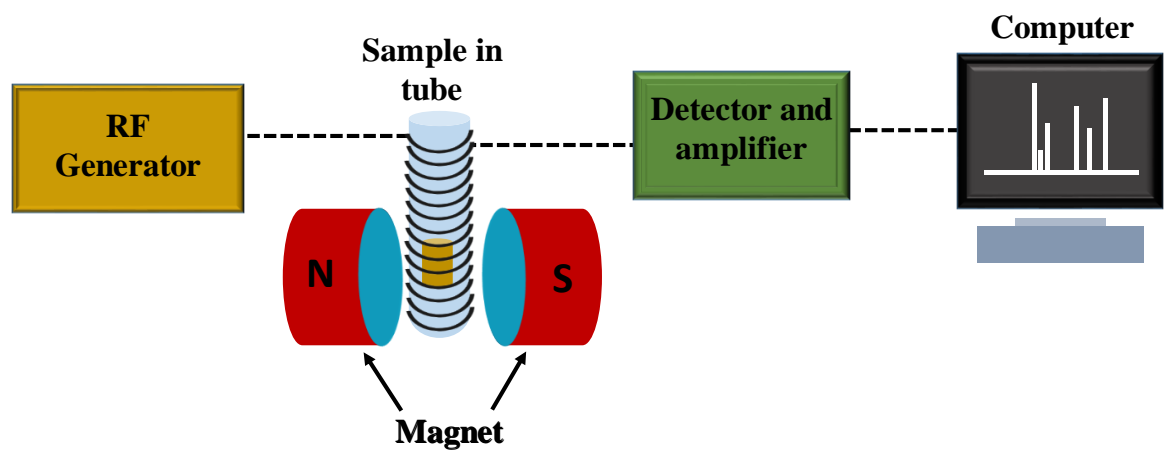

***Figure 14:*** *The schematic diagram of the nuclear magnetic resonance spectroscopy.*

## 3.3 Device characterization

### 3.3.1 Current density vs. voltage ($J - V$) characterization

The fundamental way to test a solar cell's performance is by measuring the current *vs.* voltage ($I - V$) characteristics. The current ($I$) depends on the active area ($A$) of the solar cells; hence, the current density ($J = I/A$) is an important factor to study the efficiency performance of the device. All the photovoltaic measurements were carried out under ambient conditions. Photocurrent density *vs.* applied voltage (*J-V*) measurement has been carried out using Keithley 4200 SCS and LED solar simulator (ORIEL: LSH-7320) after calibrating through a standard Si solar cell (RERA SYSTEMS-860 reference cell) **Figure 15**. The *J-V* measurement was performed with a scan rate of 100 mV/s and a hold time of 10 ms. For the intensity-dependent $J - V$ measurement, we used a set of ND filters with different optical densities (this will be explained in detail in Chapter 6).

The power conversion efficiency (PCE) of a device is calculated by using the following formula[30]

$$PCE = J_{SC} \times V_{OC} \times FF \qquad (3.14)$$

Where, $J_{SC}$ is the short circuit photocurrent density, $V_{OC}$ is the open circuit voltage, and $FF$ is the fill factor **Figure 15**. The FF is defined by the maximum squareness of the $J - V$ curve, and it can be represented as,

$$FF = \frac{J_{max} \times V_{max}}{J_{SC} \times V_{OC}} \qquad (3.15)$$

The solar cell is a diode and the dark current is similar to the p-n junction diode which increases exponentially with the applied forward voltage as

$$J = J_0 \exp(qV/nkT) \qquad (3.16)$$

Where $J_0$ is the reverse saturation current density, $q$ is the elementary charge, $n$ is the ideality factor, $k$ is the Boltzmann constant, and $T$ is temperature.

Upon applying the 1-Sun light, the photogenerated charges take place, and the total current in the device can be written as [31]

$$J = -J_L + J_0 \left[ \exp\left( \frac{V - JR_s}{\frac{nkT}{q}} \right) - 1 \right] + \frac{V - JR_s}{R_{Sh}} \qquad (3.17)$$

Where $J_L$ is the illuminated current, $R_S$ and $R_{Sh}$ are the series and shunt resistance shown in the equivalent circuit diagram in **Figure 16**.

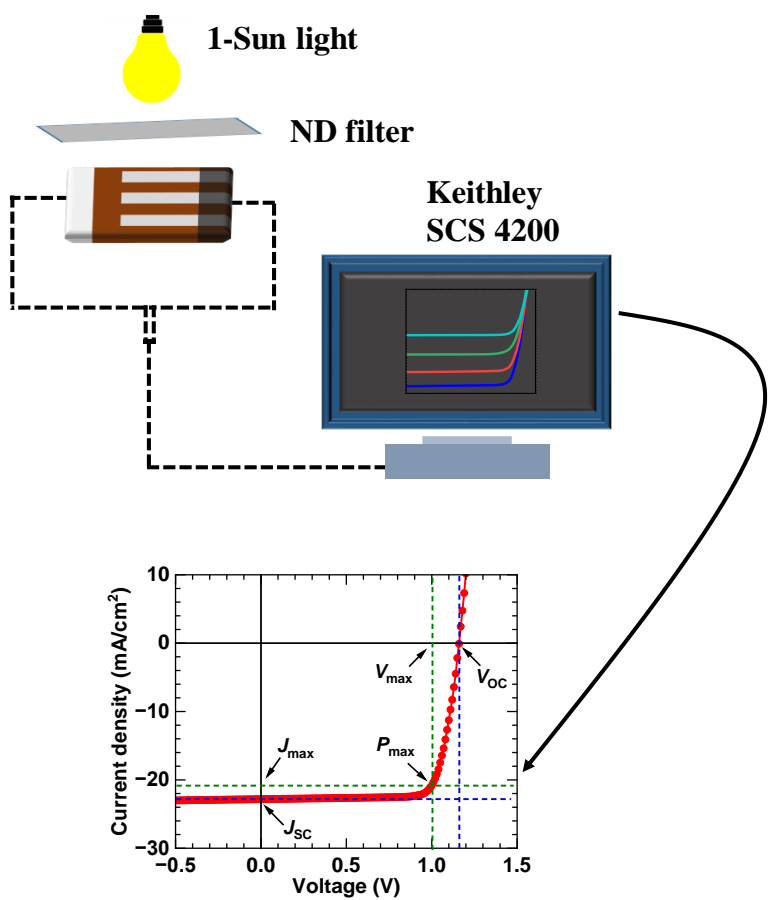

**Figure 15**: *Schematic representation of I-V measurement and the current density* vs. *voltage* $(J - V)$ *characteristics with the PV parameters for a typical perovskite solar cell.*

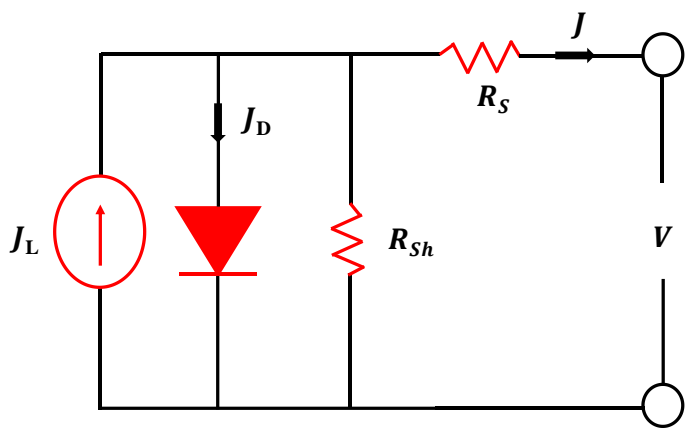

**Figure 16**: *Schematic representation of the equivalent circuit diagram under illumination conditions including the series and shunt resistance with a single diode model.*

## 3.3.2 Incident photon to current efficiency

The incident photon to current efficiency (IPCE) is a powerful technique to characterize a solar cell and helps in the optical engineering of the solar cell device. The IPCE technique is used to estimate the spectral responsivity (SR), external quantum efficiency (EQE), and internal quantum efficiency (IQE). The IPCE spectrum of all the perovskite solar cells was measured using the Bentham PVE 300 system. The schematic diagram of the IPCE spectrum measuring set-up is shown in **Figure 17.** In the Bentham PVE 300 set-up, there are two lamps: (i) 75W Xenon lamp and (ii) 100W Quartz halogen lamp (QTH), which are used to make a white light source to produce photons. The intensity of the lamp is less than 1 mW/cm$^2$. The light passes through a monochromator to get the desired wavelength. The wavelength selection and IPCE measurement are done with BenWin+ software in the Bentham PVE 300 system. The measurement can be done in three modes (DC, AC, and transformer), and for the AC or transformer mode, the DC light is chopped with a chopper at a frequency of 390 Hz and is considered as a reference signal for the lock-in amplifier. The chopped light passes through optical lenses and focuses on the device. The light spot should be lower than the active area of the device (should fall inside the active area) and for that different slits are used for different active area devices. The Bentham PVE 300 set-up is calibrated every time before starting the measurement using a standard reference Si diode (300-1100 nm)

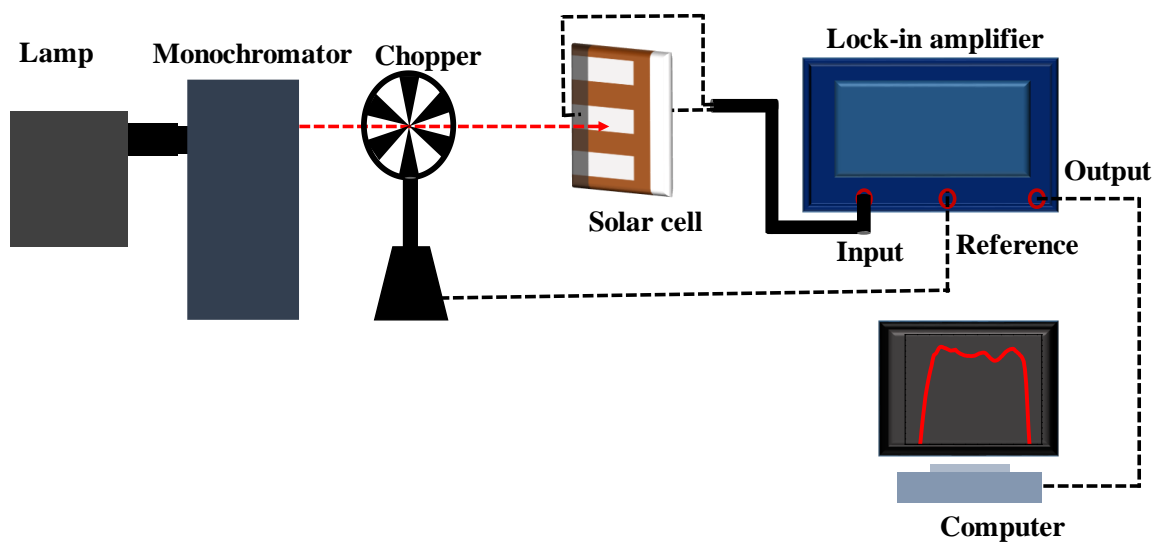

**Figure 17:** *Schematic diagram of the incident photon to current efficiency (IPCE) measurement set-up.*

The EQE of a solar cell is defined as the ratio of the number of charge carriers collected by the solar cell to the number of incident photons (of a given energy) on the device.[32]

$$EQE\ (\lambda) = \frac{number\ of\ charge\ carriers\ collected\ (\lambda)}{number\ of\ incident\ photons\ (\lambda)}$$

The Bentham PVE 300 set-up has an integrating sphere to measure the reflection R (%) and transmission T (%) spectra. The IQE spectra are obtained from the EQE spectra by using the reflection and transmission spectra as follows[33]

$$IQE\ (\%) = \frac{EQE(\%)}{1 - 0.01 \times R(\%) - 0.01 \times T(\%)} \tag{3.18}$$

### 3.3.3 Electroluminescence quantum efficiency measurement

The steady state electroluminescence quantum efficiency ($QE_{EL}$) is measured in an open atmosphere and in the dark condition to avoid any randomly scattered light. To do the $QE_{EL}$ measurement, we used a 100 mm$^2$ Si-photodetector which is connected with a 12V DC supply, and a Keithley 2400 multimeter (for photovoltage measurement) as shown in **Figure 18**. The device was injected a current with a Keithley 2000 source meter. Both the Keithley 2400 and Keithley 2000 are connected in series using National Instrument (NI) GPIB to GPIB cable. Both the Keithleys are interfaced using the LabView program to record the real-time current-voltage-luminescence ($J - V - L$). The EL spectra are collected using an optical fiber and Ocean optics spectrometer (FLAME-T-XR1-ES).

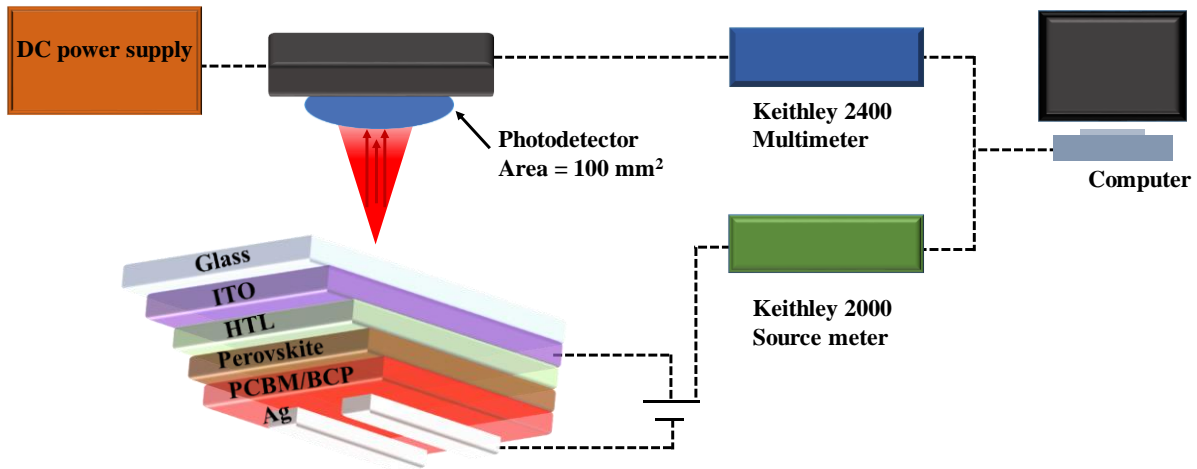

**Figure 18:** *Schematic diagram of the electroluminescence quantum efficiency ($QE_{EL}$) measurement set-up.*

### 3.3.4 Transient photovoltage/photocurrent measurement

Transient photovoltage (TPV) is an optoelectronic measurement technique widely used in the organic solar cell community, where the steady-state device is kept in an open circuit condition (along with background 1-Sun white light bias) using a high external resistance across the device and a short-lived laser perturbation is applied to the device.[34] The schematic of the TPV measurement is shown in **Figure 19**. For the TPV measurement, the termination resistance used is 1MΩ. The solar cell is illuminated by 1-sun DC white light using a THORLABS white lamp S/N M00304198. A 490 nm TOPTICA diode laser is used to perturb the device. The diode laser is modulated with a frequency of 1 kHz and duty cycle of 0.05% (for 500 ns pulse) using an ArbStudio 1104 function generator. The perturbed laser intensity is chosen such that the perturbed voltage $\Delta V_{OC} \approx 10$ mV while the background DC light intensity was 1-sun.[35] The perturbed signal is measured using a digital oscilloscope Tektronix DPO 4104B. The intensity of the background DC white light was changed using a set of neutral density (ND) filters for the intensity dependent TPV measurement. The intensity-dependent TPV analysis will be discussed in detail in Chapter 6. For the TPC measurement, the termination resistance used is 50Ω. In this case, the generated perturbed charge carriers leave the device due to short circuit conditions.[36,37] The TPC analysis will be discussed in good detail in appendix A.

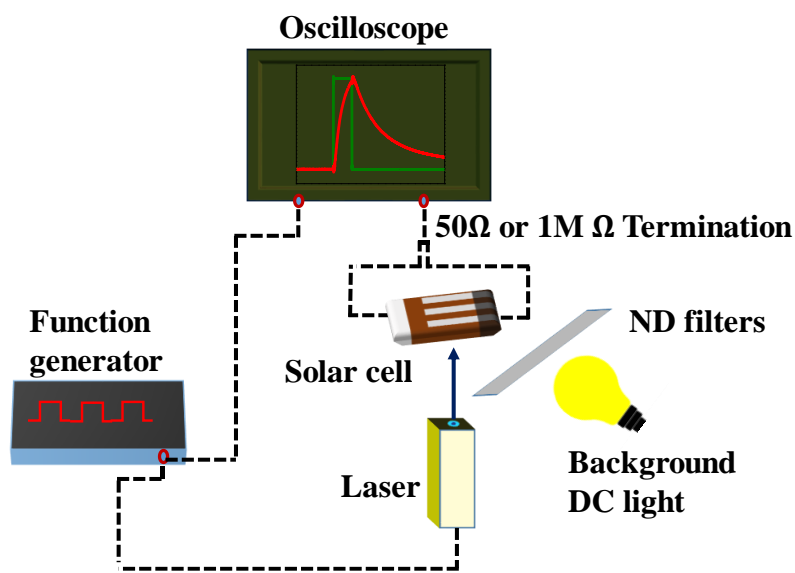

*Figure 19:* Schematic diagram of the transient photovoltage/photocurrent measurement set-up.

### 3.3.5 Frequency dependent photocurrent measurement

Frequency-dependent photocurrent measurement is a highly promising technique to understand the defect states and dielectric relaxation process in solar cell devices.[38] A frequency modulated laser is applied to the solar cell device and the corresponding photocurrent is recorded in a lock-in amplifier. A 490 nm TOPTICA diode laser is used whose ON/OFF frequency (duty cycle 50%) is modulated using an ArbStudio 1104 function generator. We used an SR830 DSP lock-in amplifier whose reference channel frequency range is 1 mHz to 102 kHz hence we modulated the laser ON/OFF frequency from 200 Hz to 102 kHz. The photogenerated current is frequency dependent and has real and imaginary parts. By analyzing the imaginary part, the dielectric relaxation is determined for perovskite solar cells.[39] This technique will be discussed in good detail in chapter 4.

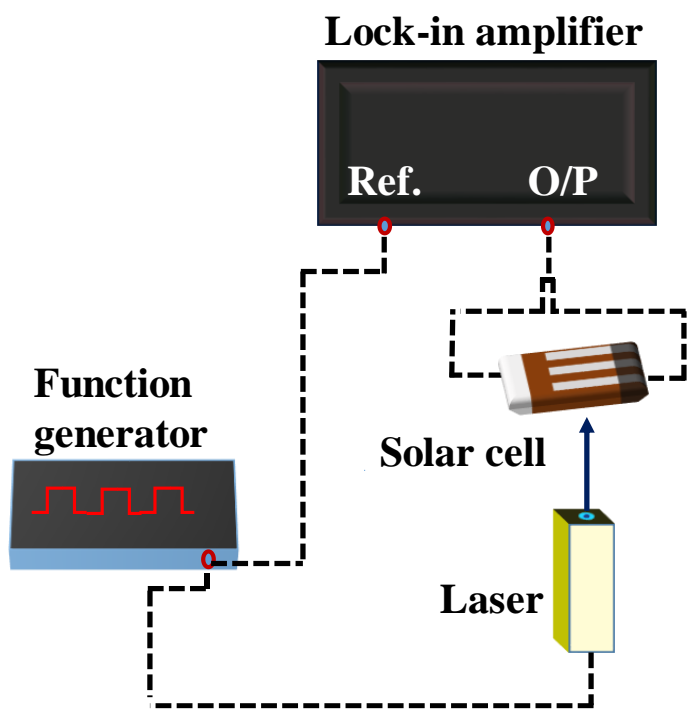

*Figure 20: Frequency-dependent photocurrent measurement set-up.*

Time of Excess Carriers in Si and CIGS Solar Cells by Modulated Electroluminescence Technique," Phys. Status Solidi Appl. Mater. Sci. **215**(2), 1700267 (2018).

# CHAPTER 4

# Role of Monovalent Cation on the Dielectric Relaxation Processes in Hybrid Metal Halide Perovskite Solar Cells and Correlation with Thermal Stability

# CHAPTER 4

# Role of Monovalent Cation on the Dielectric Relaxation Processes in Hybrid Metal Halide Perovskite Solar Cells and Correlation with Thermal Stability

## Abstract

Perovskite solar cells (PSCs) are the fastest growing photovoltaic devices in the solar cells community and offer a bright future for cheap solar electricity. In the last few years, it has been observed that the efficiency and stability of the PSCs can be enhanced by introducing multi-cations into the perovskite crystal structure. Herein, we have examined the triple monovalent cation based PSC $(FA_{0.83}MA_{0.17})_{0.95}Cs_{0.05}Pb(I_{0.90}Br_{0.10})_3$ (abbreviated as CsFAMA) over single monovalent cation based PSC $MAPbI_3$ (MAPI) through frequency dependent photocurrent and dielectric measurements in terms of the dielectric relaxation process. The dielectric relaxation time constant ($\tau_d$) is lower for the CsFAMA based PSC as compared to the MAPI based PSC. The lower $\tau_d$ is attributed to a lower dielectric constant using classical semiconductor physics. Unprecedentedly, the relaxation process is correlated with the presence of a monovalent cation of CsFAMA *vs* MAPI as an absorber in PSCs, which is well correlated with the presence of relative defect density. We note that defect formation under illumination conditions reduces $J_{SC}$ as a major component in PV parameters, which is significant in MAPI PSCs. This can be hypothesized based on relative ion density accumulation at the interface of charge extracting contact as per available literature. This study provides a unique, in-depth knowledge towards the role of monovalent cations in the dielectric relaxation process and their connection with the defects and thermal stability of halide perovskite semiconductors based solar cells.

## 4.1 Introduction

Organic-inorganic halide perovskite solar cells (PSCs) have enticed much interest in the semiconductor community and promise to deliver the next generation of highly cost-effective photovoltaic devices. The PSCs are the fastest growing photovoltaic devices in the solar cell community and reached the certified efficiency from 3.8% in 2009[1] to >25.6%.[2,3] There are various engineering techniques reported in the perovskite research community to optimize the active and charge transport layers, such as additive engineering, anti-solvent treatment, interface, and surface modification, etc. to improve the power conversion efficiency (PCE) and the stability of the PSCs.[4–6] In general, the enhanced performance of the PSC is attributed to the improved morphology and suppressed non-radiative recombination. However, in-depth insight into the stability of the PSC is still being explored in the community, which involves both intrinsic and extrinsic factors. One needs to start at the intrinsic properties of the absorbing semiconductor as a pre-requisite and then at solar cell device physics.[7–9]

It is widely known that PCE and stability are two important factors that are essential for the commercialization of PSCs. The PCE of PSCs has reached close to the conventional commercialized single junction Si solar cells[2], but the stability is still under optimization.[10–12] The stability of the PSCs depends on various factors, for example, moisture resistivity, oxygen induced degradation**,** thermal degradation or ion migration, etc.[13–15] The moisture and oxygen induced degradations can be reduced by advanced encapsulation techniques.[16,17] However, thermal degradation is crucial and needs to be addressed for future stable PSCs. MAPbI$_3$ (MAPI) perovskite is the most widely explored material in the perovskite community in terms of photophysics and device applications. However, single monovalent cation based PSCs possess numerous defects and lead to lower PCE and stability.[18–20] These defects act as recombination centers for the charge carriers and act as migration channels for the ions under illumination.[21] The defects in the perovskite can be reduced by changing the cation and anion or both simultaneously. Hence, instead of MA+, thermally stable FA+ ion-based perovskites are introduced.[22] It is found that FAPbI$_3$ (FAPI) is potentially more suitable than the conventional MAPI perovskite, but it has also been reported that the FAPI trigonal α phase is quite sensitive to humidity and easily transforms into a non-photo active hexagonal δ-phase at room temperature.[23] It is observed that the partial introduction of Cs cation in the FAPI perovskite further increases the structural and thermal stability of the photoactive cubic α-phase.[24,25] From the first principle calculation, it is reported that the formation energy of the point defects can be maximized by mixing the different A cations in the ABX$_3$ perovskite, and correspondingly the point defects reduce, as we discussed in Chapter **2**.[26] Later on, it is found that the triple cation (CsFAMA) and mixed halide (IBr) based perovskite thin films are used to fabricate highly efficient single junction PSCs.[7,27,28] It is reported that the mixing of different A cations leads to a higher tolerance factor (*t*) value through lattice strain relaxation and thus it minimizes the defects to achieve a relatively stable crystal structure.[29,30] In order to have more clarity behind the stability issue of these ionic perovskite semiconductors is still being explored in the community as an important research problem. Perovskite is being debated to be a ferroelectric material, however, it is proven to have a frequency dependent dielectric polarization as expected.[31–33] A distinct change in the performance of halide perovskites as a function of A-site is evident and being explained using first principle calculations, however, there is no experimental macroscopic study which connects this stability issue to the intrinsic semiconductor property of absorber. Semiconducting absorbers are known to have their dielectric relaxation time ($\tau_d$) depends on dielectric constant and conductivity.[34] These two macroscopic quantities have been reported however, a direct link of them to $\tau_d$ is missing. Recently, Singh et. al. reported a correlation between the $\tau_d$ to charge transport properties in MAPI films.[35] Whether $\tau_d$ can be linked to the defects in the perovskite and connects to the device stability for these ionic character semiconductors has not explored so far as per our knowledge.

In this work, we fabricated conventional *p-i-n* architecture based monovalent single cation MAPI and triple cation $(FA_{0.83}MA_{0.17})_{0.95}Cs0_{.05}Pb(IBr)_3$ (CsFAMA) based PSCs to study the dielectric relaxation process and correlate with the defects studies. A frequency dependent photocurrent ($I_{ph}(f)$) measurement has been carried out to study the dielectric relaxation time constant ($\tau_d$). We observed that the $\tau_d$ for CsFAMA perovskite is lower than that for MAPI. To validate the $I_{ph}(f)$ measurement, we carried out conductivity studies and frequency dependent capacitance measurements for CsFAMA and MAPI based perovskites. We carried out various optoelectronic measurements to compare the defects present in the MAPI and CsFAMA perovskite and found that the CsFAMA perovskite has lower defects than MAPI perovskite in confirmation with literature.[28,29] Interestingly, the relatively lower defect states for CsFAMA perovskite can be attributed to measured lower capacitance. We also measured the stability of MAPI and CsFAMA based PSCs and found the stability of CsFAMA is much higher than that of MAPI based PSCs. We correlated the $I_{ph}(f)$ with the polarization of A-cation in terms of defect mediated dielectric relaxation process to validate the enhanced stability of CsFAMA-based PSCs.

## 4.2 Experimental Section

**Materials:** Methylammonium iodide (MAI), lead iodide(PbI$_2$), formamidinium iodide (FAI), cesium iodide (CsI), lead bromide (PbBr$_2$) all were purchased from TCI chemical and used as received. Phenyl-C61-butyric acid methyl ester (PC$_{61}$BM) were purchased from Solenne BV and used as received. The poly(9,9-bis(3'-(N,N-dimethyl)-N-ethylammoinium-propyl-2,7-fluorene)-alt-2,7-(9,9-dioctylfluorene))dibromide (PFN-Br) ordered from 1-Material and used as received. Poly (triaryl amine) (PTAA) and bathocuproine (BCP) were purchased from Sigma Aldrich and used as received

**Solution preparation:** (FA$_{1-X}$MA$_X$)$_{0.95}$Cs$_{0.05}$PbI$_3$ solution prepared by mixing of FA$_{0.95}$Cs$_{0.05}$PbI$_3$ and MA$_{0.95}$Cs$_{0.05}$PbI$_3$ in different volume ratios. FA$_{0.95}$Cs$_{0.05}$PbI$_3$ perovskite precursor prepared by mixing of 212 mg of FAI, 17 mg of CsI, and 599 mg of PbI$_2$ in 1 ml of a mixed solvent of DMF and DMSO in a 4:1 volume ratio. MA$_{0.95}$Cs$_{0.05}$PbI$_3$ perovskite precursor prepared by mixing of 196 mg of MAI, 17 mg of CsI, and 599 mg of PbI$_2$ in 1 ml of a mixed solvent of DMF and DMSO in a 4:1 volume ratio. We added FA$_{0.95}$Cs$_{0.05}$PbI$_3$ and MA$_{0.95}$Cs$_{0.05}$PbI$_3$ in the volumetric ratio to obtain X = 0.00, 0.25, 0.50, 0.75, 1.00. MAPI perovskite precursor solution was prepared by mixing 581 mg PbI$_2$ and 209 mg MAI in 1 ml of a mixed solvent of DMF and DMSO in a 4:1 volume ratio and stirred at room temp for 4 hours before spin coating. The CsFAMA perovskite precursor was prepared by mixing of 16.9 mg of CsI, 33.1 mg of MAI, 71.6 mg of PbBr$_2$, 176.6 mg of FAI, and 509.4 mg of PbI$_2$ in 1 ml of a mixed solvent of DMF and DMSO in 4:1 volume ratio and stirred overnight at room temperature before spin coating. 0.5 mg of PFN-Br is dissolved in 1 ml of methanol and stirred overnight at room temperature. 1.5 mg of PTTA is dissolved in 1 ml of toluene and stirred overnight at room temperature. 20 mg PC$_{61}$BM is dissolved in 1 ml of 1,2-dichlorobenzene and stirred for overnight at room temperature. 0.5 mg BCP is dissolved in 1 ml of anhydrous isopropanol and stirred overnight at room temperature and 10 minutes at 70º C before spin coating.

**Perovskite solar cells fabrication and characterization:** ITO-coated glass (10-15 Ω/square) substrate was sequentially cleaned with hellmanex soap solution, deionized (DI) water, acetone, and isopropanol for 10 minutes each. After drying the substrate with a nitrogen gun, we kept them inside the oxygen plasma ashing chamber for 10 minutes. After plasma ashing, we immediately spin-coat PTAA at 2000 rpm for 30 seconds and then annealed at 100º C for 10 minutes under a nitrogen environment. After that, we cooled down the substrates for 5 minutes, and then PFN-Br solution was spin-coated at 5000 rpm for 30 seconds, and immediately the perovskite solution spin-coated on the ITO/PTAA/PFN-Br substrates. All the perovskite precursors were filtered using 0.45 μm PTFE filter before spin coating. The $(FA_{1-x}MA_x)_{0.95}Cs_{0.05}PbI_3$ perovskite spin-coated at 4000 rpm for 30 seconds, and at the last 10 seconds, we used 150 μl of chlorobenzene anti-solvent treatment on a 1.5 cm by 1.5 cm substrate and then annealed at $100^o$ C for 30 minutes. The $MAPbI_3$ (abbreviated as MAPI) perovskite spin-coated at 4000 rpm for 30 seconds and at the last 10 seconds, we used 150 μl of chlorobenzene anti-solvent treatment on a 1.5 cm by 1.5 cm substrate and then we did 5 minutes of room temperature dry and then annealed at $100^o$ C for 10 minutes. The $(FA_{0.83}MA_{0.17})_{0.95}Cs_{0.05}Pb(I_{0.90}Br_{0.10})_3$ (abbreviated as CsFAMA) perovskite spin-coated at 5000 rpm for 30 seconds and at the last 10 seconds, we used 150 μl of chlorobenzene anti-solvent treatment on a 1.5 cm by 1.5 cm substrate and then annealed at $100^o$ C for 30 minutes. After that, $PC_{61}BM$ was spin-coated at 2000 rpm for 30 seconds and BCP was spin-coated at 5000 rpm for 20 seconds. Finally, 120 nm of Ag was deposited under a vacuum of $2 \times 10^{-6}$ mbar using a metal shadow mask. For all studied devices, the active area of the device is 17.5 mm$^2$.

All the photovoltaic measurements were carried out under ambient conditions. Photocurrent density versus applied voltage (*J-V*) measurement was carried out using Keithley 4200 SCS and an LED solar simulator (LSH-7320) after calibrating through standard Si solar cells provided by ABET, IIT Bombay. The *J-V* measurement was performed with a scan rate of 40 mV/s. EQE measurement has been carried out to measure the photoresponse as a function of wavelength using the Bentham quantum efficiency measurement system (Bentham PVE 300). The XRD measurements were carried out in a Rigaku smart lab diffractometer with Cu Kα radiation (λ=1.54Å). Θ-2Θ scan has been carried out from $10^o$-$45^o$ with step size $0.001^o$. Morphological analysis was done using field emission scanning electron microscopy (FESEM). Optical absorption spectra were carried out using a spectrometer (PerkinElmer LAMBDA 950). Steady-state PL measurement was done on thin films in a vacuum at a pressure of $10^{-3}$ mbar in a custom-made chamber. Excitation energy and wavelength were 70 nJ and 355 nm respectively. For time-resolved PL, a gated i-CCD (ICCD, Andor iStar) was used for detection with excitation laser (3rd harmonic Nd:YAG with emission wavelength of 355 nm) pulse width of 840 ps pulse with 1 kHz repetition rate. The PLQY measurement was done using a 4-inch customized integrating sphere and $Nd^{3+}$:YAG 355 nm laser and ICCD, Andor iStar-based spectrometer using a method developed earlier by de Mello et al.[36] Steady-state current-voltage–light characteristics were measured using a Keithley 2400 source meter,

Keithley 2000 multimeter, and calibrated Si photodiode (RS components). The capacitance of the devices was measured by using Novocontrol Technology, concept 80. TPV was measured by using a 490 nm TOPTICA diode laser, THORLABS white lamp S/N M00304198, ArbStudio 1104, and digital oscilloscope Tektronix DPO 4104B.

## 4.3 Results

### 4.3.1 Dielectric relaxation process in perovskites

We fabricated the inverted (*p-i-n*) heterojunction PSCs with device structure of indium tin oxide (ITO)/ poly[bis(4-phenyl)(2,4,6-trimethylphenyl)amine(PTAA)/poly(9,9-bis(3'-(N,N-dimethyl)-N-ethylammoinium-propyl-2,7-fluorene)-alt-2,7-(9,9 dioctylfluorene)) dibromide (PFN-Br)/ perovskite / phenyl-$C_{61}$-butyric acid methyl ester(PC$_{61}$BM)/ bathocuproine(BCP)/ silver (Ag) to understand the role of monovalent cations in dielectric relaxation process. **Figure 1a** represents the crystal structure of ABX$_3$ perovskite where A is a monovalent cation (MA$^+$/FA$^+$/Cs$^+$) represented by grey balls; B is a divalent cation (Pb$^{2+}$) represented by blue balls at the center of the octahedral cage, and X is the halide (I$^-$/Br$^-$) represented by pink balls at the corner of octahedral. In this work, we focus on the thermal stability of the PSCs as a function of 'A' cation and correlate it with the dielectric relaxation process. **Figure 1b** depicts the device

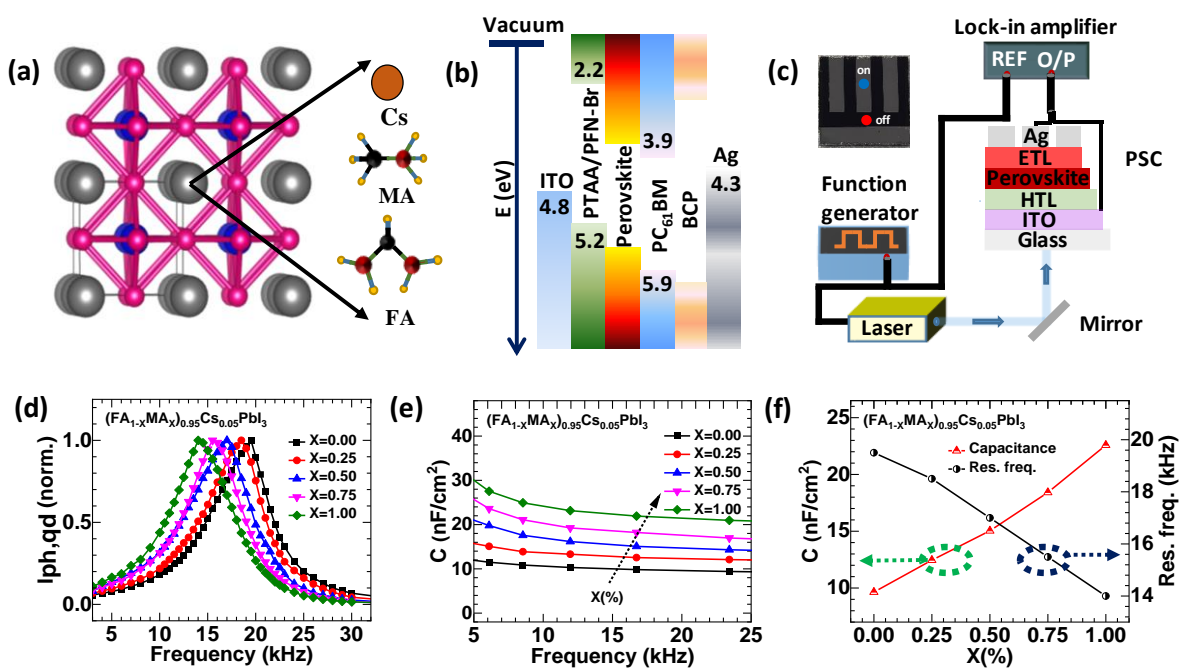

***Figure 1:*** *(a) Schematic diagram of ABX$_3$ perovskite crystal structure. (b) Energy level diagram of each layer used in the p-i-n architecture based PSC, values are taken from literature.[37,38] (c) Typical frequency dependent photocurrent measurement ($I_{ph}(f)$) set-up used*

*for this study and an inset image of actual device picture with two dots (blue on device & red off device) to show sites at which light is irradiated for $I_{ph}(f)$ measurements. **(d)** Quadrature part of the frequency dependent normalized photocurrent for $(FA_{1-X}MA_X)_{0.95}Cs_{0.05}PbI_3$, where X= 0.00, 0.25, 0.50, 0.75, and 1.00. **(e)** The frequency dependent device capacitance was measured for $(FA_{1-X}MA_X)_{0.95}Cs_{0.05}PbI_3$ perovskite absorbers. **(f)** The variation in capacitance and corresponding photocurrent resonance frequency with respect to X in $(FA_{1-X}MA_X)_{0.95}Cs_{0.05}PbI_3$ composition.*

configuration of the *p-i-n* architecture based PSCs with an energy level diagram of each layer.[37,38] Detailed information about the perovskite precursor preparation, device fabrication, and characterization are given in the experimental section above.

The $I_{ph}(f)$ measurement has been carried out to understand the dielectric relaxation process in PSCs.[39,40] **Figure 1c** represents the schematic diagram of the experimental set-up used to determine the $\tau_d$. The frequency of the input signal (490 nm diode laser) is varied from 200 Hz to 100 kHz with a duty cycle of 50% using the Arb Studio 1104 function generator. The photogenerated current is measured through the SRS830 DSP lock-in amplifier, which also provides the phase delay ($\phi$) between the device signal and the input signal. The inset figure represents the actual experimental device having 3 pixels (with Ag electrode) with an active area of each pixel is 17.5 mm$^2$. The rest of the area that is not covered by the Ag electrode is termed as outside the active area of the device. In order to study the effect of monovalent 'A' cation on the dielectric relaxation process in the ABX$_3$ perovskite, we choose a series of perovskite systems denoted by $(FA_{1-X}MA_X)_{0.95}Cs_{0.05}PbI_3$, where X varies from 0.00 to 1.00 with a step of 0.25 and the atomic concentration of Cs, Pb, & I remains constant. The crystallographic XRD study shows incorporating of smaller size MA cation by increasing X in the $(FA_{1-X}MA_X)_{0.95}Cs_{0.05}PbI_3$ perovskite system, the (110) peaks shift towards higher angle from 13.94º to 14.11º **Figure 2**.[41] Optical studies UV-vis-NIR absorption spectra and steady-state photoluminescence (SSPL) suggest that incorporating of smaller size MA cation by increasing X in the $(FA_{1-X}MA_X)_{0.95}Cs_{0.05}PbI_3$ perovskite system increases the bandgap **Figure 3**. We fabricated a series of *p-i-n* architecture based $(FA_{1-X}MA_X)_{0.95}Cs_{0.05}PbI_3$ PSCs (**Figure 1b).** The illuminated *J-V* characteristics and corresponding PV parameters of $(FA_{1-X}MA_X)_{0.95}Cs_{0.05}PbI_3$ based PSCs are shown in **Figure 4** and **Table 1** respectively.

In this chapter, the $I_{ph}(f)$ is used to determine the relaxation time of the photogenerated charge carriers in the PSC by analyzing the quadrature (imaginary) component of the photocurrent. **Figure 1d** shows a typical quadrature part of the normalized photocurrent ($I_{Ph,qd}$) as a function of frequency for $(FA_{1-X}MA_X)_{0.95}Cs_{0.05}PbI_3$ perovskite systems. The in-phase ($I_{Ph,in}$) part of the normalized photocurrent ($I_{Ph,qd}$) as a function of frequency is shown in **Figure 4b**. It is observed that $I_{ph,qd}(f)$ slowly increases with an increase in frequency and then attains a maximum photocurrent value at a specific frequency (termed as resonance frequency). After the resonance frequency, the photocurrent starts decreasing and attains zero value at a higher frequency. It is observed that the resonance frequency decreases with an increase in X in composition $(FA_{1-X}MA_X)_{0.95}Cs_{0.05}PbI_3$. Further, to underpin the trend in resonance frequency for different compositions of X in the $I_{ph}(f)$ measurement, a frequency dependent capacitance measurement is carried out for the $(FA_{1-X}MA_X)_{0.95}Cs_{0.05}PbI_3$ based PSCs, and shown in **Figure 1e**. The frequency dependent capacitance increases with an increase in X. It suggests that the capacitance is also a function of the A-site cation.[42] The variation in the resonance frequency and the capacitance with respect to the X (%) for $(FA_{1-X}MA_X)_{0.95}Cs_{0.05}PbI_3$ based PSCs is shown in **Figure 1f**. The effect of A-site cation on the resonance frequency will be explained in the discussion part in good detail.

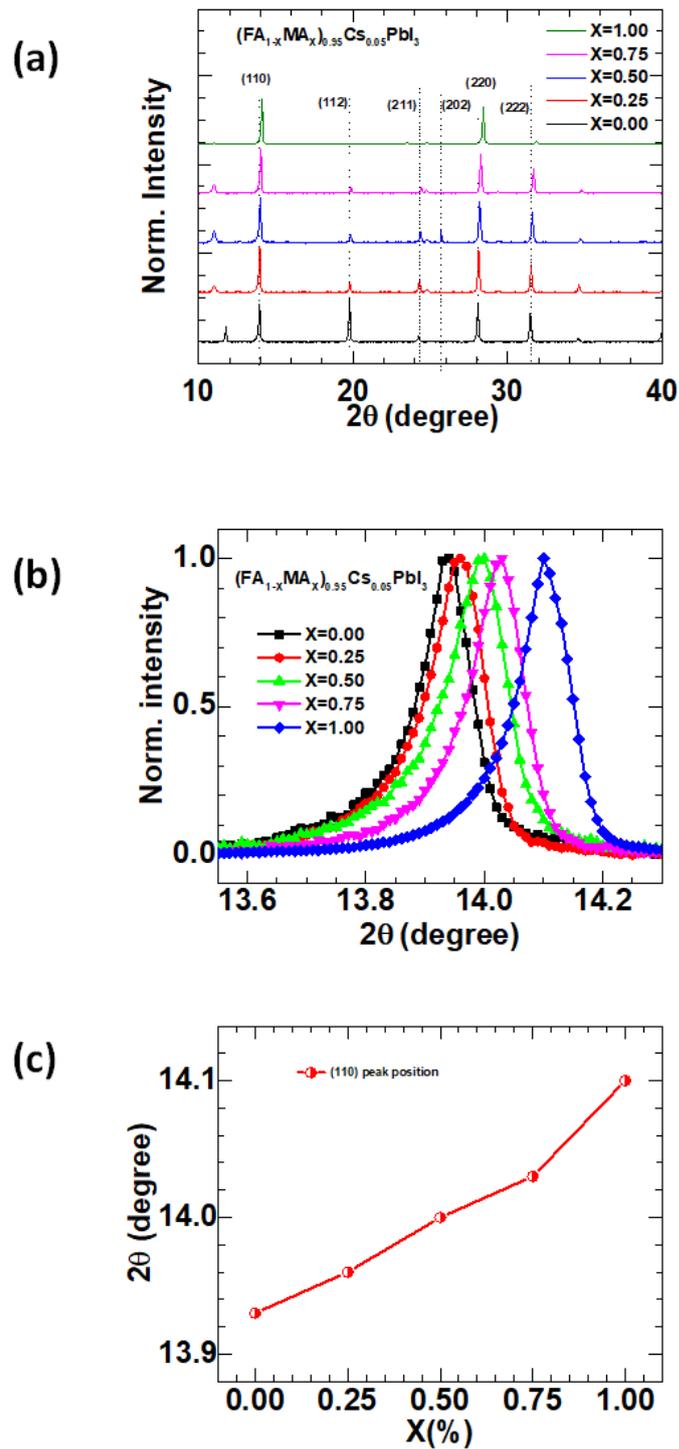

**Figure 2:** *(a) Normalized XRD pattern of (FA_{1-X}MA_X)_{0.95}Cs_{0.05}PbI_3 perovskite films where X= 0.00, 0.25, 0.50, 0.75, and 1.00. (b) Peak position of (110) plane of (FA_{1-X}MA_X)_{0.95}Cs_{0.05}PbI_3 perovskite films. (c) the shift of (110) peak in 2θ degree with respect to X composition.*

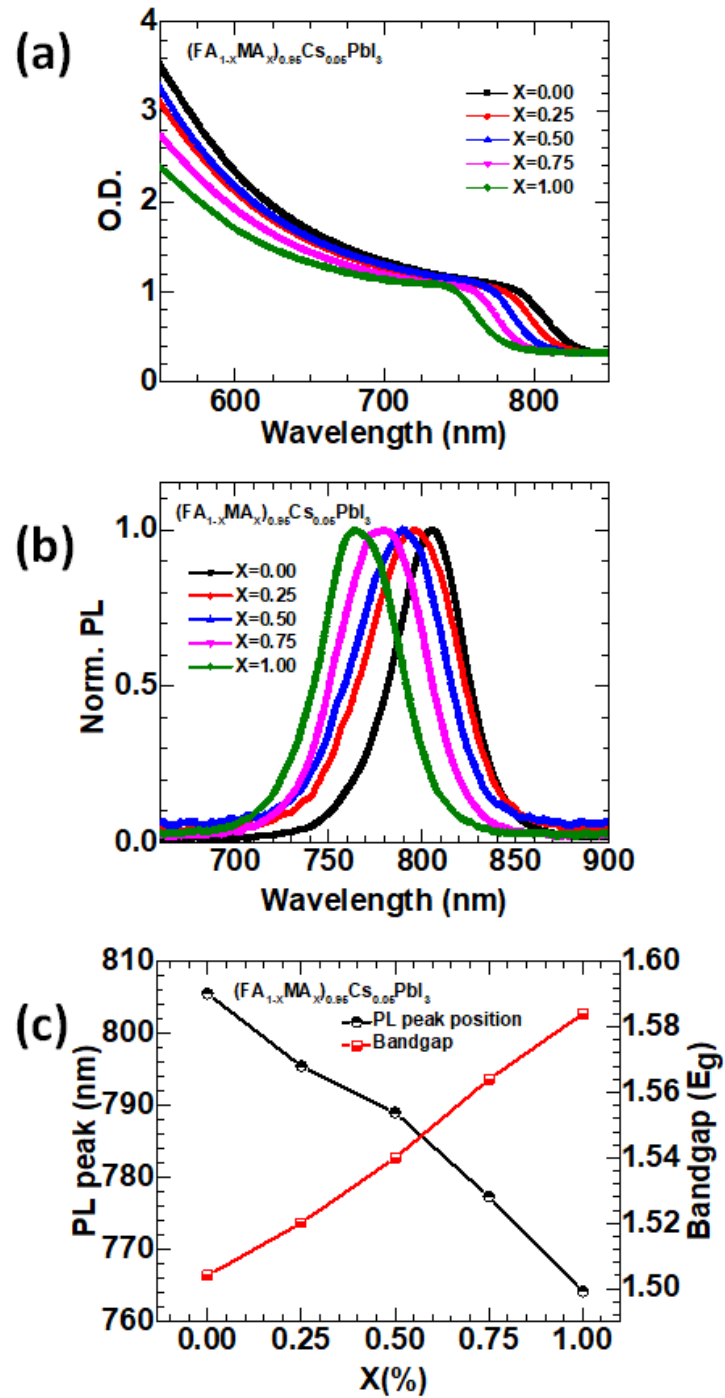

***Figure 3:*** (***a***) *Absorption,* (***b***) *steady-state normalized PL spectra for the (FA₁₋ₓMAₓ)₀.₉₅Cs₀.₀₅PbI₃ perovskite films where X= 0.00, 0.25, 0.50, 0.75, and 1.00.* (***c***) *PL peak position and bandgap for the (FA₁₋ₓMAₓ)₀.₉₅Cs₀.₀₅PbI₃ perovskite films.*

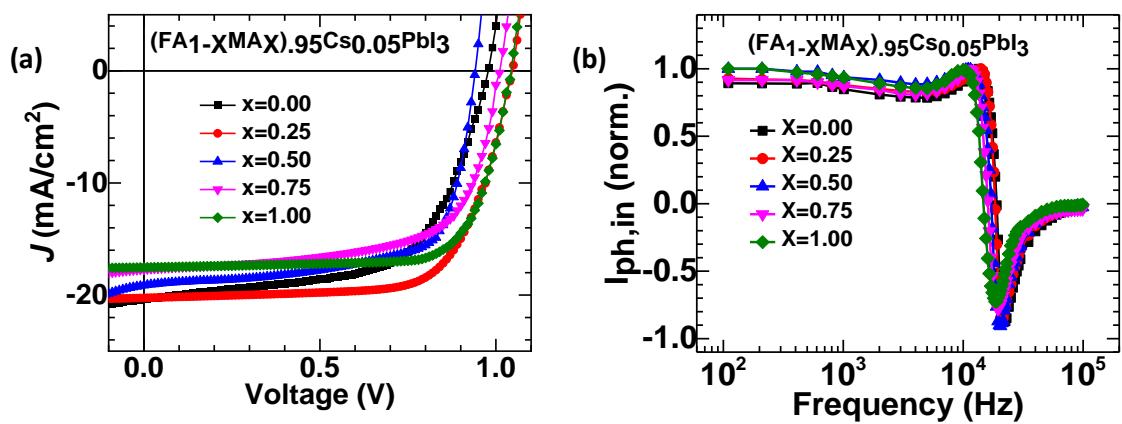

**Figure 4:** *(a) Illuminated J-V characteristics and (b) in-phase part of the frequency dependent photocurrent of the $(FA_{1-X}MA_X)_{0.95}Cs_{0.05}PbI_3$ PSCs, where X= 0.00, 0.25, 0.50, 0.75, and 1.00.*

**Table 1:** *The PV parameters of the $(FA_{1-X}MA_X)_{0.95}Cs_{0.05}PbI_3$ PSCs, where X= 0.00, 0.25, 0.50, 0.75, and 1.00.*

| Device | $J_{SC}$ (mA/cm$^2$) | $V_{OC}$ (V) | $FF$ | PCE (%) |
|---|---|---|---|---|
| X= 0.00 | 20.56 | 0.98 | 0.60 | 12.12 |
| X=0.25 | 20.25 | 1.05 | 0.63 | 13.47 |
| X=0.50 | 19.12 | 0.94 | 0.69 | 12.44 |
| X=0.75 | 17.68 | 1.01 | 0.66 | 11.74 |
| X=1.00 | 17.53 | 1.04 | 0.74 | 13.56 |

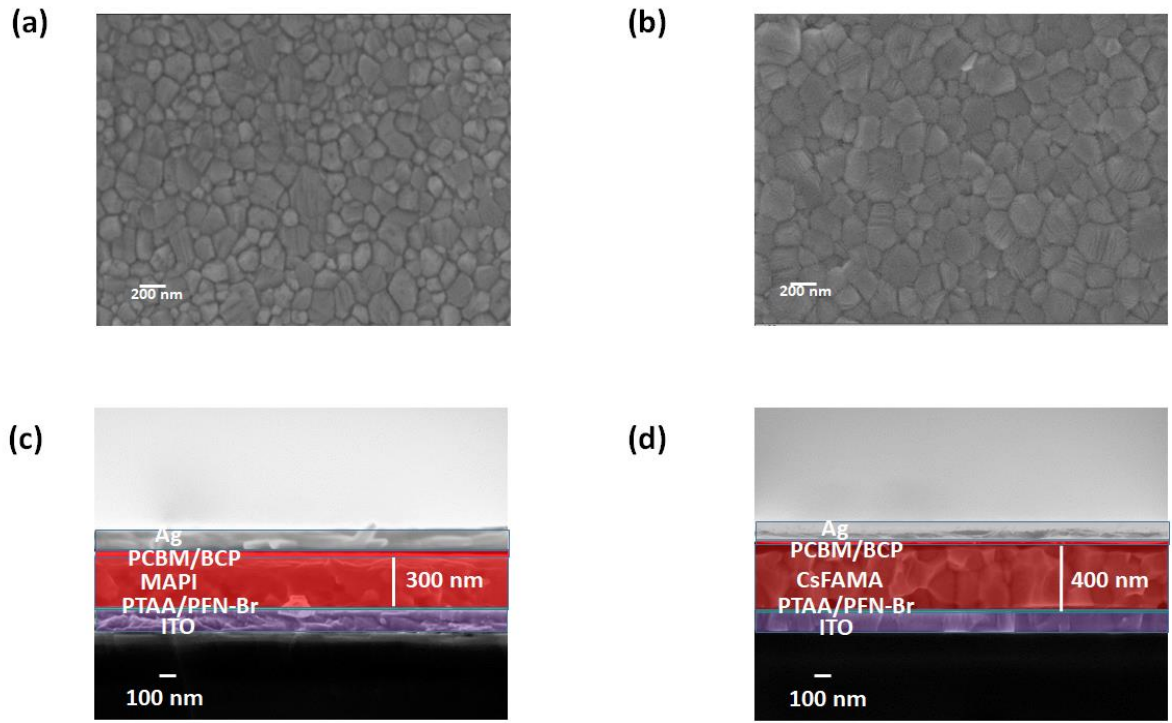

***Figure 5:*** *Top-view FESEM images of (**a**) MAPI and (**b**) CsFAMA perovskite films. The cross-section FESEM images of the complete device of (**c**) MAPI and (**d**) CsFAMA PSCs.*

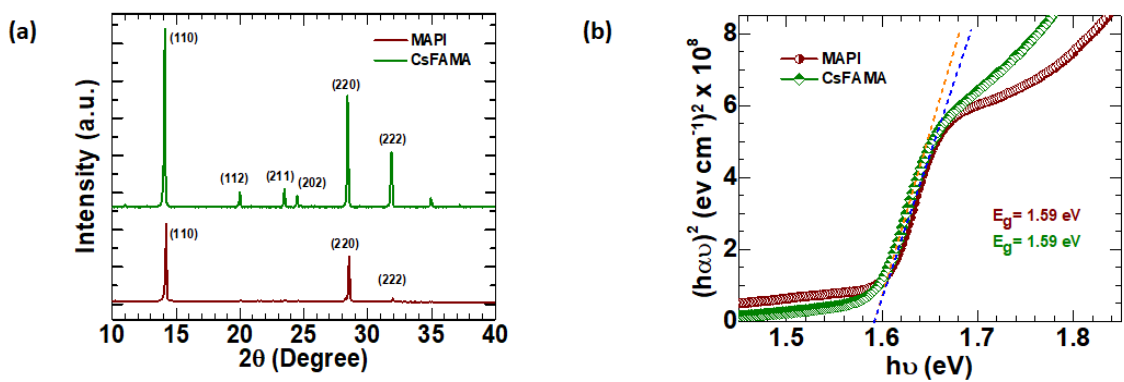

**Figure 6: (a)** *X-ray diffraction (XRD) pattern of MAPI and CsFAMA perovskite films.* **(b)** *The energy bandgap is calculated from the absorption spectra.*

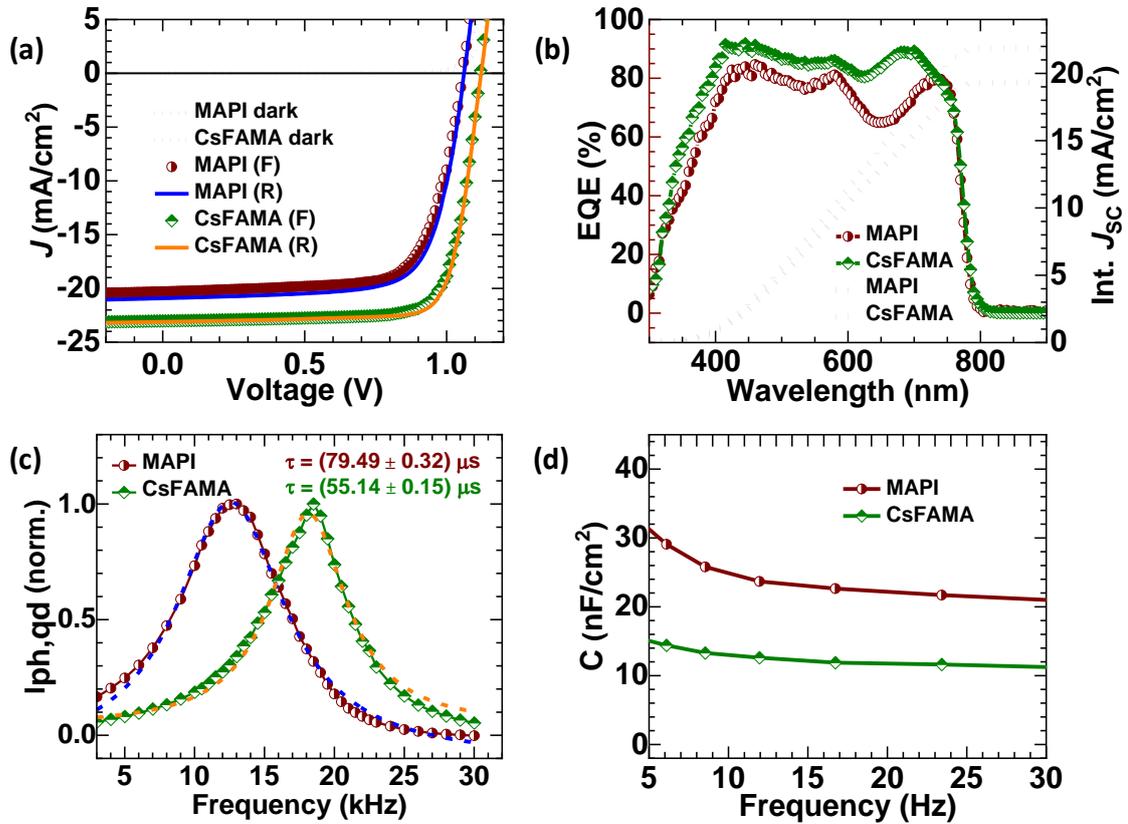

**Figure 7: (a)** *Dark and illuminated J-V characteristics of MAPI and CsFAMA based PSCs in forward and reverse scan.* **(b)** *EQE spectra of MAPI and CsFAMA based PSCs and the integrated $J_{SC}$ according to the corresponding EQE spectra (Dip in EQE is due to reflection losses, see Figure 9).* **(c)** *Quadrature part of the frequency dependent normalized photocurrent for MAPI and CsFAMA based PSCs fitted with Lorentzian peak function.* **(d)** *Frequency dependent capacitance of MAPI and CsFAMA based PSCs.*

**Table 2:** *PV parameters of the best performed and averaged over 24 PSCs under 1-sun illumination conditions in forward (F) and reverse (R) scan directions.*

| Device | $J_{SC}$ (mA/cm$^2$) | $V_{OC}$ (V) | FF | PCE (%) | Int. $J_{SC}$ (mA/cm$^2$) | HI index (%) |
|---|---|---|---|---|---|---|
| MAPI (F) | 20.30 | 1.06 | 0.72 | 15.49 | | |
| | 18.36 ± 1.63 | 1.02 ± 0.03 | 0.71 ± 0.04 | 13.98 ± 0.81 | 19.34 | 4.32 |
| MAPI (R) | 20.92 | 1.06 | 0.73 | 16.19 | | |

|  | 19.59 ± 1.57 | 1.03 ± 0.03 | 0.72 ± 0.03 | 14.50 ± 0.79 |  |  |
|---|---|---|---|---|---|---|
| CsFAMA (F) | 23.02 | 1.12 | 0.78 | 20.11 |  |  |
|  | 22.04 ± 0.81 | 1.10 ± 0.02 | 0.75 ± 0.03 | 18.15 ± 0.89 | 21.88 | 1.67 |
| CsFAMA (R) | 23.11 | 1.12 | 0.79 | 20.45 |  |  |
|  | 22.53 ± 0.88 | 1.10 ± 0.02 | 0.76 ± 0.03 | 18.76 ± 0.84 |  |  |

Further, we extended the frequency dependent photo current study to the widely explored MAPI and CsFAMA based PSCs to understand the dielectric relaxation process in these two wonderful and interesting perovskite systems.[43,44] These devices were fabricated in *p-i-n* configuration (Figure 1b). It is found that both the films are compact and grain size is relatively bigger in CsFAMA film in comparison to MAPI film (**Figure 5**). **Figure 7a** represents the dark and illuminated current density *vs* voltage (*J-V*) characteristics in forward and reverse scan direction. The illuminated *J-V* characteristics are carried out under 1-Sun (100 mW/cm$^2$) condition, and all the photovoltaic parameters for the champion MAPI and CsFAMA based PSCs are listed in **Table 2**. The illuminated PV parameters for the MAPI based PSCs are, short circuit current density ($J_{SC}$) = 20.92 mA/cm$^2$, open-circuit voltage ($V_{OC}$) = 1.06 V, fill factor (*FF*) = 0.73 and PCE = 16.19 %. The illuminated PV parameters for the CsFAMA based PSCs are, $J_{SC}$ = 23.11 mA/cm$^2$, $V_{OC}$ = 1.12 V, *FF* = 0.79 and PCE = 20.45 %. The dark *J-V* characteristics in the semi-log scale for both devices suggest three times lower leakage current density (at the bias of - 0.5V) in CsFAMA than MAPI based PSC (**Figure 8b**).[45] The higher value of $J_{SC}$ for CsFAMA based PSCs may be attributed to the higher built-in potential $V_{bi}$= 0.89 V than MAPI $V_{bi}$ = 0.84 V measured from dark *J-V* characteristics (**Figure 8c**). The higher value of $J_{SC}$ in the CsFAMA based PSC is in good agreement with the external quantum efficiency (EQE) spectrum (**Figure 7b**). EQE spectrum shows a dip around 600 nm, similar to previous reports.[46,47] In order to understand the origin of these PSCs EQE spectral profiles, reflection spectra were carried out using an integrating sphere in Bentham PVE 300. We observed the dip in the EQE spectrum correlates well with the hump in the reflection spectrum of the PSCs **Figure 9(a) and (b).** Hence, it is purely a micro-cavity effect for a multilayer thin-film device with back metal reflecting contact. The integrated $J_{SC}$ over the EQE spectrum (**Figure 7b**) is 19.34 mA/cm$^2$ and 21.88 mA/cm$^2$ for MAPI and CsFAMA PSC respectively **Table 2**.

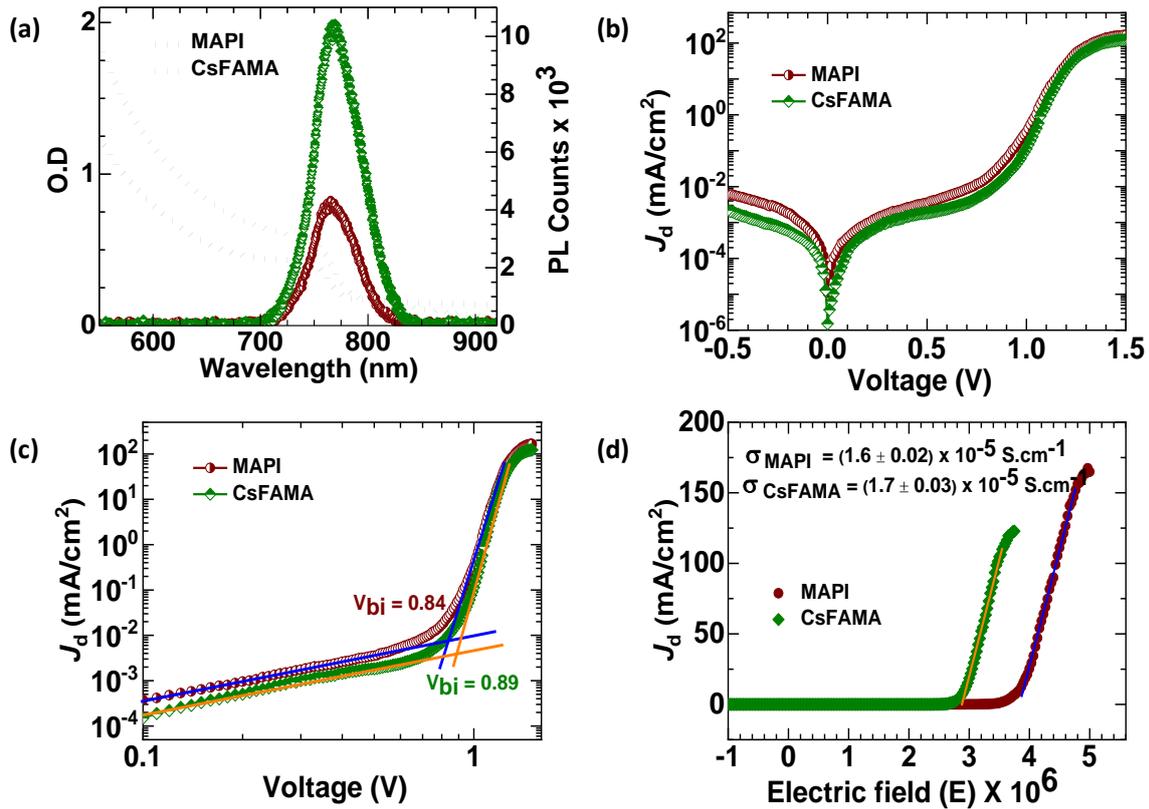

***Figure 8:*** (***a***) *Absorption and PL spectra of MAPI and CsFAMA perovskite films.* (***b***) *Dark J-V characteristics of MAPI and CsFAMA PSCs.* (***c***) *Built-in potential of MAPI and CsFAMA based PSCs.* (***d***) *Conductivity of MAPI and CsFAMA based PSCs.*

The slightly lower value ($\sim 5\%$) of integrated $J_{SC}$ compared to the $J_{SC}$ obtained from illuminated *J-V* measurement can be attributed to either the edge effect from the active area (17.5 mm$^2$) of the devices or ion migration or pre-bias measurement condition.[48,49] The lower value of $J_{SC}$, $V_{OC}$ and $FF$ of MAPI based PSC compared to the CsFAMA based PSC is obtained, which will be discussed in the subsequent part. The hysteresis Index (HI)[50] of a solar cell indicates the ionic migration in the solar cell and can be calculated as

$$\text{HI} = \frac{PCE(reverse) - PCE(forward)}{PCE(reverse)} \qquad (4.1)$$

The HI for MAPI based PSC is 4.32 % whereas for the CsFAMA based PSC it is 1.67 %. This can be correlated with the relatively higher degree of ion migration in MAPI based PSC due to the presence of defect states.[51,52] Here, we draw attention to the fact that the devices are not treated with any kind of passivating agents purposefully in order to have an in-depth insight into the relaxation process in PSCs based on their crystal structure.

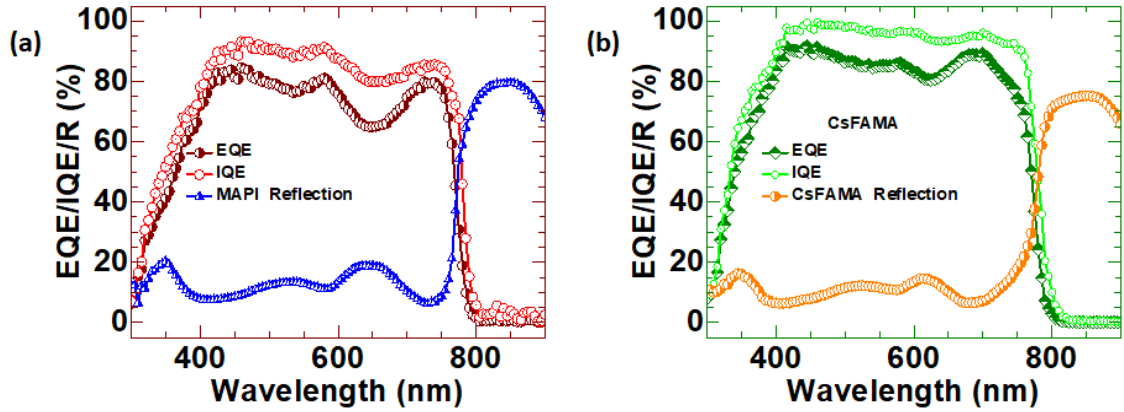

***Figure 9:*** *(**a**) EQE, IQE and reflection spectroscopy (R) of (a) MAPI and (**b**) CsFAMA based PSCs.*

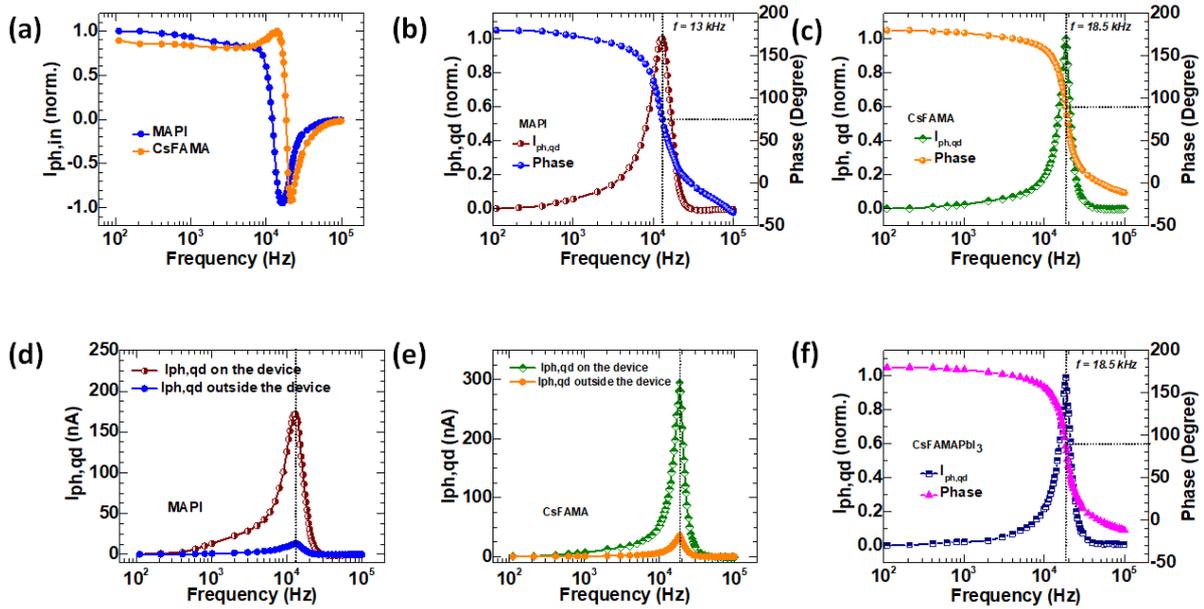

***Figure 10:*** *(**a**) Normalized in-phase part of the frequency dependent photo current of MAPI and CsFAMA PSCs. Normalized quadrature part of the frequency dependent photo current and phase delay of (**b**) MAPI and (**c**) CsFAMA PSC. Normalized quadrature part of the frequency dependent photo current on the active area and outside the active area of (**d**) MAPI and (**e**) CsFAMA PSC. (**f**) Normalized quadrature part of the frequency dependent photo current and phase of CsFAMAPbI₃ PSC.*

The $I_{ph}(f)$ measurement has been carried out to understand the dielectric relaxation processes in MAPI *vs* CsFAMA based PSCs. **Figure 7c** shows a typical frequency dependent normalized $I_{Ph,qd}$ profile for MAPI and CsFAMA based PSCs . The normalized $I_{Ph,in}$ is shown in **Figure 10a**. The full scale (200 Hz to 102 kHz) of the $I_{Ph,qd}$ along with phase for MAPI and CsFAMA based PSCs is shown in **Figure 10b** and **10c** respectively. The resonance frequency for MAPI based PSC is at 13 kHz whereas, for CsFAMA based PSC, it is at 18.5 kHz.[35] We observe that the FWHM of the $I_{ph,qd}(f)$ profile is narrower for CsFAMA than MAPI. However, we are still looking for the specific reason behind the narrow FWHM of the $I_{ph,qd}(f)$ profile for CsFAMA. Again, to validate the frequency depending on photocurrent results, we measured the frequency dependent capacitance of MAPI and CsFAMA based PSCs **Figure 7d**. We observed that the capacitance is lower for CsFAMA than MAPI based PSC which is in good agreement with the results obtained earlier for $(FA_{1-X}MA_X)_{0.95}Cs_{0.05}PbI_3$ compositions (**Figure 1e**).

We used $I_{ph,qd}(f)$ response to calculate the dielectric relaxation time constant ($\tau_d$) of the photogenerated charge carriers in PSCs. The $\tau_d$ represents the response time in the dielectric material with respect to the externally applied sinusoidal electric field of 490 nm wavelength (electromagnetic) EM wave. Here, we have used the Lorentz peak fitting function to extract $\tau_d$ from experimental results (**Figure 7 c**).

$$I_{ph,qd}(f) = I_{ph,s-\infty} + \frac{2\Omega\delta}{4\pi(f - f_{max})^2 + \delta^2} \qquad (4.2)$$

Where $I_{ph,s-\infty}$ is the photocurrent at very high or low (static) frequency, $\Omega$ = area under the plot, $\delta$ = width of the plot, $f$ = applied frequency, $f_{max}$ = frequency at which the $I_{ph,qd}(f)$ is maximum termed as the resonance frequency.

The calculated $\tau_d$ for MAPI and CsFAMA is (79.49 ± 0.32 ) µs and (55.14 ± 0.15 ) µs respectively. Hence, the relaxation process is faster in CsFAMA based PSC than that in MAPI. It indicates that any optically perturbed charge carrier in the MAPI based PSC will take more time to come back to its equilibrium position than the CsFAMA based PSC, a representative of higher defect density in the former one. The $I_{ph,qd}(f)$ profile on the device and outside of the active area of the device for MAPI and CsFAMA based PSCs shows a similar resonance peak position (**Figure 10d, e**). Thus, it indicates that the photocurrent profile is a material property and does not depend on the spatial position of the device which is in good agreement with our earlier publication.[35] It is observed that the $I_{ph,qd}(f)$ profile outside the active area of the device is low compared to the $I_{ph,qd}(f)$ profile on the active area of the device. This can be understood by the lateral photovoltaic effect.[53] We do notice that CsFAMA based perovskite system contains Br⁻ and I⁻ both. In order to validate the effect of 'A' cation on $I_{ph,qd}(f)$ profile, we fabricated $(FA_{0.83}MA_{0.17})_{0.95}Cs_{0.05}PbI_3$ based PSC with only I⁻ as halide. The frequency dependent $I_{ph,qd}(f)$ measurement did not show any significant shift in the resonance frequency of $(FA_{0.83}MA_{0.17})_{0.95}Cs_{0.05}PbI_3$ based PSC in comparison to (FA$_{0.83}$MA$_{0.17}$)$_{0.95}$Cs$_{0.05}$Pb(I$_{0.90}$Br$_{0.10}$)$_3$ based PSC (**Figure 10f**). However, it is acknowledged that we do not observe any significant shift in resonance frequency because the ratio of Br$^-$ was quite less (10%) in CsFAMA based perovskite. The complete replacement of I$^-$ by Br$^-$ may cause shift in the resonance frequency, and it can be a separate study.

### 4.3.2 Experimental validation of the frequency dependent photocurrent measurement

Resonance frequency determined by a direct frequency dependent photocurrent is correlated with an independent estimation of capacitance plus conductivity studies, which is found to be in good agreement.

The classical dielectric relaxation time can be expressed as[34]

$$\tau_d = \frac{\varepsilon_0 \varepsilon_s}{\sigma} \qquad (4.3)$$

The capacitance of a parallel plate capacitor can be expressed as

$$\varepsilon_0 \varepsilon_s = \frac{C.d}{A} \qquad (4.4)$$

Thus, the dielectric relaxation time constant can be written as

$$\tau_d = \frac{C.d}{\sigma.A} \qquad (4.5)$$

Where, C= capacitance, d= thickness, A= device area, $\sigma$= conductivity, $\varepsilon_s$= dielectric constant, $\varepsilon_0$= dielectric permittivity in vacuum.

The thickness of MAPI is d$_{MAPI}$ = 300 nm, and for CsFAMA is d$_{CsFAMA}$= 400 nm (**Figure 5**).

The conductivity of MAPI is $\sigma_{MAPI}$ = 1.6 x 10$^{-5}$ S. cm$^{-1}$ and for CsFAMA is $\sigma_{CsFAMA}$ = 1.7 x 10$^{-5}$ S. cm$^{-1}$ (**Figure 8d**).

The capacitance of MAPI is C$_{MAPI}$ ~22 nF/cm$^2$ and for CsFAMA is C$_{CsFAMA}$ ~12 nF/cm$^2$ (**Figure 7d**).

The area of both the devices is the same A = 17.5 mm$^{2.}$

The ratio of dielectric relaxation time constants using equation (4.5) is,

$$\frac{\tau_{CsFAMA}}{\tau_{MAPI}} = \frac{C_{CsFAMA}.\ d_{CsFAMA}}{\sigma_{CsFAMA}} . \frac{\sigma_{MAPI}}{C_{MAPI}.d_{MAPI}}$$

$$\frac{\tau_{CsFAMA}}{\tau_{MAPI}} = 0.68$$

Further, the dielectric relaxation time constants from the frequency-dependent photocurrent measurement are $\tau_{d,CsFAMA} = (55.14 \pm 0.15)$ µs and $\tau_{d,MAPI} = (79.49 \pm 0.32)$ µs and the ratio of the time constant measured from the frequency-dependent photocurrent measurement is

$$\frac{\tau_{CsFAMA}}{\tau_{MAPI}} = 0.69$$

Thus, the ratio of the time constant measured from two different experiments is almost similar, which validates our measurement that experimentally determined resonance frequency can provide a direct connection with semiconductor material properties.[34]

**Table 3:** *The ratio of dielectric relaxation time measured from photocurrent response $I_{ph,qd}$ (f) and C-f combined with J-V measurement.*

| Measurement technique | $\tau_{d, CsFAMA}/\tau_{d, MAPI}$ |
|---|---|
| Photocurrent $I_{ph,qd}$ (f) | 0.69 |
| C-f and dark J-V measurement | 0.68 |

### 4.3.3 Defects in perovskite solar cells

Further, optical studies are carried out to gain deep insight into the defect states in MAPI versus CsFAMA based perovskite thin films. It is observed that the steady-state photoluminescence (PL) intensity of CsFAMA is ~2.38 times higher than that of MAPI based perovskite film (**Figure 8a**) under identical conditions of excitation and measurement set-up. The higher PL intensity of the CsFAMA film could be due to either of two reasons: (a) higher thickness or (b) lower defect density than MAPI (Figure 5). To validate the higher PL intensity of CsFAMA, the time-resolved (TR) PL measurement is carried out for both films. **Figure 11a** represents the TRPL spectra for both films. The charge carrier lifetime for MAPI and CsFAMA based perovskite films is 20.36 ns and 49.06 ns respectively. Hence, the average charge carrier lifetime in CsFAMA film is in good agreement with steady-state PL intensity results (**Figure 8a**). The reason for the higher TRPL lifetime for CsFAMA perovskite is due to higher photoluminescence quantum yield (PLQY). The PLQY of MAPI and CsFAMA perovskite films on the glass substrate is 4.06% and 9.78% respectively. Interestingly the PLQY ratio is 2.41 which correlates with the steady-state PL.[36,54]

The $V_{OC}$ of a solar cell can be expressed as [34]

$$Voc = \frac{nkT}{q}\ln(I) + constant$$

$$Voc = \frac{nkT}{q} \times 2.303 \times \log(I) + constant \qquad (4.6)$$

Where $n$ = ideality factor $k$ = Boltzmann constant, $T$ = temperature, $q$ = charge, $I$ = intensity of the light.

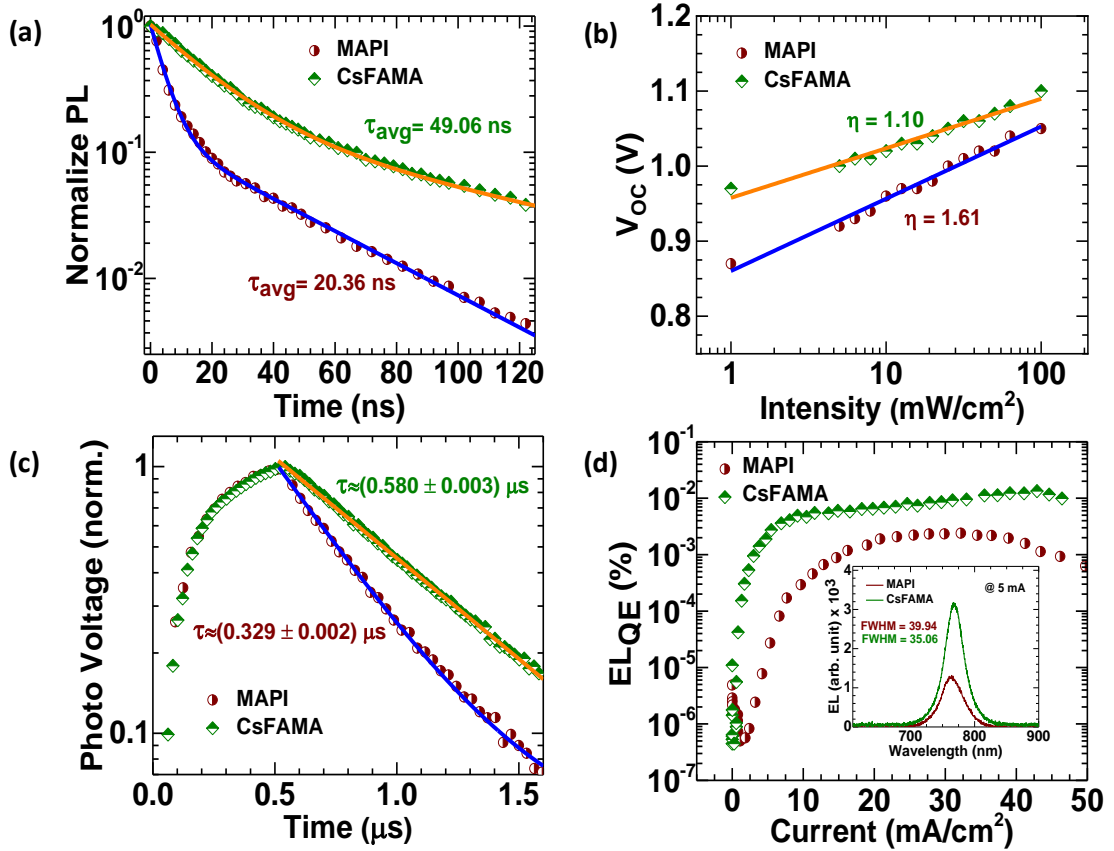

***Figure 11: (a)*** *Time-resolved photoluminescence (TRPL) decay profile of MAPI and CsFAMA perovskite films, solid line is bi-exponential fit to the experimental data.* ***(b)*** *Intensity dependent $V_{OC}$ of MAPI and CsFAMA PSCs, the solid line fits the Shockley diode equation.[55]* ***(c)*** *Transient photovoltage (TPV) decay profile of MAPI and CsFAMA PSCs under 1-Sun illumination condition.* ***(d)*** *Electroluminescence (EL) quantum efficiency as a function of injected current density for MAPI and CsFAMA PSCs, and the inset represents the normalized EL spectra of MAPI and CsFAMA PSCs at an injected current of 5 mA.*

**Figure 11b** represents the intensity dependent f $V_{OC}$ of MAPI and CsFAMA PSCs. To calculate the ideality factor, we measured the slope of $V_{OC}$ *vs.* log(I) in **Figure 11(b)** and used equation (4.6) to calculate the ideality factor. It is observed that the ideality factor (*n*) for MAPI is *n*=1.61 whereas for CsFAMA *n*=1.10. The higher value of *n* for MAPI indicates the presence of a higher number of trap-assisted recombination centers.[56,57] **Figure 11c** represents the transient photovoltage (TPV) response for MAPI and CsFAMA based PSCs.[58] The calculated lifetime (τ) of perturbed charge carriers for MAPI and CsFAMA are 0.33 μs and 0.58 μs respectively. The higher value of τ for CsFAMA represents lower defect density and is responsible for higher $V_{OC}$ despite of having the same bandgap of semiconductor (**Figure 6b**).[59] **Figure 11d** represents the electroluminescence (EL) quantum efficiency (EL_{QE}) of MAPI and CsFAMA PSCs. The inset of **Figure 11d** represents the EL spectra for MAPI and CsFAMA at 5 mA constant current injection. It is observed that EL intensity is $\sim$ 2.49 times higher in the case of CsFAMA than MAPI based PSCs (**Figure 11d**). Further, we found that the EL$_{QE}$ is dependent on injected current density up to $J_{inj} = 20$ mA/cm$^2$ in the case of MAPI based PSCs whereas it is largely independent beyond $J_{inj} \sim 5$ mA/cm$^2$ for CsFAMA PSCs.[60]

### 4.3.4 Thermal stability of the perovskite solar cells

**Figure 12a** represents the histogram of the MAPI and CsFAMA based PSCs over 24 devices. It is observed that the average PCE of MAPI based PSCs is 14.69%, whereas for CsFAMA based PSC the PCE is 18.83%. So, replacing the MAPI with CsFAMA perovskite, there is an increase in PCE of $\sim$ 28% even though the optical bandgap edge is the same. Further, the thermal stability of MAPI and CsFAMA based PSCs is investigated (**Figure 12**). The unencapsulated devices were kept under the continuous 1-Sun illumination for 14 hours, and the humidity was $\sim$ 40%. The illuminated *J-V* characteristic is measured for both devices at regular time intervals. It is observed that MAPI based PSC retains 47.11% of its initial $J_{SC}$, 105.56% of its initial $V_{OC}$, 97.34% of its initial *FF*, and 49.28% of its initial PCE after 14 hours of continuous 1-Sun illumination **Figure 12b**. However, the CsFAMA based PSC retains 85.89% of its initial $J_{SC}$, 97.20% of its initial $V_{OC}$, 94.17% of its initial *FF*, and 78.61% of its initial PCE after 14 hours of continuous 1-Sun illumination **Figure 12c**. The retained PV parameters after 14 hours of continuous 1-Sun illumination are listed in **Table 4**. Thus, continuous illumination of the PSCs under the 1-Sun condition deteriorates the performance of the PSCs. The major change in the PCE of MAPI and CsFAMA based PSCs after 14 hours of continuous 1-Sun illumination is associated with the $J_{SC}$. The $J_{SC}$ decreases for both MAPI and CsFAMA PSCs with time under 1-Sun continuous illumination, however, it is higher for MAPI based PSC. Reduction in $J_{SC}$ and a slight increase in $V_{OC}$ under continuous illumination for MAPI based PSCs have been observed earlier too by Joshi et al.[61] MAPI based PSC showed the $V_{OC}$ starts increasing after 6 hours of illumination and eventually, after 14 hours of illumination, it 105.56% of its initial $V_{OC}$. Whereas for CsFAMA, it is continuously decreasing, and eventually after 14 hours it remains 97.20 % of its initial $V_{OC}$. Even though the $V_{OC}$ of MAPI based PSC is increasing with time during continuous illumination but the $J_{SC}$ is degrading very significantly. Hence, the overall PCE remains lower for MAPI based PSC after degradation.

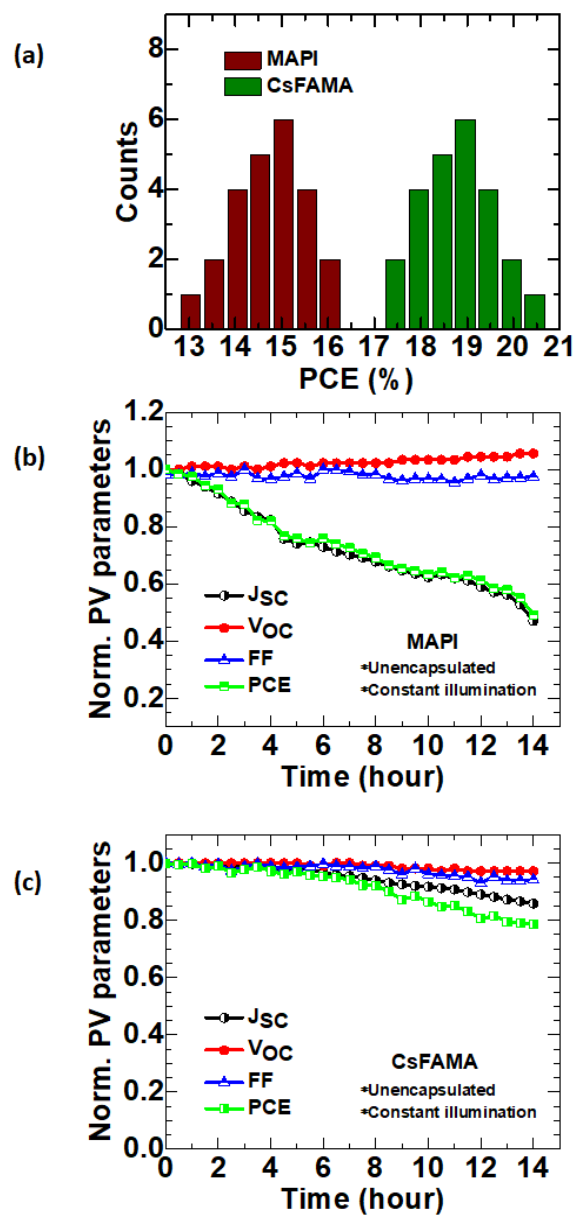

**Figure 12: (a)** *Histogram of the PCE of MAPI and CsFAMA based PSCs over 24 devices.* **(b)** *Thermal stability of PSCs under the continuous 1-Sun illumination for 14 hours of (a) MAPI and (c) CsFAMA based PSCs.*

**Table 4:** *The kinetics of PV parameters in percentage (%) of its initial value (i.e. 100%) for MAPI and CsFAMA based unencapsulated PSCs after 14 hours of continuous 1-Sun illumination.*

| Device | $J_{SC}$ (%) | $V_{OC}$ (%) | $FF$ (%) | PCE (%) |
|--------|--------------|--------------|----------|---------|
| MAPI | 47.11 | 105.56 | 97.34 | 49.28 |
| CsFAMA | 85.89 | 97.20 | 94.17 | 78.61 |

## 4.4 Discussion

**Figure 1d** suggests the role of FA *vs.* MA on dielectric resonance frequency while keeping absorber layer thickness fixed at ~ 500nm. However, the device structure is not optimized for optimal PCE. We note that there is not a linear relationship or a correlation between the PCE *vs.* resonance frequency based on figure 1d and Table 1. This can be attributed to the multiple factors which govern the solar cell efficiency, like; compactness, i.e., the morphology of the thin-film, crystallinity and domain size, thickness optimization of the absorber, and electronic coupling with charge extracting layers.[7,62] However, for an optimized PSC structure, dielectric relaxation studies consistently explain the influence on resonance frequency due to the presence of defects which are ions. Herein, we show a unique set of results which correlates the $\tau_d$ with the defects present in MAPI and CsFAMA PSCs. Further, in semiconductor physics, it is widely known that the dielectric relaxation time constant ($\tau_d$) is proportional to the resistivity ($\rho$) and dielectric permittivity ($\varepsilon_s$) i.e. $\tau_d = \rho\varepsilon_s$.[34] Two independent methods are used to determine the ratio of $\tau_{d,CsFAMA}/\tau_{d,MAPI}$, $I_{ph,qd}(f)$ and *C-f* combined with conductivity, which is found to be almost similar 0.69 and 0.68 respectively **Table 3**. This phenomenon can be understood by considering the dipoles oscillating in an external electric field.[40] When the *1/f* $\gg \tau_d$, the dipoles have enough time to respond to the electric field, and thus it shows some positive number of current. As we further increase the frequency and *1/f* becomes comparable to $\tau_d$, the resonance condition occurs and $I_{ph,qd}(f)$ attains a maximum value. Further, an increase in frequency results in a decrease in $I_{ph,qd}(f)$ (relaxation process) due to dipole lag behind the electric field. Perovskite materials are known to show ferroelectric behavior.[32] It is observed that ferroelectric dipoles of CsFAMA show a better response to an electric field than MAPI (**Figure 7c**). Hence, the relaxation process is also faster in the CsFAMA ($\tau_d = 55.14 \pm 0.15$ μs) than MAPI ($\tau_d = 79.49 \pm 0.32$ μs). This indicates that if an additional charge is perturbed in the system, it will come back to the ground state faster in CsFAMA than MAPI based PSCs. The faster relaxation process in CsFAMA is attributed to lower defect density, which is in good agreement with other optoelectronic measurements. The higher steady-state PL intensity (Figure 8a), higher charge carrier lifetime (TRPL and TPV) (Figure 11a, c), and the injection current density independent higher EL$_{QE}$ behavior (Figure 11d) is evidence of lower defect density in CsFAMA based perovskite film.

When light interacts with absorbing semiconductor layers at different modulated frequencies, like *1/f* $\gg \tau_d$, then the dipoles have enough time to respond to the alternating electric field. We note that any perturbed charge carrier in the MAPI-based PSC takes a longer time to come back into its equilibrium position than the CsFAMA-based PSC. This suggests that at the microscopic level dipoles are interacting with defects, which are relatively hindering the dipole oscillations in MAPI based devices. Thermal stability studies suggested that there is a significant decrease in the $J_{SC}$ which can be attributed to prominent ion migration in MAPI based PSCs than CsFAMA based PSCs as studied by Joshi *et. al.* using capacitance studies as a function of illumination dose and irradiation time duration.[61,63] These two combined studies connect well macroscopic dielectric parameters to microscopic level properties of the semiconductor film in the form of the presence of relative ionic defects density. A significant decrease in $J_{SC}$ and a slight increase in $V_{OC}$ is an established key feature of ion migration.[61,64] Under constant illumination, a relative enhancement of $V_{OC}$ in MAPI based PSC can be understood by the presence of relatively more ionic defects. It has also been established previously by theoretical and experimental studies.[61,64] Based on DFT calculation it is proven that due to presence of only 3 H-bonds from A-site cation with octahedral I⁻ can cause the release of I⁻ ions.[65] Whereas these H-bonding sites can be increased with the addition of FA+ ion and this results in relatively better control over ion migration issues in pristine halide perovskites, i.e., CsFAMA studied here.[66] Hence, we believe that the low ionic/structural defect density in CsFAMA bulk due to the presence of relatively more number of H-bonding between cation to halide ions is key for stable absorber and hence PSCs than conventional MAPI based PSCs.

## 4.5 Conclusions

In conclusion, we have presented a comparative study on the dielectric relaxation process in MAPI *vs* CsFAMA in terms of $I_{ph}(f)$ measurement. The $I_{ph}(f)$ measurements allow us to determine $\tau_d$, and it is lower for CsFAMA than MAPI, which suggests the lower defect density in the former one. We also compared the defects in MAPI and CsFAMA by various optoelectronic measurements and observed that CsFAMA possesses lower defects than MAPI. We validated our measured $\tau_d$ from $I_{ph}(f)$ measurement and $\tau_d = \rho\varepsilon$ and it is in good agreement. The longer frequency response of CsFAMA (lower $\tau_d$) dipoles to the electric field in comparison to the MAPI is attributed to lower defect density in the CsFAMA bulk. We also showed that low $\tau_d$ in multi-cation CsFAMA is connected for stable absorber under illumination conditions and can be used as a tool to screen other ionic semiconductors for stable solar cell application.

## 4.6 Postscript

Overall, this chapter deals with the compositional engineering of hybrid organic-inorganic halide perovskite semiconductors and introduces the study of defects using various optoelectronic measurements. From the first principle calculation studies, it is shown that the intrinsic defects in the perovskite can be minimized by using multications in the perovskite crystal structure (discussed in Chapter **2**). In the literature reports, the single monovalent cation perovskite MAPbI$_3$ has higher defects and lower stability compared to the FAPbI$_3$ or triple cation CsFAMA based perovskite, but the direct correlation was not established. In this chapter, we used frequency dependent photocurrent measurement technique and studied the dielectric relaxation process. The faster relaxation time for the CsFAMA based PSCs indicates lower defects, which are later verified using various optoelectronic measurements. Further, the CsFAMA-based perovskite solar cells show higher thermal stability, which is correlated with the halide ion migration and hydrogen bonding with the monovalent cation. More hydrogen bonding in the CsFAMA perovskite crystal structure is responsible for faster dielectric relaxation and higher thermal stability, hence the correlation established. However, we demonstrated the thermal stability studies under continuous 1-Sun illumination and relative humity of 40%, of our unencapsulated PSCs. The stability can be increased by encapsulating the device and carrying the stability measurement under N$_2$ atmosphere. In addition, the reproducible efficiency and thermal stability of the PSCs can be further improved by choosing a self-assembled monolayer (SAM) based HTLs that facilitates good energy band alignment with the perovskite layer of bandgap 1.6 eV **Figure 13**.[67] We used a phosphonic acid anchored carbazol SAM based HTL and thermally stable triple cation CsFAMA perovskite (as discussed above) in the inverted architecture-based solar cells. The work function of the SAM is modified using a polyelectrolyte polymer and well matched with the perovskite work function, resulting in excellent interfacial charge transport quality. This work will be discussed in next Chapter 5.

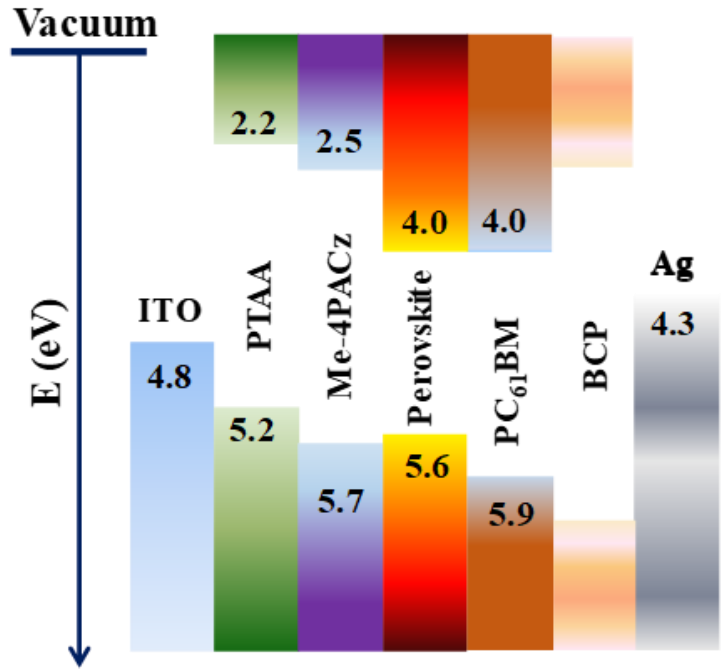

**Figure 13.** Energy level band diagram of the *p-i-n* architecture based perovskite solar cell where PTAA and Me-4PACz SAM are used as HTLs and CsFAMA perovskite (bandgap 1.6 eV) as an absorber. The energy level values are taken from the literature.[68–70]

# CHAPTER 5

# Resolving the Hydrophobicity of Me-4PACz Hole Transport Layer for Highly Reproducible and Efficient (>20%) Inverted Perovskite Solar Cells

# CHAPTER 5

# Resolving the Hydrophobicity of Me-4PACz Hole Transport Layer for Highly Reproducible and Efficient (>20%) Inverted Perovskite Solar Cells

## Abstract

[4-(3,6-Dimethyl-9H-carbazol-9-yl)butyl]phosphonic acid (Me-4PACz) self-assembled monolayer (SAM) has been employed in perovskite single junction and tandem devices demonstrating high efficiencies. However, a uniform perovskite layer does not form due to the hydrophobicity of Me-4PACz. Here, we tackle this challenge by adding a conjugated polyelectrolyte poly(9,9-bis(3'-(N,N-dimethyl)-N-ethylammonium-propyl-2,7-fluorene)-alt-2,7-(9,9dioctylfluorene))dibromide (PFN-Br) to the Me-4PACz in a specific ratio, defined as Pz:PFN. With this mixing engineering strategy of Pz:PFN, the PFN-Br interacts with the A-site cation is confirmed via solution-state nuclear magnetic resonance studies. The narrow full width at half maxima (FWHM) of diffraction peaks of perovskite film revealed improved crystallization on the optimal mixing ratio of Pz:PFN. Interestingly, the mixing of PFN-Br additionally tunes the work function of the Me-4PACz, as demonstrated by Kelvin probe force microscopy and built-in-voltage estimation in solar cells. Devices employing optimized Pz:PFN mixing ratio deliver open-circuit voltage ($V_{OC}$) of 1.16V and efficiency >20% for perovskites with a bandgap of 1.6 eV with high reproducibility and concomitant stability. Considering significant research on Me-4PACz SAM, our work highlights the importance of obtaining a uniform perovskite layer with improved yield and performance.

## 5.1 Introduction

In the previous chapter 4, we have demonstrated that the compositional engineering of the organic cations in the hybrid organic –inorganic metal halide perovskite is helpful in minimizing the defect states. The mixing the monovalent cations results in improved device performance and concomitant stability. However, in Chapter 4, we used PTAA as a hole transport layer (HTL). PTAA is hydrophobic in nature, and hence deposition of a uniform perovskite layer is challenging. Therefore, we used PFN-Br as an interlayer to deposit a uniform perovskite layer, which restrains direct contact of perovskite with the PTAA layer. In addition, the PTAA shows a band offset by 200-300 meV for ~1.6 eV perovskite, which further hampers charge collection efficiently.[1] Hence, we used another HTL of self-assembled monolayer (SAM) in the *p-i-n* device architecture of perovskite solar cells (PSCs). So, in this chapter 5, we used the mixed cation perovskite CsFAMA as an absorber due to its enhanced performance and stability, as discussed in Chapter 4, and PCBM as an electron transport layer (ETL) in the *p-i-n* device architecture. The inverted or *p-i-n* device architecture is of significant interest using low-temperature processed charge transport layers (CTLs) such as poly(3,4-ethylenedioxythiophene) polystyrene sulfonate (PEDOT: PSS), poly (triaryl amine) (PTAA), fullerene ($C_{60}$) are employed.[2–6] Recently, Al-Ashouri *et al.*, used various phosphonic acid group anchored carbazole-based self-assembled monolayers (SAMs) such as 2PACz, MeO-2PACz, and Me-4PACz as hole transport layers (HTL), demonstrating superior device performance than the widely employed PTAA as HTL.[1,7,8] In particular, by employing Me-4PACz SAM as HTL, the same group, reported one of the highest efficiency in single junction devices and a certified PCE of ~29% in silicon-perovskite two-terminal tandem solar cells.[1] In comparison to other SAMs, the Me-4PACz-based device is reported to demonstrate the lowest density of interface traps, indicating suppressed non-radiative recombination at the perovskite/Me-4PACz interface.[9] These salient features make Me-4PACz SAM a promising HTL for PSCs. Motivated by this, we fabricate PSCs with Me-4PACz as HTL. Unfortunately, we (and others) observed poor perovskite layer formation on the Me-4PACz and is in line with recent reports.[10–13] Strategies such as the incorporation of an $Al_2O_3$ insulating layer at the interface of Me-4PACz and perovskites and deposition of Me-4PACz by evaporation have been reported to improve the perovskite layer coverage on Me-4PACz.[10,11,14] These strategies are promising, but developing a strategy to deposit Me-4PACz by solution process and simultaneously obtain a uniform perovskite layer with improved device performance with a proper understanding remains a challenge.

Until recently, the interface engineering approach has been applied to address the issue of poor wetting of perovskite on underlying hydrophobic polymeric HTLs. For instance, PTAA is known to be a hydrophobic HTL and modification of its surface with poly(9,9-bis(3'-(N,N-dimethyl)-N-ethylammoinium-propyl-2,7-fluorene)-alt-2,7-(9,9dioctylfluorene))dibromide (PFN-Br), poly(methyl methacrylate) (PMMA):2,3,5,6-tetrafluoro-7,7,8,8-tetracyanoquinodimethane (F4-TCNQ), polyvinyl oxide (PEO) and tetra-n-propylammonium bromide (TPAB), phenylethylammonium iodide (PEAI), etc. have been reported.[5,15–19] Among others, modification with PFN-Br has been widely employed as it helps in obtaining a uniform perovskite layer with augmented device efficiency. In the previous chapter 4, we have also employed PFN-Br modified PTAA and reported PCE of >20%.[2] In addition to polymeric HTLs, modification of SAMs has also been reported.[20,21] Recently, Li *et al*., employed a mixture of MeO-2PACz and 2PACz to improve the charge extraction and reported a PCE of 25% for all-perovskite tandem solar cells.[21] Deng *et al*., showed that co-assembled monolayers (a mixture of SAM and alkylammonium containing SAM) can help in simultaneously suppressing the non-radiative recombination and surface functionalization and the resultant device showed a PCE of 23.59%.[20] Very recently, Al-Ashouri added 1,6-hexylenediphosphonic acid with Me-4PACz to improve the perovskite layer deposition; however, a detailed investigation has not been performed.[13] Until present, the modification of carbazole-based SAMs has been limited to co-assemble with phosphonic or carboxylic acid anchoring group which might limit the exploration of SAM. Therefore, strategies need to be developed beyond co-assembly with SAM, to improve the perovskite layer coverage on Me-4PACz SAM.

In this chapter, we present a mixing engineering strategy by combining Me-4PACz SAM with conjugated polyelectrolyte PFN-Br to obtain a highly reproducible and improved performance of perovskite solar cells. With the aid of mixed Me-4PACz:PFN-Br, a uniform perovskite thin film is obtained and the device with optimized Me-4PACz:PFN-Br mixing ratio showed a PCE of >20% in 0.175 cm$^2$ device active area and 19% over an active area of 0.805 cm$^2$. To the best of our knowledge, the obtained device efficiency is one of the highest values reported for Me-4PACz with a triple cation perovskite composition having a bandgap ($E_g$) of 1.6 eV. More importantly, with the help of solution nuclear magnetic resonance (NMR), X-ray diffraction (XRD) and Kelvin probe force microscopy (KPFM) outcomes, we elucidate the triple role of PFN-Br. The PFN-Br interacts with the A-site cation and improves the crystallization. This also elevates the valence band position of Me-4PACz, leading to better interfacial energy level alignment with perovskite having a bandgap of 1.6 eV. We note that this improved performance could be achieved due to reduced interfacial localized states (low series resistance ($R_S$)) and bulk defects in the absorber layer (ideality factor $n \sim 1$), which result not only in high fill factor (*FF*), short-circuit current density ($J_{SC}$) and open-circuit voltage ($V_{OC}$) but also provide an excellent yield and stability for optimal PSCs. The unencapsulated device showed enhanced stability by retaining the initial device performance at $T_{95}$ even after 3000 hours when measured in $\sim$ 40% humidity. Further, the operational stability at the 1-Sun condition and thermal stability at 85° C of the un-encapsulated device under $N_2$ environment retains > 90% of the initial efficiency after 300 and 600 hours respectively.

## 5.2 Experimental Section

**Materials:** ITO-coated glass substrates (15 $\Omega$/sq) were purchased from Lumtech. Lead iodide ($PbI_2$), formamidinium iodide (FAI), cesium iodide (CsI), lead bromide ($PbBr_2$), and Me–4PACz, all were purchased from TCI chemical and used as received. Methyl ammonium Bromide (MABr) was purchased from Greatcell Solar. Phenyl-C61-butyric acid methyl ester ($PC_{61}BM$) was purchased from Lumtech and was used as received. The poly(9,9-bis(3'-(N,N-dimethyl)-N-ethylammoinium-propyl-2,7-fluorene)-alt-2,7-(9,9-dioctylfluorene))dibromide (PFN-Br) ordered from Solarmer Material Inc. and used as received. Bathocuproine (BCP) was purchased from Sigma Aldrich and used as received. The list of all chemicals, companies, CAS, and product numbers is as follows.

| Chemical | Company | CAS number | Product number |
|----------|---------|------------|----------------|
| **MABr** | Great cell Solar | 6876-37-5 | MS301000 |
| **PbBr₂** | TCI | 10031-22-8 | L0288 |
| **FAI** | TCI | 879643-71-7 | F0974 |
| **PbI₂** | TCI | 10101-63-0 | L0279 |
| **CsI** | TCI | 7789-17-5 | C2205 |
| **Me-4PACz** | TCI | 2747959-96-0 | M3359 |

| PFN-Br | Solarmer Material Inc. | 889672-39-6 | ZJ710A |
|---|---|---|---|
| PCBM | Lumtec. | 160848-22-6 | LT-S905 |
| BCP | Sigma | 4733-39-5 | 699152 |
| DMF | Sigma | 68-12-2 | 227056 |
| DMSO | Sigma | 67-68-5 | 276855 |
| Methanol | Sigma | 67-56-1 | 322415 |
| Chlorobenzene | Sigma | 108-90-7 | 284513 |
| IPA | Sigma | 67-63-0 | 19516 |
| Ag | Local company – Parekh Industry Limited | Silver wire – 1 mm diameter | Purity 99.99% |
| ITO coated glass | Lumtec. | ITO Non-Patterned $15\Omega$ | LT – G001 |
| DMSO d6 | ACROS Organic | 2206-27-1 | 166291000 |

**Solution preparation:** To make triple cation $(FA_{0.83}MA_{0.17})_{0.95}Cs_{0.05}Pb(I_{0.83}Br_{0.17})_3$ perovskite solution, first we added 22.5 mg of MABr, 73.5 mg of $PbBr_2$, 172 mg of FAI, and 507.5 mg of PbI2 in 1 ml of DMF: DMSO (4:1) and stirred at room temperature for 2 hours to make a premixed solution. Separately we made 1.5 (M) CsI solution in DMSO i.e. 100 mg of CsI in 257 µl of DMSO and stirred for 2 hours. We filtered the premixed solution with a PTFE 45 mm filter in a separate vial. Finally, we added 950 µl of premixed solution and 50 µl of CsI solution to get the final triple cation$(FA_{0.83}MA_{0.17})_{0.95}Cs_{0.05}Pb(I_{0.83}Br_{0.17})_3$ perovskite solution and stirred for 1 hour. To obtain mixed Me–4PACz: PFN-Br solution, first, we prepared 0.4 mg/ml Me-4PACz solution in anhydrous methanol and 0.4 mg /ml PFN-Br solution in anhydrous methanol and stirred overnight. Finally, one hour before spin-coating, we mix the Me–4PACz and PFN-Br solution in the desired volume ratio e.g. 6:4, 7:3, 8:2, 9:1, and 9.5:0.5.

20 mg PCBM is dissolved in 1 ml of chlorobenzene and stirred overnight. 0.5 mg BCP is dissolved in 1 ml of anhydrous isopropanol and stirred overnight at room temperature and 10 minutes at 70° C just before spin coating.

**Perovskite solar cells fabrication:** ITO-coated glass (10-15 $\Omega$/square) substrate was patterned with Zn powder and HCl and then sequentially cleaned with soap solution, deionized (DI) water, acetone, and isopropanol for 10 minutes each. After drying the substrate with a nitrogen gun, we kept the ITO substrates at 80° C for 10 minutes and then took them inside the oxygen plasma ashing chamber for 20 minutes, and plasma ashing was done at an RF power of 18 watts. After plasma ashing, we immediately take the substrates inside the $N_2$ environment glove box ($O_2$<0.1 ppm, $H_2O$<0.1 ppm) and spin-coat Me-4PACz: PFN-Br mixed solution at 4000 rpm for 30 seconds and then annealed at 100° C for 10 minutes. After that, we cool down the substrates for 5 minutes and then perovskite solution spin-coated on the ITO/ Me-4PACz: PFN-Br substrates. The CsFAMA perovskite was spin-coated at 5000 rpm for 30 seconds and at the last 5 seconds, we used 150 µl of chlorobenzene as an anti-solvent treatment on a 1.5 cm by 1.5 cm substrate and then annealed at 100° C for 30 minutes. After that, $PC_{61}BM$ was spin-coated at 2000 rpm for 30 seconds and BCP was spin-coated at 5000 rpm for 20 seconds. Finally, 150 nm of Ag was deposited under a vacuum of $2x10^{-6}$ mbar using a metal shadow mask. Further, we fabricated PSC of the active area of 0.805 $cm^2$ (i.e., 11.5 mm x 7 mm) using ITO-coated glass substrates of sheet resistance 7 $\Omega$/square.

**Characterization:**

All the photovoltaic measurements were carried out under ambient conditions. Photocurrent density versus applied voltage ($J - V$) measurement has been carried out using Keithley 4200 SCS and LED solar simulator (LSH-7320) after calibrating through standard Si solar cells provided by ABET, IIT Bombay. The $J - V$ measurement was performed with a scan rate of 100 mV/s. EQE measurement has been carried out to measure the photo response as a function of wavelength using the Bentham quantum efficiency measurement system (Bentham PVE 300). The XRD measurements were carried out in Smartlab, Rigaku diffractometer with Cu K$\alpha$ radiation ($\lambda$=1.54Å). $\Theta$-2$\Theta$ scan has been carried out from 10°-45° with a step size of 0.001°. Morphological analysis was done using field emission scanning electron microscopy (FESEM). Optical absorption spectra were carried out using a spectrometer (PerkinElmer LAMBDA 950). Steady-state PL measurement was done on thin films in a vacuum at a pressure of $10^{-3}$ mbar in a custom-made chamber with a diode laser of wavelength 490 nm.

Steady-state current-voltage–light characteristics were measured using a Keithley 2400 source meter, Keithley 2000 multimeter, and calibrated Si photodiode (RS components). TPV was measured by using a 490 nm TOPTICA diode laser, THORLABS white lamp S/N M00304198, ArbStudio 1104, and digital oscilloscope Tektronix DPO 4104B. The contact angle measurement is done in the GBX Digidrop instrument. The KPFM measurement was carried out in Asylum/Oxford instrument, MFP3D origin. KPFM images were also recorded in the dark using the same instrument in dual-pass mode under ambient conditions. Voltage was applied to the cantilever tip. The NMR spectra measurement was carried out in JEOL, ECZR series 600 MHz NMR spectrometer.

## 5.3 Results and Discussion

### 5.3.1 Hydrophobicity of the HTL and perovskite layer deposition

Me-4PACz has been used as HTL in a single junction and tandem (with silicon) devices, demonstrating one of the highest efficiencies.[1,9,22] In an attempt to fabricate the device using Me-4PACz SAM, we observed poor perovskite layer formation. The schematic illustration of perovskite deposition on the Me-4PACz coated ITO substrate and the photographic image of a perovskite thin film is shown in **Figure 1a**. The poor perovskite layer coverage can be attributed to the presence of non-polar groups such as methyl (–CH$_3$) and long-alkyl chain (C$_4$H$_8$) in Me-4PACz SAM, which are in general responsible for hydrophobic nature.[23–27]

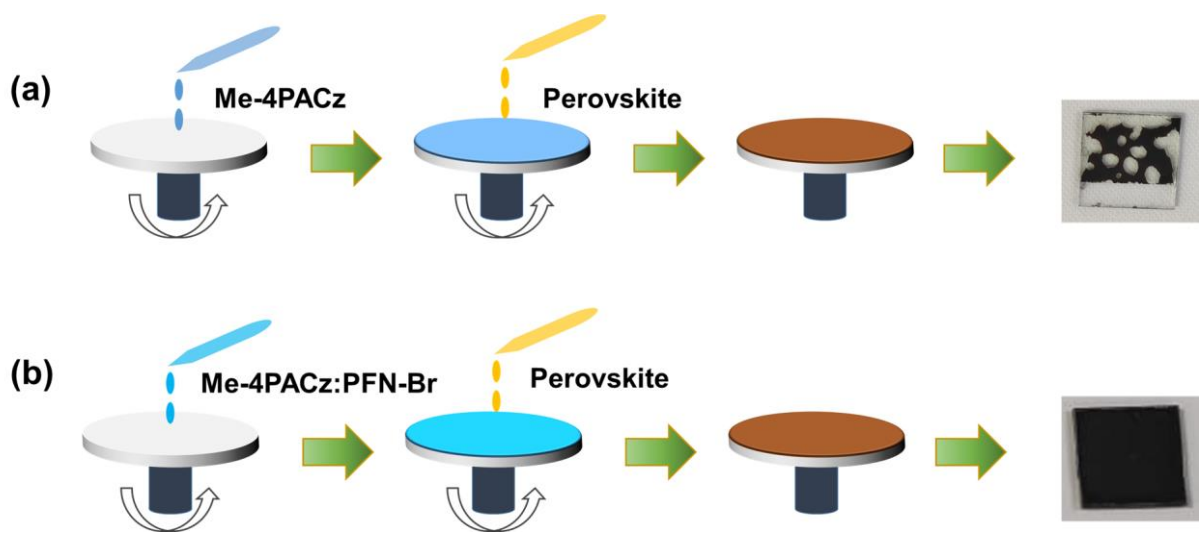

***Figure 1:*** *Schematic of the thin film deposition of the Me-4PACz, Me:4PACz:PFN-Br (abbreviated as Pz:PFN), and perovskite layer. (**a**) Me-4PACz and perovskite layer deposition. (**b**) Deposition of mixed Me-4PACz:PFN-Br and perovskite layers.*

To understand this more clearly, we performed water contact angle measurements of Me-4PACz and compared it with widely used hydrophobic polymer and small organic molecules based HTLs such PTAA, poly[2,6-(4,4-bis-(2-ethylhexyl)-4H-cyclopenta [2,1-b;3,4-b′]dithiophene)-alt-4,7(2,1,3-benzothiadiazole)] (PCPDTBT), poly(3-hexylthiophene-2,5-diyl) (P3HT) and 4,4′,4″-Tris[phenyl(m-tolyl)amino]triphenylamine (MTDATA). **Figure 2** shows the molecular structure of the aforementioned HTLs and their respective water contact angle in comparison with Me-4PACz. As expected, the water contact angle for Me-4PACz is high and in the same range as compared to other HTLs, evidencing its hydrophobic nature.

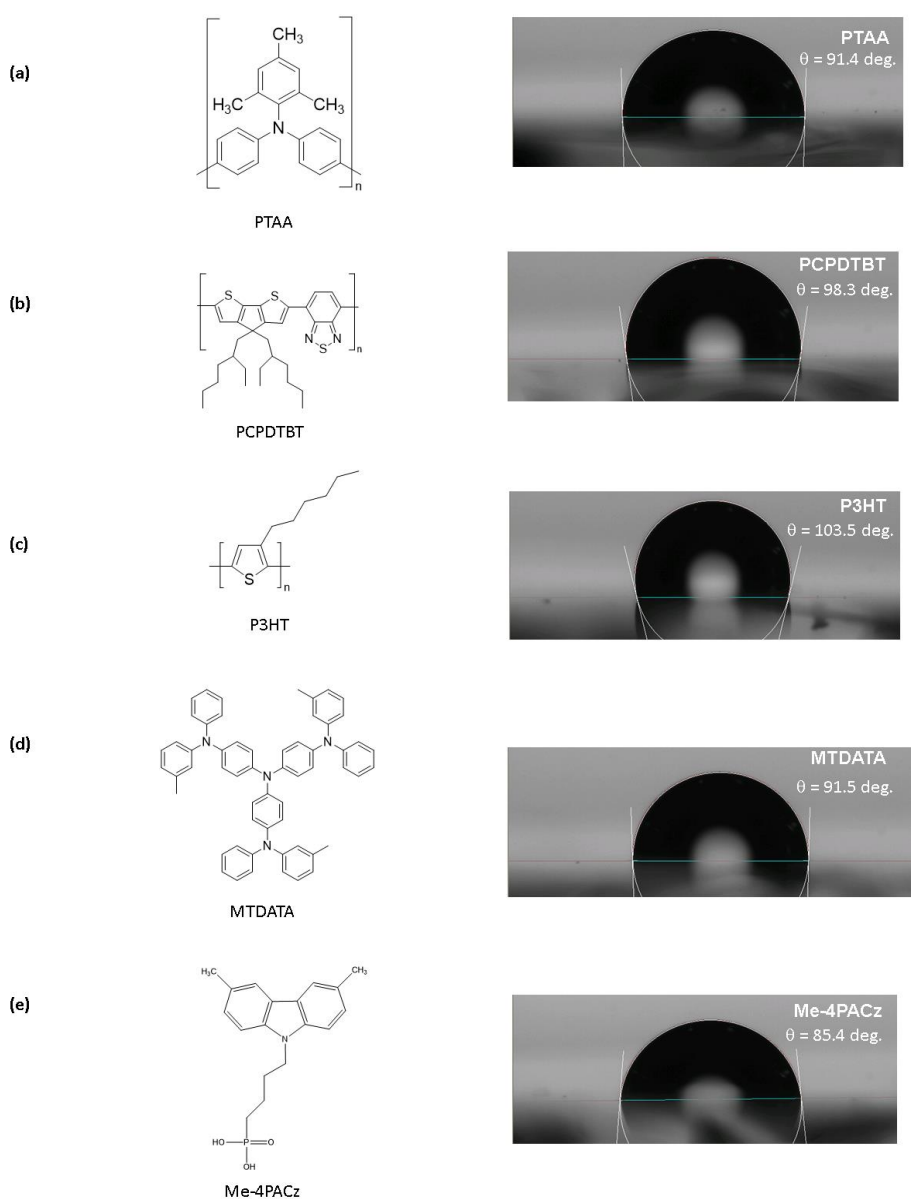

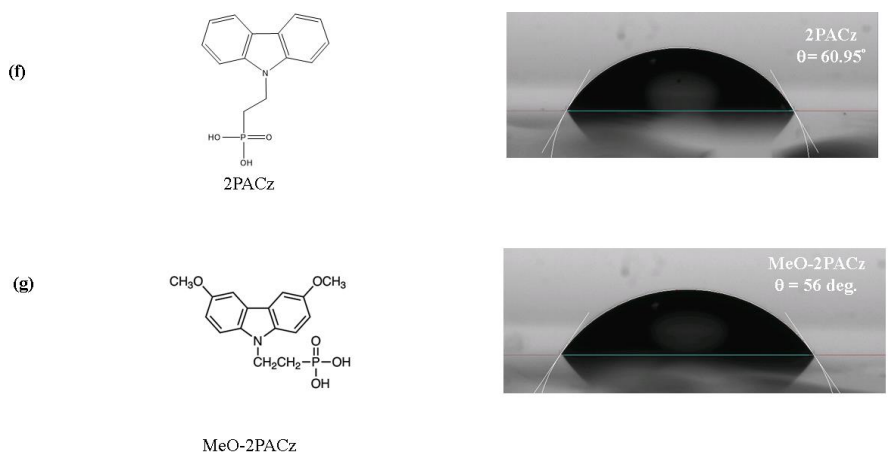

**Figure 2** : *Molecular structure and corresponding contact angle of (**a**) Poly[bis(4-phenyl)(2,4,6-trimethylphenyl)amine (PTAA), (**b**) poly[2,6-(4,4-bis-(2-ethylhexyl)-4H-cyclopenta [2,1-b;3,4-b']dithiophene)-alt-4,7(2,1,3-benzothiadiazole)] (PCPDTBT), (**c**) poly(3-hexylthiophene-2,5-diyl) (P3HT) (**d**) 4,4',4''-Tris[phenyl(m-tolyl)amino]triphenylamine (MTDATA), (**e**) [4-(3,6-Dimethyl-9H-carbazol-9-yl)butyl]phosphonic Acid (Me-4PACz), (**f**) [2-(9H-Carbazol-9-yl)ethyl]phosphonic Acid (2PACz) and (**g**) [2-(3,6-Dimethoxy-9H-carbazol-9-yl)ethyl]phosphonic Acid (MeO-2PACz)*

We additionally washed the Me-4PACz coated substrate with methanol (solvent to dissolve SAM), in an attempt to reduce the hydrophobicity and obtain a uniform perovskite film. However, the perovskite layer showed non-uniform coverage, as shown in **Figure 3**. This indicates that the phosphonic acid group strongly binds with ITO, and a strong monolayer fingerprint is present even after the washing step, which is in line with previous reports.[7]

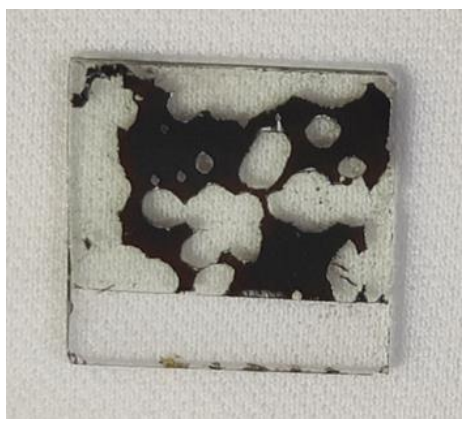

**Figure 3:** *Photographic image showing non-uniform perovskite thin film being deposited on methanol washed ITO/Me-4PACz substrate.*

To further confirm the role of non-polar groups that are present in Me-4PACz in preventing the perovskite layer formation, we fabricated device by employing 2PACz SAM as HTL. Because of the absence of the –CH$_3$ group and long alkyl chain, not only a uniform perovskite layer was obtained but the resultant device showed a PCE of ~ 20%, this further confirms that other than the SAM all other layers of the device stack are working as expected **Figure 4**. The current density ($J$) – voltage ($V$) curve of the best-performing device employing 2PACz SAM is shown in **Figure 4**. These results evidently suggest that Me-4PACz is sufficiently hydrophobic to prevent the formation of a uniform perovskite layer, and strategies need to be developed to improve the perovskite layer coverage on the Me-4PACz SAM.

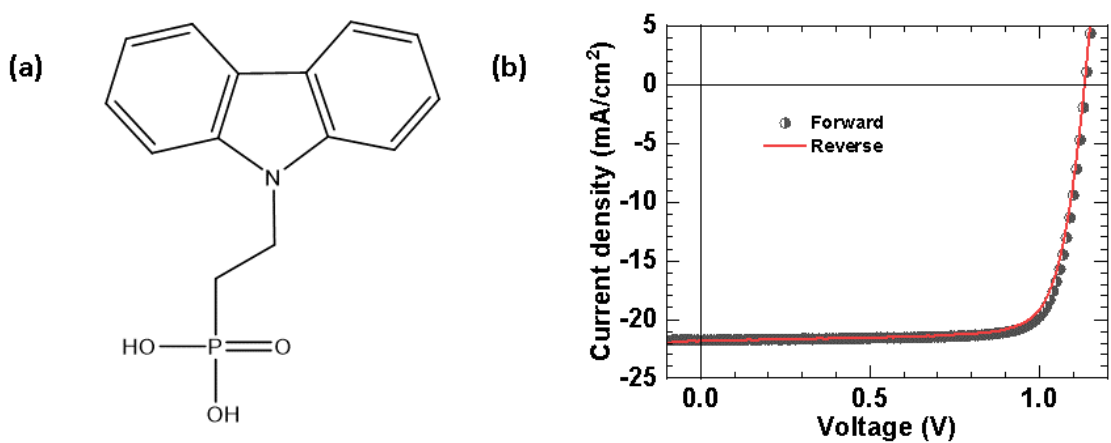

***Figure 4:*** *(**a**) molecular structure of 2PACz hole transport material (HTL). (**b**) $J - V$ characteristic curve of perovskite device employing 2PACz as HTL*

***Table 1:*** *Photovoltaic device parameters of perovskite device employing 2PACz as HTL.*

| Scan direction | $J_{SC}$ (mA/cm$^2$) | $V_{OC}$ (V) | $FF$ (%) | PCE (%) |
|:---:|:---:|:---:|:---:|:---:|
| Forward | 21.75 | 1137 | 80.31 | 19.86 |
| Reverse | 21.78 | 1133 | 78.62 | 19.40 |

Typically, researchers have incorporated PFN-Br as an interlayer to modify the interface of hydrophobic polymeric HTLs and obtain a uniform perovskite layer.[15,28] We initially modified the surface of Me-4PACz with PFN-Br and obtained a uniform perovskite layer **Figure 5**.

However, the device performance was not significantly high and reproducible and showed a PCE of 15.54% and 16.52% in forward and reverse scans. The J − V curve and box plot of device parameters show the degree of non-reproducibility, as shown in **Figure 5** and **Figure 6** respectively. We initially assumed that the PFN-Br solution (in methanol) might corrode the Me-4PACz underlayer, leading to low device efficiency. However, the washing step of Me-4PACz with methanol solvent (as discussed earlier) suggests that Me-4PACz/PFN-Br (abbreviated as Pz/PFN) dual layer might not be suitable to achieve high device efficiency, unlike PTAA/PFN-Br as discussed in chapter 4.[29]

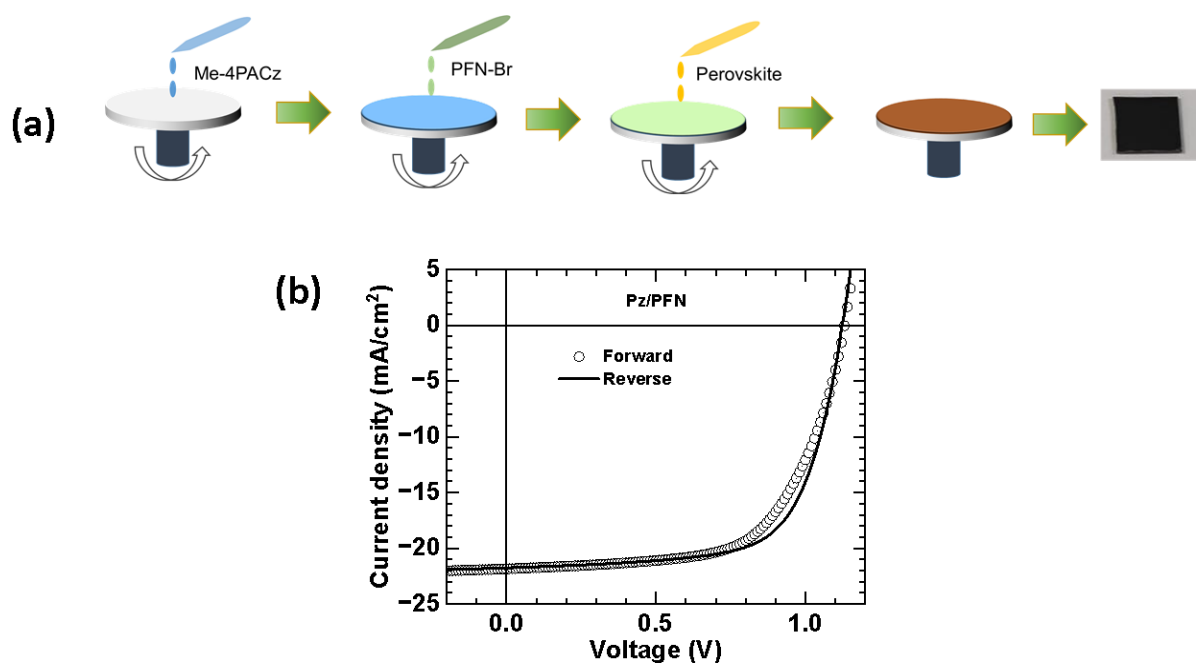

***Figure 5:*** *(**a**) PFN-Br interlayer deposition in between Me-4PACz and perovskite layers. (**b**) J − V characteristic curves of the device with Me-4PACz and PFN-Br dual layer.*

***Table 2:*** *Photovoltaic device parameters of perovskite device employing Me-4PACz and PFN-Br dual layer.*

| Scan direction | $J_{SC}$ (mA/cm$^2$) | $V_{OC}$ (V) | $FF$ (%) | PCE (%) |
|---|---|---|---|---|
| Forward | 21.84 | 1130 | 62.97 | 15.54 |
| Reverse | 21.75 | 1122 | 67.7 | 16.52 |

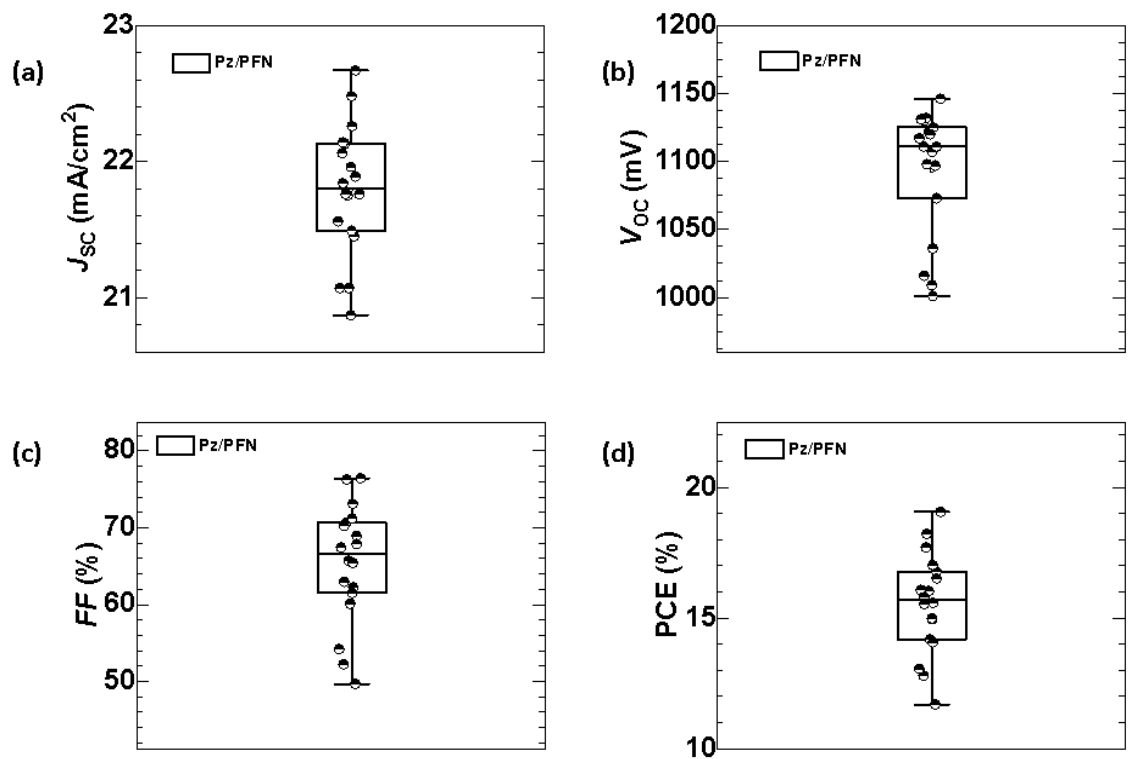

***Figure 6:*** *Histogram of the PV parameters over 18 devices employing Me-4PACz and PFN-Br dual layer.*

Previously, Levine et al. reported the lowest density of interface traps for Me-4PACz SAM.[9] This implies that high device performance can be obtained when perovskite is directly interfaced with Me-4PACz. However, in the Me-4PACz/PFN-Br interlayer modification, the perovskite is not directly interfaced with Me-4PACz SAM, which could be the reason for low device efficiency. Therefore, to make a direct interface between the Me-4PACz and perovskite, we performed a mixing engineering strategy.

### 5.3.2 Resolving the hydrophobicity and mixing engineering strategy

The mixing engineering strategy steps involving device stack layers deposition are schematically shown in **Figure 1b**. The Me-4PACz and PFN-Br were mixed in 6:4, 7:3, 8:2, 9:1, and 9.5:0.5 ratios followed by the perovskite layer deposition by one-step method (please see the experimental section above for more details). From now onwards, for our convenience, we term Me-4PACz:PFN-Br as Pz:PFN. Irrespective of all the mixing ratios, a uniform perovskite layer was obtained as shown in Figure 1b and **Figure 7**. Top surface scanning electron microscope (SEM) images, see **Figure 8**, showed no significant difference in the perovskite layer morphology with respect to the different Pz:PFN ratios.

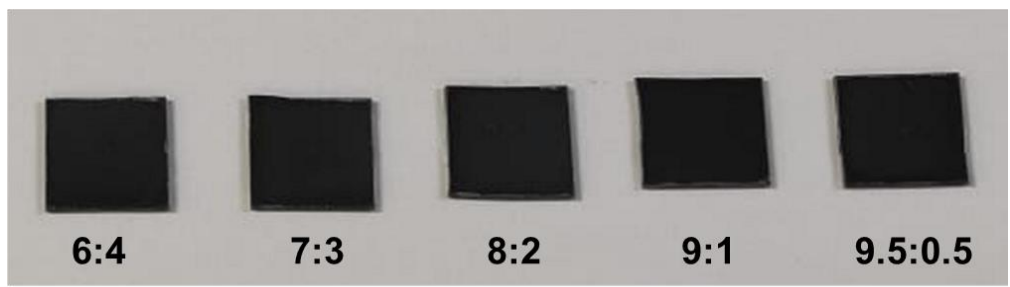

***Figure 7:*** *Photographic image of perovskite films deposited on Pz:PFN mixed HTLs with different mixing ratios such as, Pz:PFN (6:4), Pz:PFN (7:3), Pz:PFN (8:2), Pz:PFN (9:1), and Pz:PFN (9.5:0.5).*

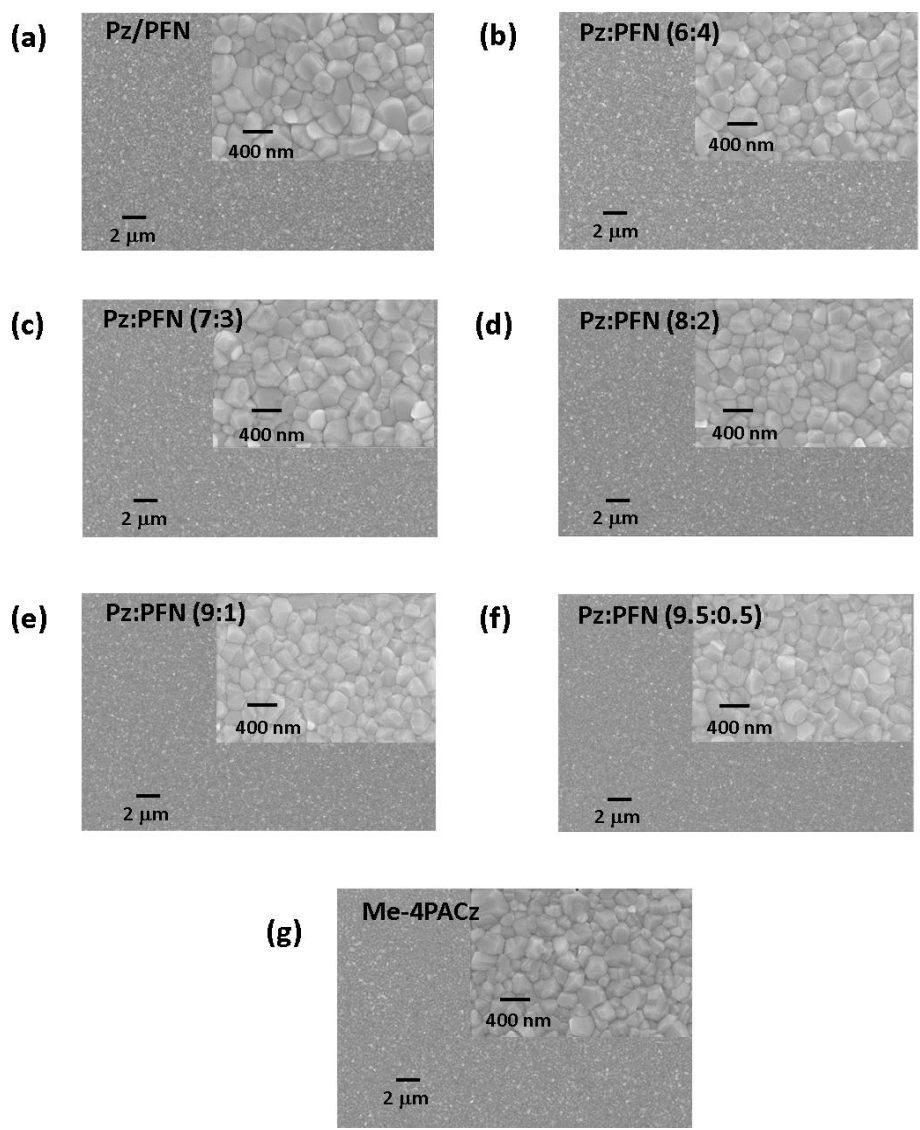

***Figure 8:*** *Top surface SEM images of perovskite thin film deposited on **(a)** Pz/PFN interlayer, **(b)** to **(f)** at different mixing ratios of Pz:PFN, **(g)** Me-4PACz.*

Moreover, we observed indistinct perovskite layer morphology on both the Me-4PACz/PFN-Br (term as Pz/PFN) (**Figure 8a**) dual layer and Me-4PACz (where perovskite layer is deposited on the substrate Figure 1a) (**Figure 8g**) compared to the mixing ratio cases. This further indicates that the perovskite layer crystallization is not influenced by the PFN-Br mixing or interlayer modification. The X-ray diffraction (XRD) diffractograms, as shown in **Figure 9**, showed no traces of residual lead iodide and no significant difference in the perovskite crystalline layer. The combined outcomes of SEM and XRD indicate that perovskite crystallizes similarly in all cases without any significant changes.

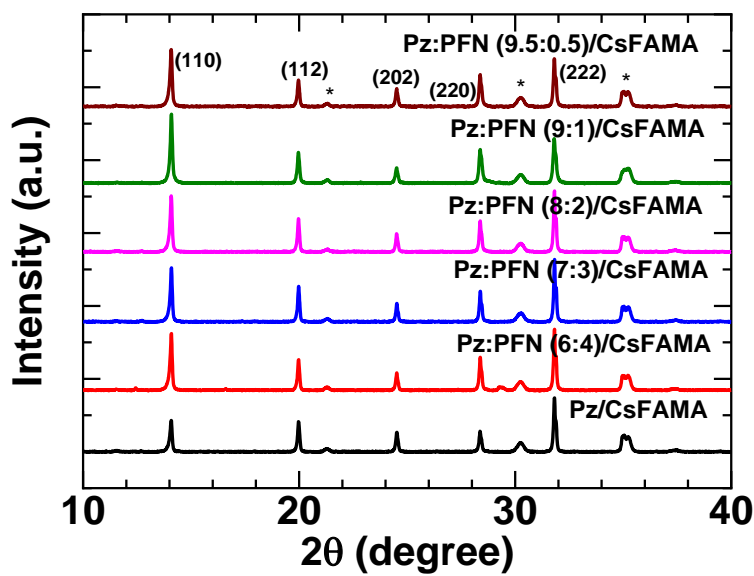

***Figure 9:*** *XRD diffractograms of perovskite thin films deposited on Me-4PACz (without and with different ratios of PFN-Br) coated ITO substrates.. Me-4PACz and PFN-Br are referred to as Pz and PFN respectively. The XRD peaks due to the indium tin oxide layer of ITO-coated glass substrates are marked with a star (\*).*

After the perovskite layer deposition, phenyl-$C_{61}$-butyric acid methyl ester (PCBM) and bathocuproine (BCP) were deposited via the solution process as an electron transport layer (ETL) and buffer layer respectively, and the silver electrode was thermally evaporated to complete the cell. **Figure 10a** depicts the best-performing device $J − V$ curves in forward bias under 1 Sun illumination for all the mixed Pz:PFN based perovskite solar cells. The $J − V$ curves under forward and reverse scan directions are shown in **Figure 11b** and the device parameters are tabulated in **Table 3**. With a change in mixing ratio, the device performance first increased and then decreased. We note that there is a change in $J_{SC}$ (also integrated $J_{SC}$ from IPCE) with respect to the compositional ratio of HTL, which will be discussed later along with KPFM results. The device obtained from Pz:PFN with a 9:1 ratio showed the highest PCE of 20.67% with a $J_{SC}$ of 22.54 mA/cm$^2$, $V_{OC}$ of 1.16 V, $FF$ of 79.12% with slight hysteresis (please see **Table 3**). Note that the obtained efficiency is one of the highest values reported with Me-4PACz based 1.6 eV bandgap PSCs device **Figure 13, Table 4.** The photovoltaic device performance over 0.805 cm$^2$ active area (11.5 mm x 7 mm) PSC device is 19.06% with $J_{SC}$ = 22.25 mA/cm$^2$, $V_{OC}$ = 1.145V, and $FF$ = 74.81% **Figure 11a**. **Figure 10b** shows the incident photon to current efficiency (IPCE) spectrum of our best-performing device along with the reflection (R), and transmission (T) spectrum. The IPCE spectra of the devices based on different Pz:PFN ratios are shown in **Figure 12**. The dip in the EQE spectrum correlates well with the hump in the reflection spectra. The loss in the current density due to reflection can be calculated from the external quantum efficiency (EQE) and internal quantum efficiency (IQE) spectrum using equations (5.3) and (5.4). The integrated $J_{SC,EQE}$ is calculated by integrating the EQE spectrum over the 1-Sun spectrum. There is a current density mismatch by less than 5% of the $J_{SC,EQE}$ to $J_{SC}$ measured from the J − V measurement, and this can be understood by edge effect from the active area (17.5 mm$^2$ inset figure 11c) of the device or pre-bias measurement condition.[30,31]

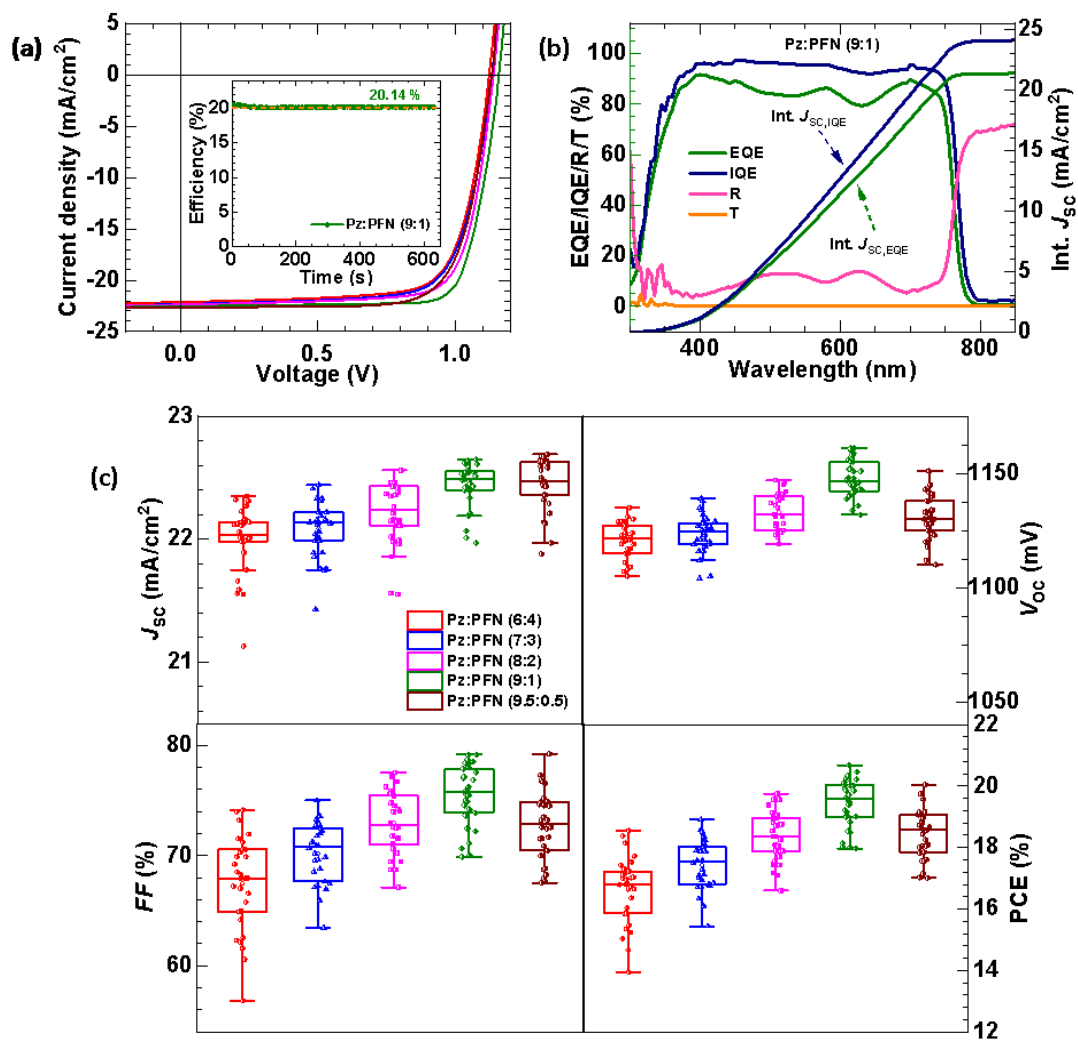

**Figure 10:** (Colour line figures – Pz:PFN (6:4) is red colour, Pz:PFN (7:3) is blue colour, Pz:PFN (8:2) is magenta colour, Pz:PFN (9:1) is olive colour, Pz:PFN (9.5:0.5) is wine colour). **(a)** Current density vs. Voltage ($J-V$) characteristics of the photovoltaic devices under 1-Sun (100 mW/cm²) condition in the forward scan direction. The forward and reverse $J-V$ scans of the representative devices are shown in Figure 11b. The inset figure represents the power output of the Pz:PFN (9:1) PSC device taken under maximum power point tracking with a stabilized efficiency is 20.14% (at 970 mV). **(b)** IPCE spectrum of the best performed Pz:PFN (9:1) PSC including the reflection (R), transmission (T), and integrated current density (Int. $J_{SC}$). **(c)** Boxplot of short circuit current density ($J_{SC}$), open circuit voltage ($V_{OC}$), fill factor (FF), and power conversion efficiency (PCE%) over 30 devices.

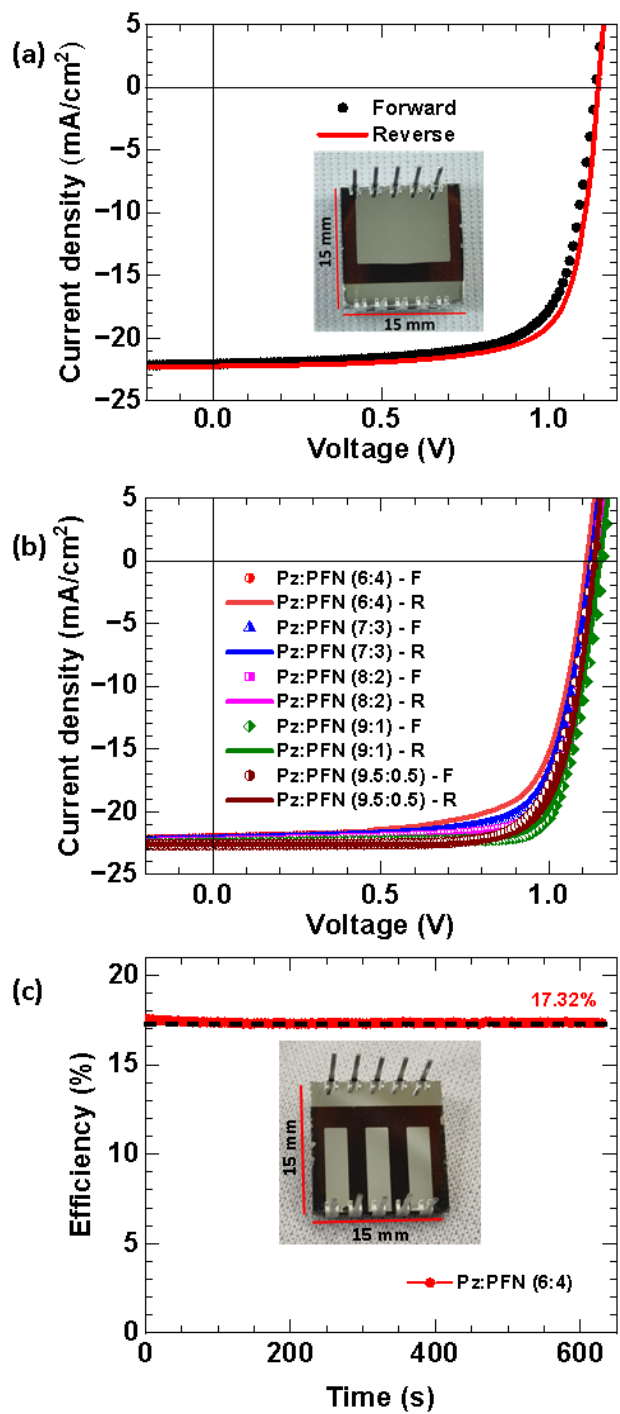

**Figure 11:** *(a)* $J-V$ *characteristics of the photovoltaic devices under 1-Sun (100 mW/cm$^2$) condition in the forward and reverse scan direction for 0.805 cm$^2$ active area (11.5 mm x 7 mm) PSC device. The inset figure represents the PSC of active area 0.805 cm$^2$. (b) $J-V$ characteristic curves of photovoltaic devices device under AM1.5G illumination conditions employing different mixing ratios of Pz:PFN HTL, under forward and reverse scan for 0.175 cm$^2$ active area (2.5 mm x 7 mm) PSC device. (c) The power output of the PSCs taken under maximum power point tracking for Pz:PFN (6:4) with a stabilized efficiency of 17.32% (at 920 mV). The inset figure represents the PSC of active area 0.175 cm$^2$.*

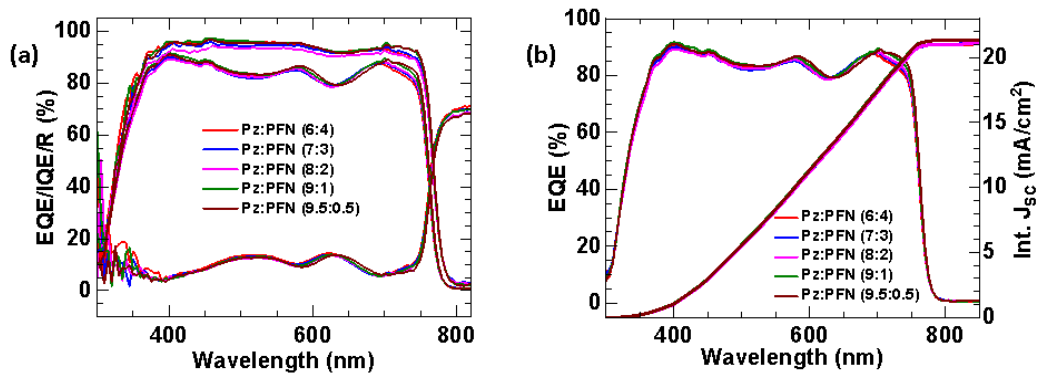

***Figure 12:*** *(**a**) EQE/IQE & reflectance (R) spectra and (**b**) EQE & integrated $J_{SC}$ of PSC devices employing different mixing ratios of Pz:PFN HTLs.*

Following are the equations used to calculate the integrated current density (Int. $J_{SC}$) from the EQE and IQE spectra to calculate the loss in the current due to reflection.

$$IQE(\%) = \frac{EQE\ (\%)}{1 - 0.01 * R(\%) - 0.01 * T(\%)} \qquad (5.1)$$

$$IQE(\%) = \frac{EQE\ (\%)}{1 - 0.01 * R(\%)} \qquad (5.2)$$

$$J_{SC}(EQE) = q \int_{300\ nm}^{850\ nm} \phi(\lambda)EQE(\lambda)d\lambda \qquad (5.3)$$

$$J_{SC}(IQE) = q \int_{300\ nm}^{850\ nm} \phi(\lambda)IQE(\lambda)d\lambda \qquad (5.4)$$

where T (%) and R (%) are the transmission and reflection spectra in percentage. $\phi(\lambda)$ represents the 1-Sun spectrum, and $q$ is the elementary charge.

T (%) is <1%, so it can be neglected in the equation (5.1) and represents equation (5.2).

The int. $J_{SC,EQE}$ is 24.28 mA/cm$^2$ and int. $J_{SC,EQE}$ is 21.41 mA/cm$^2$. The loss in the current density ($J_{SC,EQE}$) is due to reflection from the opaque device. This is a significant loss that we can overcome by the anti-reflection (ARC) coating.

**Table 3:** *Photovoltaic device parameters, integrated $J_{SC}$ (measured from IPCE), and hysteresis index of Pz:PFN HTLs over the active area of 0.175 cm² and 0.805 cm² derived from Figure 11 and Figure 12. Where F and R represent forward and reverse scan directions respectively.*

| HTL | $J_{SC}$ (mA/cm²) | $V_{OC}$ (V) | FF (%) | PCE (%) | Int. $J_{SC}$ (mA/cm²) | Hysteresis index |
|---|---|---|---|---|---|---|
| Pz:PFN (6:4) –F | 22.12 | 1123 | 73.99 | 18.38 | 21.11 | 6.92 |
| Pz:PFN (6:4) –R | 22.01 | 1111 | 70.30 | 17.19 | | |
| Pz:PFN (7:3) –F | 22.33 | 1134 | 73.33 | 18.57 | 21.14 | 3.05 |
| Pz:PFN (7:3) –R | 22.06 | 1123 | 72.74 | 18.02 | | |
| Pz:PFN (8:2)–F | 22.37 | 1137 | 76.23 | 19.39 | 21.13 | 0.87 |
| Pz:PFN (8:2) –R | 22.38 | 1139 | 76.73 | 19.56 | | |
| Pz:PFN (9:1) –F | 22.54 | 1159 | 79.12 | 20.67 | 21.41 | 2.22 |
| Pz:PFN (9:1) –R | 22.65 | 1151 | 77.56 | 20.22 | | |
| Pz:PFN (9.5:0.5) –F | 22.64 | 1134 | 74.67 | 19.17 | 21.43 | 2.89 |
| Pz:PFN (9.5:0.5) –R | 22.67 | 1133 | 76.85 | 19.74 | | |
| Pz:PFN (9:1) –F (0.805 cm²) | 22.06 | 1137 | 72.12 | 18.09 | 21.25 | 5.09 |
| Pz:PFN (9:1) –R (0.805 cm²) | 22.25 | 1145 | 74.81 | 19.06 | | |

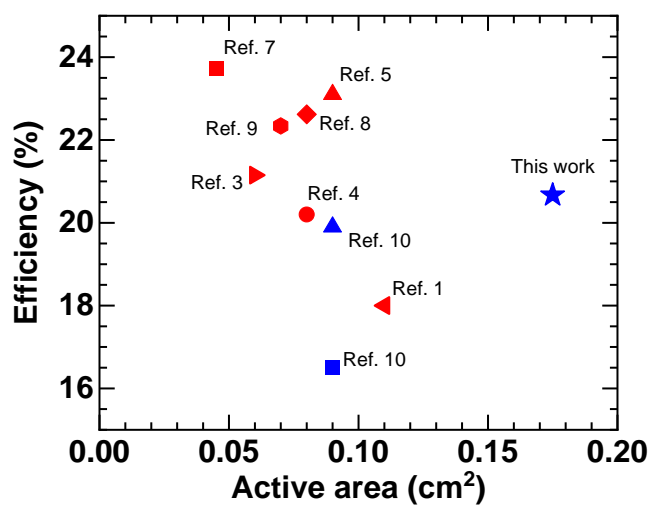

**Figure 13:** *The performance of the solar cells corresponds to the active area of the device and has a bandgap of the perovskite layer of $E_g \sim (1.60 \pm 0.04)$ eV.*

**Table 4:** *List of the PV parameters of PSCs for bandgap $E_g \sim (1.60 \pm 0.04)$ eV with corresponding active area of the device and the hole transport layer used.*

| SL. #. | Bandgap (eV) | *HTLs* | $J_{SC}$ (mA/cm²) | $V_{OC}$ (V) | *FF* (%) | PCE (%) | Active area (cm²) | ref. in this chapter | ref. in the Figure 13 |
|--------|--------------|--------|-------------------|--------------|----------|---------|-------------------|---------------------|----------------------|
| 1. | ~ 1.60 | PEDOT:PSS | 20.11 | 1.11 | 80.6 | 18.0 | 0.11 | [32] | 1 |
| 2. | ~ 1.60 | P3CT-Rb | 21.67 | 1.144 | 82.78 | 20.52 | - | [33] | 2 |
| 3. | ~ 1.60 | PTAA | 20.72 | 1.23 | 83 | 21.15 | 0.06 | [34] | 3 |
| 4. | ~ 1.60 | PTAA | 22.61 | 1.091 | 82 | 20.20 | 0.08 | [35] | 4 |
| 5. | ~ 1.59 | PTAA/PEAI | 24.51 | 1.15 | 82.1 | 23.1 | 0.09 | [36] | 5 |
| 6. | ~ 1.62 | PTAA/PEAI | 21.00 | 1.167 | 82 | 20.1 | - | [37] | 6 |
| 7. | ~ 1.60 | PTAA | 24.13 | 1.162 | 84.6 | 23.72 | 0.045 | [5] | 7 |
| 8 | ~ 1.58 | PTAA | 22.89 | 1.169 | 84.55 | 22.62 | 0.073 | [38] | 8 |
| 9 | ~ 1.56 | PTAA | 23.86 | 1.143 | 82 | 22.34 | 0.07 | [39] | 9 |
| 10 | ~ 1.58 | Me-4PACz/Al₂O₃ | 23.0 | 1.09 | 79.4 | 19.9 | 0.09 | [40] | 10 |
| 11 | ~ 1.58 | Me-4PACz/ PFN-Br | 20.50 | 1.08 | 74.7 | 16.5 | 0.09 | [40] | 10 |
| 12 | ~ 1.60 | Me-4PACz:PFN-Br | 22.54 | 1.159 | 79.12 | 20.67 | 0.175 | This work | This work |

To verify the reproducibility of the device performance with Pz:PFN mixing engineering strategy, 30 devices of each mixing ratio were fabricated using the device procedure outlined in the experimental section above. **Figure 10(c)** summarizes the distribution of photovoltaic parameters under forward scan. The average performance improved from ~17% for Pz:PFN (6:4) to >20% for Pz:PFN (9:1) ratio case with the most narrow distribution among 30 PSCs, i.e., highly reproducible efficient PSCs. The performance improvement is attributed to the increase in $J_{SC}$, $V_{OC,}$ and $FF$. The stabilized efficiency under maximum power point tracking for the Pz:PFN (9:1) PSC device is 20.14% (at 970 mV) inset **Figure 10(a)** ( and for Pz:PFN (6:4) PSC device is 17.32% (at 920 mV) **Figure 11 (c)** )

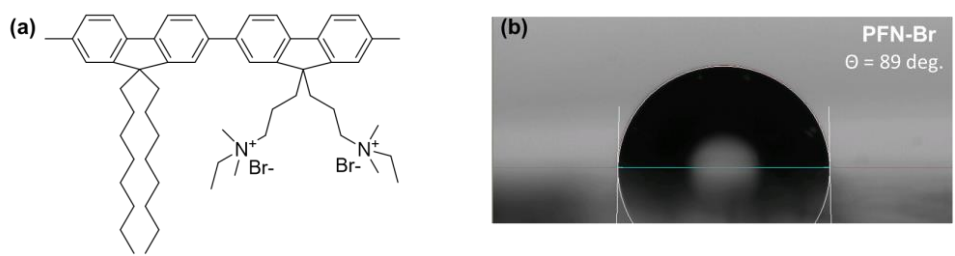

***Figure 14:*** *(**a**) Structure of PFN-Br showing non-polar groups responsible for hydrophobic nature and (**b**) water contact angle showing hydrophobicity of PFN-Br.*

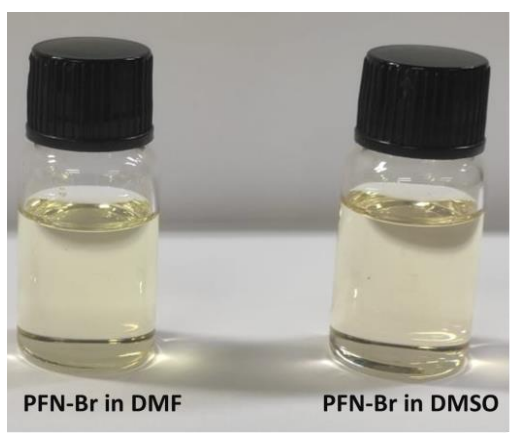

***Figure 15:*** *Solubility of PFN-Br in DMF and DMSO solvents.*

### 5.3.3 Understanding of the buried interface of perovskite solar cells

Numerous research reports have employed PFN-Br as an interlayer to overcome the hydrophobicity of polymeric HTLs such as PTAA to obtain a uniform perovskite film and even as an additive in the perovskite solution to improve its performance.[15,28,41] On the other hand, by looking at the structure of PFN-Br (**Figure 14a**), one can see the presence of a non-polar long alkyl chain. This can cause severe wetting issues and, therefore, raises similar concerns about forming a non-uniform perovskite layer. Moreover, the high water contact angle of PFN-Br, as shown in **Figure 14b**, corroborates the hypothesis. However, a uniform perovskite layer was formed after the incorporation of PFN-Br. This raises the point of how the perovskite layer is formed when a highly hydrophobic PFN-Br is introduced. In literature, the role of PFN-Br in helping to form a uniform perovskite layer has not been focused on yet. For instance, Wang *et al.* incorporated PFN-Br interlayer in between the PTAA HTL and perovskite absorber layer and assumed the possibility of PFN-Br redissolution during the deposition of the perovskite layer, thereby improvement in the device performance.[41] Therefore, it is of utmost importance to understand the underlying mechanism of a uniform perovskite layer formation on PFN-Br interlayer or mixed Pz:PFN HTL. We noticed that PFN-Br is readily soluble in DMF and DMSO solvents (**Figure 15**) and allows us to make good quality films as shown in **Figure 8** and domain analysis in **Figure 16,** with almost similar average domain sizes. Further, we deposited MAPbI$_3$ (without Br$^-$) independently on MeO-2PACz and Pz:PFN-coated glass substrate. The MAPbI$_3$ perovskite film deposited on MeO-2PACz and Pz:PFN (9:1) is indistinctly uniform in **Figure 17**.

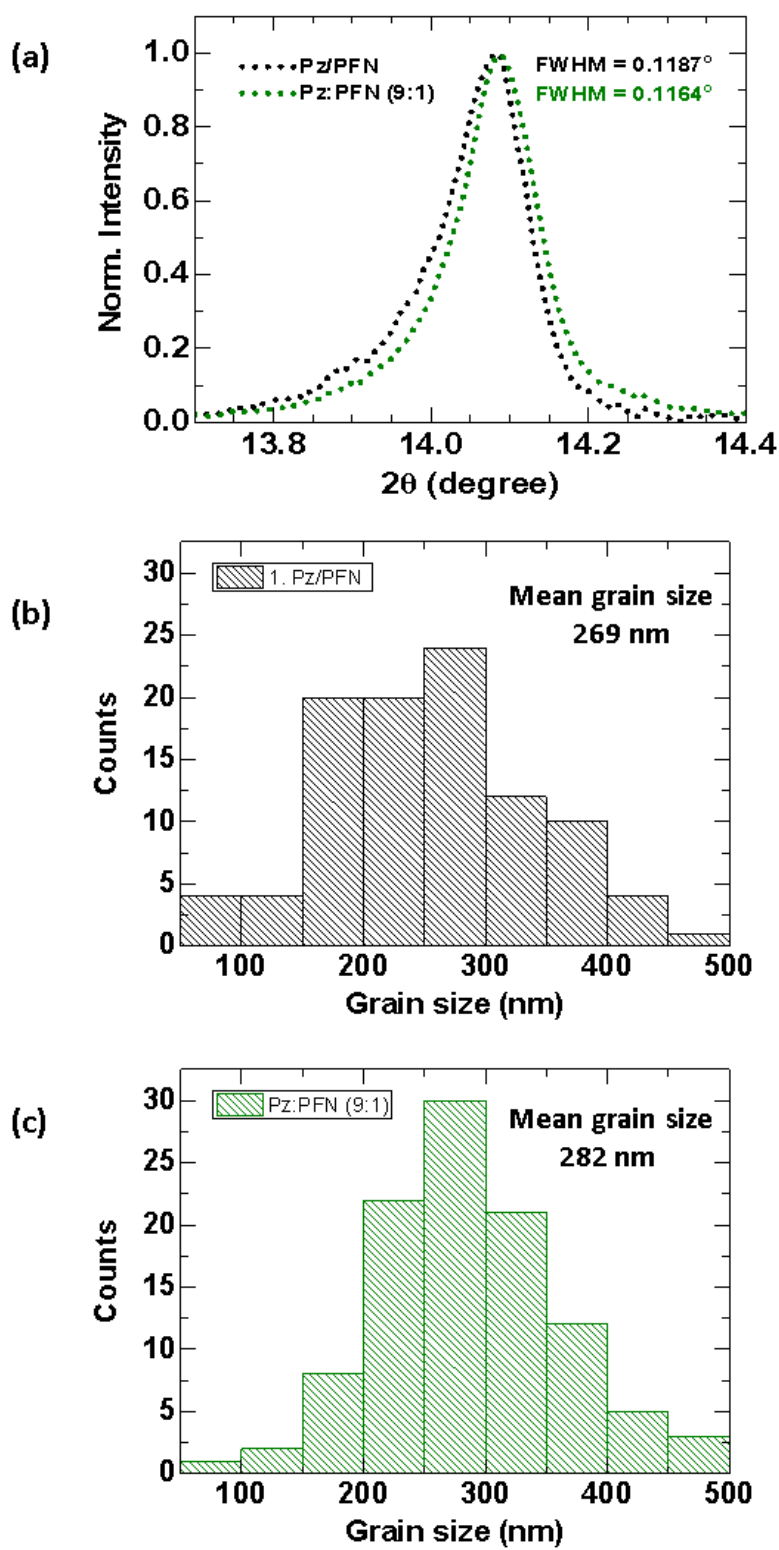

**Figure 16: (a)** *XRD pattern of normalized (110) peak of the CsFAMA perovskite film deposited on bilayer Pz/PFN and Pz:PFN(9:1) coated substrates. The grain size was calculated from the FESEM images of the perovskite films (using Image-J software) deposited on bilayer* **(b)** *Pz/PFN and* **(c)** *Pz:PFN(9:1) coated substrates.*

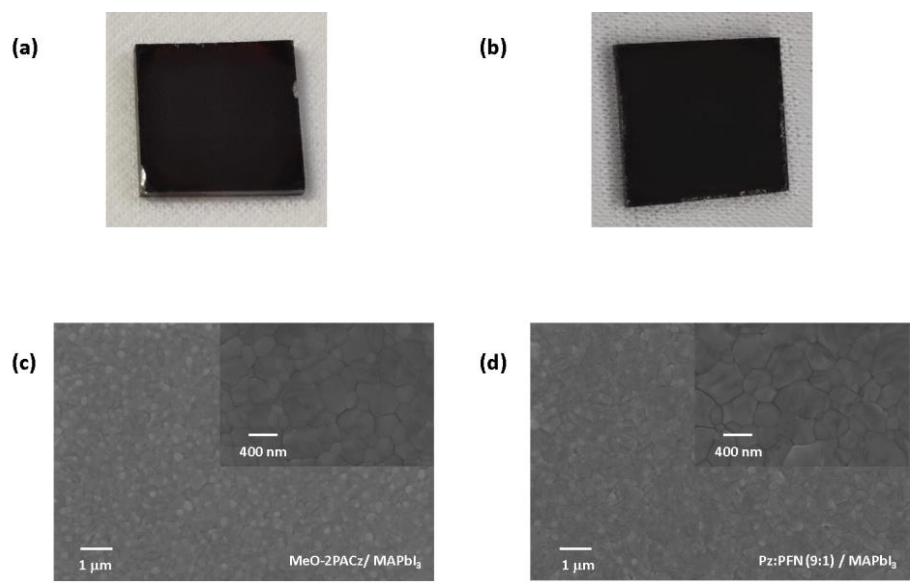

***Figure 17:*** *Photographic image of the MAPbI₃ perovskite film deposited on the (**a**) MeO-2PACz and (**b**) Pz:PFN (9:1) HTLs coated glass substrates. FESEM images of the MAPbI₃ perovskite film deposited on the (**c**) MeO-2PACz and (**d**) Pz:PFN (9:1) HTLs coated glass substrates.*

The XRD diffractograms were carried out for both films as shown in **Figure 18a**. The XRD pattern of MAPbI₃ deposited on Pz:PFN HTL showed a peak with lower full width at half maximum (FWHM). A slight shift (~0.0024°) in XRD peak is observed between MAPbI₃ film deposited on MeO-2PACz coated glass substrate *vs* MAPbI₃ film on Pz:PFN (9:1) coated glass substrate. The reason for this small shift is investigated via spectroscopic technique as the spectroscopic probe is known to be more sensitive than the structural probe. As per previous reports by us and others, micro-structure synchrotron studies were found to be less conclusive than spectroscopic studies, where spectroscopy could provide essential insight to correlate with solar cells performance.[42,43] These samples are tested for PL spectroscopy. The raw PL data showed small peak-shift and FWHM differences. FWHM of MAPI film being broad on MeO-2PACz substrates could be explained based on the FWHM of the XRD peak. However, peak-shift which is found to be red-shifted in raw data, interestingly overlapped once an O.D. (or self-absorption) correction is introduced **Figure 18 (b)** and **Figure 19.**[44]

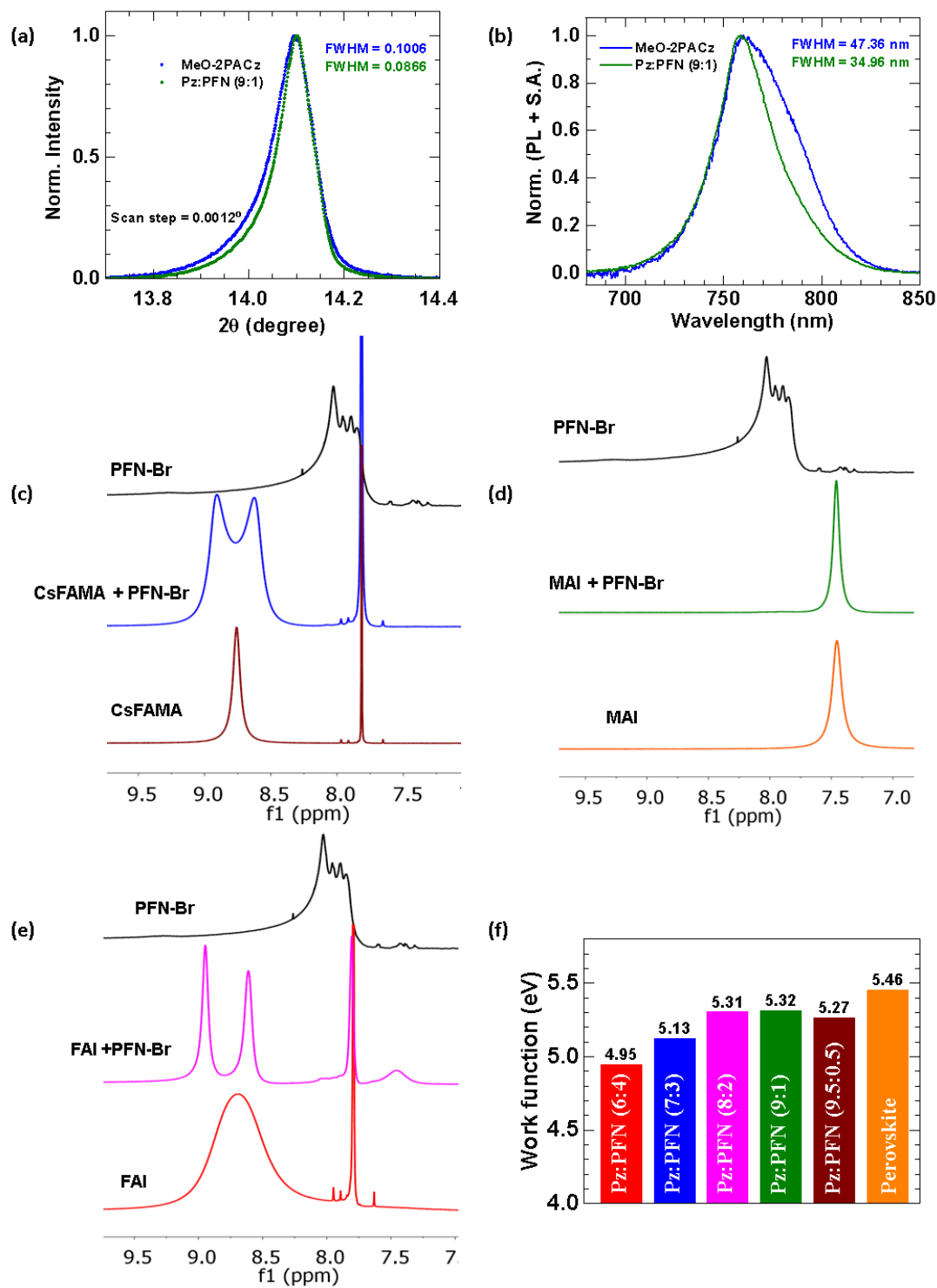

**Figure 18: (a)** *XRD pattern and* **(b)** *PL spectra (self-absorption corrected) of the MAPbI₃ perovskite thin film ( thickness ~ 250 nm ) deposited on the MeO-2PACz and Pz:PFN(9:1) coated glass substrates.*[44] **(c)** *¹H NMR of the CsFAMA, CsFAMA +PFN-Br and PFN-Br solution prepared in DMSO d6 solvent.* **(d)** *¹H NMR of the MAI, MAI +PFN-Br and PFN-Br solution prepared in DMSO d6 solvent.* **(e)** *¹H NMR of the FAI, FAI +PFN-Br, and PFN-Br solution prepared in DMSO d6 solvent.* **(f)** *The work function of the Pz:PFN HTLs was measured using KPFM study.*

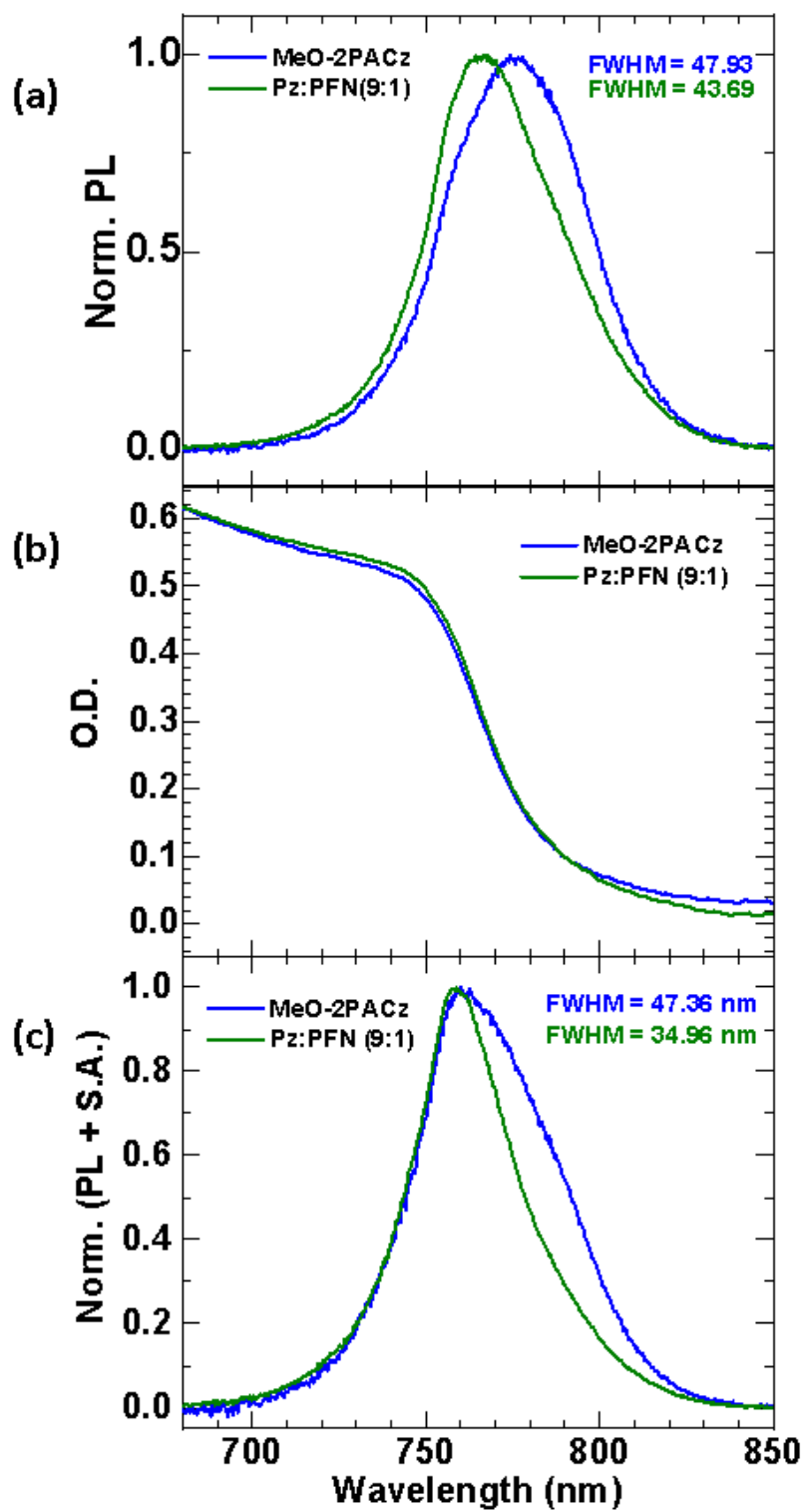

**Figure 19:** *(a) The photoluminescence (PL) spectra and the (b) absorption spectra of the MAPbI₃ perovskite film deposited on the MeO-2PACz and Pz:PFN (9:1) HTL. (c) The PL spectra after self-absorption correction of the MAPbI₃ perovskite films deposited on the MeO-2PACz and Pz:PFN (9:1) HTLs.*

This suggests that bromine containing Pz:PFN is not incorporate the Br$^-$ ion into the absorber layer within the sensitivity limits of used structural and spectroscopic probes. A small peak shift on the higher degree side in the XRD peak falls near the resolution limit of the instrument, which is 0.0012°.

To further understand the interaction between PFN-Br and perovskite, a series of liquid-state $^1$H nuclear magnetic resonance (NMR) measurements in solution (deuterated DMSO) were performed, and the results are shown in **Figure 18c – e.** Initially, an $^1$H NMR spectrum of triple-cation perovskite solution dissolved in deuterated DMSO was recorded (**Figure 18c**), showing a resonance peak at around 9 ppm, which corresponds to protons bound to the nitrogen atoms of formamidinium (HC(NH$_2$)$_2^+$ or FA$^+$) and methylammonium (CH$_3$NH$_3^+$ or MA$^+$). Upon addition of PFN-Br into the perovskite solution in deuterated DMSO, the peak at around 9 ppm splits into two new peaks due to interaction between cationic HC(NH$_2$)$_2^+$ (FA$^+$)/CH$_3$NH$_3^+$ (MA$^+$) and PFN-Br. To further examine the interaction between MA$^+$ and PFN-Br, an additional $^1$H NMR experiment of only MAI without and with PFN-Br was carried out. As shown in **Figure 18d**, no peaks are split and shifted, indicating that the peak split in the perovskite solution with PFN-Br addition (Figure 18c) is related to the change in the FA$^+$ cation in the presence of PFN-Br. To further confirm this conclusion, we recorded a $^1$H NMR spectrum of only FAI without and with PFN-Br. As shown in **Figure 18e**, this peak around 9 ppm of FAI solution with PFN-Br additive gets split into two peaks in a way similar to the splitting observed in the perovskite solution with PFN-Br additive. The splitting of FA$^+$ protons into two new signals has also been observed by other research groups and is ascribed to the formation of hydrogen bond complexes of the amidinium moiety of the FA$^+$ cation.[45–48] Considering all NMR results, it is thus proposed that a hydrogen bond between FA$^+$ and PFN-Br is existed, which might help in perovskite layer formation during its spin coating step and crystallization and might also remain present in the final perovskite film and can be beneficial for prolong device stability. In addition to the perovskite layer formation with an improved interfacial property, we hypothesized that the addition of PFN-Br might influence the electronic properties of Me-4PACz. Therefore, we measured the work function of Me-4PACz (without and with PFN-Br different ratios), and perovskite using Kelvin Probe Force Microscopy (KPFM) technique. The work function of the films is determined using gold as a reference and the results are shown in **Figure 18f, Figure 20, and Figure 21**.

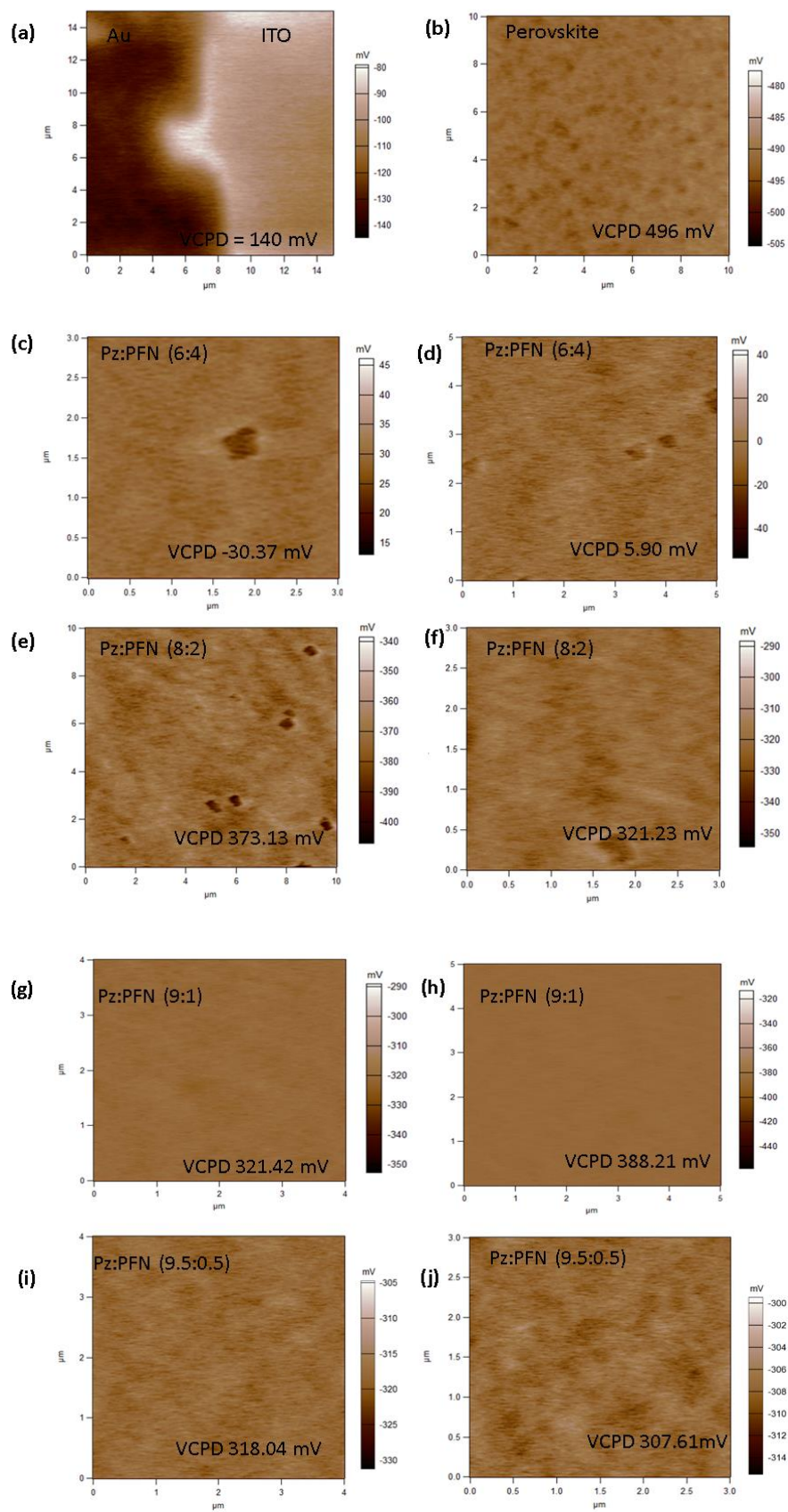

***Figure 20:*** ***(a)*** *The KPFM scanning image of a standard gold (Au) sample deposited on ITO substrate.* ***(b)*** *KPFM scanning image of perovskite film deposited on the ITO substrate. The*

*work function of the Pz:PFN HTL at two different locations and the average of them are listed in Table S5. All the Pz:PFN HTL samples were deposited on the ITO substrates. (**c**) and (**d**) are the KPFM scanning image at two different points for the Pz:PFN (6:4). (**e**) and (**f**) are the KPFM scanning image at two different points for the Pz:PFN (8:2). (**g**) and (**h**) are the KPFM scanning image at two different points for the Pz:PFN (9:1). (**i**) and (**j**) are the KPFM scanning images at two different points for the Pz:PFN (9.5:0.5).*

The work function of the samples is calculated using the following formula.

$$V_{CPD} = \frac{WF_{tip} - WF_{sample}}{-e} \qquad (5.5)$$

Where $WF_{tip}$ is the work function of the tip and $WF_{sample}$ is the work function of the sample. $V_{CPD}$ is the contact potential difference.[49,50]

***Table 5:*** *List of the average of the contact potential difference ($V_{CPD}$) for the Au, perovskite, Pz:PFN HTLs, and corresponding work functions.*

| Substrate | Average $V_{CPD}$ (mV) | Work function (eV) |
|---|---|---|
| ITO/Au | 140 | 5.1 |
| ITO/Perovskite | 496 | 5.46 |
| ITO/Pz:PFN (6:4) | -12 | 4.95±0.018 |
| ITO/Pz:PFN(8:2) | 347 | 5.307±0.026 |
| ITO/Pz:PFN(9:1) | 355 | 5.315±0.034 |
| ITO/Pz:PFN(9.5:0.5) | 313 | 5.273±0.005 |

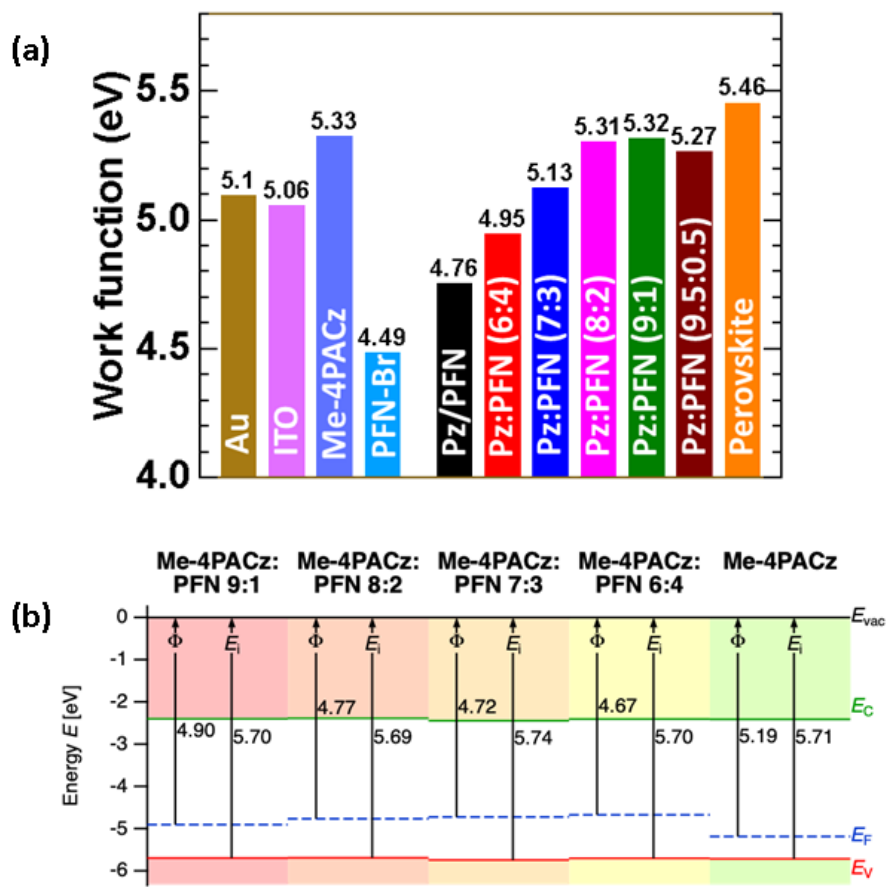

***Figure 21:*** *(**a**) Work function of Me-4PACz without and with PFN-Br interlayer and with different Pz:PFN mixing ratios, perovskite, PFN-Br, ITO, and Au metal electrode measured using KPFM. (**b**) The valence band maximum and the Fermi-level of the Me-4PACz and PFN-Br mixed HTLs are measured with respect to the vacuum using ultraviolet photoelectron spectroscopy (UPS).*

Though the pristine Me-4PACz shows a better energy level alignment with the perovskite layer (**Figure 21a**). However, forming the perovskite layer on pristine Me-4PACs is a challenge. Because of the effect of interface dipoles caused by the electrolyte moiety present in PFN-Br, the work function of Me-4PACz showed dramatic changes with the mixing ratios. For the particular Pz:PFN (9:1) case, the energy level alignment matches closely with the Me-4PACz (without any PFN-Br) and also with the work function of the perovskite layer, implying better hole extraction efficacy. Moreover, we used ultraviolet photoelectron spectroscopy (UPS) on the above mentioned HTLs and found that the Fermi level values are well aligned with KPFM results, as shown in **Figure 21b**. Further, the built-in potential difference (($V_{bi}$) calculated from the dark current characteristics of diodes show a clear trend in reduced barrier-voltage for the case of Pz:PFN (9:1) composition as compared to the other mixing ratio composition, as seen in **Figure 22.** We observed a slight relative change in the EQE and corresponding $J_{SC}$ (whether integrated or measured using an AM1.5G solar simulator light source), which can be seen modulating with respect to ($V_{bi}$ values of diodes. ($V_{bi}$ values modulate with respect to different Fermi-levels of ITO/(Pz:PFN::X:Y) layer, as ETL is common in all diodes (Figure 21).[51] This also explains the reason behind the high device efficiency Pz:PFN (9:1) case compared to the other cases despite similar perovskite crystal structure and top surface morphology.

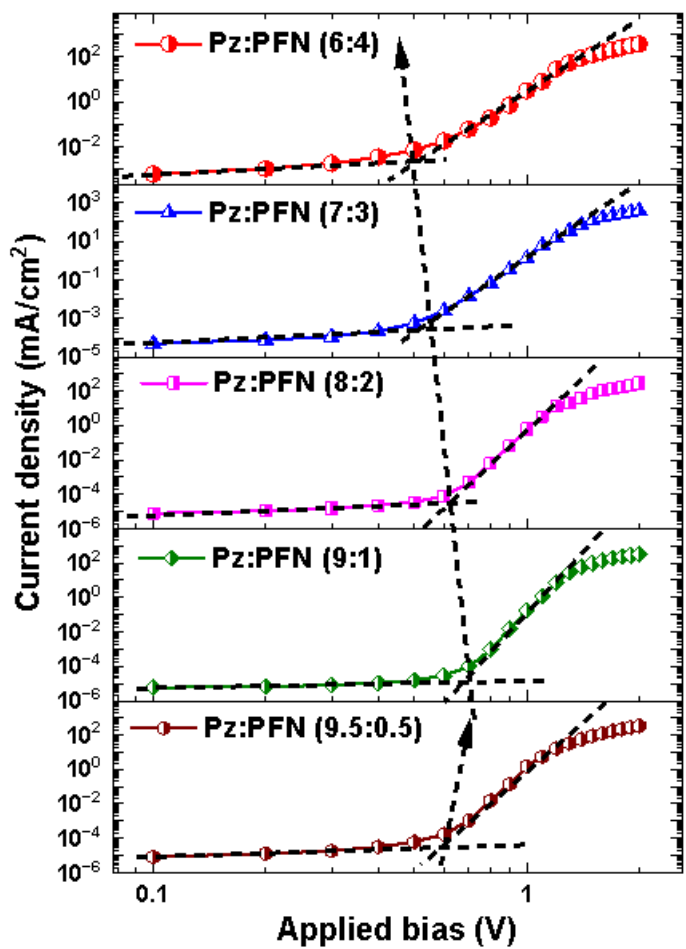

**Figure 22:** *The built-in potential ($V_{bi}$) is estimated from the dark current-voltage characteristics for the diode of ITO/Pz:PFN(X:Y)/Perovskite/PCBM/BCP/Ag.*

**5.3.4 Device physics of perovskite solar cells**

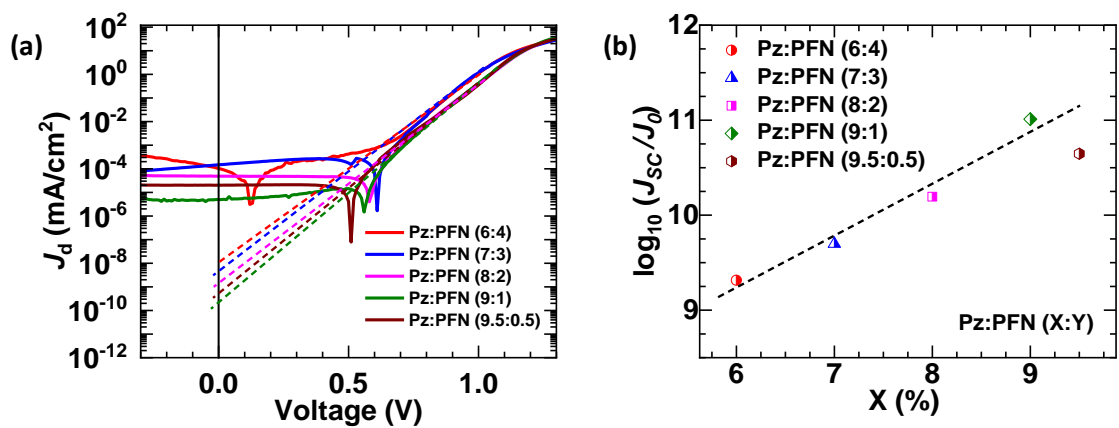

***Figure 23:*** *(a) The dark current density plotted in semi-log scale as a function of applied bias. (b) The log ($J_{SC}/J_0$) vs. mixing ratio Pz:PFN (X:Y), where X and Y represent the volume ratio for Me-4PACz and PFN-Br respectively.*

***Table 6:*** *The $J_d$ and $J_0$ values of all the devices with different Pz:PFN mixing ratios derived from Figure 23a.*

| HTL | Dark current $J_d$ @ -0.3 (mA/cm$^2$) | $J_0$ (mA/cm$^2$) |
|---|---|---|
| Pz:PFN (6:4) | 3.72 X 10$^{-4}$ | 1.07 X 10$^{-8}$ |
| Pz:PFN (7:3) | 8.04 X 10$^{-5}$ | 4.47 X 10$^{-9}$ |
| Pz:PFN (8:2) | 5.10 X 10$^{-5}$ | 1.43 X 10$^{-9}$ |
| Pz:PFN (9:1) | 5.28 X 10$^{-6}$ | 2.20 X 10$^{-10}$ |
| Pz:PFN (9.5:0.5) | 2.04 X 10$^{-5}$ | 5.10 X 10$^{-10}$ |

The dark current measurement can also give insights into the recombination at the perovskite and Pz:PFN interface by an estimation of reverse saturation current density $J_0$.[52] Therefore, we measured the dark current of the devices with different ratios of Pz:PFN and observed the lowest dark current density $J_d$ at –0.3 V of $5.28 \times 10^{-6}$ mA/cm$^2$ and the reverse saturation current density $J_0$ of $2.20 \times 10^{-10}$ mA/cm$^2$ for the Pz:PFN (9:1) (**Figure 23**) whereas a typical $J_0$ value for established Silicon photovoltaics is $10^{-10} \sim 10^{-9}$ mA/cm$^2$.[53,54] The $J_d$ and $J_0$ values of all the devices with different Pz:PFN mixing ratios are tabulated in **Table 6**.

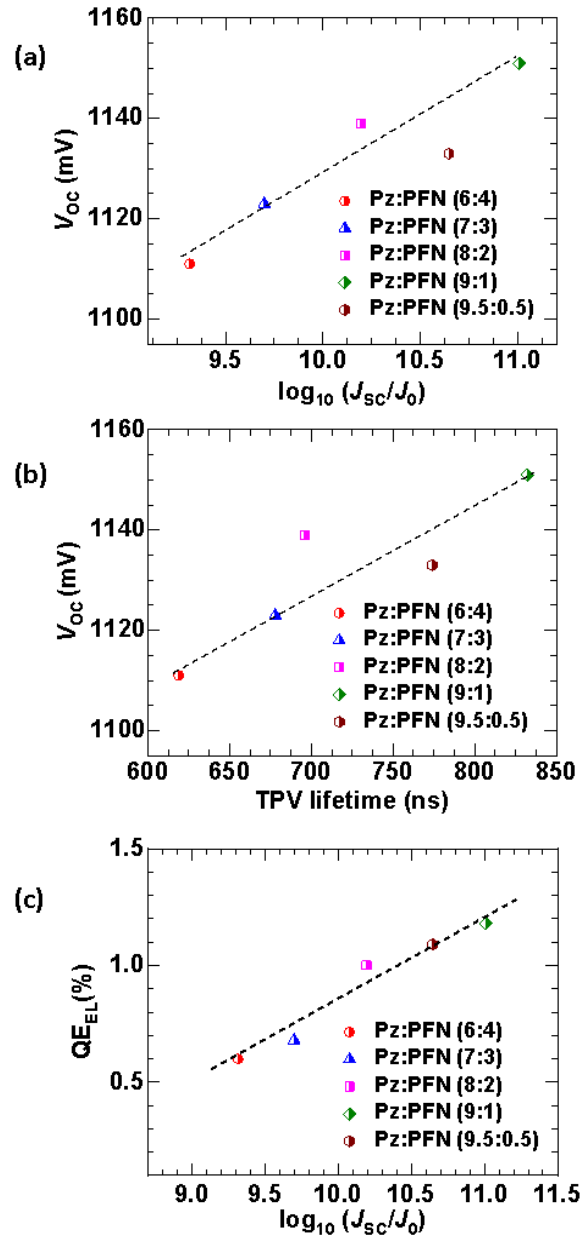

**Figure 24:** (a) The $V_{OC}$ of the representative devices as a function of short circuit ($J_{SC}$) and reverse saturation current density ($J_0$). (b) The $V_{OC}$ of the representative devices as a function of the perturbed charge carrier lifetime measured from TPV measurement at 1-Sun illumination condition. (c) Electroluminescence (EL) quantum efficiency plotted as a function of $J_{SC}$ and $J_0$.

The lowest $J_d$ and $J_0$ values for the 9:1 case imply suppressed recombination at the perovskite/ Pz:PFN interface. **Figure 24a** represents the variation in $V_{OC}$ *vs.* $\log_{10}(J_{SC}/J_0)$, where $J_{SC}$ and $J_0$ are calculated from the illuminated and dark currents explained above. The $\log_{10}(J_{SC}/J_0)$ is higher for the Pz:PFN (9:1) (**Figure 23b**) and the corresponding $V_{OC}$ is higher as per the basic diode equation.[55]

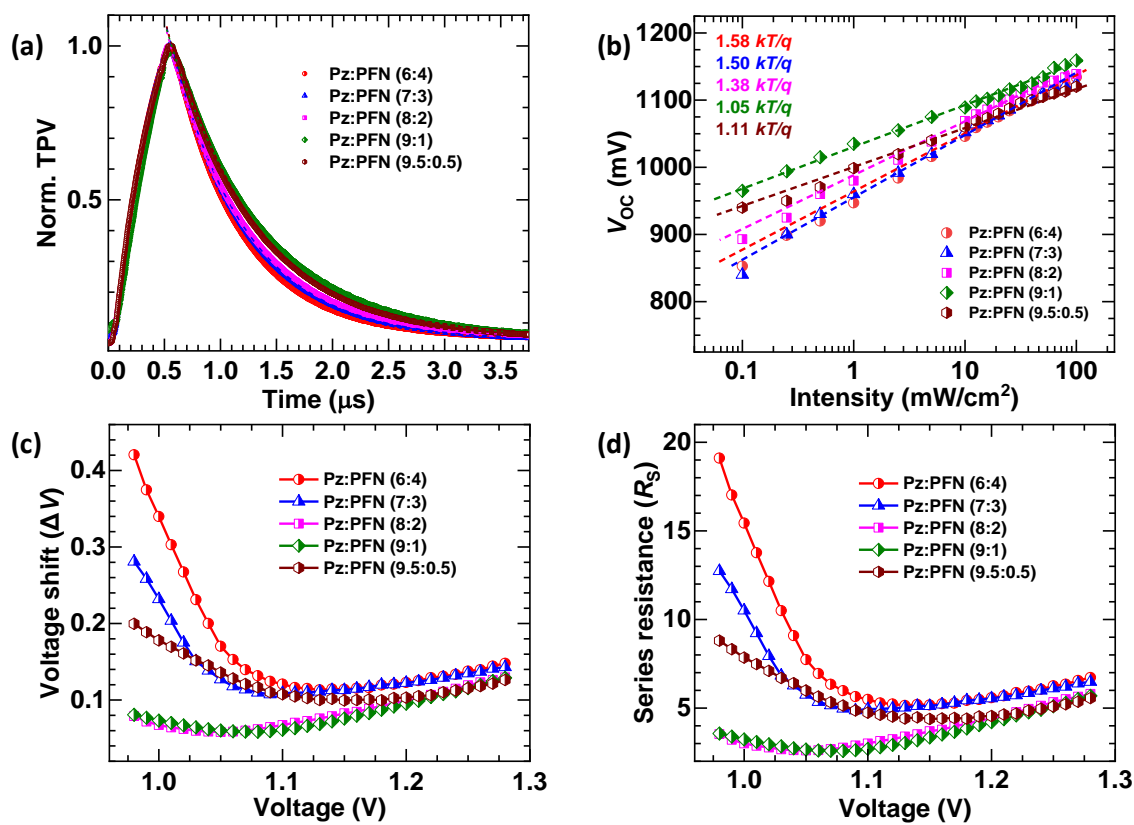

**Figure 25: (a)** *Normalized transient photovoltage (TPV) decay profile of the Pz:PFN based PSCs under 1-Sun light bias.* **(b)** *Intensity dependent $V_{OC}$ (please see Figure 26) of Pz:PFN HTL based PSCs, solid line is fit to the Shockley diode equation.[55]* **(c)** *The voltage shift (ΔV) (please see Figure 27) of the illuminated current and dark shifted current.* **(d)** *The series resistance calculated from the voltage shift by $R_S = \frac{\Delta V}{J_{SC}}$.* [56,57]

***Table 7:*** *The charge carrier lifetime ($\tau$) derived from TPV studies and ideality factor (n) are calculated from the intensity dependent studies $V_{OC}$ (see Figure 25a, & b)*

| HTL | $\tau$ TPV (ns) @ 1-Sun | $n$ (Ideality factor) |
|---|---|---|
| Pz:PFN(6:4) | $619.21 \pm 0.92$ | 1.58 |
| Pz:PFN(7:3) | $678.14 \pm 1.03$ | 1.50 |
| Pz:PFN(8:2) | $696.82 \pm 0.80$ | 1.38 |
| Pz:PFN(9:1) | $832.30 \pm 1.61$ | 1.05 |
| Pz:PFN(9.5:0.5) | $774.42 \pm 1.34$ | 1.11 |

Transient photovoltage (TPV) decay measurements were carried out to determine the charge carrier lifetime and to quantify the recombination process. [58,59] The normalized TPV decay profile, fitted with a mono-exponential decay and measured for all the cases, is shown in **Figure 25a**. Among all the cases studied, the TPV decay profile of the device with Me-4PACz:PFN-Br (9:1) showed a slow decay and a high perturbed charge carrier lifetime ($\tau$) (please refer to **Table 7**), implying suppressed non-radiative recombination.[60,61] This can be further correlated with the device $V_{OC}$. The $V_{OC}$ of the device is calculated as a function of the perturbed charge carrier lifetime ($\tau$) and high $V_{OC}$ as in the case of 9:1 case (**Figure 24b**); implying the lower recombination. The dependence of $V_{OC}$ on the incident light fluence was investigated for all the Pz: PFN-based devices to get further insight into recombination processes. The incident light varies from 100 mW/cm$^2$ (1 Sun) to 1 mW/cm$^2$ (0.01 Sun) with a set of neutral density (ND) filters. The raw data of the intensity dependent $V_{OC}$ for all the Pz:PFN HTLs are shown in **Figure 26.** The legends ND 0.0, 0.1, 0.2, 0.3, and so on indicate that the 1-Sun light is falling on the device through different ND filters of optical density ND 0.0, 0.1, 0.2, 0.3, and so on. The ND 0.0 indicates that there is no ND filter i.e. zero optical density and in that case, the intensity of light is 1-Sun and lower of higher NDs.

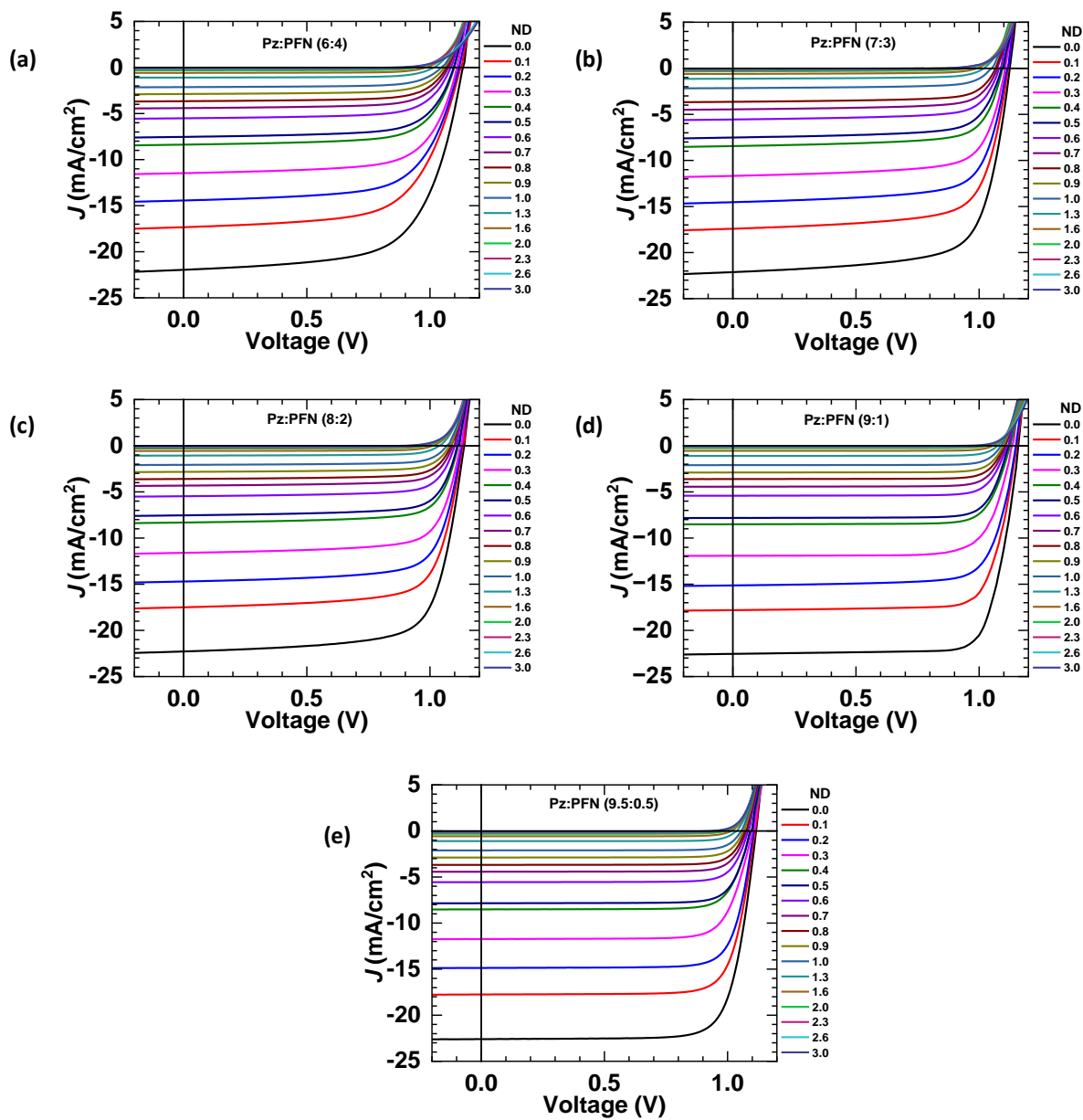

**Figure 26:** *Raw data plot of Figure 25b. Intensity dependent J − V characteristics of Pz:PFN HTL based PSCs of (**a**) Pz:PFN (6:4), (**b**) Pz:PFN (7:3), (**c**) Pz:PFN (8:2), (**d**) Pz:PFN (9:1), and (**e**) Pz:PFN (9.5:0.5). The ND 0.0 indicates 1-Sun light intensity and the rest are indicating 1-Sun light is passing through ND filters of optical density 0.1, 0.2,0.3, etc.*

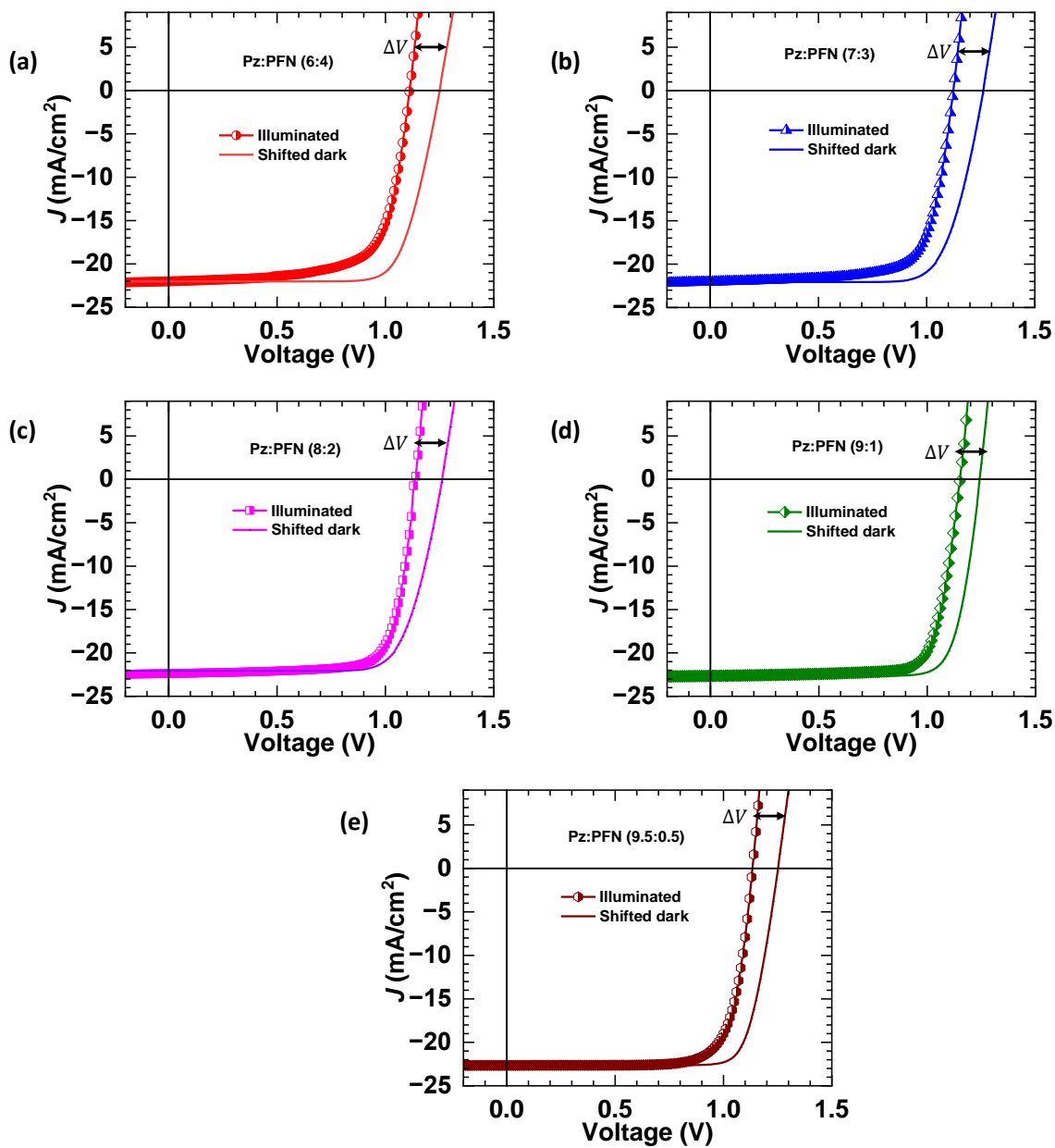

**Figure 27:** *Raw data plot of Figure 25c. The scattered point with the line indicates the illuminated current, and the solid line plot indicates the dark current shifted by $J_{SC}$ and merged with the illuminated $J_{SC}$ for the Pz:PFN HTL based PSCs of (a) Pz:PFN (6:4), (b) Pz:PFN (7:3), (c) Pz:PFN (8:2), (d) Pz:PFN (9:1), and (e) Pz:PFN (9.5:0.5). The difference between the voltage for the same current is represented by $\Delta V$ (see Figure 25c) and used to calculate the series resistance (see Figure 25d) from the voltage shift by $R_S = \frac{\Delta V}{J_{SC}}$.* [56,57]

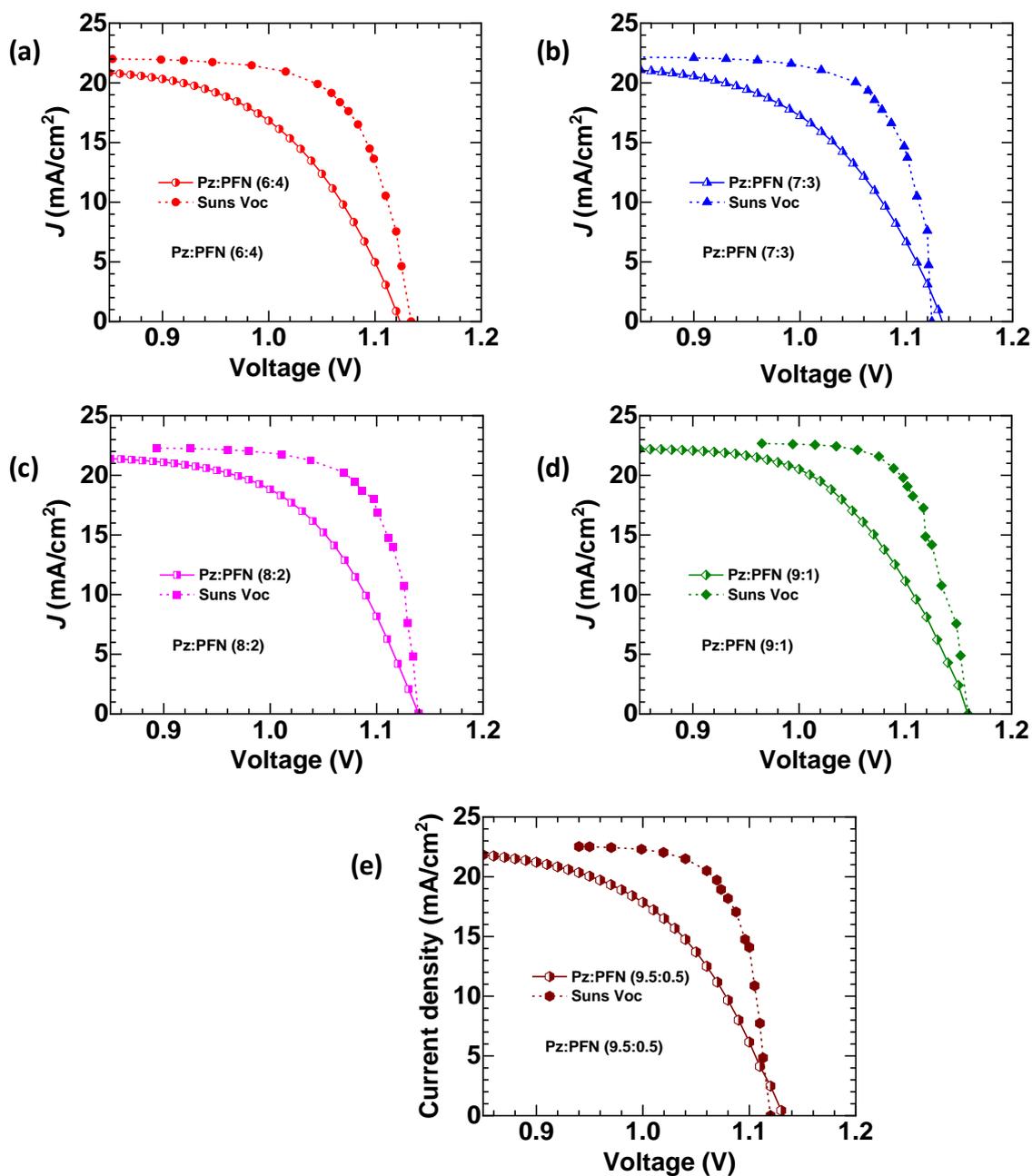

**Figure 28:** *The plots represent the pseudo J − V (solid point) measured from Suns-$V_{OC}$ and real J − V (half solid points) at the 1-Sun condition for the Pz:PFN HTL based PSCs of (**a**) Pz:PFN (6:4), (**b**) Pz:PFN (7:3), (**c**) Pz:PFN (8:2), (**d**) Pz:PFN (9:1), and (**e**) Pz:PFN (9.5:0.5).*[62]

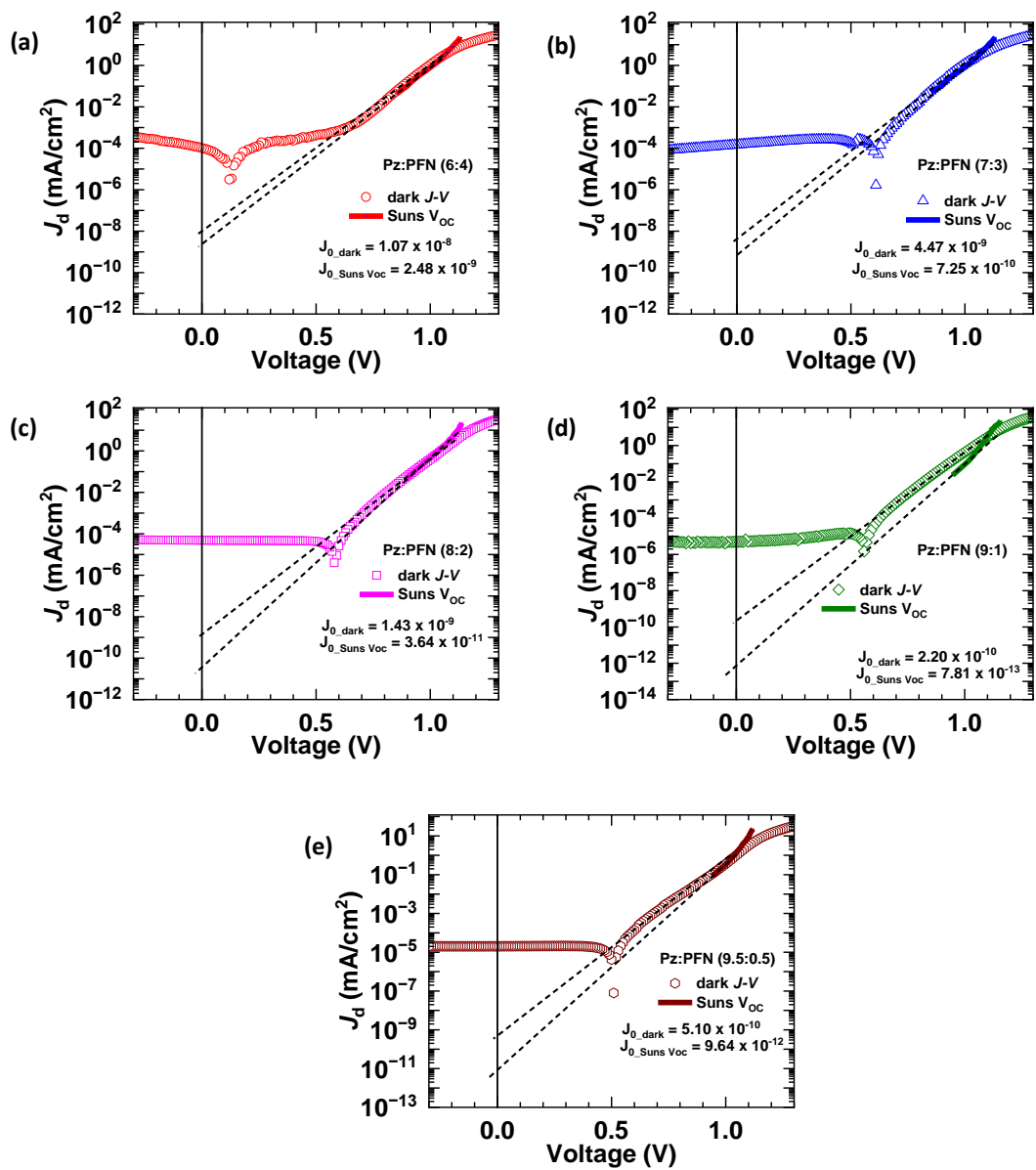

**Figure 29:** *The reverse saturation current density is calculated from the dark current ($J_d$) i.e. $J_{0,d}$ and from pseudo $J − V$ i.e. $J_{0,sv}$ which is plotted in a semi-log scale along with the dark current. The pseudo $J − V$ is measured from the Suns-$V_{OC}$ measurement for the Pz:PFN HTL based PSCs of (**a**) Pz:PFN (6:4), (**b**) Pz:PFN (7:3), (**c**) Pz:PFN (8:2), (**d**) Pz:PFN (9:1), and (**e**) Pz:PFN (9.5:0.5). The $J_{0,d}$ and $J_{0,sv}$ are tabulated in Table 8.* [62–64]

**Table 8:** *The reverse saturation current density calculated from the dark $J_{0,d}$ current Figure 23 and from the Suns-$V_{OC}$ measurement $J_{0,sv}$ Figure 29.*

| Devices | $J_{0,d}$ (mA/cm$^2$) | $J_{0,sv}$ (mA/cm$^2$) |
|---------|-----------------------|------------------------|
| Pz:PFN (6:4) | $1.07 \times 10^{-8}$ | $2.48 \times 10^{-9}$ |
| Pz:PFN (7:3) | $4.47 \times 10^{-9}$ | $7.25 \times 10^{-10}$ |
| Pz:PFN (8:2) | $1.43 \times 10^{-9}$ | $3.64 \times 10^{-11}$ |
| Pz:PFN (9:1) | $2.20 \times 10^{-10}$ | $7.81 \times 10^{-13}$ |
| Pz:PFN(9.5:0.5) | $5.10 \times 10^{-10}$ | $9.64 \times 10^{-12}$ |

The intensity dependent $V_{OC}$ (Figure 25b) of a photovoltaic device can be written as

$$V_{OC} = \frac{nkT}{q} \ln(I) \qquad (5.6)$$

$$V_{OC} = \frac{nkT}{q} \times 2.303 \times \log(I) \qquad (5.7)$$

Where $n$ is the ideality factor, $k$ is the Boltzmann constant, $T$ is the temperature, $q$ is the electronic charge, $I$ is the illumination intensity.[65]

**Figure 25b** plots $V_{OC}$ against intensity on a logarithmic scale, and the slope corresponds to $\frac{nkT}{q}$.[65,66] The slope of the best-performing device, that is, employing Pz:PFN (9:1) gives an $n$ value of 1.05 (Figure 25b) and is smallest when compared to other mixing ratios of Pz:PFN based devices, implying suppressed non-radiative trap assisted recombination. Ideality factor $n$ being close to 1 suggests that the absorber is free from bulk defects, and hence it provides higher $V_{OC}$.[52] Relatively, this higher $V_{OC}$ (free from non-radiative recombination channel) further facilitates charge carrier extraction under short circuit conditions, i.e. $J_{SC}$ for optimal PSCs. The lower value of series resistance ($\sim 4\Omega$ /sq) calculated from the dark and illuminated current, also indicates higher $FF$ and $V_{OC}$ for Pz:PFN (9:1) as shown in **Figure 25 c, d (and Figure 27)**.[56,57] Generally, in practical devices there is a role of the series resistance and results in the illuminated $J - V$ plot shifted towards lower voltage (due to voltage drop) compared to the dark shifted by $J_{SC}$. The separation in the voltage difference signifies the series resistance (Figure 25c) and it is calculated as $R_S = \frac{\Delta V}{Jsc}$.[56,57] Hence, the device performances are limited by the series resistance. If the devices are free from series resistance, then the *FF* will be higher, and the reverse saturation current will be lower, which can be calculated from the Suns-$V_{OC}$ measurement as shown in **Figure 28**.[62] The Suns-$V_{OC}$ method is a characterization technique of solar cells where the effect of series is avoided and the pseudo $J - V$ characteristics can be generated. This technique more properly can be termed as illumination $V_{OC}$.[67] Therefore, we obtained the Suns-$V_{OC}$ pseudo plot from the intensity dependent $J - V$ measurement. The measurement technique and pseudo $J - V$ estimation well explained in chapter 2. Further, the pseudo $J - V$ estimated from the Suns-$V_{OC}$ is shifted towards the dark current by $J_{SC}$ to estimate the reverse saturation current (as previously shown). [56,57]The reverse saturation current density $J_0$ is estimated by fitting the curve using the basic diode equation and its lower for the pseudo $J - V$ plot **Figure 29**. The Suns $V_{OC}$ based pseudo $J - V$ curve showed more than two orders of magnitude reduction in $J_0$ to a value of sub-pA, i.e. 0.78 pA/cm$^2$ **Table 8**.[64]

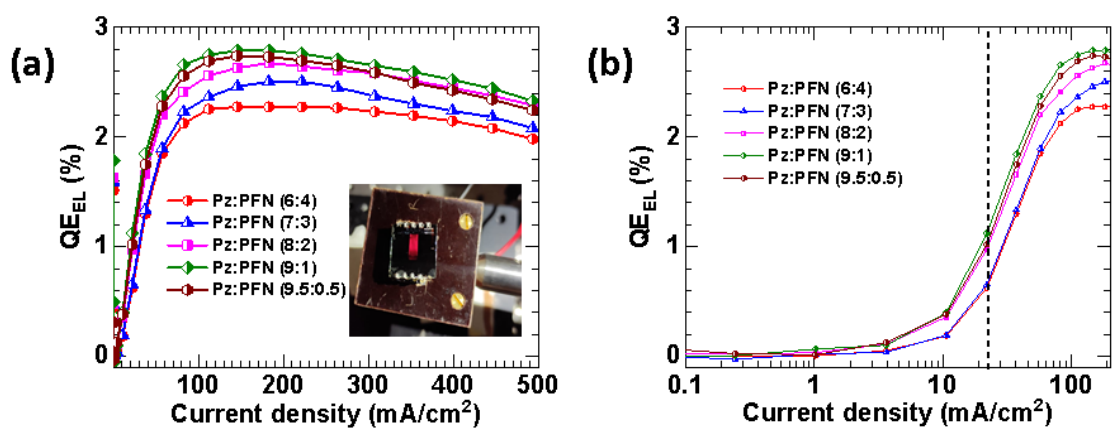

**Figure 30:** **(a)** *The electroluminescence quantum efficiency (QE_EL) is a function of injected current density and the inset represents the digital image of the device during a 5 mA injected current.* **(b)** *The QE_EL at the injected $J_{SC}$ of the representative PSCs.*

**Table 9:** *The QE_EL @ injected current density ($J_{SC}$) is calculated from Figure 30b.*

| HTL | $J_{SC}$ (mA/cm$^2$) | QE_EL @$J_{SC}$ |
|---|---|---|
| Pz:PFN(6:4) | 22.12 | 0.60 |
| Pz:PFN(7:3) | 22.33 | 0.68 |
| Pz:PFN(8:2) | 22.37 | 1.00 |
| Pz:PFN(9:1) | 22.54 | 1.18 |
| Pz:PFN(9.5:0.5) | 22.64 | 1.09 |

We measured the electroluminescence of all the Pz:PFN HTL based PSCs. For electroluminescence measurement, a current is injected into the device, and due to recombination, photons are emitted from the device. The inset of **Figure 30a** represents a device operation under electroluminescence measurement. The electroluminescence quantum efficiency (QE$_{EL}$) was found to be higher for the Pz:PFN(9:1) as shown in **Figure 30** and **Table 9**. **Figure 24c** depicts the (QE$_{EL}$) plotted as a function of log$_{10}$ $(J_{SC}/J_0)$ and it is higher for the Pz:PFN(9:1). This again indicates that the Pz:PFN(9:1) HTL-based PSC interface has relatively lesser trap states which also verified from higher TPV lifetime (indicating the presence of lower trap states).[68,69] A comparison of the PV parameters for the different mixing ratios of Me-4PACz and PFN-Br are tabulated in **Table 10**.

***Table 10:*** *The different mixing ratios of the Me-4PACz and PFN-Br based PSC and corresponding PV parameters e.g. short circuit current $J_{SC}$, shunt resistance $R_{sh}$ (k$\Omega$), and $V_{OC}$ calculated from illuminated $J-V$ plot Figure 10a. The ideality factor (n) was calculated from intensity dependent $J-V$ plot (Figures 25b and 26). TPV lifetime calculated from Figure 25a. QE$_{EL}$ at injected $J_{SC}$ calculated from Figure 30b.*

| HTL | $J_{SC}$(mA/cm$^2$) | $R_{sh}$ (k$\Omega$) | $V_{OC}$ (V) | $n$ (Ideality factor) | $\tau$ TPV (ns) @ 1-Sun | QE$_{EL}$ @$J_{SC}$ |
|---|---|---|---|---|---|---|
| Pz:PFN(6:4) | 22.12 | 1.05 | 1123 | 1.58 | 619.21 | 0.60 |
| Pz:PFN(7:3) | 22.33 | 1.22 | 1134 | 1.50 | 678.14 | 0.68 |
| Pz:PFN(8:2) | 22.37 | 1.45 | 1137 | 1.38 | 696.82 | 1.00 |
| Pz:PFN(9:1) | 22.54 | 2.63 | 1159 | 1.05 | 832.30 | 1.18 |
| Pz:PFN(9.5:0.5) | 22.64 | 2.51 | 1134 | 1.11 | 774.42 | 1.09 |

This table includes all the device parameters to compare the device performance for all the Pz:PFN mixed HTLs-based PSCs. The highest shunt resistance (R$_{sh}$), highest $V_{OC}$ measured from illuminated $J-V$, The lowest ideality factor (*n*) calculated from intensity dependent $J-V$, highest TPV lifetime @1-Sun, highest QE$_{EL}$ @injected $J_{SC}$ for Pz:PFN (9:1) over the other Pz:PFN ratio mixed HTL indicates that the Pz:PFN (9:1) HTL based PSC has suppressed non-diative channels at the interface and as a result improved device performance. Further, the stability measurement of the Pz:PFN (9:1) device was carried out under various conditions.

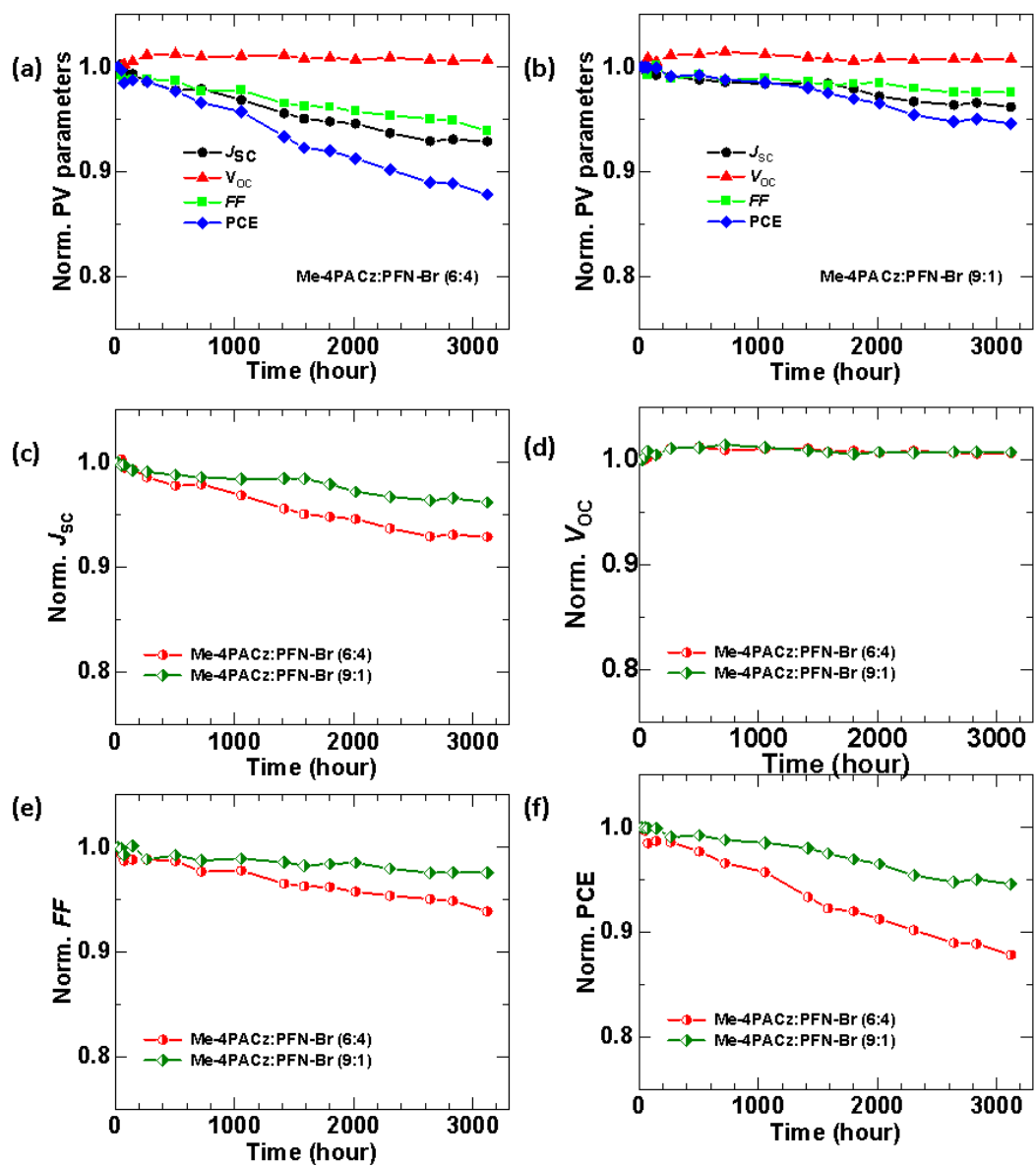

***Figure 31:*** *The J − V scans were taken periodically of the unencapsulated devices, measured in RH ∼ 40% for >3000 hours. The normalized J − V parameters of the (**a**) Pz:PFN (6:4) and (**b**) Pz:PFN (9:1) based PSCs over 3000 hours were plotted. The individual parameters are compared in (**c**) $J_{SC}$ (**d**) $V_{OC}$ (**e**) FF (**f**) PCE.*

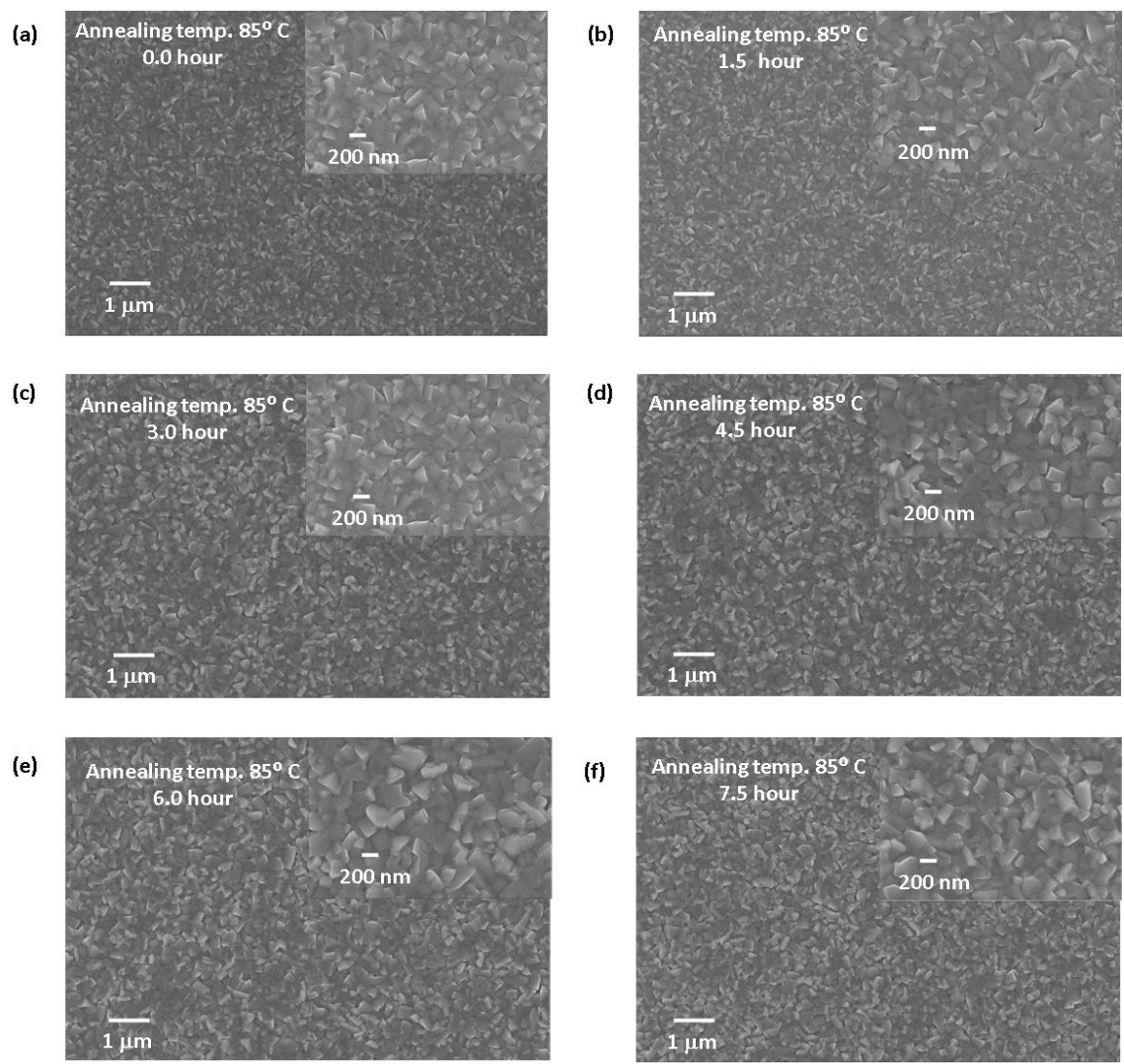

**Figure 32:** *Top surface SEM images of perovskite thin films deposited on Pz:PFN(9:1) coated ITO substrates. The perovskite films were annealed at 100º C for 30 minutes and cooled down to room temperature. Further, the perovskite films were annealed at 85º C for (**a**) 0.0 hours, (**b**) 1.5 hours, (**c**) 3.0 hours (**d**) 4.5 hours (**e**) 6.0 hours, and (**f**) 7.5 hours respectively.*

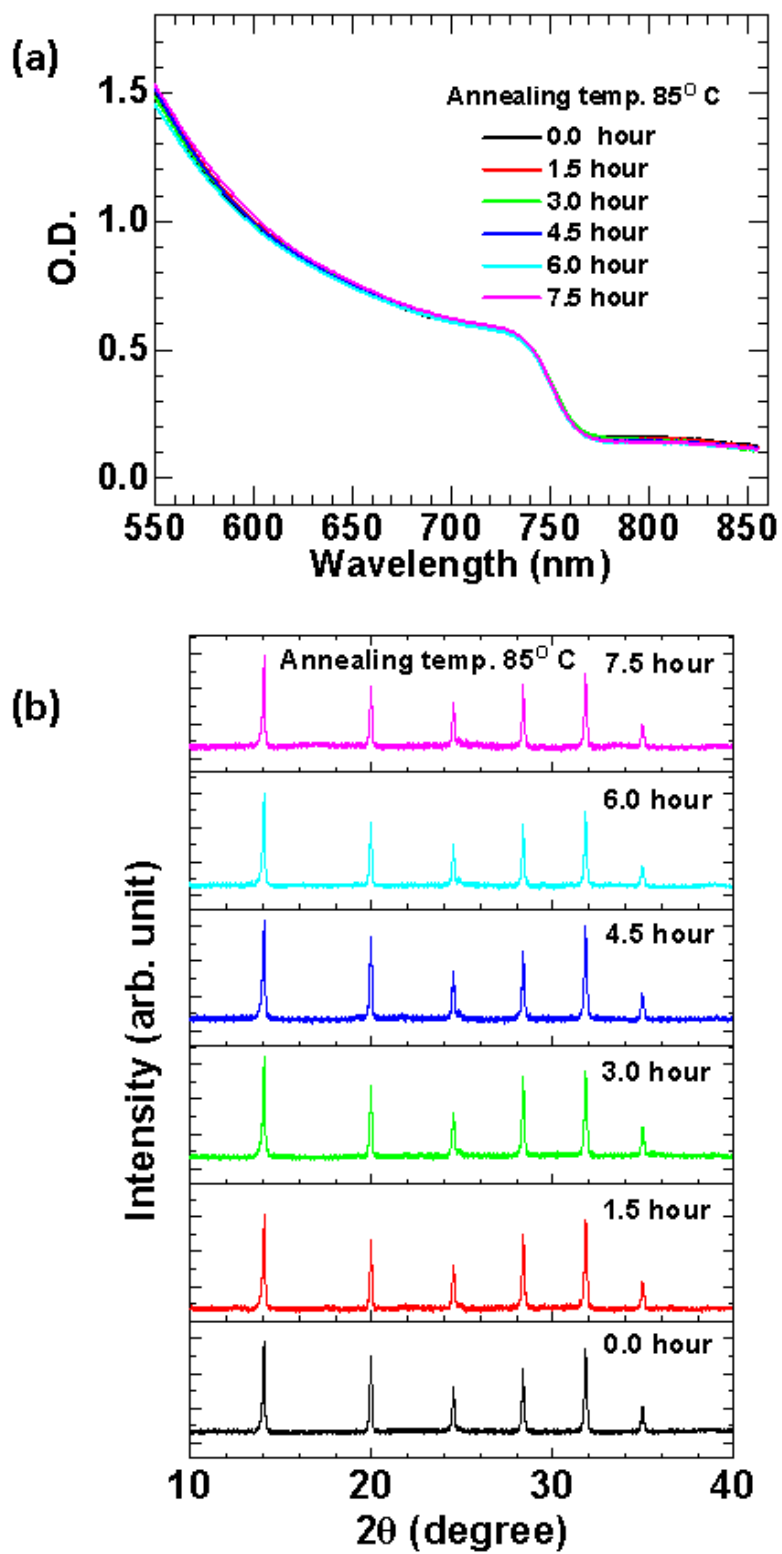

**Figure 33:** (**a**) *Absorption spectra and* (**b**) *crystallographic XRD plot of the 85º C annealed perovskite films at 0.0 hour, 1.5 hours, 3.0 hours, 4.5 hours, 6.0 hours, and 7.5 hours respectively.*

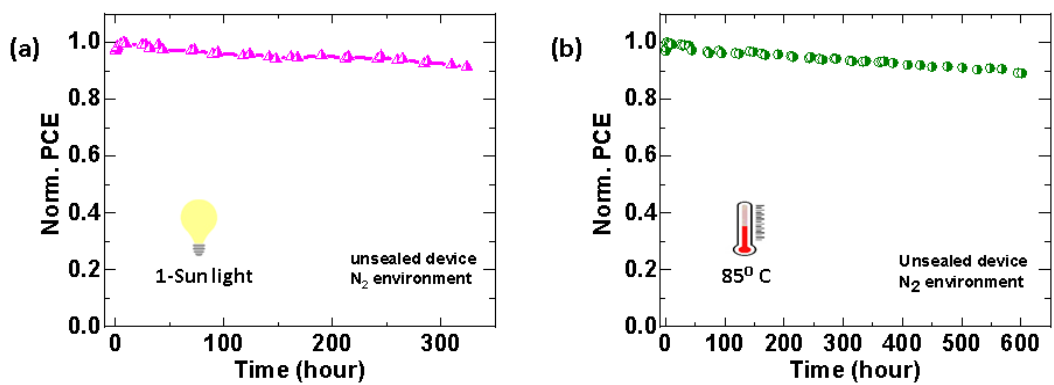

**Figure 34:** *(a) The operational stability of the un-encapsulated device was measured under N₂ environment and kept under constant 1-Sun illumination. (b) For thermal stability, we kept the device at 85º C and measured the photovoltaic performance. We kept the device at 85º C for approximately 8 hours and then kept it at room temp. Again, on the next day, we kept the device at 85º C and measured the PV device performance and so on.*

We investigated the long-term stability of our best performing Pz:PFN (9:1) based device in comparison with Pz:PFN (6:4).We kept the devices under $N_2$ environment in the dark, and the $J - V$ scans were taken periodically of the unencapsulated devices. The device performance was measured under relative humidity in a range RH ∼ 40-50%. We performed the stability of the Pz:PFN(9:1) and Pz:PFN(6:4) devices, and we observed that the Pz:PFN(9:1) is retaining 95% of its initial efficiency even after >3000 hours **Figure 31**. Further, the morphological stability was carried out at 85º C of the unencapsulated perovskite films. The perovskite films deposited on Pz:PFN(9:1) coated ITO substrates were annealed at 85º C for (**a**) 0.0 hours, (**b**) 1.5 hours, (**c**) 3.0 hours (**d**) 4.5 hours (**e**) 6.0 hours and (**f**) 7.5 hours respectively **Figure 32.** The top surface SEM images of annealed perovskite thin films were taken and we did not observe any significant change in the morphology. Further, we checked the absorption spectra and XRD pattern of the annealed films, which also indicate the crystal structural stability of the perovskite films **Figure 33**. To study the operational stability we kept the device under constant 1-Sun illumination in the $N_2$ environment, and the $J - V$ scans were taken periodically. The best performing Pz:PFN(9:1) device retains > 90% of the initial efficiency after 300 hours **Figure 34a**. For the thermal stability measurement, we kept the device on a hot plate at 85º C in $N_2$ environment and periodically measured the photovoltaic performance. We kept the device at 85º C for approximately 8 hours and then kept it at room temp. Again, on the next day, we kept the device at 85º C and measured the PV device performance and so on. The best performing Pz:PFN(9:1) device retains > 90% of the initial efficiency after 600 hours **Figure 34b**. This improved device yield and stability can be ascribed not only to the uniform perovskite layer coverage on the Me-4PACz but also to the interfacial interaction of PFN-Br with the perovskite.

## 5.4 Conclusion

To summarize, we present a mixing engineering strategy of combining Me-4PACz SAM with the conjugated polyelectrolyte PFN-Br polymer. The mixing of the PFN-Br with the Me-4PACz, facilitates uniform deposition of the perovskite layer on the top of hydrophobic Me-4PACz HTL. The uniform deposition of the perovskite layer on the top of PFN-Br mixed Me-4PACz happened due to Me-4PACz:PFN-Br interaction with A-site cation, which was confirmed via solution-state NMR. This further facilitates an improved crystallization of the perovskite layer, which was confirmed via narrow diffraction in XRD, and narrow PL peaks result. In addition to this, the KPFM result reveals that mixing with PFN-Br tunes the work function of Me-4PACz, and for the optimized 9:1 mixing ratio, the energy level alignment of Pz:PFN matches well with the perovskite. As a result of this, the perovskite device demonstrates high and reproducible efficiency of over 20% concomitant with high stability for $T_{95}$ >3000 hours when measured in relative humidity of ~ 40%. We believe that the mixing engineering of SAM Me-4PACz with electrolyte polymer PFN-Br will not only open new doors to tackle hydrophobic SAMs in solution-processable efficient photovoltaic devices but also will allow designing new electrolyte-based polymers and/or small molecules that can be combined with SAMs, thereby quilting a better interface.

## 5.5 Postscript

Overall, this chapter helps to understand the origin of the hydrophobicity of carbazol acid-anchored self-assembled monolayer (Me-4PACz) HTL and the non-uniform perovskite layer deposition. A conjugated polyelectrolyte polymer PFN-Br can be used in a bi-layer or mixing engineering strategy to deposit a uniform perovskite layer. We observed that the mixing engineering technique is highly efficient in a particular volume ratio (9:1) due to the well matched surface potential of composite HTL Pz:PFN and the perovskite layer. The PFN-Br in the mixed Pz:PFN(9:1) facilitates improved crystallization and defect passivation at the interface. This improvement in the interface results in the lowest leakage current and improves photovoltaic parameters ($V_{OC}, J_{SC}, FF$), hence improved efficiency with concomitant stability. Thus, 10% of the conjugated polymer (PFN-Br) mixing in the Me-4PACz results in stable, reproducible, and efficient >20% perovskite solar cells. Interestingly, we have observed that the ideality factor estimated from the intensity dependent $V_{OC}$ for the Pz:PFN(9:1) based PSC is ~1.05. This indicates that the device might be operating near the radiative limit. Hence, we studied the radiative bimolecular recombination process for the Pz:PFN(9:1) based PSCs using steady-state and transient optoelectronic measurements in the next chapter 6. In this regard, we developed a new set of characterization schemes that can be used to study the dominant recombination mechanism in solar cells. The details of characterization and analysis will be discussed in the next Chapter 6.

Furthermore, it is crucial to note that the leakage current for the Pz:PFN(9:1) based PSC is very low (sub-pA/cm$^2$). Such a low current is immensely useful in the photodetector application. The photodetector figure of merit (FOM) parameters are inversely related to the dark current and hence expect better detectivity (detection of the faint signal) for lower dark current devices. Therefore, we conducted a detailed study of the Pz:PFN(9:1) HTL based PSC as a photodetection diode. The photodetector study will be discussed in Appendix A.

# CHAPTER 6

# Perovskite solar cells dominated by bimolecular recombination- how far is the radiative limit?

# Chapter 6

# Perovskite solar cells dominated by bimolecular recombination- how far is the radiative limit?

**Abstract**

In this chapter we present an experimental demonstration of perovskite solar cells dominated by bimolecular recombination and critically analyse their performance against radiative limits. To this end, we first establish a set of quantitative benchmark characteristics expected from solar cells limited by bimolecular recombination. Transient as well as steady state intensity dependent measurements indicate that our solar cells indeed operate at such limits with interface passivation comparable to the champion c-Si technology. Further, we identify characterization schemes which enable consistent back extraction of recombination parameters from transient optoelectrical and electroluminescence measurements. Remarkably, these parameters predict important features of dark current density *vs.* voltage characteristics ($J - V$) and Suns-$V_{OC}$ measurements, thus validating the estimates and the methodology. Uniquely, this work provides a consistent and coherent interpretation of diverse experimental trends ranging from dark $J - V$, Suns-$V_{OC}$, steady state and transient intensity dependent measurements to electroluminescence quantum yield. As such, insights shared in this chapter could have significant implications towards fundamental electronic processes in perovskite solar cells and further efficiency optimization towards Shockley-Queisser limits.

## 6.1 Introduction.

In the previous Chapter 5, we showed the mixing engineering strategy to resolve the hydrophobicity of Me-4PACz HTL and deposit a uniform perovskite layer. We used different mixing ratios of Me-4PACz with PFN-Br and tailored the work function of the HTL. The interface engineering using the mixed Pz:PFN HTL for the CsFAMA (bandgap 1.6 eV) based PSC showed fantastic device performance with augmented stability for a specific mixing ratio Pz:PFN (9:1). The interface engineering study suggests the clean interface for the Pz:PFN (9:1) based PSC. The Pz:PFN (9:1) based PSC showed the lowest dark current, higher charge carrier lifetime, and ideality factor close to unity (estimated from intensity dependent $V_{OC}$ study), compared to the other mixing ratios. The unity ideality factor indicates that the device might be operating near the radiative limit. Hence, we studied the dominant recombination mechanisma for the Pz:PFN(9:1) based PSCs using steady-state and transient optoelectronic measurements in this chapter.

The literature survey of perovskite solar cells indicates the progress in the PSC field is so impressive that performance close to the radiative or the Shockley-Quiesser (SQ) limit[1] could possibly be achieved in the near future. Indeed, solar cells limited by radiative recombination are perceived as epitomes of perfection in Photovoltaics. Practical solar cells involve multiple materials and fabrication steps with different thermal budgets. As a result, practical solar cell efficiencies are usually much lower than SQ limits – mainly due to the presence of defects in the bulk of the active material and imperfect interfaces.[1–3] Trap states or defects in a semiconductor [4–7] play a significant role in the solar cell efficiency as they directly contribute to non-radiative recombination mechanisms.[8,9]

In this context, we present an experimental demonstration of detailed balance limit operation of perovskite solar cells, dominated by bimolecular recombination, under 1 Sun illumination. To establish this claim, we identified the ideal characteristics of a solar cell dominated by bimolecular recombination at detailed balance limit (i.e., the benchmark criteria) under illumination intensity dependent measurements (steady state and transient). Remarkably, our perovskite solar cells perform well in accordance with the theoretical benchmarks, indicating that our device operation is dominated by bimolecular recombination . These benchmark criteria also allow coherent and consistent back extraction of the recombination parameters, which in turn predicts important features like the reverse saturation current density obtained through dark current density *vs.* voltage characteristics $(J - V)$ and Suns-$V_{OC}$ methods. In addition, we present excellent passivation of the interfaces between perovskite and the contact layers in the finished device – which are comparable to the state-of-the-art c-Si technology. Further, we used Electroluminescence measurements to explore the physical origin of such non-radiative bimolecular recombination which allows us to directly estimate the radiative recombination coefficient. Interestingly, in contrast to the optical spectroscopy-based characterization of thin films, our proposed methodology for estimation of recombination parameters relies only on device level terminal current-voltage characteristics.

Below, we first provide experimental results which indicate that the performance of our solar cells is comparable to the state of the art (for comparable band gap), followed by theoretical analysis on benchmark criteria to establish radiative limit performance and its implications.

## 6.2 Experimental section

**Materials:** ITO-coated glass substrates (15 $\Omega$/sq) were purchased from Lumtech. Lead iodide (PbI$_2$), formamidinium iodide (FAI), cesium iodide (CsI), lead bromide (PbBr$_2$), and Me–4PACz, all were purchased from TCI chemical and used as received. Methyl ammonium Bromide (MABr) was purchased from Greatcell Solar. Phenyl-C61-butyric acid methyl ester (PC$_{61}$BM) was purchased from Lumtech and was used as received. The poly(9,9-bis(3'-(N,N-dimethyl)-N-ethylammoinium-propyl-2,7-fluorene)-alt-2,7-(9,9-dioctylfluorene))dibromide (PFN-Br) ordered from Solarmer Material Inc. and used as received. Bathocuproine (BCP) was purchased from Sigma Aldrich and used as received.

**Solution preparation:** To make triple cation (FA$_{0.83}$MA$_{0.17}$)$_{0.95}$Cs$_{0.05}$Pb(I$_{0.9}$Br$_{0.1}$)$_3$ perovskite solution, first we added 22.5 mg of MABr, 73.5 mg of PbBr$_2$, 172 mg of FAI, and 507.5 mg of PbI$_2$ in 1 ml of DMF: DMSO (4:1) and stirred at room temperature for 2 hours to make a premixed solution. Separately we made 1.5 (M) CsI solution in DMSO i.e. 100 mg of CsI in 257 µl of DMSO and stirred for 2 hours. We filtered the premixed solution with PTFE 45 mm filter in a separate vial. Finally, we added 950 µl of premixed solution and 50 µl of CsI solution to get the final triple cation(FA$_{0.83}$MA$_{0.17}$)$_{0.95}$Cs$_{0.05}$Pb(I$_{0.9}$Br$_{0.1}$)$_3$ perovskite solution and stirred for 1 hour. To obtain mixed Me–4PACz: PFN-Br solution, first, we prepared 0.4 mg/ml Me–4PACz solution in anhydrous methanol and 0.4 mg /ml PFN-Br solution in anhydrous methanol and stirred overnight. Finally, one hour before spin-coating we mix the Me–4PACz and PFN-Br solution in 9:1 volume ratio.[10] 20 mg PC$_{61}$BM is dissolved in 1 ml of chlorobenzene and stirred overnight. 0.5 mg BCP is dissolved in 1 ml of anhydrous isopropanol and stirred overnight at room temperature and 10 minutes at 70º C just before spin coating.

**Perovskite solar cells fabrication:** ITO-coated glass (15 $\Omega$/square) substrate was patterned with Zn powder and HCl and then sequentially cleaned with soap solution, deionized (DI) water, acetone, and isopropanol for 10 minutes each. After drying the substrate with a nitrogen gun, we keep the ITO substrates at 80$^o$C for 10 minutes and then took them inside the oxygen plasma ashing chamber for 20 minutes, and plasma ashing was done at an RF power of 18 watts. After plasma ashing, we immediately take the substrates inside the $N_2$ environment glove box ($O_2$<0.1 ppm, $H_2O$<0.1 ppm) and spin-coat Me-4PACz: PFN-Br mixed solution at 4000 rpm for 30 seconds and then annealed at 100$^o$C for 10 minutes. After that, we cool down the substrates for 5 minutes and then perovskite solution spin-coated on the ITO/ Me-4PACz: PFN-Br substrates. The CsFAMA perovskite was spin-coated at 4000 rpm for 30 seconds and at the last 5 seconds, we used 150 ml of chlorobenzene as an anti-solvent treatment on a 1.5 cm × 1.5 cm substrate and then annealed at 100$^o$C for 30 minutes.  After that, $PC_{61}BM$ was spin-coated at 2000 rpm for 30 seconds and BCP was spin-coated at 5000 rpm for 20 seconds. Finally, 150 nm of Ag was deposited by thermal evaporation (integrated inside the $N_2$ filled glove box) under a vacuum of 2x10$^{-6}$ mbar using metal shadow mask with device active area of 7 mm × 2.5 mm, i.e., 17.5 mm$^2$ and of 7 mm × 11.5 mm, i.e., 80.5 mm$^2$.

**Characterization:**

All the photovoltaic measurements were carried out under ambient conditions. Photocurrent density *vs.* applied voltage (*J-V*) measurement has been carried out using Keithley 4200 SCS and LED solar simulator (ORIEL: LSH-7320) after calibrating through a standard Si solar cell (RERA SYSTEMS-860 reference cell). The *J-V* measurement was performed with a scan rate of 100 mV/s and hold time 0.01 second.  EQE measurement has been carried out to measure the photo response as a function of wavelength using the Bentham quantum efficiency measurement system (Bentham PVE 300). TPV was measured by using a 490 nm TOPTICA diode laser, THORLABS white lamp S/N M00304198, ArbStudio 1104, and digital oscilloscope Tektronix DPO 4104B.[11] The intensity of the background DC light changed using a set of neutral density (ND) filters. The details of the intensity dependent steady-state and transient measurements are explained later in this chapter (section 6.3.2).

**List of the variables:**

The list of the variables used in this chapter is provided in Appendix B.

## 6.3 Results

### 6.3.1 Solar cell performance

To study the detail balance limit in a perovskite solar cell, we fabricated state-of-the-art perovskite solar cells with the active material $(FA_{0.83}MA_{0.17})_{0.95}Cs_{0.05}Pb(I_{0.83}Br_{0.17})$ of bandgap 1.6 eV (as we discussed in Chapter 5). Detailed fabrication and characterization methodology are provided in the experimental sections above. It is worth mentioning that, we purposely decreased the perovskite deposition spin speed from 5000 rpm to 4000 rpm, to slightly increase the thickness of the perovskite absorber without compromising the fill factor ($FF$) of the PSC device. Our devices show an efficiency of $\eta = 21.54\%$ under 1-Sun illumination with $J_{SC} = 23.35$ mA/cm$^2$, $V_{OC} = 1.160\ V$, $FF = 79.52\%$, over an active area of 0.175 cm$^2$. (see **Figure 1a** and **Table 1**). The steady state efficiency at maximum power point tracking under 1-Sun condition is 20.97% **Figure 2a**. A cross-sectional SEM image of the device (glass/ITO/Pz:PFN(9:1)/Perovskite/PCBM/BCP/Ag) is shown in the inset of Figure 1a. The performance of the device over an active area of 0.805 cm$^2$ is $\eta = 19.30\%$ with $J_{SC} = 22.54$ mA/cm$^2$, $V_{OC} = 1.144\ V$, $FF = 74.85$ (Figure 1a, Table 1) .**Figure 1b** represents the EQE, IQE spectrum with reflection spectra (R) and the integrated current densities (Int. $J_{SC}$) over the AM 1.5G spectra for a device with an active area of 0.175 cm$^2$ (large area 0.805 cm$^2$ device's EQE, reflection and integrated $J_{SC}$ results are in **Figure 2b**). We note that the integrated $J_{SC}$ over the EQE spectrum is slightly lower than the $J_{SC}$ measured from the $J - V$ scan, which can be due to the pre-bias condition or edge effect from the active area of the device, as previously identified in the literature.[12–14] **Figure 1c** compares the performance of our devices against the state-of-the-art solar cells from literature with active material of comparable band gap. Open circuit voltage and efficiency of our devices compare well with the best reported results. It is well known that the performance of perovskite solar cells not only depends on the bandgap[15] but also on the active area of the device for a particular bandgap.[16–18] This can explain the slightly higher reported $V_{OC}$ and efficiency from literature as compared to our smaller (0.175 cm$^2$) devices and is consistent with a slight reduction of these parameters in our own scaled up (up to 0.805 cm$^2$) devices. Further, a  statistical study of ~30 devices is shown in **Figure 3** with an average efficiency of 20.21%. As our devices are comparable to the state-of-the-art (for a given band gap), we further tested their performance against the theoretical benchmark criteria expected from perovskite solar cells at detailed balance limits (will be discussed later in this chapter)

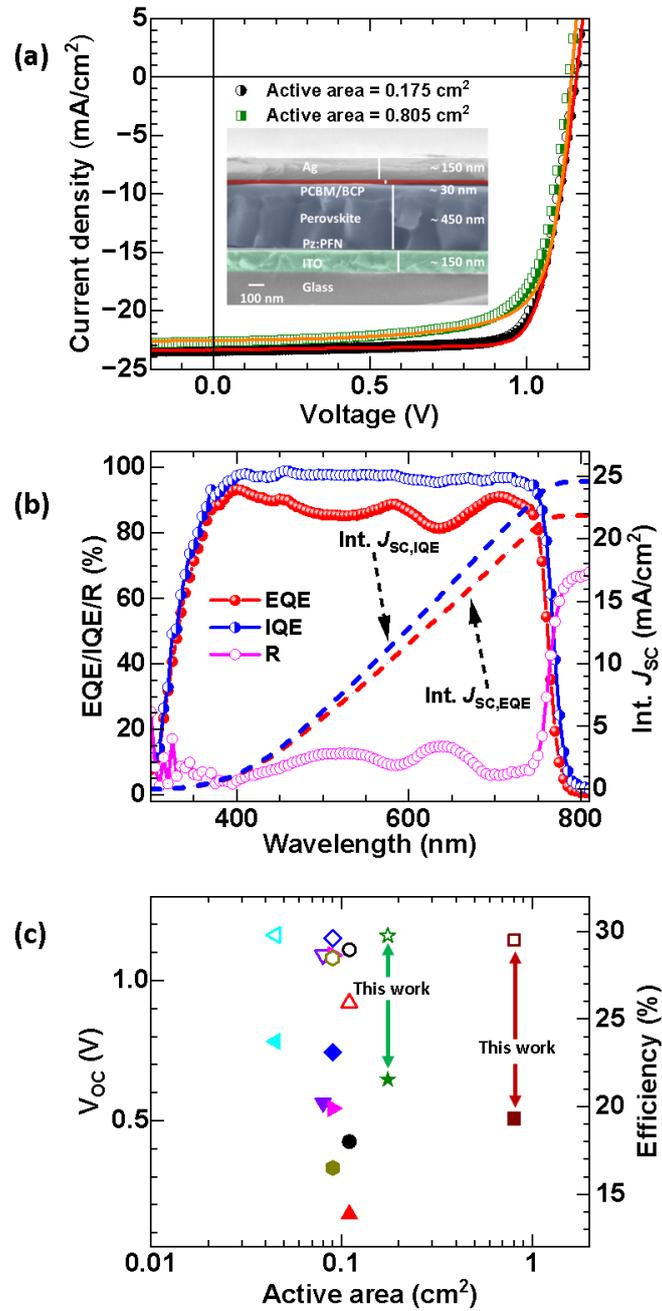

**Figure 1**: *Current density vs. Voltage (J-V) characteristics of the photovoltaic devices under 1-sun (100 mW/cm2) illumination conditions in the forward and reverse scan direction. The active area of the devices used are 0.175 cm$^2$ and 0.805 cm$^2$. The inset figure represents the cross-sectional view of the device (glass/ITO/Pz:PFN/Perovskite/PCBM/BCP/Ag). (b) EQE, IQE spectrum with reflection spectra (R), and the integrated current densities (Int. J$_{SC}$) over the AM 1.5G spectra. (c) Comparison of our devices against the state-of-the-art from literature. The open-circuit voltage (V$_{OC}$) denoted by the open symbols and the efficiency (solid symbols) are plotted against the active area of the inverted architecture-based PSCs having a bandgap of E$_g$ = 1.6 ± 0.02 eV ( please see Table 2 for references, etc.). Colour symbols:: ○-black circle: ref.1, △-red up-pointing triangle: ref.2, ▽- violet down-pointing triangle: ref.3, ◇-blue diamond: ref.4, ◁-cyan left-pointing triangle: ref.5, ▷- magenta right-pointing triangle:*

*ref.6,* ○-*dark yellow hexagon: ref.7,* ☆- *olive star: this work on the active area of 0.175 cm²,* □- *wine square: this work on the active area of 0.805 cm².*

**Table 1**: *Photovoltaic (PV) parameters, integrated $J_{SC}$ (measured from EQE spectra), and hysteresis index.*

| Scan direction | $J_{SC}$ (mA/cm²) | $V_{OC}$ (V) | *FF* (%) | PCE (%) | Int. $J_{SC}$ (mA/cm²) | Hysteresis index (%) |
|---|---|---|---|---|---|---|
| **0.175 cm² - FS** | 23.40 | 1.157 | 77.82 | 21.06 | 21.95 | 2.23 |
| **0.175 cm² - RS** | 23.35 | 1.160 | 79.52 | 21.54 | | |
| **0.805 cm² - FS** | 22.66 | 1.138 | 72.09 | 18.59 | 21.37 | 3.68 |
| **0.805 cm² - RS** | 22.54 | 1.144 | 74.85 | 19.30 | | |

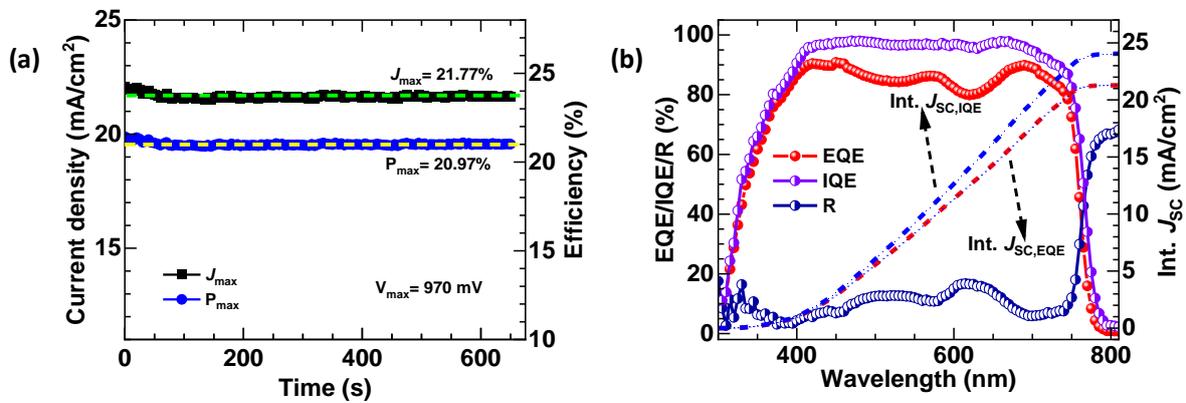

**Figure 2**: **(a)** *The steady state efficiency at maximum power point tracking under 1-Sun condition for 0.175 cm² active area PSC.* **(b)** *EQE, IQE spectrum with reflection spectra (R), and the integrated current densities (Int. $J_{SC}$) over the AM 1.5G for 0.805 cm² PSC.*

**Table 2**: *List of the PV parameters of PSCs for bandgap $E_g \sim (1.60 \pm 0.02)$ eV with corresponding active area of the device and the hole transport layer used.*

| SL. #. | Bandgap (eV) | HTLs | $J_{SC}$ (mA/cm²) | $V_{OC}$ (V) | FF (%) | PCE (%) | Active area (cm2) | References in this chapter | References in the Figure 1c |
|---|---|---|---|---|---|---|---|---|---|
| 1. | ~ 1.60 | PEDOT:PSS | 20.11 | 1.11 | 80.6 | 18.0 | 0.11 | [19] | 1 |
| 2. | ~1.59 | PEDOT:PSS | 19.29 | 0.92 | 77.26 | 13.86 | 0.11 | [20] | 2 |
| 3. | ~ 1.60 | PTAA | 22.61 | 1.091 | 82 | 20.20 | 0.08 | [21] | 3 |
| 4. | ~1.59 | PTAA/PEAI | 24.51 | 1.15 | 82.1 | 23.1 | 0.09 | [22] | 4 |
| 5. | ~ 1.60 | PTAA | 24.13 | 1.162 | 84.6 | 23.72 | 0.045 | [23] | 5 |
| 6 | ~ 1.58 | Me-4PACz/Al₂O₃ | 23.0 | 1.09 | 79.4 | 19.9 | 0.09 | [24] | 6 |
| 7 | ~ 1.58 | Me-4PACz/ PFN-Br | 20.50 | 1.08 | 74.7 | 16.5 | 0.09 | [24] | 7 |
| 8. | ~ 1.60 | P3CT-Rb | 21.67 | 1.144 | 82.78 | 20.52 | - | [25] | 8 |
| 9. | ~ 1.60 | Me-4PACz:PFN-Br | 23.35 | 1.160 | 79.52 | 21.54 | 0.175 | This work | This work |
| 10. | ~ 1.60 | Me-4PACz:PFN-Br | 22.54 | 1.144 | 74.85 | 19.30 | 0.805 | This work | This work |

The list of the performance of perovskite solar cells with an $E_g \sim (1.60 \pm 0.02)$ eV **Table 2**. Please note that there is only one report as of now with Me-4PACz used as a hole transport layer in the perovskite solar cells having a bandgap of ~1.58 eV, which is included in Table 2 with reference number 24 in this chapter.

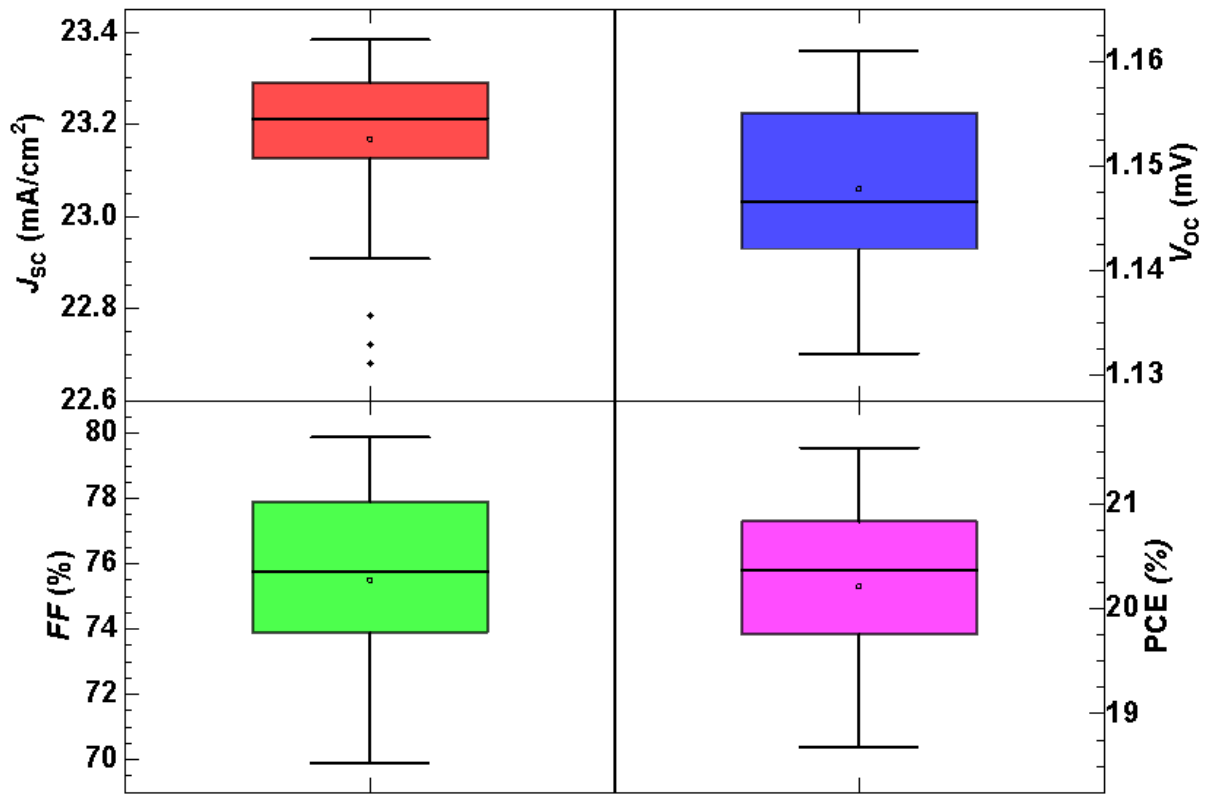

***Figure 3***: *Boxplot histogram of short circuit current density ($J_{SC}$), open circuit voltage ($V_{OC}$), fill factor (FF), and power conversion efficiency (PCE%) over 30 devices of device architecture glass/ITO/Pz:PFN(9:1)/Perovskite/PCBM/BCP/Ag.*

### 6.3.2 Intensity dependent steady-state and transient measurements

To study the perovskite solar cells at detail balance dominant with dominant bimolecular recombination, we have carried out the steady-state and transient measurements of the perovskite solar cell device. Both the characterizations are done separately on the same device to keep consistency in the result. First, we measured the steady state $J - V$ at different intensities and obtained the ideality factor. If the ideality factor is unity, a signature of dominant bimolecular recombination, but it's not necessarily true always (exceptions are discussed later explicitly). To check the dominant bimolecular recombination at the device level, we carried out intensity dependent steady state $J - V$ and transient photovoltage measurement. The detailed of the characterization techniques are discussed below.

### 6.3.2.1 Intensity dependent steady-state $J - V$ characteristics measurement

Intensity dependent current *vs.* voltage characteristics study is a steady-state measurement where the solar cell is kept under 1-sun condition and the corresponding current *vs.* voltage is recorded in Keithley **Figure 4a.** We used Keithley 4200 SCS to record the current *vs.* voltage characteristics, and the 1- Sun light is made using a LED solar simulator (LSH-7320). The intensity of the 1-sun light is varied by using a set of neutral density (ND) filters whose optical densities are O.D. = 0.1, 0.2, 0.3, and so on. The ND 0.1 means the DC background 1-sun light passing through ND filter of optical density O.D.= 0.1 and the resultant intensity to the device is $(100/10^{O.D.} \text{ mW/cm}^2)$ 79.43 mW/cm$^2$. The intensity dependent $J - V$ characteristics at different intensities are shown in **Figure 4b.** As the ND filter's optical density increases the DC background 1-sun light intensity decreases and as a result the $J_{SC}$ and $V_{OC}$ both decrease. The intensity dependent $J_{SC}$ is shown in **Figure 4c** and the intensity dependent $V_{OC}$ is shown in **Figure 9a**. The intensity dependent $J_{SC}$ shows linear dependency ($J_{SC} = I^{\alpha}$) over the intensity with exponent $\alpha = 0.99$. The intensity dependent $V_{OC}$ shows linear dependency over the intensity with slope $= 1.05 \times kT/q$ **Figure 9a**.

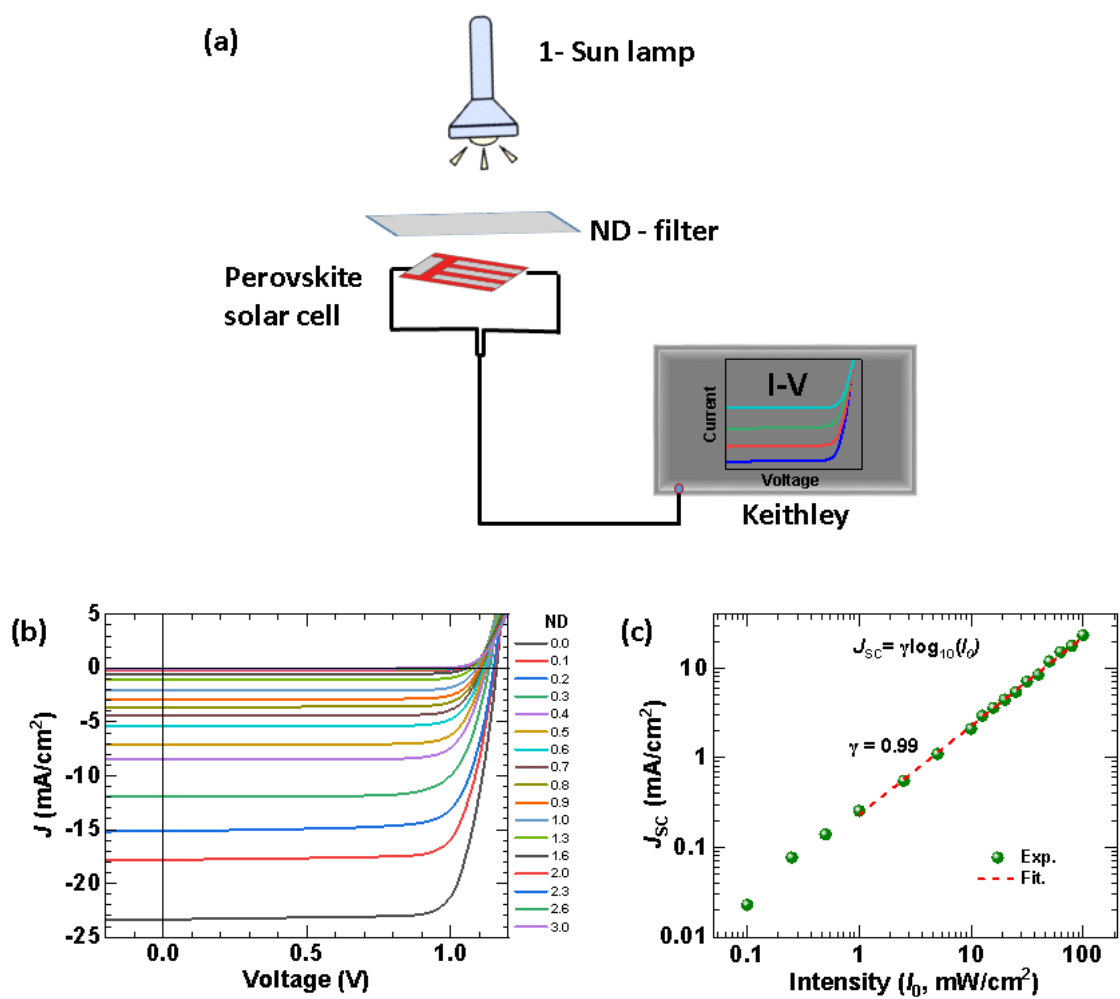

**Figure 4: (a)** *Schematic of the experimental set-up of steady-state Current* vs. *Voltage characteristics measurement at different intensities.* **(b)** *Current density vs. Voltage* $(J-V)$ *characteristics of the photovoltaic devices under varied illumination conditions. The intensity of the 1-sun light changed using a set of neutral density (ND) filters. The legends in the figure indicate that the DC light is passing through different ND filters of optical density 0.1, 0.2, 0.3, and so on. For example, ND 0.1 means the DC background 1-sun light passing through ND filter of optical density 0.1 and the resultant intensity to the device is* $(100/10^{O.D.} \ mW/cm^2)$ *79.43 mW/cm². **(c)** The variation in* $J_{SC}$ *with respect to different illumination intensities.*

**6.3.2.2 Intensity-dependent transient photovoltage (TPV) measurements**

The transient photovoltage (TPV) is an optoelectronic measurement technique where the steady-state (using background -1 sun white light illumination condition) device was kept in an open circuit condition using a high external resistance across the device and a short-lived laser perturbation is applied to the device. At an open-circuit voltage ($V_{OC,0}$) condition the generated charge carriers cannot flow through the circuit. Further, the application of the short-lived laser perturbation leads to generating small charges within the devices. Since the device is at open circuit condition, the generated perturbed charges cannot leave the device instead they create a small voltage ($\Delta V_{OC,max}$) above the open circuit voltage ($V_{OC,0}$) due to accumulation across the electrodes. So, during perturbation for a short time, the resultant voltage across the device becomes ($V_{OC,0} + \Delta V_{OC,max}$). But immediately after the perturbation is turned off, the generated small charges decay due to recombination (radiative + non-radiative) in the presence of background steady-state charges created by DC light. As a result, the generated voltage again comes back from ($V_{OC,0} + \Delta V_{OC,max}$) to $V_{OC,0}$ over a finite time called the lifetime of the perturbed charge carriers. In the intensity-dependent transient measurement, we kept the short-lived laser intensity fixed where the background DC light changed. Initially, we kept the DC light background at the 1-sun condition and the laser intensity was chosen such that the generated perturbed voltage $\Delta V_{OC,max} \approx 10$ mV which is much smaller than the $V_{OC,0}$ (~1.16 V) of the device at 1-sun condition. The condition of TPV measurement is that the perturbation signal should be much smaller so that the perturbation remains in the first-order regime that forces the decay signal to be mono-exponential, which is easy to analyse and interpret the results.[26,27] In the intensity-dependent TPV measurement, the background DC light intensity is changed with a set of neutral density (ND) filters. As the background DC light intensity changes, the recombination dynamics of the perturbed charge carrier change. At higher DC background light intensity, the steady-state background charge densities are higher. Hence, the recombination of the short-lived perturbed charge carrier with the background charges is more prominent at higher intensities and results in a shorter perturbed charge carrier lifetime.

The schematic of the TPV measurement is shown in **Figure 5a**. A perovskite solar cell was illuminated by 1-sun DC light using a THORLABS white lamp S/N M00304198. A 490 nm TOPTICA diode laser is used to perturb the device. The diode laser is modulated with a frequency of 1 kHz and duty cycle of 0.05% using an ArbStudio 1104 function generator. The perturbed laser intensity is chosen such that the perturbed voltage $\Delta V_{OC} \approx 10$ mV while the background DC light intensity was 1-Sun. The perturbed signal is measured using a digital oscilloscope Tektronix DPO 4104B. The intensity of the DC white light was changed using a set of neutral density (ND) filters. As the DC background light intensity changes while keeping the perturbed AC signal intensity constant, the generated perturbed voltage changes. The lower DC background light intensity results in lower background charges and as a result lower recombination and hence higher generated AC perturb voltage **Figure 5b**. The normalized decay profile of the TPV signal is shown in **Figure 5c.**

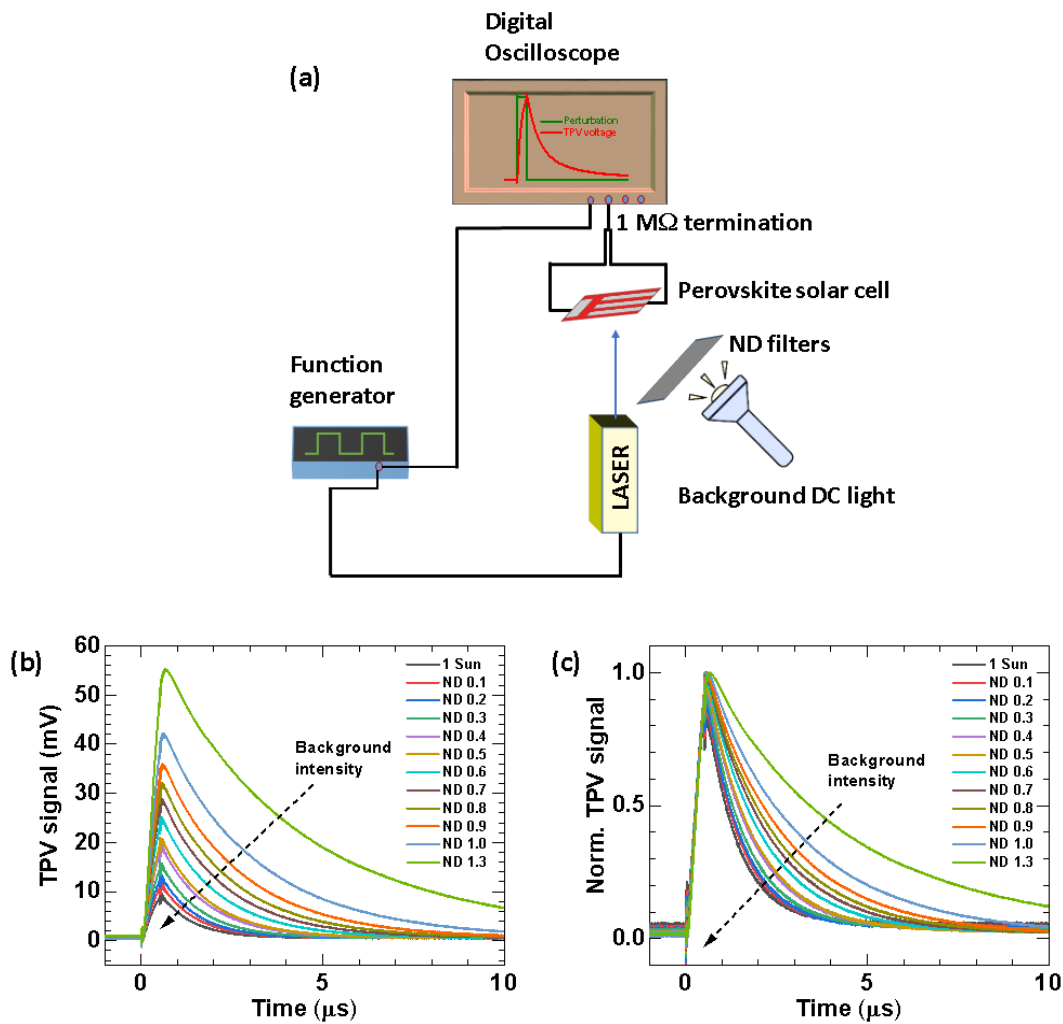

***Figure 5:(a)*** *Schematic of the experimental set-up of transient photovoltage (TPV) measurement. Decay profile of the transient photovoltage at different DC background intensities while the perturb excitation laser pulse (ON time 500 ns) intensity.* ***(b)*** *The TPV signal amplitude* ($\Delta V_{OC,max}$) *decay profile of a perovskite solar cell with respect to different DC background illumination. The legends indicate that the DC light is passing through different ND filters of optical density 0.1, 0.2, 0.3, and so on. ND 0.1 means the DC background 1-sun light passing through ND filter of optical density 0.1 and the resultant intensity to the*

*device is $(100/10^{O.D.}\ mW/cm^2)\ 79.43\ mW/cm^2$ . (c) The normalized TPV signal at different background DC light intensities.*

### 6.3.2.3 Working range of transient photovoltage (TPV) measurements

To confirm the working range of the measurement i.e. whether the transient photovoltage measurement is in the perturbation limit or not, we measure the ratio of the intensity of the laser perturbation light and the background DC light. The calculation is done as follows,

During the TPV measurement, the perturbation laser pulse power was kept constant while the background DC light intensity changed to obtain the intensity dependent transient photovoltage characteristics. The laser power 10 mW (spot diameter ~ 1 mm$^2$) with a pulsed width of 500 ns and repetition rate 1 kHz was kept remain constant throughout the measurement, whereas the background DC light intensity changed from 100 mW/cm$^2$ to 5.01 mW/cm$^2$. The RMS short circuit current due to pulsed laser (500 ns duty cycle, frequency 1kHz) is 0.97 μA and the short circuit current due to background DC light changes from a few mA to sub-mA **Figure 6**. The current corresponding to the light intensity of the pulsed laser and background DC light is such that the ratio of the pulsed laser intensity and background DC intensity is always less than 1%. This indicates that the measurement is always lying in the perturbation regime

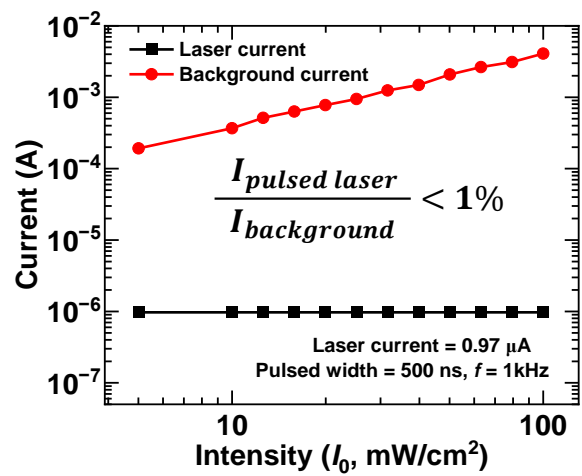

**Figure 6**: *Comparison of the perturbation light and the background light. During the measurement, the laser intensity was kept fixed at 10 mW with pulsed with 500 ns and repetition rate 1 kHz whereas the background intensity changed from 100 mw/cm² to 5.01 mw/cm².*

**6.3.2.4 Effect of capacitance on transient photovoltage (TPV) measurements**

Further, it is worth noting that, as per literature,[28,29] the charge carrier recombination lifetime from transient photovoltage (TPV) measurement may be affected by the capacitance of the device.

As per the literature, the capacitance affected TPV lifetime can be expressed as[28,29]

$$\tau_{TPV} = (\tau_{B,eff} + R_0 C_0) \left[ 1 + \left( \frac{1}{R_{Sh}} + \frac{1}{R_L} \right) R_0 \right] \qquad (6.1)$$

Where $\tau_{TPV}$ is the measured carrier recombination lifetime from transient photovoltage measurement (TPV), $\tau_{B,eff}$ is the effective carrier recombination lifetime from TPV, $C_0 = \frac{\epsilon \epsilon_0}{d}$ is the geometric capacitance per unit area, $R_{Sh}$ is the device shunt resistance, $R_L$ is the load resistance and $R_0$ is the differential resistance, which can be expressed as

$$R_0 = \left( \frac{kT}{q} \right) \left( \frac{n}{J_0} \right) \times exp \left( \frac{-q V_{OC}}{nkT} \right) \qquad (6.2)$$

Where $n$ is the ideality factor.

$\frac{kT}{q} = 25.7$ meV thermal voltage.

To find the effect of capacitance on the carrier recombination lifetime, we used the following

$\epsilon_0 = 20$ for perovskite material.[30]

$d = 450$ nm, the thickness of the perovskite layer.

$R_{Sh} = 16$ k$\Omega$ for our perovskite solar cell device.

$R_L = 1$M$\Omega$ for used Oscilloscope – DPO 4000B.

$n = 0.98$ experimentally measured (please see Figure 9b).

$J_0 = 3.80 \times 10^{-19}$ mA/cm$^2$. Experimentally measured (please see Figure 9b).

By considering the above parameters, the capacitance effects time $R_0C_0$ is plotted along with the measured TPV lifetime **Figure 7**. The effective carrier recombination lifetime $\tau_{B,eff}$ is estimated using eq.(6.1) and shown in **Figure 7**.

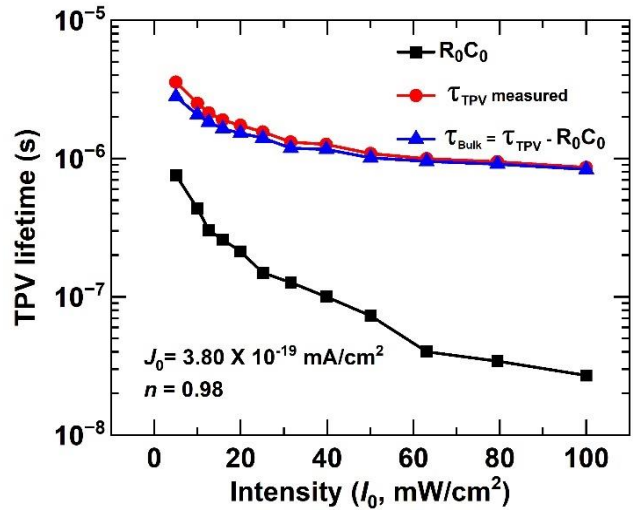

**Figure 7**: *The effect of capacitance on the charge carrier recombination lifetime from transient photovoltage measurement (TPV). The red plot represents the measured data, the black plot represents the time due to the capacitance effect, and the blue plot represents the corrected charge carrier recombination lifetime.*

Therefore, the capacitance can affect transient measurements. However, as per the literature,[28,29] the time constants estimated by our scheme is not significantly affected by the capacitance issues. The capacitance effect is slightly there at the lower intensity, but at higher intensity, this effect is insignificant.

### 6.3.3 Intensity dependent scaling laws of the steady state and transient measurements

We consider optoelectrical characterizations in which the solar cell is maintained under open circuit conditions and is subjected to a pulsed illumination (**see Figure 8**). The carrier dynamics under illumination (with intensity $I$ and corresponding photo-generation rate $G$) for an undoped sample with $n = p$ is given as

$$\frac{\partial n}{\partial t} = G - k_1 n - k_2 n^2 - k_3 n^3 \qquad (6.3)$$

where $t$ is the time, and $n$ and $p$ denote the electron and hole density respectively. The parameters $k_1$, $k_2$, and $k_3$ denote the monomolecular, bi-molecular recombination, and Auger recombination rates respectively (as discussed in Chapter 2). While $k_1$ is contributed by mid-gap recombination centers, $k_2$ could be due to radiative and non-radiative recombination mechanisms. Auger recombination is expected to play a significant role at higher carrier levels and is considered later. Under such steady-state conditions (i.e., ignoring Auger recombination), with $G = G_0$, we have $G_0 = k_1 n_0 + k_2 n_0^2$ where $n_0$ is the steady-state carrier density. With $n = n_0 + \Delta n$, through a perturbation analysis[28,29,31–34] ($\Delta G \ll G_0, \Delta n \ll n_0$) we find

$$\frac{\partial \Delta n}{\partial t} = G' - (k_1 + 2k_2 n_0)\Delta n \qquad (6.4)$$

where $G' = \Delta G$ during $t < 0$ and $G' = 0$ during $t > 0$ (**see Figure 8**). During $t > 0$, the carrier transients are given by $\Delta n(t) = \Delta n(0)e^{-t/\tau}$ with $\tau = (k_1 + 2k_2 n_0)^{-1}$ and $\Delta n(0) \approx \Delta G T_{ON}$. As $V_{OC}$ is defined as the separation between quasi-Fermi levels, we have $V_{OC} = \left(\frac{kT}{q}\right) \times \ln\left(\frac{np}{n_i^2}\right)$, where $n_i$ is the intrinsic carrier concentration. With $n = p$, and under the assumption of small perturbation in comparison to the background illumination (see Section 6.3.4 for detailed derivation) this leads to $\Delta V_{OC}(t) = \Delta V_{OC,max} e^{-t/\tau}$.

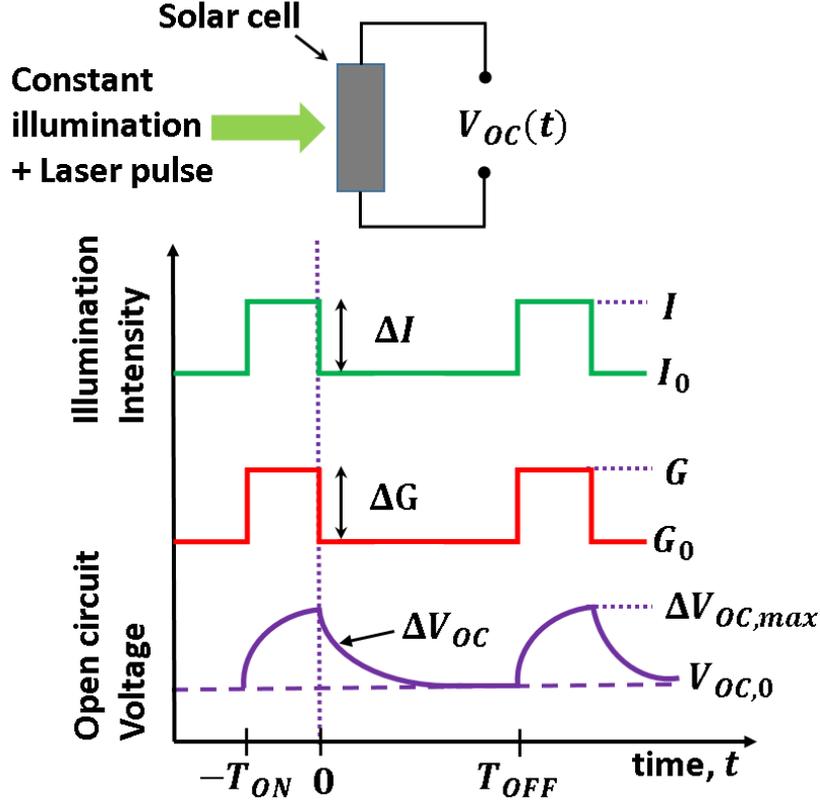

**Figure 8**: *Optoelectronic characterizations to explore radiative detailed balance limits of perovskite solar cells. The top panel shows the schematic of the experimental setup (not drawn to scale). The solar cell is subjected to a constant illumination which is modulated by laser pulses, and the resultant open circuit voltage transients are measured. The bottom panel shows the time dependence of illumination intensity (I), corresponding photo-generation rate (G) and open circuit voltage ($V_{OC}$).*

Further mathematical analysis correlates the various parameters related to the steady state and transient measurements as (Note that the subscript 0 denotes steady-state parameters.)

$$V_{OC,0} = \frac{kT}{q}\ln(G_0) - \frac{kT}{q}\ln(k_2 n_i^2) \qquad (6.5a)$$

$$\Delta V_{OC,max} \approx 2\frac{kT}{q}\frac{\Delta G T_{ON}}{\sqrt{G_0/k_2}} \qquad (6.5b)$$

$$\tau^{-1} \approx k_1 + 2\sqrt{k_2}\sqrt{G_0} \qquad (6.5c)$$

$$\ln \tau \approx -\ln(2k_2 n_i) - \frac{q}{2kT}V_{OC,0} \qquad (6.5d)$$

Let's name all these equations 6.5a, 6.5b, 6.5c, and 6.5d together as eq. 6.5. The detailed derivation of the above equations is shown below in section 6.3.4.

### 6.3.4 Theoretical analysis of intensity dependent scaling laws of the steady state and transient measurements

In this section, we provide the detailed derivations of equations 6.5a, 6.5b, 6.5c and 6.5d

The carrier dynamics under illumination (with intensity $I$ and corresponding photo-generation rate $G$) for an undoped sample with $n = p$ is given as

$$\frac{\partial n}{\partial t} = G - k_1 n - k_2 n^2$$

with $G = G_0$,

we have $G_0 = k_1 n_0 + k_2 n_0^2$

where $n_0$ is the steady-state carrier density.

With $n = n_0 + \Delta n$, through a perturbation analysis ($\Delta G \ll G_0, \Delta n \ll n_0$) we find

$$\frac{\partial (n_0 + \Delta n)}{\partial t} = (G_0 + \Delta G) - k_1(n_0 + \Delta n) - k_2\ (n_0 + \Delta n)^2$$

$$\frac{\partial \Delta n}{\partial t} = \ (G_0 + \Delta G) - \ k_1 n_0 - k_1 \Delta n - k_2(n_0^2 + 2n_0 \Delta n + \Delta n^2)$$

$$\frac{\partial \Delta n}{\partial t} = \ G_0 - (k_1 n_0 + k_2 n_0^2) + \Delta G - k_1 \Delta n - 2k_2 n_0 \Delta n$$

$$\frac{\partial \Delta n}{\partial t} = \ \Delta G - k_1 \Delta n - 2k_2 n_0 \Delta n$$

$$\frac{\partial \Delta n}{\partial t} = \ G' - (k_1 + 2k_2 n_0)\Delta n$$

where $G' = \Delta G$ during $t < 0$ and $G' = 0$ during $t > 0$ (**see Figure 8**).

During $t > 0$,

$$\frac{\partial \Delta n}{\partial t} = \ 0 - (k_1 + 2k_2 n_0)\Delta n$$

$$\frac{\partial \Delta n}{\Delta n} = \ -(k_1 + 2k_2 n_0)\partial t$$

$$\ln(\Delta n) = \ -(k_1 + 2k_2 n_0)t + c$$

$$\Delta n(t) = \Delta n(0)e^{-(k_1 + 2k_2 n_0)t}$$

$$\Delta n(t) = \Delta n(0)e^{-t/\tau}$$

$$\tau = \ 1/(k_1 + 2k_2 n_0)$$

Assuming negligible recombination during $0 < t < T_{ON}$, we have $\Delta n(0) \ \approx \Delta G T_{ON}$.

Now, we have

$$G_0 = k_1 n_0 + k_2 n_0^2$$

For dominant bimolecular recombination,

$$G_0 \approx k_2 n_0^2$$

$$n_0 \approx \sqrt{G_0/k_2}$$

**Derivation of eq. 6.5a**

.

At the dominant bimolecular recombination, with $n = p$ and $k_2 np = G_0$, we have $n^2 \approx \frac{G_0}{k_2}$

The split in Quasi-Fermi levels indicate $V_{OC} = \left(\frac{kT}{q}\right) \times \ln\left(\frac{np}{n_i^2}\right)$

$$V_{OC} = \left(\frac{kT}{q}\right) \times \ln\left(\frac{n^2}{n_i^2}\right)$$

$$V_{OC,0} = \left(\frac{kT}{q}\right) \times \ln\left(\frac{G_0}{k_2 n_i^2}\right)$$

$$V_{OC,0} = \frac{kT}{q}\ln(G_0) - \frac{kT}{q}\ln(k_2 n_i^2)$$

**Derivation of eq. 6.5b**

$$V_{OC} = \left(\frac{kT}{q}\right) \times \ln\left(\frac{np}{n_i^2}\right)$$

With $n = p$, and under the assumption of small perturbation in comparison to the background illumination,

$$V_{OC} = \left(\frac{kT}{q}\right) \times \ln\left(\frac{n^2}{n_i^2}\right)$$

$$V_{OC} = \frac{2kT}{q} \, \ln\left(\frac{n}{n_i}\right)$$

$$V_{OC,0} + \Delta V_{OC}(t) = \frac{2kT}{q} \, \ln\left(\frac{n_0 + \Delta n}{n_i}\right)$$

$$V_{OC,0} + \Delta V_{OC}(t) = \frac{2kT}{q} \, \ln\left(\frac{n_0}{n_i}\right) + \frac{2kT}{q} \, \ln\left(1 + \frac{\Delta n}{n_0}\right)$$

$$\Delta V_{OC}(t) = \frac{2kT}{q} \, \ln\left(1 + \frac{\Delta n}{n_0}\right)$$

$$\Delta V_{OC}(t) \approx \frac{2kT}{q}\left(\frac{\Delta n}{n_0}\right)$$

Now, $\Delta n = \Delta n(0)e^{-t/\tau}$

$$n_0 = \sqrt{G_0/k_2}$$

$$\Delta n(0) = \Delta G T_{ON}$$

$$\Delta V_{OC}(t) \approx \frac{2kT}{q} \times \frac{\Delta n(0)e^{-t/\tau}}{\sqrt{G_0/k_2}}$$

$$\Delta V_{OC}(t) \approx \frac{2kT}{q}\left(\frac{\Delta G T_{ON}}{\sqrt{G_0/k_2}}\right)e^{-t/\tau}$$

$$\Delta V_{OC,max} \approx \frac{2kT}{q}\left(\frac{\Delta G T_{ON}}{\sqrt{G_0/k_2}}\right)$$

$$\Delta V_{OC}(t) = \Delta V_{OC,max}\, e^{-t/\tau}$$

**Derivation of eq. 6.5c**

We have,

$$\tau = 1/(k_1 + 2k_2 n_0)$$

$$\tau^{-1} = k_1 + 2k_2 n_0$$

$$\tau^{-1} = k_1 + 2k_2\sqrt{G_0/k_2}$$

$$\tau^{-1} = k_1 + 2\sqrt{k_2}\sqrt{G_0}$$

**Derivation of eq. 6.5d**

With dominant bimolecular recombination, we have

$$\tau^{-1} \approx 0 + 2k_2 n_0$$

$$\frac{1}{\tau} \approx 2k_2 n_0$$

$$2k_2 n_0 \approx \frac{1}{\tau}$$

$$n_0 \approx \frac{1}{2k_2\tau}$$

$$V_{OC,0} = \frac{2kT}{q}\, ln\left(\frac{n_0}{n_i}\right)$$

$$V_{OC,0} = \frac{2kT}{q}\, ln\left(\frac{1}{2k_2\tau\, n_i}\right)$$

$$\ln(2k_2 n_i \tau) = -\frac{qV_{OC,0}}{2kT}$$

$$2k_2 n_i \tau = e^{-qV_{OC,0}/2kT}$$

$$\tau = \frac{1}{2k_2 n_i} e^{-qV_{OC,0}/2kT}$$

$$\ln \tau = \ln\left(\frac{1}{2k_2 n_i}\right) - qV_{OC,0}/2kT$$

$$\ln \tau = -\ln(2k_2 n_i) - \frac{q}{2kT}V_{OC,0}$$

**Intensity dependent carrier generation rate**

$J_{SC} = q \times G \times W$ at all conditions (assuming negligible recombination)

$J_{SC,AM\ 1.5G} = q \times G_{AM\ 1.5G} \times W$

Where, $J_{SC,AM\ 1.5G} =$ current density at the 1-Sun condition

$\quad\quad G_{AM\ 1.5G} =$ charge carrier generation at the 1-Sun condition

$q$ and $W$ are the electronic charge and the thickness of the device respectively.

$G_{AM\ 1.5G}$ is the generation of charge carriers at $AM\ 1.5G$ condition i.e. illumination light intensity $I = 100$ mW/cm$^2$

Let $G_0$ is the generation of charge carriers at illumination light intensity $I_0$ (in units of mW/cm$^2$)

Then, we have $\frac{G_{AM\ 1.5G}}{G_0} = \frac{100}{I_0}$

$$G_0 = G_{AM\ 1.5G} \times \frac{I_0}{100}$$

Now, $J_{SC,AM\ 1.5G} = q \times G_{AM\ 1.5G} \times W$

$$G_{AM\ 1.5G} = \frac{J_{SC,AM\ 1.5G}}{q \times W}$$

Therefore,

$$G_0 = G_{AM\ 1.5G} \times \frac{I_0}{100}$$

$$G_0 = \frac{J_{SC,AM\ 1.5G}}{q \times W} \times \frac{I_0}{100}$$

$$G_0 = M \times I_0$$

Where $M$ is a constant term as,

$$M = \frac{J_{SC,AM1.5G}}{q \times W \times 100}$$

**The further derivation of eq. 6.5a is the following**

$$V_{OC,0} = \left(\frac{kT}{q}\right) \times \ln\left(\frac{G_0}{k_2 n_i^2}\right)$$

$$V_{OC,0} = \left(\frac{kT}{q}\right) \times \ln\left(\frac{MI_0}{k_2 n_i^2}\right)$$

$$V_{OC,0} = \left(\frac{kT}{q}\right) \times \ln(I_0) + \left(\frac{kT}{q}\right) \ln\left(\frac{M}{k_2 n_i^2}\right)$$

$$V_{OC,0} = \left(\frac{kT}{q}\right) \times \ln(I_0) + \left(\frac{kT}{q}\right) \ln\left(\frac{J_{SC,AM1.5G}}{q \times W \times 100 \times k_2 n_i^2}\right)$$

### 6.3.5 Benchmarks for a solar cell dominated by bimolecular recombination

With $G \propto I$, a perovskite solar cell at detailed balance limit with dominant bimolecular recombination should satisfy the following benchmarks in terms of the background illumination intensity $I_0$ :

**(a)** light ideality factor is close to 1 (as per eq. 6.5a),

**(b)** $\Delta V_{OC,max}$ varies as $I_0^{-0.5}$ (as per eq. 6.5b),

**(c)** $\tau^{-1}$ varies as $I_0^{0.5}$ (as per eq. 6.5c), and

**(d)** the slope of $\ln(\tau)$ *vs.* $V_{OC,0}$ will be $\frac{q}{2kT}$ (as per eq. 6.5d).

**(e)** consistent back extraction of parameters like $k_1$, $k_2$, and $n_i$ from transient measurements. Further, eq. 6.3 indicates that mono-molecular recombination dominates at low carrier densities. Accordingly, we expect that the device is dominated by monomolecular recombination under dark conditions and by bimolecular recombination under illumination. Such a device should exhibit the following additional characteristics.

**(f)** dark ideality factor is close to 2.

**(g)** ideality factor obtained using the Suns-$V_{OC}$ based pseudo-$J - V$ should be 1.

**(h)** The recombination parameters estimated using the illumination dependent steady state and transient measurements should anticipate/predict the reverse saturation current densities obtained from dark $J - V$ and the Suns-$V_{OC}$ based pseudo-$J - V$ characteristics. Further, we expect the above analysis to lead to

**(i)** a consistent explanation for the external quantum efficiency of electroluminescence of the same solar cell under dark conditions (i.e., when configured as a Light Emitting Diode). And,

**(j)** this should enable reliable estimates for the coefficient of radiative recombination, which in turn compares well with SQ analysis.

It is evident that the above list demands accurate and self-consistent analysis of multiple characterization techniques. This is clearly significant and relevant to the community as the above is expected from device level characterizations and not thin-film or material characterization. Hence, if successful, this could lead to the self-consistent characterization of multiple phenomena based on terminal $J - V$ characteristics. To this end, below, we provide a summary of the experimental results and then the required theoretical analysis.

### 6.3.6 Experimental validation of theoretical predictions

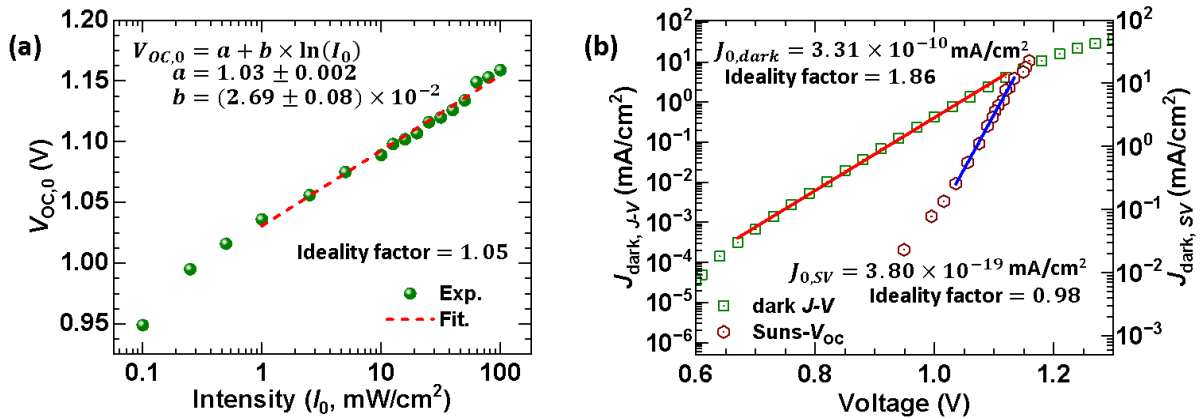

**Figure 9**: *Intensity dependence of $V_{OC}$ and dark $J - V$ of perovskite solar cells. (a) Variation of the open-circuit voltage ($V_{OC,0}$) with steady state illumination intensity. The ideality factor obtained[35] from $V_{OC,0}$ vs. $ln(I_0)$ is 1.05, which indicates that the device could be operating close to the radiative detailed balance limit. (b) The dark current characteristics measured from $J - V$ scans and Suns $V_{OC}$ measurements.[36–38] The $J_{dark,J-V}$ on the left y-axis indicates measurement from dark $J - V$ scans whereas the $J_{dark,SV}$ on the right y-axis indicates the dark current estimated using Suns $V_{OC}$ method.[39,40] Solid lines indicate numerical fits to obtain parameters like reverse saturation current density and ideality factor.*

The results from steady-state and transient optoelectrical measurements are shown in **Figure 9** and **Figure 10** respectively. Remarkably, the experimental trends are well in accordance with the predictions of eq. 6.5. Note that $\Delta I$ is kept a constant in our measurements (see **Figure 8**), as per literature.[26,27] The key trends are as follows:

(i)     light ideality factor is 1.05 (see **Figure 9a**) which compares well with the predictions of eq. 6.5a for devices dominated by bimolecular recombination.

(ii)    $\Delta V_{OC,max}$ varies as $I_0^\alpha$ (see **Figure 10a**, left panel) with $\alpha = -0.57$. The power exponent ($\alpha$) improves from $-0.57$ to $-0.54$ if the fit excludes the transients with $\Delta V_{OC,max} < 15\ mV$. This compares well with the theoretical predictions of eq. 6.5b.

(iii)   $\tau^{-1}$ scales as $I_0^\beta$ (see **Figure 10a**, right panel) with $\beta = 0.48$. The power exponent ($\beta$) improves from $0.48$ to $0.51$ if the fit excludes the transients with $\Delta V_{OC,max} < 15\ mV$ - in accordance with the predictions of eq. 6.5c.

(iv)    **Figure 10c** shows that $\tau$ varies exponentially with $V_{OC,0}$ with two distinct slopes. Indeed, for a broad range of measurements with $\Delta V_{OC,max} > 15\ mV$, we find that $\ln(\tau)$ varies linearly with $V_{OC,0}$ with a slope of $-19.94$ which compares well with $-q/2kT$, as per eq. 6.5d.

(v)     **Figure 9b** indicates that the ideality factor associated with dark $J - V$ is close to 2 while the ideality factor associated with pseudo $J - V$ based on Suns-$V_{OC}$ measurements is close to 1. These observations clearly indicate that monomolecular recombination dominates the dark current while bimolecular recombination dominates under illumination. Further,

(vi)    As the experimental trends are in accordance with theoretical predictions, it is desirable to self-consistently back extract recombination parameters. For example, eq. 6.5 allows such back extraction of parameters like $k_1$, $k_2$, and $n_i$ from transient measurements. If robust, then these estimates should anticipate key features of dark current, like the reverse saturation current density. Such a self-consistent estimation of recombination parameters from finished devices is provided in the next section.

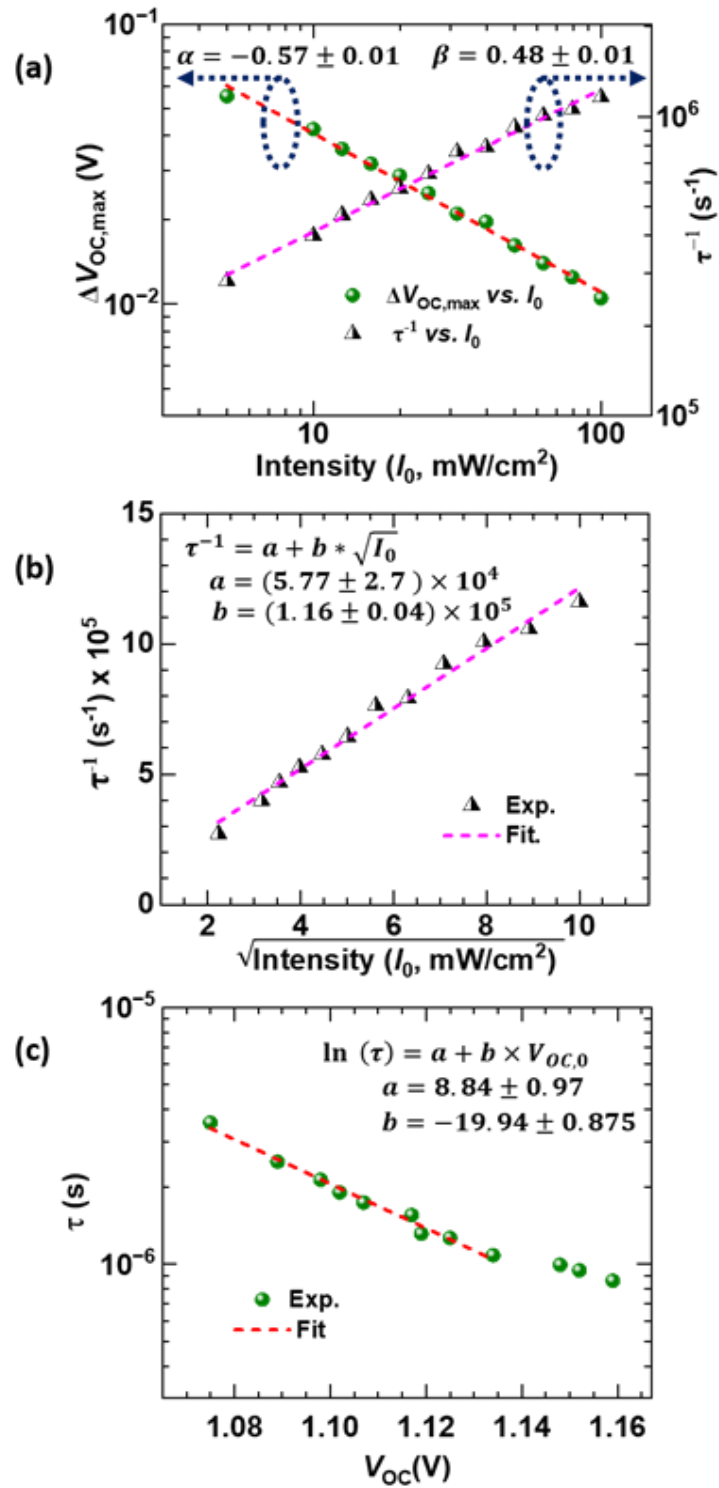

***Figure 10****: Comparison of experimental results against theoretical benchmark criteria (eq. 6.5) for perovskite solar cells. (**a**) $\Delta V_{OC,max}$ and $\tau^{-1}$ variation with $I_0$ are in accordance with eq. 6.5b,c (**b**) Direct extraction of $k_1$ and $k_2$ using $\tau^{-1}$ vs. $\sqrt{I_0}$ plot, from the intercept and the slope respectively (i.e., using eq. 6.5c). (**c**) $ln(\tau)$ varies linearly with $V_{OC,0}$ with slope close to $-q/2kT$, as anticipated by eq. 6.5d. These trends indicate that the device is limited by bimolecular recombination.*

**6.3.7 Self-consistent back extraction of recombination parameters**

The results shown in **Figures 9-10** clearly indicate that our device is indeed operating at detailed balance limit with dominant bimolecular recombination, as the experimental trends compare well against all the benchmark criteria identified by eq. 6.5. In addition, eq.6.5 also allows self-consistent back extraction of various recombination parameters. For example, as per eq. 6.5c, **Figure 10b** allows back extraction of $k_2$ as $(1.04 \pm 0.07) \times 10^{-10}$ cm$^3$s$^{-1}$ (from the slope) while the intercept leads to $k_1$ as $(5.77 \pm 2.7) \times 10^4$ s$^{-1}$ (see **Table 3**). Importantly, an independent estimate for $k_2$ is obtained using eq.6.5b as $1.12 \times 10^{-10}$ cm$^3$s$^{-1}$ (**Figure 11**) which compares well with the previous estimate. This independent scheme for the estimation of $k_2$ is immensely useful for scenarios in which the direct estimation of $k_1$ and $k_2$ using eq. 6.5c could be limited by the noise in the measured data. From the intercept of **Figure 10c** and using eq. 6.5d, $n_i$ is estimated as $(0.25 - 1.98) \times 10^6$ cm$^{-3}$. Note that these estimates are obtained from transient measurements. Another estimate for $n_i$ can be obtained from steady-state measurements using eq. 6.5a and **Figure 9a** as the y-intercept of $V_{OC,0}$ *vs.* $\ln(I_0)$ is related to $-\ln(k_2 n_i^2)$. Using the back-extracted value for $k_2$ and the y-intercept of **Figure 9a** (See **Table 3** for details), we find that $(1.03 - 1.19) \times 10^6$ cm$^{-3}$. Note that both these estimates for $n_i$ rely on the value of $k_2$. An independent estimate for $n_i$ can be obtained as follows: Specifically, $k_2 n_i^2$ can be obtained using eq. 6.5a and **Figure 9a** while $k_2 n_i$ can be obtained using eq. 6.5d and **Figure 10c**. This leads to an independent estimate for $n_i$ as $1.76 \times 10^6$ cm$^{-3}$. Interestingly and reassuringly, the estimate for $n_i$ obtained independently through respective steady state and transient measurements compare very well. Thus, through a combination of steady-state and transient illumination dependent measurements, here we obtain a coherent as well as consistent set of estimates for $k_1$, $k_2$, and $n_i$ (**Table 3** for details). The estimated recombination parameters are well in accordance with literature **Table 4**. It is worth noting that the estimates of $k_1$, $k_2$, and $n_i$ reported in the literature are typically based on the optical characterization of the thin films. Here, we estimated and validated the recombination parameters through a consistent set of device level experiments relying only on terminal characteristics, both steady state and transient. The parameters estimated from the steady-state and transient measurements compare well with each other and with the values reported in the literature (see **Table 4**).

**Back extraction of the recombination parameters:**

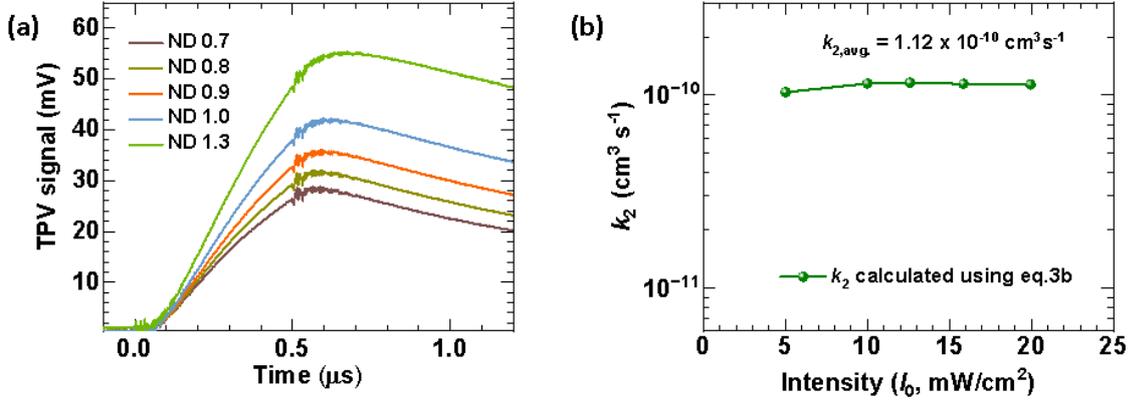

***Figure 11:*** *(**a**) The transient photovoltage (TPV) signals a rising profile for different background intensities. (**b**) The estimation of $k_2$ from eq. 6.5b for different background intensities.*

$$\Delta V_{OC,max} \approx \frac{2kT}{q} \left( \frac{\Delta G T_{ON}}{\sqrt{G_0/k_2}} \right)$$

Where $\Delta G$ is the photons incident per unit time per unit area

$$\Delta G = \frac{laser \; powere \; (P)}{A \times W \times \frac{hc}{\lambda}} \times f_{in}$$

where, laser power $P = 10$ mW, wavelength $\lambda = 490$ nm, Active area of the device $A = 0.175$ cm$^2$, Thickness of the device $W = 450$ nm $f_{in} \approx 0.85$ is the fraction of light incident to the device after reflection loss $R_L \approx 15\%$ (**Figure 1b**) Here, note that the generation rate is calculated against the device area (A) although the spot size (~0.01 cm$^2$) of the laser is much smaller than A. This is well justified as follows: Under open circuit conditions, although the carriers are generated under the influence of the localized laser pulse, they quickly spread over the entire device area. The typical time taken is in the order of $t_{diff} = \frac{W^2}{D}$, where D is the diffusion coefficient. With carrier mobilities of the order of 10 cm$^2$/Vs, we find $t_{diff} \approx 7.88 \; ns$. As this time is much shorter than the recombination time of the order of $\sim \mu$s, the effective generation can be assumed to be uniform over the entire device area (and not over the area of the laser spot).

Accordingly, we find, $\Delta G = 2.66 \times 10^{21}$ cm$^{-3}$s$^{-1}$

$T_{ON} = 500$ ns

$$G_0 = M \times I_0$$

Where $M$ is a constant term as,

$$M = \frac{J_{SC,AM1.5G}}{q \times W \times 100}$$

$$\Delta V_{OC,max} = 183.4 \times \sqrt{\frac{100 \times k_2}{I_0 \times J_{SC,AM\,1.5G}}}$$

$$k_2 = 2.97 \times 10^{-5} \times \left(\frac{I_0}{100}\right) \times J_{SC,AM\,1.5G} \times \left(\Delta V_{OC,max}\right)^2$$

The average value of $k_2$ is estimated by using the above equations. At lower background intensities, the photovoltage increase with time is quite linear (i.e., negligible recombination), whereas the trends at higher intensities show the influence of recombination (and hence non-linear variation). We used five data points to calculate the $k_2$ at each background intensity, and the average is shown in **Figure 11b**.

***Table 3***: *List of the recombination coefficients $k_1, k_2$ and $n_i$ measured from benchmark eq. 6.5,* ***Figure 9****, and* ***Figure 10***.

| SL.# | Parameters | Equation and Figure used | Calculations used | Values |
|---|---|---|---|---|
| (i) | $k_1$ | eq. 6.5c and Figure 10b. | $\tau^{-1} \approx k_1 + 2\sqrt{k_2}\sqrt{G_0}$ <br><br> $intercept = k_1 = 5.77 \times 10^4$ | $k_1 = (5.77 \pm 2.7) \times 10^4 \, \mathrm{s^{-1}}$ |
| (ii) | $k_2$ | eq. 6.5c and Figure 10b. | $\tau^{-1} \approx k_1 + 2\sqrt{k_2}\sqrt{G_0}$ <br><br> $\tau^{-1} \approx k_1 + 2\sqrt{\frac{k_2 \times J_{SC,AM\,1.5G}}{qW \times 100}}\sqrt{I_0}$ | $k_2 = (1.04 \pm 0.07) \times 10^{-10} \, \mathrm{cm^3 s^{-1}}$ |
| | | | $slope = 2\sqrt{\dfrac{k_2 \times J_{SC,AM\,1.5G}}{qW \times 100}}$ $= (1.16 \pm 0.04) \times 10^5$ | |
| (iii) | $k_2$ | Eq.6.5b and Figure 11. | $\Delta V_{OC,max} \approx \dfrac{2kT}{q}\left(\dfrac{\Delta GT_{ON}}{\sqrt{G_0/k_2}}\right)$ $k_2 = 2.97 \times 10^{-5} \times \left(\dfrac{I_0}{100}\right) \times J_{SC,AM\,1.5G}$ $\times \left(\Delta V_{OC,max}\right)^2$ | $k_2 = 1.12 \times 10^{-10}$ cm$^3$s$^{-1}$ |
| (iv) | $n_i$ | Eq. 6.5d, Figure 10c and using $k_2$ | $\ln\tau \approx -\ln(2k_2 n_i) - \dfrac{q}{2kT}V_{OC,0}$ $intercept = -\ln(2k_2 n_i)$ $= (8.84 \pm 0.97)$ | Using the value of $k_2$ (from Figure 10b) $k_2 = (1.04 \pm 0.07) \times 10^{-10}$ cm$^3$s$^{-1}$ The range of $n_i$ is $n_i = (0.25 - 1.98) \times 10^6$ cm$^{-3}$ <br><br> Using the value of $k_2$ (from Figure 11) $k_2 = 1.12 \times 10^{-10}$ cm$^3$s$^{-1}$ The range of $n_i$ is $n_i = (0.24 - 1.71) \times 10^6$ cm$^{-3}$ |

| (v) | $n_i$ | Eq. 6.5a and Figure 9a and using $k_2$ | $V_{OC,0} = \dfrac{kT}{q}\ln(G_0) - \dfrac{kT}{q}\ln(k_2 n_i^2)$ $V_{OC,0} = \dfrac{kT}{q} \times \ln(I_0) + \dfrac{kT}{q} \times \ln\left(\dfrac{M}{k_2 n_i^2}\right)$ $intercept = \dfrac{kT}{q} \times \ln\left(\dfrac{M}{k_2 n_i^2}\right)$ $intercept$ $= \dfrac{kT}{q}\ln\left(\dfrac{J_{SC,AM1.5G}}{q \times W \times 100 \times k_2 n_i^2}\right)$ $= 1.03 \pm 0.002$ | Using the value of $k_2$ (from Figure 10b) $k_2 = (1.04 \pm 0.07) \times 10^{-10}$ cm³s⁻¹ The range of $n_i$ is $n_i = (1.03 - 1.19) \times 10^6$ cm⁻³ |
| | | | | Using the value of $k_2$ (from Figure 11) $k_2 = 1.12 \times 10^{-10}$ cm³s⁻¹ The range of $n_i$ is $n_i = (1.03 - 1.11) \times 10^6$ cm⁻³ |
| (vi) | $n_i$ | Eq. 6.5a and Eq. 6.5d | $V_{OC,0} = \dfrac{kT}{q}\ln(G_0) - \dfrac{kT}{q}\ln(k_2 n_i^2)$ $\ln\tau \approx -\ln(2k_2 n_i) - \dfrac{q}{2kT}V_{OC,0}$ $\dfrac{k_2 n_i^2}{k_2 n_i} = \dfrac{127.5}{7.24 \times 10^{-5}}$ | $n_i = 1.76 \times 10^6$ cm⁻³ |

It is worth noting that the estimates of $k_1$, $k_2$, and $n_i$ in the literature are based on the characterization of thin films and not from completed devices. This chapter provides the first estimates for the recombination parameters from the device level characterizations, and the parameters obtained from the steady-state and transient measurements are consistent with each other. Further, these estimates compare well with the literature reports based on thin film studies (see **Table 4**).

**Table 4**: The list of recombination parameters for the perovskite semiconductor (literature survey).

| List of references | $k_1$ (s⁻¹) | $k_2$ (cm³s⁻¹) | $k_{2,rad}$ (cm³s⁻¹) | $k_3$ (cm⁶ s⁻¹) | $n_i$ (cm⁻³) | Ref. |
|---|---|---|---|---|---|---|
| A[41] | - | - | $5 \times 10^{-11}$ | - | $8 \times 10^4$ | [41] |
| B[41] | - | $(8.77 \pm 0.79) \times 10^{-10}$ | $(4.78 \pm 0.43) \times 10^{-11}$ | $(8.83 \pm 1.57) \times 10^{-29}$ | - | [41] |
| C[42] | - | $(8.77 \pm 0.79 \times 10^{-10}$ | $(4.78 \pm 0.43) \times 10^{-11}$ | | $(3.46 \pm 0.47) \times 10^5$ | [42] |
| D[43] | - | - | $(4.6 \pm 0.2) \times 10^{-11}$ | $(8.0 + 2.1) \times 10^{-29}$ | - | [43] |
| E[43] | - | - | $(6.3 \pm 0.3) \times 10^{-11}$ | $(9.7 + 1.4) \times 10^{-29}$ | - | [43] |
| F[44] | - | $6.8 \times 10^{-10}$ | - | - | - | [44] |
| G[45] | $5 \times 10^6$ | $8.1 \times 10^{-11}$ | $7.2 \times 10^{-11}$ | $1.1 \times 10^{-28}$ | - | [45] |
| H[45] | $5 \times 10^6$ | $7.9 \times 10^{-11}$ | $5.6 \times 10^{-11}$ | $1.8 \times 10^{-28}$ | - | [45] |
| I[45] | $2.5 \times 10^6$ | $7.0 \times 10^{-11}$ | $5.4 \times 10^{-11}$ | $6 \times 10^{-29}$ | - | [45] |
| J[46] | $1.7 \times 10^5$ | $2.3 \times 10^{-10}$ | - | $5.4 \times 10^{-28}$ | - | [46] |
| K[47] | | $5 \times 10^{-11}$ | | $4.4 \times 10^{-29}$ | $n_i = 8.05 \times 10^4$ | [47] |
| L[48] | - | $1.1 \times 10^{-10}$ | - | $3 \times 10^{-29}$ | - | [48] |
| M[49] | - | $(0.6 - 1.1) \times 10^{-10}$ | - | | - | [49] |
| N[50] | 0 | $(2 - 4) \times 10^{-11}$ | - | - | - | [50] |
| O[51] | - | $4.75 \times 10^{-10}$ | - | $10^{-28}$ | | [51] |
| P[52] | $10^5 - 10^9$ | $10^{-12} - 10^{-7}$ | - | $10^{-29} - 10^{-26}$ | - | [52] |
| Q[53] | $2 \times 10^6$ | $3 \times 10^{-11}$ | - | - | | [53] |
| R[54] | $4 \times 10^6$ | $3.8 \times 10^{-10}$ | $(3 \pm 0.3) \times 10^{-11}$ | - | $1.25 \times 10^5$ | [54] |

### 6.3.8 Validation of the back extracted recombination parameters

As a further validation, the back extracted parameters can be used to predict the diode reverse saturation currents obtained from dark $J - V$ and Suns-$V_{OC}$ measurements. For this we use $k_1 = 5.77 \times 10^4$ s$^{-1}$, $k_2 = 10^{-10}$ cm$^3$s$^{-1}$, $n_i = 0.5 \times 10^6$ cm$^{-3}$, and $W = 450$ nm, where W is the thickness of the perovskite active layer. Note that the value of $n_i$ chosen is consistent with literature[42,47,54] and within the broad range estimated from our experiments. The ideality factor of dark current is close to 2 which indicates that the monomolecular recombination is dominant under such conditions (see **Figure 9b**). The reverse saturation current density measured from the dark $J - V$ is $3.31 \times 10^{-10}$ mA/cm$^2$, whereas the corresponding theoretical estimate for trap assisted recombination ($J_{0,SRH} = qk_1n_iW$) is $2.08 \times 10^{-10}$ mA/cm$^2$. The ideality factor of dark pseudo $J - V$ characteristics is close to 1 which indicates that bimolecular recombination dominates under illuminated conditions **Figure 12a**. The corresponding reverse saturation current density obtained experimentally is $3.80 \times 10^{-19}$ mA/cm$^2$ which compares well with the theoretical estimate $J_{0,BB} = qk_2n_i^2W = 1.87 \times 10^{-19}$ mA/cm$^2$ (**see Figure 12b and Table 5**).[54]

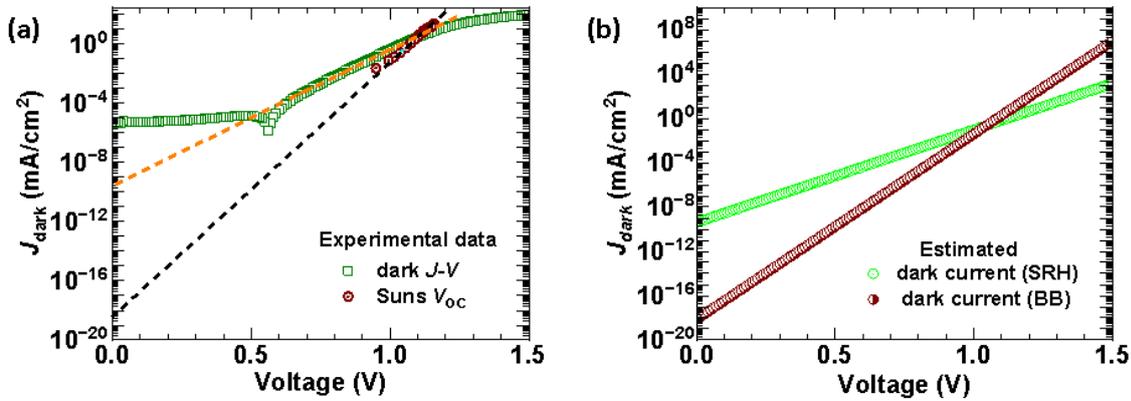

***Figure 12:*** *(**a**) The dark current measured (experimental) from the dark $J - V$ characteristic and from the Suns $V_{OC}$ measurement. The Suns $V_{OC}$ plot is obtained from the intensity dependent $J - V$ characteristics.[37,38] The dark current from the Suns $V_{OC}$ is obtained by shifting the pseudo $J - V$ by $J_{SC}$ towards the dark current.[39,40] (**b**) The dark current is estimated from the achievable limit calculation by using $J_{dark,SRH} = qk_1n_iWe^{qV/2kT}$ and $J_{dark,BB} = qk_2n_i^2We^{qV/kT}$.*

.

**Table 5**: *The reverse saturation current density* $J_0$ *(mA/cm²) and the ideality factor calculated from the dark current, Suns-* $V_{OC}$ *measurement and compared it with the estimates for trap assisted (SRH) and bimolecular recombination using* $J_{dark,SRH} = qk_1 n_i W e^{qV/2kT}$ *and* $J_{dark,BB} = qk_2 n_i^2 W e^{qV/kT}$ *respectively.*

| | Technique | $J_0$ (mA/cm²) | Ideality factor |
|---|---|---|---|
| **Dark $J - V$** | Experimental | $3.31 \times 10^{-10}$ | 1.86 |
| **dark current (SRH)** $k_1 = 5.77 \times 10^4$ **s⁻¹** , $n_i = 0.5 \times 10^6$ **cm⁻³** | Estimated | $2.08 \times 10^{-10}$ | 2 |
| **Suns $V_{OC}$** | Experimental | $3.80 \times 10^{-19}$ | 0.98 |
| **dark current (BB)** $k_2 = 1.04 \times 10^{-10}$ **cm³s⁻¹** $n_i = 0.5 \times 10^6$ **cm⁻³** | Estimated | $1.87 \times 10^{-19}$ | 1 |

Further, with the measured value of the band gap ($E_g = 1.6 \ eV$), and $n_i \approx 0.5 \times 10^6$ cm⁻³ with the assumption $N_C = N_V$, we find that $N_C \approx 1.65 \times 10^{19}$ cm⁻³ were $N_C, N_V$ are the effective density of states of conduction and valence band respectively. It is well known that a solar cell remains limited by bimolecular recombination for all carrier densities $n_{BB} > k_1/k_2$. Using the back extracted values of $k_1$ and $k_2$, we find the corresponding value of $n_{BB} = 5.55 \times 10^{14}$ cm⁻³. For 1 Sun illumination, the carrier density under open circuit conditions is given as $n_0 = \sqrt{G_0/k_2} = \sqrt{J_{SC}/qWk_2}$ ($W$ is the thickness of the active layer, $W = 450$ nm) which evaluates to $n_0 = 5.58 \times 10^{15}$ cm⁻³. Hence, our device is dominated by bimolecular recombination for illumination intensities as low as 0.01 Sun. The carrier density at which Auger recombination dominates is given as $n_{Aug} > k_3/k_2$, where $k_3$ is the coefficient of Auger recombination[55–58]. For $k_3 = 10^{-28}$ cm⁶s⁻¹, we find $n_{Aug} = 10^{18}$ cm⁻³ which is much larger than $n_0$. This indicates that the relative contribution of Auger recombination under 1 Sun conditions is negligible. All these factors contribute to ensure that our devices remain limited by bimolecular recombination over a large range of illumination intensities explored in **Figure 9.**

### 6.3.9 Analysis of the achievable limits (ALs)

The SQ limit parameters[59,60] (i.e., with only radiative recombination), for a solar cell with a band gap $\sim 1.6\ eV$ and thickness $W \approx 500$ nm are $\eta \approx 30\%$, $J_{SC,SQ} \approx 26$ mA/cm$^2$, $V_{OC,SQ} \approx 1.3$ V, $FF_{SQ} \approx 90\%$. The recombination limited $J - V$ characteristics of a corresponding solar cell is given as

$$J_{AL} = -J_{SC,SQ} + qk_1 n_i W e^{qV/2kT} + qk_2 n_i^2 W e^{qV/kT} + qk_3 n_i^3 W e^{3qV/2kT} \qquad (6.6)$$

Here, the first term on the RHS denotes the maximum $J_{SC}$ while the rest of the terms denote mono-molecular, bimolecular, and Auger recombination respectively (with the assumption that $n = p$ over the entire active layer). With the parameters $k_1 = 10^4$ s$^{-1}$, $k_2 = 10^{-10}$ cm$^3$s$^{-1}$, $k_3 = 10^{-28}$ cm$^6$s$^{-1}$, $n_i = 0.5 \times 10^6$ cm$^{-3}$, $W = 450$ nm, and $J_{SC,SQ} = 26$ mA/cm$^2$ the achievable limits (AL) of performance of such a cell is $\eta_{AL} = 27.76\%$, $V_{OC,AL} = 1.192\ V$, $FF_{AL} = 89.57\%$ (**Figure 13c** and **Table 6**). With $k_1 = k_3 = 0$ (i.e., under detailed balance limit with only bimolecular recombination), the above estimate improves only marginally (**Figure 13d** and **Table 6**). The gap between the achievable limit of efficiency with that of the SQ limit is of fundamental importance. Similar aspects could be valid for other perovskite systems.[61–64] Hence, exploration of the various electronic states and processes involved in the bimolecular recombination process is very relevant for identifying the achievable limit of efficiency and scope for further optimization.

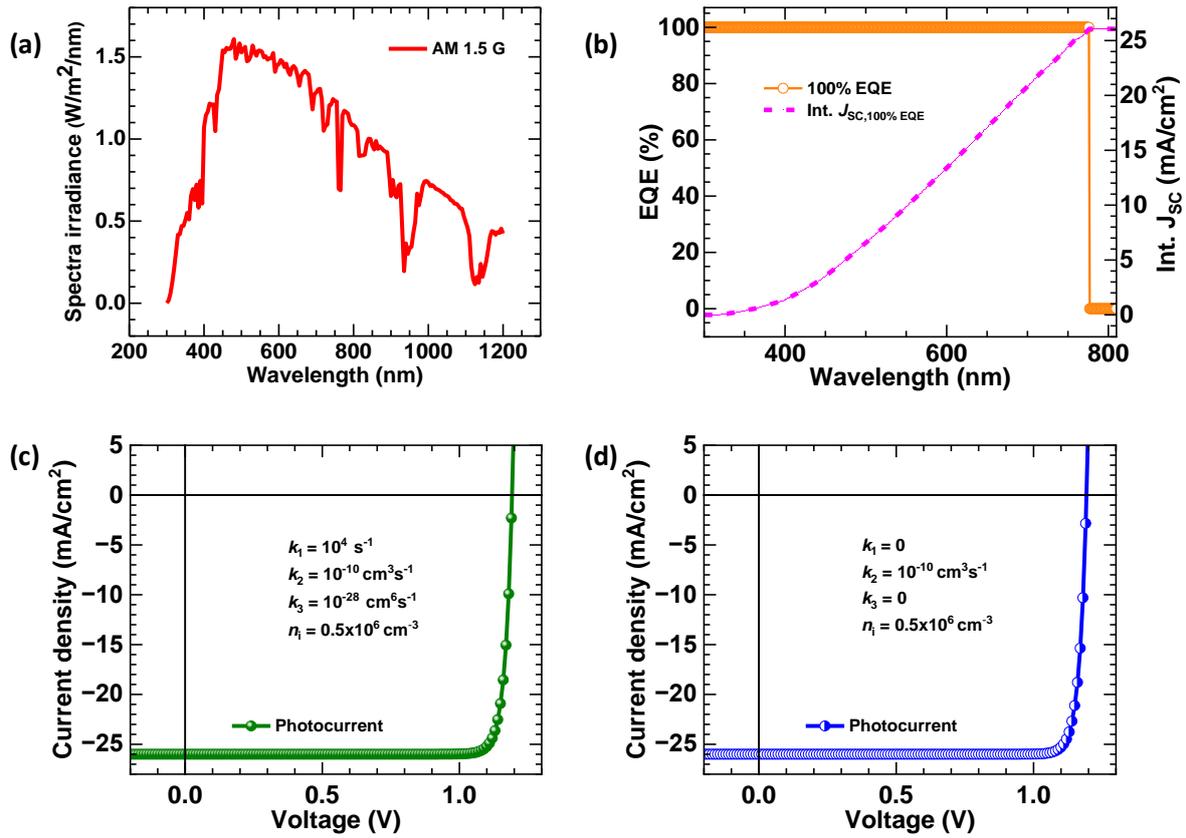

***Figure 13:*** (***a***) *AM 1.5G spectrum is taken from the PV lighthouse, which is used to calculate the integrated current density over the EQE spectrum of the PSCs.*[65] (***b***) *The ideal EQE spectrum (no loss) i.e. 100% EQE is considered, and the corresponding integrated $J_{SC}$ is calculated for $1.6\ eV$ $(300\ to\ 776\ nm)$ solar cells. (**c**) the achievable limit of the PV parameters calculated using integrated $J_{SC,SQ} \approx 26\ mA/cm^2$ (calculated from **Figure 13b**) and $k_1 = 10^4\ s^{-1}$ $k_2 = 10^{-10} cm^3 s^{-1}$, $k_3 = 10^{-28}\ cm^6 s^{-1}$, and $n_i = 0.5 \times 10^6\ cm^{-3}$ in equation (6.6). (**d**) The achievable limit of the PV parameters calculated using integrated $J_{SC,SQ} \approx 26\ mA/cm^2$ (calculated from **Figure 13b**) and revised values of $k_1 = 0$, $k_2 = 10^{-10}\ cm^3 s^{-1}$, $k_3 = 0$, and and $n_i = 0.5 \times 10^6\ cm^{-3}$ in equation (6.6).*

**Table 6**: *The list of the PV parameters estimated from illuminated $J-V$ at 1-sun condition, Suns $V_{OC}$ measurement and achievable efficiency limit calculation.*

|  | $J_{sc}$ (mA/cm²) | $V_{OC}$ (V) | FF (%) | PCE (%) |
|---|---|---|---|---|
| **achievable limit**<br>$k_1 = 10^4$ s⁻¹<br>$k_2 = 10^{-10}$ cm³s⁻¹<br>$k_3 = 10^{-28}$ cm⁶s⁻¹<br>$n_i = 0.5 \times 10^6$ cm⁻³ | 26 | 1.192 | 89.57 | 27.76 |
| **achievable limit**<br>$k_1 = 0$<br>$k_2 = 10^{-10}$ cm³s⁻¹<br>$k_3 = 0$<br>$n_i = 0.5 \times 10^6$ cm⁻³ | 26 | 1.193 | 89.72 | 27.83 |

### 6.3.10 Physical mechanisms that contribute to bimolecular recombination

In general, the bimolecular recombination rate could be due to radiative and non-radiative processes. Accordingly, we have

$$k_2 = k_{2,rad} + k_{2,nonrad} \qquad (6.7)$$

where $k_{2,rad}$ is the rate of band-band radiative recombination while $k_{2,nonrad}$ denotes non-radiative recombination processes[45,66,67] which vary as $n^2$. The origin of such non-radiative bimolecular processes is not well understood. Nevertheless, classical literature on trap assisted recombination indicates that shallow traps could lead to such effects[68]. Specifically, defects at an effective energy level $E_T$ could contribute to $k_2$ under the conditions $n_i e^{(E_T-E_i)/kT} > n_0$. With the back extracted value of $n_i = 0.5 \times 10^6$ cm⁻³ and $n_0 = 5.58 \times 10^{15}$ cm⁻³, we find that such traps should have $E_T - E_i > 0.59$ eV – rather the shallow traps could be with in 0.2 eV from the conduction/valence band edge for our active material with band gap of 1.6 eV. Preliminary transient photo-current (TPC) measurements of our devices (**Figure 14**) indeed indicate the presence of shallow traps[69–71] in accordance with the above arguments. A device dominated by such shallow traps could result in a light ideality factor of 1. However, under dark conditions, the expected ideality factor due to such shallow traps is 1. Experimental observation of dark ideality factor closes to 2 (see **Figure 9b**) indicates that our devices have both mid-gap recombination centres and shallow traps. As a result, the dark ideality factor is 2 and the light ideality factor could be close to 1. Numerical simulation results shown in **Figure 15** clearly support this inference.[72–75] Even though the hypothesis of shallow trap limited performance consistently explains the experimental observations on dark $J - V$, transient photo-voltage, and transient photo-current, it is evident that more experimental characterizations are needed to further establish and quantify role of shallow traps.

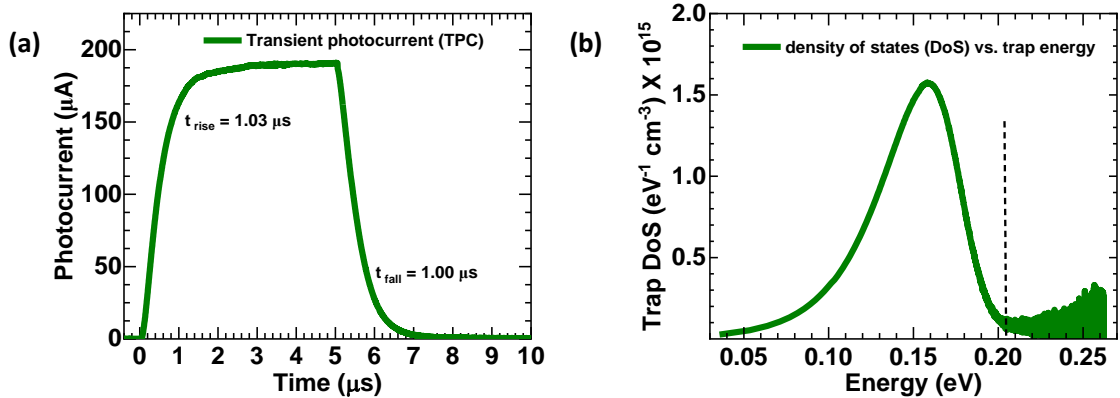

***Figure 14***: *(**a**) Transient photocurrent (TPC) using a 490 nm modulated laser of pulse duration 5 μs, and repetition rate 10 KHz with duty cycle 5%. (**b**) Trap density of states (DoS) as a function of demarcation energy calculated from transient photocurrent (TPC).*

The trap density is calculated using transient photocurrent (TPC) measurement **Figure 14a.** When the photocurrent is dominated by detrapping from a distribution of trap states, the density of trap states is given by[69,70]

$$N(E) = \frac{I(t).t}{q.V.f.k.T} \qquad (6.8)$$

Where *I(t)* is the photocurrent as a function of time, *q* is the elementary charge, *V* is the volume of the sample, *f* is the fraction of states filled at each energy level, *k* is the Boltzmann constant, and *T* is the temperature. It is assumed that *f=1*, all the states are filled at each energy level.

The above relation assumes that at time *t* the carriers at energy state E are fully detrapped. The detrapping energy is given by

$$E(t) = kTln(v_0 t) \qquad (6.9)$$

where the $v_0$ is the attempt to escape frequency which is in the range of ~$10^9$ Hz for perovskite solar cells.[71,76]

By using the above equations, the trap density of states (DoS) and location of the traps are calculated **Figure 14b**.

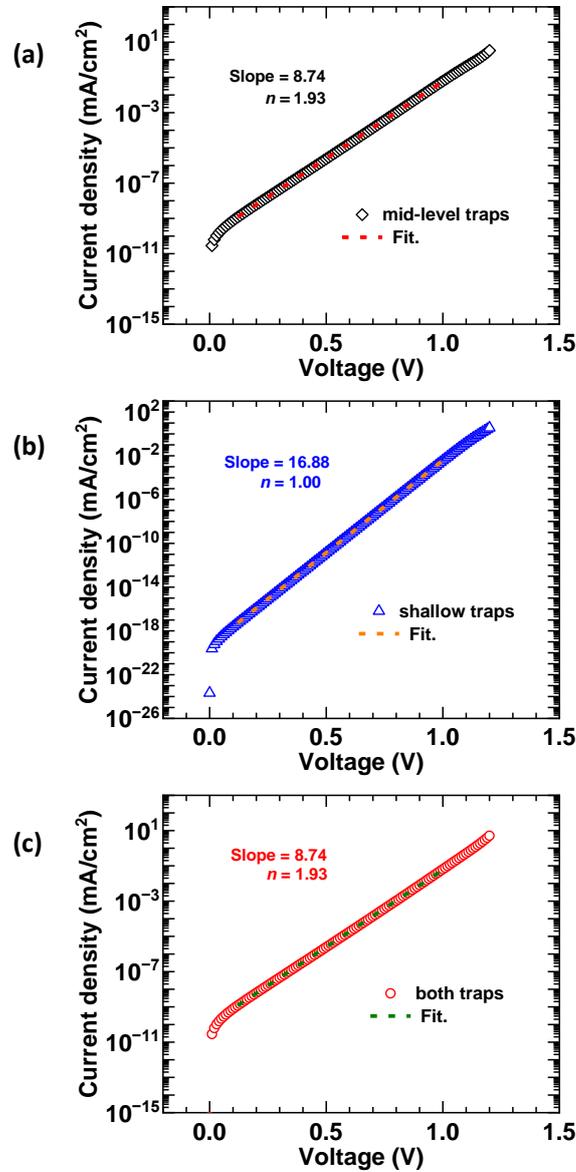

**Figure 15**: *Simulated dark J − V characteristic considering the (**a**) shallow traps (**b**) mid-level traps and (**c**) both shallow and mid-level traps. The shallow traps have $E_T − E_I = 0.6\ eV$, where $E_T$ is the effective trap level, $E_g \sim 1.6eV$. The simulation results were obtained through numerical solutions of non-linear coupled Poisson and drift-diffusion simulations. Methodology and calibration of simulation scheme are discussed in our prior literature.*[72–75,77–80]

**6.3.11 Electroluminescence and estimation of $k_{2,rad}$**

It is evident that non-radiative bimolecular recombination limits the performance of our solar cell. Hence, it is desirable to further characterize the parameters $k_{2,rad}$ and $k_{2,nonrad}$. This could be achieved through Electroluminescence ($EL$) measurements[81] with the same devices being used as LEDs. Under forward bias conditions, the LED current could be dominated by recombination and is given by $J \approx qW(k_1 n + k_2 n^2 + k_3 n^3)$. The internal quantum efficiency of EL is given as

$$IQE_{EL} = \frac{k_{2,rad}n^2}{k_1 n + k_2 n^2 + k_3 n^3} \qquad (6.10)$$

The maximum value of $IQE_{EL}$, as per the above equation, occurs at $n_{EL} = \sqrt{k_1/k_3}$. With $k_1 = 10^5 \mathrm{s}^{-1}$ and $k_3 = 10^{-28}$ cm$^6$s$^{-1}$, we have $n_{EL} \approx 3 \times 10^{16}$ cm$^{-3}$. At such carrier densities, it can be shown that bimolecular recombination with $k_2 = 10^{-10}$ cm$^3$s$^{-1}$ dominates both monomolecular and Auger recombination. Hence, further simple analysis indicates that the maximum external quantum efficiency of $EL$ is given as

$$EQE_{EL,max} \approx f \frac{k_{2,rad}}{k_2} \qquad (6.11)$$

where $f$ is the out-coupling factor. eq. 6.11 allows direct estimation of $k_{2,rad}$ from the EQE$_{EL}$ measurements. The measured $EQE_{EL,max}$ as a function of the dark current of our devices (see **Figure 16**) is 2.67%. With $f \approx 0.25$, we find $k_{2,rad} \approx 10^{-11}$ cm$^3$s$^{-1}$. With this estimate of $k_{2,rad}$, and using the equation $J = -J_{SC,SQ} + qk_{2,rad}n_i^2 W e^{qV/kT}$, the radiative limit performance parameters of our devices with back extracted parameters ($k_{2,rad} = 10^{-11}$ cm$^3$s$^{-1}$, $n_i = 0.5 \times 10^6$ cm$^{-3}$, $W = 450$ nm) are $\eta_{AL} = 29.34\%$, $V_{OC,AL} = 1.25\,V$, $FF_{AL} = 90.13\%$ (**Figure 17** and **Table 7**). These estimates are very close to the SQ limit performance for a solar cell of comparable bandgap. Interestingly, the above estimate is limited by uncertainties in the parameter combination $k_{2,rad}n_i^2 W$ which explains why the predicted $V_{OC}$ is lower than that of SQ limit of ~ 1.3 V (for the same bandgap). Further accurate estimates for $k_{2,rad}n_i^2 W$ could improve the $V_{OC}$ estimates thus approaching radiative detailed balance limits (i.e., $V_{OC,SQ} \approx 1.3\,V$).

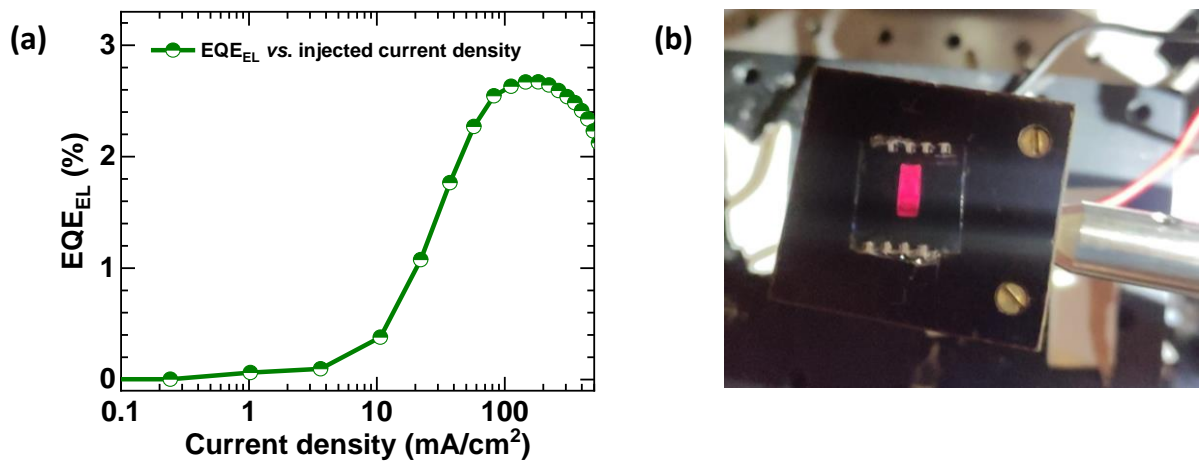

***Figure 16***: *The solar cell is used as a light-emitting diode (LED). (**a**) Electroluminescence external quantum efficiency of the solar cell device as a function of injected current density. (**b**) The photographic image of the device during LED operation.*

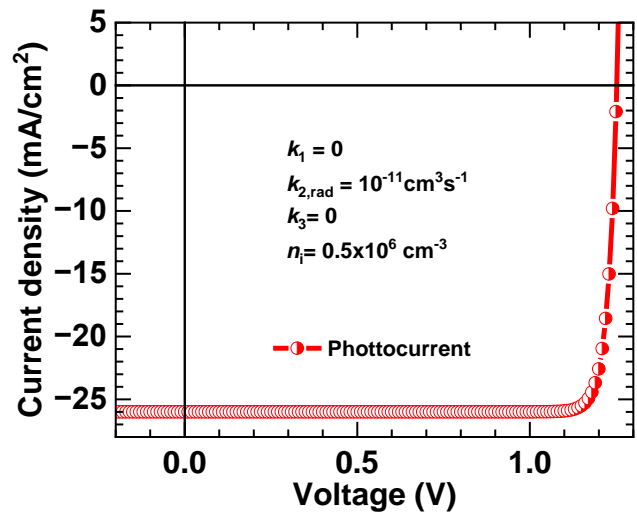

***Figure 17***: *(**a**) The achievable limit of the PV parameters calculated using integrated $J_{SC,SQ} \approx$ 26 mA/cm² (calculated from **Figure 13b**), $k_1 = 0$, $k_{2,rad} = 10^{-11}$ cm³s⁻¹, $k_3 = 0$, and $n_i = 0.5 \times 10^6$ cm⁻³. The PV parameters are listed in table 7.*

**Table 7**: *The list of the PV parameters estimated from achievable efficiency limit calculation using $k_1 = 0$, $k_{2,rad} \approx 10^{-11}$ $cm^3 s^{-1}$, $k_3 = 0$, and $n_i = 0.5 \times 10^6$ $cm^{-3}$.*

|  | $J_{SC}$ (mA/cm²) | $V_{OC}$ (V) | *FF* (%) | PCE (%) |
|---|---|---|---|---|
| **achievable limit** <br> $\boldsymbol{k_1 = 0}$ <br> $\boldsymbol{k_{2,rad} = 10^{-11} \text{ cm}^3\text{s}^{-1}}$ <br> $\boldsymbol{k_3 = 0}$ <br> $\boldsymbol{n_i = 0.5 \times 10^6 \text{ cm}^{-3}}$ | 26 | 1.25 | 90.13 | 29.34 |

## 6.3.12 Loss analysis in the efficiency limit.

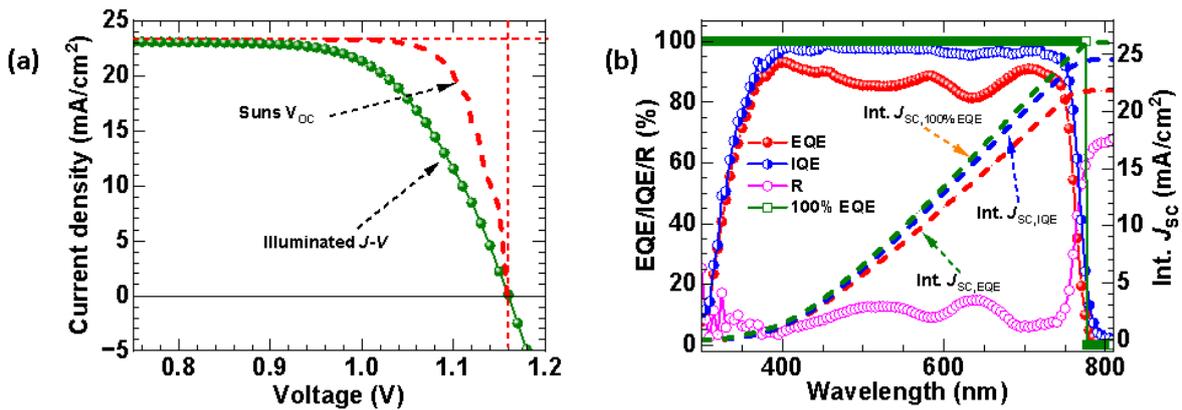

**Figure 18: (a)** *Current density vs. Voltage* (J-V) *characteristics under* $1 - sun$ *Illumination condition and pseudo* $J - V$ *measured from Suns* $V_{OC}$ *measurement.* (**b**) *EQE & IQE spectrum of the PSC including the reflection (R) spectra measured from the glass/ITO side. EQE represents the external quantum efficiency, R represents the reflection and IQE represents the internal quantum efficiency spectrum. The integrated current densities are estimated by integrating the EQE spectrums over AM 1.5G spectrum.*

Following equations are used to estimate IQE and integrated current densities.

$$IQE(\%) = \frac{EQE\ (\%)}{1 - 0.01 * R(\%)} \qquad (6.12)$$

$$J_{SC}(EQE) = q \int_{300\ nm}^{850\ nm} \phi(\lambda) EQE(\lambda) d\lambda \qquad (6.13)$$

$$J_{SC}(IQE) = q \int_{300\,nm}^{850\,nm} \phi(\lambda) IQE(\lambda) d\lambda \qquad (6.14)$$

where $\phi(\lambda)$ represents the AM 1.5G spectrum and $q$ is the elementary charge.

***Table 8***: *The list of the PV parameters estimated from illuminated $J-V$ at 1-sun condition, Suns $V_{OC}$ measurement and achievable efficiency limit calculation.*

| | $J_{SC}$ (mA/cm²) | $V_{OC}$ (V) | *FF* (%) | PCE (%) |
|---|---|---|---|---|
| $J-V$ **scans** | 23.35 | 1.160 | 79.52 | 21.54 |
| **Suns** − $V_{OC}$ | 23.30 | 1.159 | 89.02 | 24.04 |
| **achievable limit** <br> $k_1 = 10^4$ **s⁻¹** <br> $k_2 = 10^{-10}$ **cm³s⁻¹** <br> $k_3 = 10^{-28}$ **cm⁶s⁻¹** <br> $n_i = 0.5 \times 10^6$ **cm⁻³** | 26 | 1.192 | 89.57 | 27.76 |
| **achievable limit** <br> $k_1 = 0$ <br> $k_2 = 10^{-10}$ **cm³s⁻¹** <br> $k_3 = 0$ <br> $n_i = 0.5 \times 10^6$ **cm⁻³** | 26 | 1.193 | 89.72 | 27.83 |
| **achievable limit** <br> $k_1 = 0$ <br> $k_{2,rad} = 10^{-11}$ **cm³s⁻¹** <br> $k_3 = 0$ <br> $n_i = 0.5 \times 10^6$ **cm⁻³** | 26 | 1.25 | 90.13 | 29.34 |

## 6.4. Discussions

Our experimental results are well in accordance with the theoretical benchmarks identified and hence conclusively establish that our solar cells operate at detailed balance limit with dominant bimolecular recombination. There are several novelties associated with this work.

**(i)** It has been long established that the light ideality factor under steady-state conditions should be close to 1 for solar cells at dominated by bimolecular recombination.[82] Similarly, $ln(\tau)$ is also known to vary linearly[28] with $V_{OC,0}$ with a slope of $-q/2kT$. However, in isolation, these are not unique and sufficient conditions to claim operation dominated by bimolecular recombination at balance limit. For instance, under low-level injection with an effective doping density $N_A$ and limited by trap assisted recombination, the $V_{OC,0}$ could vary as $kT/q \ln(G_0/k_1 n_i) + kT/q \ln(N_A/n_i)$ – which exhibits a light ideality factor of 1. However, as compared to eq. 6.5, it can be shown that such a solar cell would have distinctly different trends in transient characterizations and do not consistently anticipate the reverse saturation current densities estimated from dark $J-V$ and Suns-$V_{OC}$ methods. Hence, this work, for the first time, identifies a coherent set of benchmarks (i.e., not just a single criterion) expected from solar at detailed balance limit and dominated by bimolecular recombination.

**(ii)** The low value for $k_1$ indicates that our devices have excellent carrier lifetime in the active layer and interfaces are well passivated. We remark that the light $J-V$ hysteresis is not significant in our devices (hysteresis index $= 2.23\%$, **Figure 1a**, **Table 1**) - which also indicates good passivation of interfaces.[83] Under open circuit conditions with a negligible electric field, we have $k_1 = 1/\tau_B + 2S/W$, where $\tau_B$ is the trap assisted carrier lifetime in the active layer and $S$ denotes the recombination velocity at the interfaces with contact layers. Given $W = 450$ nm and the back extracted value of $k_1 = 5.77 \times 10^4$ s⁻¹, it is evident that the $\tau_B \geq 17$ μs and $S \leq 1$ cm/s. Such low values of $S$ are comparable to c-Si technology[84,85] while the $S$ reported[86–88] for perovskite thin films are of the order of $10^3$ cm/s.

**(iii)** The performance parameters of our solar cells under 1-Sun illumination are $\eta = 21.54\%$ with $J_{SC} = 23.35$ mA/cm², $V_{OC} = 1.160\ V$, $FF = 79.52\%$, over an active area of $0.175$ cm² and the bandgap of the active material used is 1.6 $e$V (see **Figure 1a** and **Table 1**). The performance of our device compares well with literature reports on solar cells with photoactive material of similar bandgap (see **Figure 1c** and **Table 2**). In comparison with the SQ or AL limits, it is evident that our devices differ in $J_{SC}$ and $FF$ due to reflection loss by the multilayer thin film devices and limited by the series resistance of the ITO – coated glass substrates respectively (see **Table 8**). The Suns $V_{OC}$ measurement offers a pseudo $J - V$ which is not affected by series resistance effects.[37,38] The FF estimated from the Suns $V_{OC}$ (89.02%) and achievable limit (89.57%) are similar **Figure 18 & Table 8**. The loss in $J_{SC}$ of the devices is due to reflection loss by the micro-cavity effect of the multi-layer thin film device **Figure 18b**.[89] Comparison with the achievable limits suggests areas of improvement. Evidently, there is much scope for improvement in the optical engineering of our solar cells (for example, the $J_{SC}$ can be further increased by using $MgF_2$ anti-refection coating towards the illumination glass side.)[90,91] which ensures optimum $J_{SC}$ for lower $W$ could be of interest. The FF gap indicates that resistive loss could be improved significantly.

**(iv)** Our devices operate at the detailed balance limit with dominant non-radiative bimolecular recombination. Hence significant improvement in $V_{OC}$ is possible only if the shallow trap density is reduced. Indeed, these devices could result in near radiative limit performance if the shallow trap density can be reduced by one order of magnitude. Future research in this direction could be of broad interest.

**(v)** Importantly, this work enables, for the first time, consistent estimation for $k_1, k_2$, and $n_i$ from transient and steady state characteristics of finished devices. To this end, we employed novel characterization schemes and theoretical analysis. We have done similar calculations for a different set of device operating in the dominant bimolecular recombination regime (not included in this chapter), and the associated analysis yields similar conclusions and parameters. Our estimates compare very well with the existing literature (**Table 4**).[50,92] Further, this work identifies shallow traps as one possible cause of rather low $EQE_{EL}$ of otherwise good solar cells. Reduction in shallow trap density is indeed expected to improve the performance of both LEDs and Solar cells.

## 6.5 Conclusions

In summary, here we presented the first ever conclusive experimental demonstration of perovskite solar cells dominated by bimolecular recombination operating at detailed balance limit. This claim is supported by multiple experimental characterizations and detailed theoretical analysis. We proposed and validated coherent schemes for back extraction of recombination parameters from transient and steady state electro-optical characterizations along with EL measurements which anticipate important parameters of dark $J - V$ and Suns-

$V_{oc}$ measurements. Indeed, this work identifies achievable efficiency limits for perovskite solar cells with implications for both fundamental physics and process/device optimization.

## 6.6 Postscript

Overall, this chapter explains how steady-state and transient measurements help in characterizing a solar cell device dominated by bimolecular recombination.[93] There are various theoretical and experimental techniques reported in the literature where the perovskite films are used to estimate the recombination parameters (e.g. using the transient absorption spectroscopy measurement). So, developing a characterization scheme to study the complete solar cell device at the radiative limit is essential. Therefore, we developed a consistent characterization scheme at the device level and validated the coherent scheme from the back extraction of recombination parameters. Further, the bimolecular recombination parameter can be the sum of radiative or non-radiative recombination. The origin of such non-radiative bimolecular processes is not well understood. The electroluminescence quantum efficiency is used to correctly estimate the radiative bimolecular recombination. We observe that the non-radiative bimolecular recombination limits the device performance which originates from the shallow trap states. The estimated recombination parameters have a specific range comparable to the literature reports. Further improvement in the device could lead to the SQ limit parameters in the perovskite solar cell device; hence, we titled our work perovskite solar cells dominated by bimolecular recombination – how far is the radiative limit?

However, the efficiency is limited in the single junction solar cell due to the limited access to the solar spectrum by a single absorber. A single absorber in a solar cell leads to both the below bandgap loss and thermalization loss, as discussed in Chapter **2**. Hence, to further enhance the efficiency of solar cells, it is essential to use multiple bandgap absorbers in a single device, i.e., a tandem solar cell architecture. Therefore, we studied a four-terminal (4T) tandem solar cell architecture with silicon (Si) and cadmium telluride (CdTe) solar cells, which will be discussed in the next Chapter 7.

# CHAPTER 7
# Highly Efficient Four Terminal Tandem Solar Cells and Optical Loss Analysis

# CHAPTER 7

# Highly Efficient Four Terminal Tandem Solar Cells and Optical Loss Analysis

## Abstract:

Four-terminal (4T) tandem solar cells have emerged as a promising technology for significantly enhancing the efficiency of photovoltaic devices. By stacking multiple sub-cells with varying bandgaps, tandem solar cells can effectively utilize a broader range of the solar spectrum, achieving higher power conversion efficiencies than traditional single-junction solar cells. The top high bandgap perovskite solar cell should be semi-transparent for a 4T tandem solar cell. We have fabricated inverted architecture semi-transparent perovskite solar cells (PSCs) with 17.61% and 15% efficiency over the active area of 0.175 cm$^2$ and 0.805 cm$^2$ respectively. The transparency of the semi-transparent PSC cell of bandgap 1.6 eV device in the NIR range (>750) is ~ 45%. The transmitted NIR photons go to the bottom low bandgap Si solar cells. We have used two different kinds of Si solar cells: (i) monocrystalline PERC Si solar cell and (ii) silicon heterojunction (SHJ) solar cell. The 4T efficiencies of the Si –perovskite tandem solar cell are 27.15% and 28.02% over the active area of 0.175 cm$^2$ (perovskite solar cells) using monocrystalline PERC and silicon heterojunction solar cells respectively. Further, the 4T tandem efficiencies are slightly lower for the larger area (0.805 cm$^2$ perovskite solar cells) solar cells. In addition, we worked on the Cadmium telluride (CdTe) and perovskite 4T solar cells. The efficiency of the single junction CdTe solar cell used is 19.48%, and the efficiency improves to 23.08% in the 4T tandem solar cell structure.

## 7.1 Introduction

In Chapter 4, we discussed compositional engineering and observed that the mixing of monovalent organic cations is helpful in defect passivation and results in improved efficiency with concomitant stability for the CsFAMA based PSC. Further, in Chapter 5, we showed the interface engineering at the perovskite-HTL interface of a *p-i-n* architecture PSC. We observed that mixed Pz:PFN(9:1) HTL based PSC is highly reproducible, efficient, and stable. In Chapter 6, we showed the Pz:PFN(9:1) HTL-based PSC shows an efficiency of 21.54% and is operating under dominant bimolecular recombination regime.[1] However, the efficiency is limited in the single junction solar cell due to the limited access to the solar spectrum by a single absorber. A single absorber in a solar cell leads to both the below bandgap loss and thermalization loss, as discussed in Chapter **2**. Hence, to further enhance the efficiency of solar cells, it is essential to use multiple bandgap absorbers in a single device, i.e., a tandem solar cell architecture. Therefore, we studied a four-terminal (4T) tandem solar cell architecture of perovskite with silicon (Si) and cadmium telluride (CdTe) solar cells in this chapter, which is discussed below.

The four-terminal (4T) tandem solar recently got significant attention due to its high efficiency performance.[2–5] Traditional single junction solar cells have a fundamental limitation in their ability to convert sunlight into electricity efficiently. This limitation arises from the mismatch between the bandgap of the semiconductor material used in the solar cell and the solar spectrum, which consists of a wide range of wavelengths. As a result, only a fraction of the incoming solar radiation can be effectively converted into electrical energy. 4T tandem solar cells offer a promising solution to this challenge by combining multiple sub-cells with varying bandgaps in a stacked configuration.[6–9] Each sub-cell should be optimized to absorb a specific portion of the solar spectrum, allowing more efficient utilization of sunlight. By effectively capturing a broader range of wavelengths, these tandem solar cells can achieve higher power conversion efficiencies compared to their single-junction counterparts. To make a 4T tandem solar cell, the top high bandgap solar cell should be semi-transparent.[10] Semi-transparent perovskite solar cells (PSCs) offer the unique advantage of being able to transmit a portion of light while simultaneously converting a significant portion of incident light into electricity. This characteristic allows for their integration into various applications, such as windows, building facades, and portable electronic devices, without obstructing the passage of light.[11–14] Perovskite solar cells have rapidly progressed in power conversion efficiency (PCE).

The efficiency of semi-transparent perovskite solar cells has reached values above 20%, and there is ongoing research and development to enhance their performance further.[15,16]

In this work, we showed an inverted architecture semi-transparent PSC of efficiency of 17.61% over an active area of 0.175 cm$^2$ and 15 % over an active area of 0.805 cm$^2$. Upon placing the semi-transparent PSC on the top of a PERC Si monocrystalline solar cell, an efficiency of 23% increases to 27.15%, and using a SHJ Si cell, an efficiency of 26.81% increases to 28.02%. Furthermore, using the same semi-transparent PSC over a CdTe solar cell, with an efficiency of 19.1%, results in an improved efficiency of 23.08% over a 0.175 cm$^2$ area solar cell. The improvement in the efficiency in the 4T structures is slightly lower for the larger area (0.805 cm$^2$) devices.

## 7.2 Experimental Section

**Materials:**

ITO-coated glass substrates (15 Ω/sq) were purchased from Lumtech. Lead iodide (PbI$_2$), formamidinium iodide (FAI), cesium iodide (CsI), lead bromide (PbBr$_2$), and Me–4PACz, all were purchased from TCI chemical and used as received. Methyl ammonium Bromide (MABr) was purchased from Greatcell Solar. Phenyl-C61-butyric acid methyl ester (PC$_{61}$BM) was purchased from Lumtech and was used as received. The poly(9,9-bis(3'-(N,N-dimethyl)-N-ethylammoinium-propyl-2,7-fluorene)-alt-2,7-(9,9-dioctylfluorene))dibromide (PFN-Br) ordered from Solarmer Material Inc. and used as received. Bathocuproine (BCP) was purchased from Sigma Aldrich and used as received. The list of all chemicals, companies, CAS, and product numbers is as follows.

**Solution preparation:**

To make triple cation (FA$_{0.83}$MA$_{0.17}$)$_{0.95}$Cs$_{0.05}$Pb(I$_{0.83}$Br$_{0.17}$)$_3$ perovskite solution, first we added 22.5 mg of MABr, 73.5 mg of PbBr$_2$, 172 mg of FAI, and 507.5 mg of PbI$_2$ in 1 ml of DMF: DMSO (4:1) and stirred at room temperature for 2 hours to make a premixed solution. Separately we made 1.5 (M) CsI solution in DMSO i.e. 100 mg of CsI in 257 µl of DMSO and stirred for 2 hours. We filtered the premixed solution with PTFE 45 mm filter in a separate vial. Finally, we added 950 µl of premixed solution and 50 µl of CsI solution to get the final triple cation(FA$_{0.83}$MA$_{0.17}$)$_{0.95}$Cs$_{0.05}$Pb(I$_{0.83}$Br$_{0.17}$)$_3$ perovskite solution and stirred for 1 hour. To obtain mixed Me–4PACz: PFN-Br solution, first, we prepared 0.4 mg/ml Me-4PACz solution in anhydrous methanol and 0.4 mg /ml PFN-Br solution in anhydrous methanol and stirred overnight. Finally, one hour before spin-coating we mix the Me–4PACz and PFN-Br solution in 9:1 volume ratio.[17] 20 mg $PC_{61}BM$ was dissolved in 1 ml of chlorobenzene and stirred overnight. 0.5 mg BCP was dissolved in 1 ml of anhydrous isopropanol and stirred overnight at room temperature and 10 minutes at $70^o$ C before spin coating.

**Semi-transparent perovskite solar cell fabrication**

 ITO-coated glass (15 Ω/square) substrates were patterned with Zn powder and HCl and then sequentially cleaned with soap solution, deionized (DI) water, acetone, and isopropanol for 10 minutes each. After drying the substrate with a nitrogen gun, we kept the ITO substrates on a hot plate at $80^o$ C for 10 minutes and then took them inside the oxygen plasma ashing chamber for 20 minutes, and plasma ashing was done at an RF power of 18 watts. After plasma ashing, we immediately take the substrates inside the $N_2$ environment glove box ($O_2$<0.1 ppm, $H_2O$<0.1 ppm) and spin-coat Me-4PACz: PFN-Br mixed solution at 4000 rpm for 30 seconds and then annealed at $100^o$ C for 10 minutes. After that, we cool down the substrates for 5 minutes and then perovskite solution spin-coated on the ITO/ Me-4PACz: PFN-Br substrates. The CsFAMA perovskite was spin-coated at 5500 rpm for 30 seconds, and at the last 7 seconds, we used 150 μl of chlorobenzene as an anti-solvent treatment on a 1.5 cm by 1.5 cm substrate and then annealed at $100^o$ C for 30 minutes.  After that, $PC_{61}BM$ was spin-coated at 2000 rpm for 30 seconds and BCP was spin-coated at 5000 rpm for 20 seconds. Finally, 150 nm of Ag was deposited under a vacuum of $4x10^{-6}$ mbar using a metal shadow mask for opaque cells. To fabricate the semi-transparent cell, we replaced the opaque Ag electrode with a transparent IZO electrode. Before depositing the IZO layer, a 20 nm $SnO_2$ layer is deposited by thermal atomic layer deposition (ALD) technique at $80^o$ C using Tetrakis(dimethylamido)tin(IV) in the Anric Technologies system. The top layer IZO of 200 nm is deposited by the sputtering technique at 50 W under a vacuum of $1.6x10^{-2}$ mbar using a metal shadow mask. Finally $MgF_2$ anti-reflection coating is done by thermal evaporation under a vacuum of $4x10^{-6}$ mbar.

**Characterization**

All the photovoltaic measurements were carried out under ambient conditions. Photocurrent density versus applied voltage (*J-V*) measurement was carried out using Keithley 4200 SCS and an LED solar simulator (LSH-7320) after calibration through a standard Si solar cell (RERA SYSTEMS-860 reference cell). The *J-V* measurement was performed with a scan rate of 100 mV/s with a hold time of 10 ms. EQE measurement has been carried out to measure the photo response as a function of wavelength using the Bentham quantum efficiency measurement system (Bentham PVE 300). The optical loss analysis estimation is carried out using the transfer matrix method (TMM) calculation.

## 7.3 Results

### 7.3.1 Working principle of tandem solar cells

A single junction perovskite solar cell consists of a single absorber. It is sandwiched between the electron transport layer (ETL) and hole transport layer (HTL) with respective electrodes at the end terminal of the device. When photons are absorbed by the semiconductor absorber of energy equal to or more than the energy bandgap ($E_g$), the electron and hole pairs are generated at the conduction and valence bands respectively. The incident sunlight consists of photons with energies ranging from infrared to visible to ultra-violet region. If the energy of the incident photon is less than the bandgap of the absorber, the photon will not absorb, leading to the below bandgap loss. If the incident photon's energy is higher than or equal to the bandgap of the semiconductor, it will be absorbed and create electron and hole pairs. For the photons of energy more than the bandgap, the excited electrons will go to the conduction band's higher energy states, undergo various non-radiative fast relaxation processes, and eventually arrive at the bottom of the conduction band **Figure 1a**. These fast relaxation processes happen via phonon interaction and release heat to the semiconductor, referred to as thermalization loss. Thus, for a single absorber in a single junction solar cell, there is always loss due to below bandgap loss and thermalization loss. To understand the process, let us consider a Si solar cell of bandgap energy 1.12 eV, which has lower below bandgap loss (**Figure 1b**) and higher thermalization loss. Whereas a high bandgap perovskite solar cell of energy bandgap 1.6 eV has higher below bandgap loss and lower thermalization loss (**Figure 1c**). These two losses can be minimized by combining the two semiconductors in a single device, which implies a tandem structure. In the tandem solar cell structure, different energy bandgap sub-cells are sandwiched on top of each other, and the light incident through the high bandgap absorber sub-cell only. **Figure 1d**. The high energy photons are absorbed by the top high bandgap sub-cell, and the rest of the unabsorbed low energy photons go to the bottom low bandgap sub-cell **Figure 1e**. Thus, the below bandgap and thermalization loss are significantly reduced and result in high efficiency in the tandem structure **Figure 1f**. Generally, the top high bandgap cell is thinner to pass maximum unabsorbed low energy photons to the bottom sub-cell.[18–20] Different kinds of tandem structures are available, such as two-terminal (2T), three-terminal (3T), four-terminal (4T) etc.[2,11,12,21] The sub-cells are electrically connected in the 2T or 3T devices.[22,23] In contrast, in the 4T devices, the sub-cells are mechanically stacked on top of each other. The 2T tandem solar cells are easy to handle but more complex due to current matching.[22,23] The 4T devices need more care to do optical coupling but are easy to fabricate. The important advantage of the 4T device is that if any sub-cell (especially the perovskite sub-cell) gets damaged due to stability issues, it can easily be replaced with a new one. This chapter shows the 4T tandem solar cell structure of perovskite with silicon (Si) and Cadmium telluride (CdTe) solar cells.

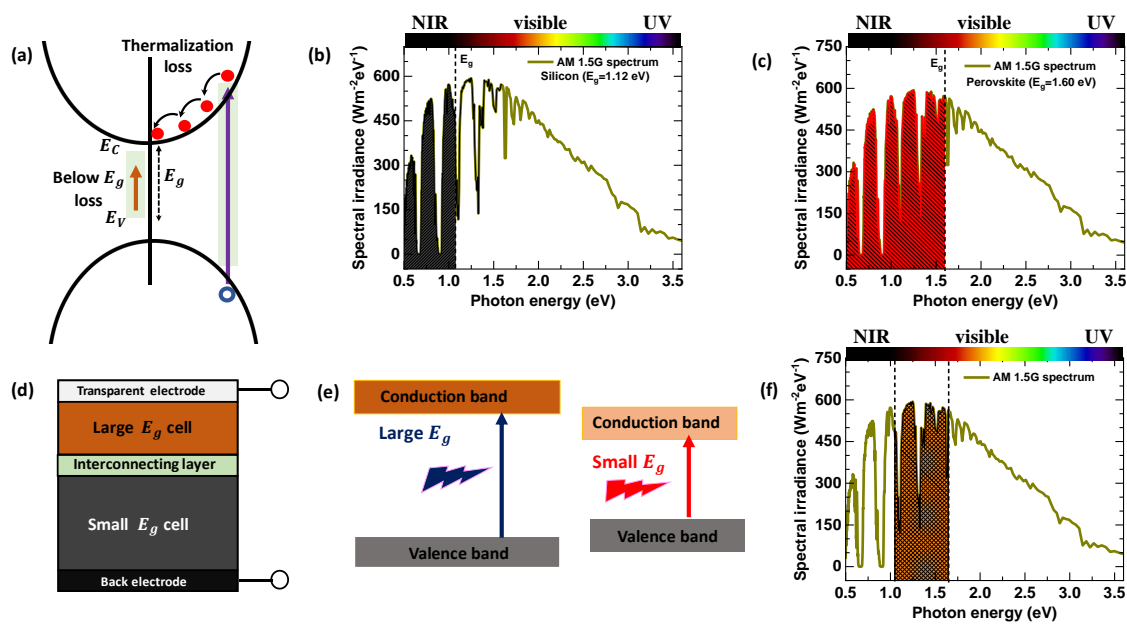

***Figure 1:*** *Schematic representation of the working principle of tandem solar cells.* ***(a)*** *Schematic representation of below bandgap loss and thermalization loss.* ***(b)*** *Silicon solar cells have lower below bandgap loss and higher thermalization loss.* ***(c)*** *Perovskite solar cells have higher below bandgap loss and lower thermalization loss.* ***(d)*** *Schematic representation of tandem solar cells using large and smaller bandgap absorber material.* ***(e)*** *Working of tandem solar cells using two semiconductors of different bandgap.* ***(f)*** *Minimizing the below bandgap loss and thermalization loss in the tandem solar cells.*

### 7.3.2 Semi-transparent perovskite solar cells fabrication

The details of fabrication and optimization of the opaque perovskite solar cells through interface engineering are discussed in Chapter 5. For making a transparent PSC, the opaque contact silver (Ag) is replaced by a semi-transparent electrode SnO$_2$/IZO/MgF$_2$. **Figure 2a** is the schematic of the semi-transparent perovskite solar cell in the inverted architecture (*p-i-n*). The top layer IZO of 200 nm is deposited by the sputtering technique at 50 W under a vacuum of $1.6 \times 10^{-2}$ mbar using a metal shadow mask. The sputtering process deals with the high-energy sputtering particles that can damage the underlying PCBM and perovskite layer. So, before depositing the IZO layer, a 20 nm $SnO_2$ layer is deposited by the thermal atomic layer deposition (ALD) technique at 80° C using Tetrakis(dimethylamido)tin(IV) in the Anric Technologies system. Finally, $MgF_2$ anti-reflection coating of 100 nm is done by thermal evaporation under a vacuum of $4 \times 10^{-6}$ mbar. The cross-sectional view of the semi-transparent perovskite solar cell is shown in **Figure 2b**. The thickness of the perovskite layer is ~400 nm, and the thickness of the rest of the layer is shown in Figure 2b.

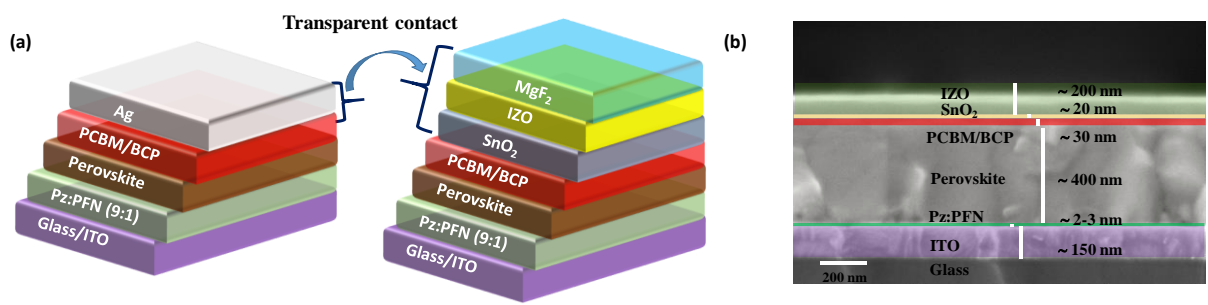

***Figure 2: (a)** Schematic of semi-transparent PSC. Glass/ITO is the transparent conductive electrode, Pz:PFN (9:1) is the hole transport layer, perovskite is the active material, PCBM/BCP acts as the electron transport layer, ALD SnO₂ acts as an electrons transport as well as sputtering damage protecting layer and IZO is the transparent back electrode. **(b)** The cross sectional view of the semi-transparent perovskite solar cells.*

### 7.3.3 Semi-transparent perovskite solar cells performance

The semi-transparent high bandgap perovskite solar cells are the front cells that absorb all the high energy photons, and the rest of the unabsorbed low energy photons go directly to the bottom low bandgap sub-cell. Hence, the transparency and performance of the semi-transparent perovskite solar cell play a key role in achieving the high efficiency 4T tandem solar cell. **Figure 3a** is the current density *vs.* voltage (*J-V*) characteristic of the semi-transparent PSC and opaque PSC devices of active area 0.175 cm² and 0.805 cm² under AM1.5G illumination conditions. The F represents the forward scan and R represents the reverse scan direction. The list of the PV parameters in the forward and reverse scan direction is given in **Table 1**. In the semi-transparent solar cells, the top electrode IZO is less conductive than silver (Ag). Hence, there is an expected loss in the *FF* of the semi-transparent PSC device compared to the opaque

PSC Table 1.[24,25] **Figures 3b** and **3c** are the photographic images of the semi-transparent PSC of the active area of 0.175 cm² and 0.805 cm² respectively. For the 0.805 cm² semi-transparent solar cells, we adopted the busbar/finger contact from the Si-solar cells technology to collect the maximum charge carriers efficiently from the complete device area.[26,27] The busbar/finger metal contacts are deposited on the top of the IZO layer by thermal evaporation using a metal mask under a vacuum of 4x10⁻⁶ mbar. The lower *FF* in the large area device can be attributed to the increased series resistance, as discussed in Chapter **2**.

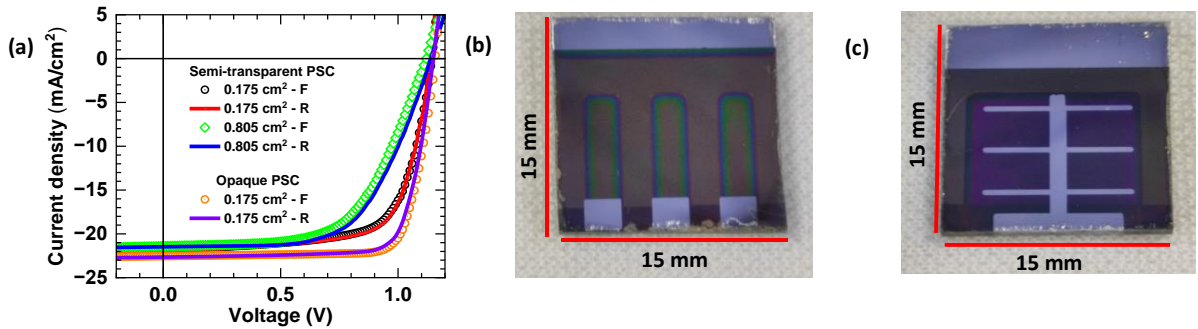

**Figure 3:** *(a) J-V characteristic curves of the semi-transparent (ST)-PSC and opaque PSC device of active area 0.175 cm² and 0.805 cm² under AM1.5G illumination conditions. (b) Photographic image of the ST-PSC of active area 0.175 cm² and (c) 0.805 cm² respectively.*

**Table 1:** *Photovoltaic device parameters of semi-transparent (ST)-PSC of active area 0.175 cm² and 0.805 cm².*

| Scan direction | $J_{SC}$ (mA/cm²) | $V_{OC}$ (V) | *FF* (%) | PCE (%) |
|---|---|---|---|---|
| ST PSC (0.175 cm²)– F | 21.78 | 1.144 | 69.51 | 17.32 |
| ST PSC (0.175 cm²)– R | 22.02 | 1.137 | 70.34 | 17.61 |
| ST PSC (0.805 cm²)– F | 21.42 | 1.120 | 59.56 | 14.29 |
| ST PSC (0.805 cm²)– R | 21.48 | 1.139 | 61.31 | 15.00 |
| Opaque PSC (0.175 cm²)– F | 22.54 | 1.159 | 79.12 | 20.67 |
| Opaque PSC (0.175 cm²)– R | 22.65 | 1.151 | 77.56 | 20.22 |

### 7.3.4 Four terminal Si-Perovskite tandem solar cells performance

The 4T solar cells comprise of a top semi-transparent perovskite solar cell and bottom Si solar cells. We used two different Si solar cells in the 4T combinations. We used a Monocrystalline PERC Si cell efficiency of 23% (**Figure 4**) and a silicon Hetero Junction (SHJ) of efficiency of 26.81% (**Figure 5**). The 4T tandem solar cell efficiency with different Si solar cells is discussed below.

### (i) Four terminal PERC Si-perovskite tandem solar cell performance

The current density *vs.* voltage ($J - V$) characteristics of the PERC cell is shown in **Figure 4a** and the EQE spectrum is shown in **Figure 4b** with a black plot. The ($J - V$) characteristics and the EQE spectrum of the semi-transparent solar cells of area 0.175 cm$^2$ and 0.805 cm$^2$ in the reverse scan direction are shown along with the Si solar cells in **Figure 4**. The current density ($J_{SC}$) of the PERC Si solar cell is 33.54 mA/cm$^2$, open circuit voltage ($V_{OC}$) is 0.732 V, fill factor ($FF$) is 82%, and the PCE is 23%. When the semi-transparent perovskite solar cell is used as a filter on the top of the Si cell, the high bandgap photons will be absorbed by the perovskite solar cells, and the low energy photons reach the bottom Si cell, which results in a current density of 15.86 mA/cm$^2$. The efficiency of the filtered Si solar cell becomes 9.54% **Table 2**. The 4T tandem efficiency of the monocrystalline PERC Si solar cells with the semi-transparent perovskite solar cell is 27.15% over 0.175 cm$^2$ and 24.54% over 0.805 cm$^2$ active area of the device respectively.

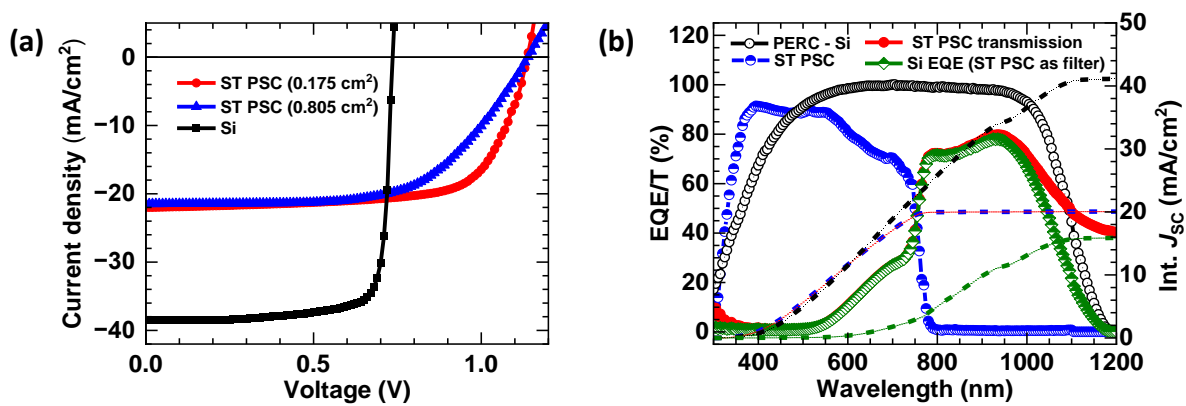

***Figure 4****: (**a**) J-V characteristic curves of the semi-transparent (ST)-PSCs device of active area 0.175 cm$^2$, 0.805 cm$^2$, and PERC Si solar cell under AM1.5G illumination conditions. (**b**) IPCE spectrum, including EQE, transmission, and integrated J$_{SC}$ of Monocrystalline PERC Si and perovskite solar cells.*

**Table 2**: *Photovoltaic device parameters of semi-transparent (ST)-PSC of active area 0.175 cm$^2$, 0.805 cm$^2$, Monocrystalline PERC Si solar cell, and 4T tandem solar cell.*

| Scan direction | $J_{SC}$ (mA/cm$^2$) | $V_{OC}$ (V) | $FF$ (%) | PCE (%) |
|---|---|---|---|---|
| ST PSC  (0.175 cm$^2$) | 22.02 | 1.137 | 70.34 | 17.61 |
| ST PSC  (0.805 cm$^2$) | 21.48 | 1.139 | 61.31 | 15.00 |
| Monocrystalline PERC Si stand-alone | 38.54 | 0.734 | 82 | 23.0 |
| ST PSC filtered Monocrystalline PERC Si | 15.86 | 0.734 | 82 | 9.54 |
| 4T with Monocrystalline PERC Si + ST PSC (0.175 cm$^2$) | - | - | - | 27.15 |
| 4T with Monocrystalline PERC Si + ST PSC (0.805 cm$^2$) | - | - | - | 24.54 |

**(ii) Four terminal SHJ Si-perovskite tandem solar cell performance**

The $J - V$ characteristics of the silicon heterojunction (SHJ)  cell is shown in **Figure 5a,** and the EQE spectrum is shown in **Figure 5b** with a wine color plot. The $J - V$ characteristics and the EQE spectrum of the SHJ Si solar cells is taken from literature by LONGi Solar.[28] The PV parameters of the SHJ Si solar cell is  $J_{SC}$ = 41.45 mA/cm$^2$, $V_{OC}$ = 0.747 V,  $FF$ =86.59%, and the PCE is 26.81%. When the semi-transparent perovskite solar cell is used as a filter on the top of the Si cell, the high bandgap photons will be absorbed by the perovskite solar cells, and the low energy photons reach the bottom Si cell, which results in a current density of 16.10 mA/cm$^2$. The efficiency of the filtered Si solar cell becomes 10.41%. The 4T tandem efficiency of the SHJ Si solar cells with the semi-transparent perovskite solar cell is 28.02% over 0.175 cm$^2$ and 25.41% over 0.805 cm$^2$ active area respectively.

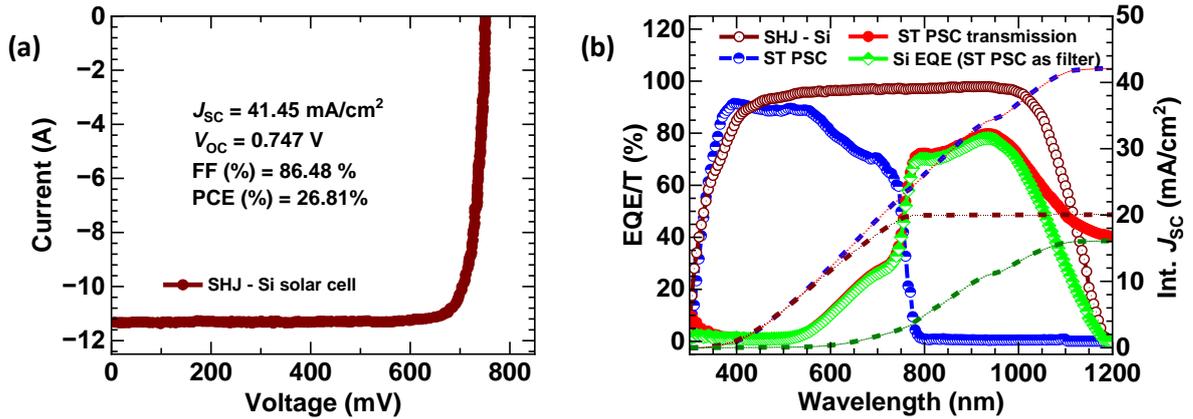

***Figure 5***: *(**a**) J-V characteristic curves Silicon Hetero Junction (SHJ) solar cell by LONGi[28] under AM1.5G illumination conditions. (**b**) IPCE spectrum including EQE, transmission, and integrated $J_{SC}$ of SHJ Si and perovskite solar cells.*

***Table 3***: *Photovoltaic device parameters of semi-transparent (ST)-PSC of active area 0.175 $cm^2$, 0.805 $cm^2$, Silicon Hetero Junction (SHJ) solar cell by LONGi,[28] and 4T tandem solar cell.*

| Scan direction | $J_{SC}$ (mA/cm$^2$) | $V_{OC}$ (V) | *FF* (%) | PCE (%) |
|---|---|---|---|---|
| ST PSC (0.175 cm$^2$) | 22.02 | 1.137 | 70.34 | 17.61 |
| ST PSC (0.805 cm$^2$) | 21.48 | 1.139 | 61.31 | 15.00 |
| SHJ stand-alone | 41.45 | 0.747 | 86.59 | 26.81 |
| ST PSC filtered SHJ | 16.10 | 0.747 | 86.59 | 10.41 |
| 4T with SHJ + ST PSC (0.175 cm$^2$) | - | - | - | 28.02 |
| 4T with SHJ + ST PSC (0.805 cm$^2$) | - | - | - | 25.41 |

### 7.3.5 Loss analysis of the four terminal SHJ Si-perovskite tandem solar cell

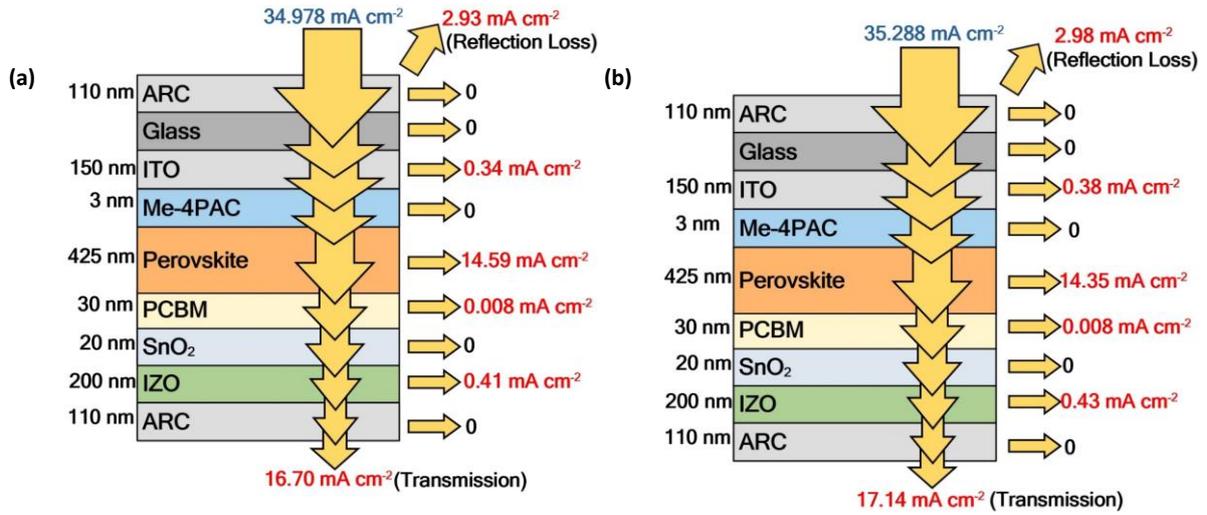

***Figure 6***: *Loss analysis by optical simulations: Loss analysis of the current density of (**a**) Monocrystalline PERC Si and (**b**) Silicon Hetero Junction (SHJ) solar cell by LONGi.*[28]

The transmission spectra of the top NIR-transparent PSCs define the amount of filtered illumination reaching the bottom low bandgap Si solar cell. The current density $J_{SC}$ of the bottom low bandgap solar cells depends on the amount of light coming the bottom cell after filtering by the top perovskite solar cell. The value of $J_{SC}$ can be calculated using the transmission spectra from the top cell (NIR-transparent PSCs), the EQE of the bottom cell, and the standard AM1.5G spectrum. Here, the EQE of two types of bottom Si solar cells is considered for the analysis: Monocrystalline PERC Si solar cell and Silicon Hetero Junction (SHJ) solar cell. The EQE of these two cells is shown in **Figures 4b** and **5b** respectively. Analysis of the optical losses in different layers of the transparent perovskite cell is carried out based on the absorption of the incident illumination in these layers. The parasitic absorption and reflection losses are calculated from the overall perovskite cell structure. The optical constants *n* & *k* values are determined experimentally by spectroscopic ellipsometry, and thickness is determined by DektakXT profilometer. We used the Transfer Matrix Method (TMM) simulations for these calculations. The analysis is carried out for the 500 -1200 nm wavelength range. The general relation between the $J_{SC}$, EQE(λ), and AM 1.5G spectrum utilized here is given as

$$J_{SC} = q . \int_{\lambda_1}^{\lambda_2} EQE(\lambda) . \varphi_{ph}^{AM1.5G}(\lambda) \ d\lambda \qquad (1)$$

Where $\varphi_{ph}^{AM1.5G}$ is the AM1.5G photon flux as a function of wavelength (λ). The equations below represent the general relations used for the calculation of absorption ($J_{abs}$) and reflection loss currents ($J_{ref}$) as well as the transmission current ($J_{trans}$) for the top cell.

$$J_{abs} = q. \int_{500\,nm}^{1200\,nm} A(\lambda). \varphi(\lambda) \qquad (2)$$

$$J_{ref} = q. \int_{500\,nm}^{1200\,nm} R(\lambda). \varphi(\lambda) \qquad (3)$$

$$J_{trans} = q. \int_{500\,nm}^{1200\,nm} T(\lambda). \varphi(\lambda) \qquad (4)$$

We started the calculation with the absorption of incident light in each layer of the top perovskite cell in our defined wavelength range. The EQE spectrum of the bottom silicon cell is incorporated in this analysis. This leads to two possible cases depending upon the type of Si solar cell used as the bottom cell in a 4T tandem structure. The current losses occurring in each layer are calculated by integrating the parasitic absorption from 500 nm-1200 nm wavelength range weighted by the AM 1.5G photon flux. Similarly, the reflection current losses are calculated by integrating the reflection spectrum from the overall cell structure weighted by the AM1.5G photon flux. For both the monocrystalline PERC Si solar cell and Silicon Hetero Junction (SHJ) solar cell by LONGi, if all the incident photons in the wavelength range of 500 to 1200 nm are converted into charge carriers, then the maximum current density from the bottom Si solar cell is ~ 35 mA/cm$^2$.

Based on our analysis, for the monocrystalline PERC Si solar cell and Silicon Hetero Junction (SHJ) solar cell, the primary sources of loss include the parasitic absorption in the perovskite absorber layer and reflection from the overall cell structure. The reflection losses amount to ~ 3 mA cm$^{-2}$. To minimize these losses, better optical coupling must be ensured between the top and bottom cells in the 4T tandem structure. The perovskite bulk layer is absorbing ~ 14.5 mA cm$^{-2}$. The absorption losses in the transport layers of the cell mainly occur in the ITO layer, amounting to ~ 0.35 mA/cm$^2$. Other significant losses arise due to the absorption within the IZO layer (TCE) with values ~0.40 mAcm$^{-2}$. Finally, after considering all the losses within the semi-transparent solar cell, the amount of current for the Si cells is 16.70 mA/cm$^2$ and 17.14 mA/cm$^2$ for the Monocrystalline PERC Si solar cell and Silicon Hetero Junction (SHJ) solar cell respectively. It is to be noted that there is ~ 0.5 mA/cm$^2$ current density mismatch from the experimental result and theoretical loss analysis estimation, which validates our experimental results as shown in **Table 4**.

***Table 4****: 4T tandem solar cells PV parameters comparison from the experimental and loss analysis estimation for the ST PSC with PERC Si and SHJ Si solar cells over an active area of 0.175 cm$^2$.*

| Scan direction | $J_{SC}$ (mA/cm$^2$) | $V_{OC}$ (V) | $FF$ (%) | PCE (%) |
|---|---|---|---|---|
| ST PSC  (0.175 cm$^2$) | 22.02 | 1.137 | 70.34 | 17.61 |
| PERC stand alone | 38.54 | 0.734 | 82 | 23.0 |
| ST PSC filtered PERC | 15.86 | 0.734 | 82 | 9.54 |
| 4T with PERC+ ST PSC | - | - | - | 27.15 |
| ST PSC filtered PERC (Loss analysis estimation) | 16.70 | 0.734 | 82 | 10.05 |
| 4T with PERC+ ST PSC (Loss analysis estimation) | - | - | - | 27.66 |
| SHJ stand alone | 41.45 | 0.747 | 86.59 | 26.81 |
| ST PSC filtered  SHJ | 16.10 | 0.747 | 86.59 | 10.41 |
| 4T with SHJ+ ST PSC | - | - | - | 28.02 |
| ST PSC filtered SHJ (Loss analysis estimation) | 17.14 | 0.747 | 86.59 | 11.09 |
| 4T with  SHJ+  ST  PSC(Loss analysis estimation) | - | - | - | 28.70 |

### 7.3.6 Four terminal CdTe - perovskite tandem solar cells performance

The Cadmium telluride (CdTe) solar cell is a kind of thin film solar cell of thickness ~ 1-2 µm. Significant progress has been observed in the CdTe solar cell field, and an efficiency of >20% has been achieved in the last decade.[29,30] The highest reported efficiency of the CdTe solar cell is 22.4% by First Solar company.[31] The efficiency of the CdTe solar cell can be further improved in the tandem solar cell structure. There is no report on the experimental demonstration of the tandem structure of CdTe and perovskite solar cells except for a theoretical estimation.[32] We are the first to demonstrate the CdTe – perovskite 4T tandem solar cell architecture. In this section, we discussed the 4T performances of CdTe with 1.6 eV bandgap perovskite solar cells. The current density *vs.* voltage ($J - V$) characteristics of the Cadmium telluride (CdTe) solar cell is shown in **Figure 7a,** and the EQE spectrum is shown in **Figure 7b** with a green plot. The PV parameters of the CdTe solar cell is $J_{SC} =$ 27.38 mA/cm$^2$, $V_{OC} = 0.867$ V, $FF =$ 82.06%, and the PCE is 19.48%. When the semi-transparent perovskite solar cell is used as a filter on the top of the CdTe cell, the high bandgap photons will be absorbed by the perovskite solar cells, and the low energy photons reach the bottom CdTe cell, which results in a current density of 7.69 mA/cm$^2$. Thus the efficiency of the filtered CdTe solar cell becomes to 5.47% **Table 5**.[33] The 4T tandem efficiency of the CdTe solar cells with the semi-transparent perovskite solar cell is 23.08% over 0.175 cm$^2$ and 20.47% over 0.805 cm$^2$ active area respectively. The obtained 4T CdTe-perovskite tandem solar cell efficiency is higher than the reported highest single junction CdTe solar cell efficiency.[31]

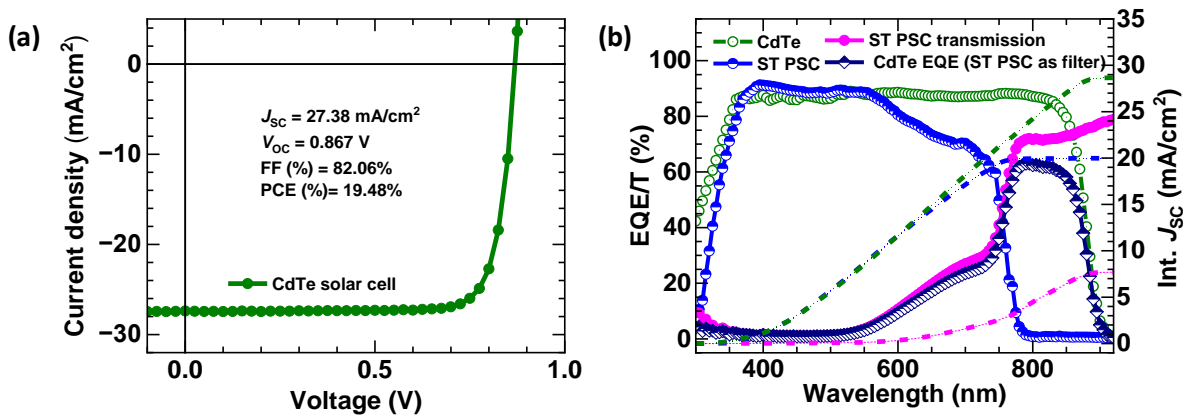

***Figure 7****: **(a)** J-V characteristic curve of CdTe solar cell under AM1.5G illumination conditions. **(b)** IPCE spectrum including EQE, transmission, and integrated $J_{SC}$ of CdTe and perovskite solar cells.*

***Table 5****: Photovoltaic device parameters of semi-transparent (ST)-PSC of active area 0.175 cm$^2$, 0.805 cm$^2$, CdTe solar cell, and 4T tandem solar cell of CdTe and perovskite solar cells.*

| Scan direction | $J_{SC}$ (mA/cm$^2$) | $V_{OC}$ (V) | $FF$ (%) | PCE (%) |
|---|---|---|---|---|
| ST PSC  (0.175 cm$^2$) | 22.02 | 1.137 | 70.34 | 17.61 |
| ST PSC (0.805 cm$^2$) | 21.48 | 1.139 | 61.31 | 15.00 |
| CdTe stand-alone | 27.38 | 0.867 | 82.06 | 19.48 |
| ST PSC filtered CdTe | 7.69 | 0.867 | 82.06 | 5.47 |
| 4T with CdTe + ST PSC (0.175 cm$^2$) | - | - | - | 23.08 |
| 4T with CdTe + ST PSC (0.805 cm$^2$) | - | - | - | 20.47 |

## 7.4 Discussion

We fabricated the inverted architecture semi-transparent perovskite solar cell, which acts as a high bandgap top cell in the 4T tandem solar cell architecture. Generally, the transparent electrode (ITO, IZO etc.) has lower conductivity than the opaque electrode (Ag), hence there is a slight drop in the $FF$. However, we have shown highly efficient semi-transparent perovskite solar cell efficiency of 17.61% and 15% over the active area of 0.175 cm$^2$ and 0.805 cm$^{2,}$ respectively. The efficiency of the larger area semi-transparent cell is slightly lower due to the increased series resistance. As discussed in Chapter **2**. The high bandgap top cell absorbs the high energy photons, and the unabsorbed photons go to the bottom low bandgap sub-cell, e.g., Si solar cell or the CdTe solar cell. The semi-transparent PSC has a transmission of ~ 45% in the NIR range (>750 nm), which is responsible for the bottom sub-cell photocurrent. It is to be noted that the photocurrent produced by the bottom low bandgap solar cell due to the NIR photons is lower than the photocurrent produced by the perovskite solar cell. However, this current mismatch can be minimized by choosing a large bandgap perovskite semiconductor. The current matching of the top perovskite and bottom Si cells is essential in the two terminal solar cells.[34,35] The goal is to ensure that the currents from all the sub-cells are well-matched, allowing for efficient power conversion and maximizing the overall performance of the tandem cell. In the 4T tandem solar cell, the current matching is not essential because, in this case, the individual cell performance is used considering the optical coupling. We achieve maximum efficiency of 28.02% and 23.08%, Si and CdTe solar cells respectively, over an active area of 0.175 cm$^2$.

## 7.5 Conclusion

We fabricated 4T tandem solar cells in the Si-Perovskite and CdTe-Perovskite architecture, which are mechanically stacked on top of each other. To make 4T of tandem solar cells, we fabricated semi-transparent perovskite solar cells with an efficiency of 17.61% and 15% over the active area of 0.175 cm$^2$ and 0.805 cm$^2$ respectively. We used two different Si solar cells (i) a monocrystalline PERC Si cell efficiency of 23% and a SHJ efficiency of 26.81%. We used the perovskite solar cells as a filter on these Si cells and the 4T configuration efficiencies are 27.15% and 28.02% respectively. The loss analysis by optical simulations was carried out to verify our experimental results, and we observed a 0.5% mismatch in the current density or efficiency. Further, we demonstrated 4T CdTe-perovskite tandem solar cell architecture with an efficiency of 23.08%.

## 7.6 Postscript

This chapter explains how tandem solar cells improve efficiency over single junction solar cells. This chapter showed the 4T tandem architecture of PSC with Si and CdTe solar cells. We used high bandgap semi-transparent perovskite solar cells as a filter on the top of Si or CdTe solar cells. The 4T tandem solar cells are easy to fabricate as they are processed individually, but it is challenging to handle individual cells together. However, the 4T solar cells benefit over the 2T solar cells because if any sub-cell gets damaged during operation, it could be easily replaceable with a new one. The 4T solar cells can be considered the initial step of 2T tandem solar cells, as optimization of the semi-transparent perovskite solar cells at the individual level is essential. Furthermore, the current matching of the Si and perovskite sub-cells using semi-transparent PSC as a filter on the top of the Si cell is the prerequisite for fabricating 2T tandem solar cells.

# CHAPTER 8

# Summary and Future Outlooks

# CHAPTER 8

# Summary and Future Outlooks

## 8.1 Summary

This thesis is focused on studying the compositional and device engineering of hybrid metal halide perovskite thin film solar cells through various optoelectronic measurements. An in-depth device physics is discussed to explore the interface engineering between the charge transport and perovskite layers. This thesis also aims to understand a new characterization scheme to characterize solar cells operating under dominant bimolecular recombination using steady-state and transient measurements. Further, the application of perovskite solar cells with silicon and CdTe in the four terminal solar cells is explored. In addition, a detailed analysis is discussed to study the scalable perovskite photodetector. Here, we summarize the thesis in detail:

1. First, we studied the role of monovalent cations in the hybrid organic-inorganic metal halide perovskite solar cells in terms of the dielectric relaxation process using frequency-dependent photocurrent measurement. The multi-monovalent cation-based perovskite has lower defect states, confirmed via various optoelectronic measurements. Further, the correlation between defects and stability is established using the concept of the first principle conclusion of the perovskite crystal structure. The increased hydrogen bonding of the FA-containing triple-cation perovskite system shows faster dielectric relaxation, which is the signature of lower defects and correlates with the stability of the perovskite solar cells.

2. Hence, we chose the multi-cation perovskite in our inverted ($p$-$i$-$n$) device architecture, and we used a new self-assembled monolayer (SAM) hole transport layer (HTL). In this chapter, we used the mixing engineering strategy of SAM with a polyelectrolyte polymer (PFN-Br). We tailored the work function of the mixed Pz:PFN HTL for the 1.6 eV bandgap based perovskite solar cell. The interface of the device is modified for different mixing ratios of HTL, and associated device physics is discussed. However, we observed that the specific mixing ratio Pz:PFN(9:1) shows excellent device performance, the lowest dark current, and the ideality factor close to the unity.

3. The unity ideality factor is a primary signature of solar cells working near the radiative limit. Hence, we studied the dominant recombination mechanism on a complete solar cell using steady-state and transient measurements. We developed a consistent characterization scheme for studying solar cells operating under dominant bimolecular recombination. Our characterization schemes are validated using multiple experiments, such as predicting the reverse saturation current density both from the dark current and pseudo $J - V$, which are comparable to the experimental results.

4. Further, we replace the opaque (Ag) electrode with a semi-transparent electrode (IZO) of the best performing Pz:PFN(9:1) solar cell to fabricate the semi-transparent solar cells. We demonstrate an efficient four-terminal tandem solar cell architecture with silicon and CdTe solar cells. We analyzed the optical loss using the transfer matrix method to support our experimentally measured results.

5. In addition, we observed the lowest dark current for the Pz:PFN(9:1) HTL based device, which is a prerequisite in photodetector applications. Therefore, we conducted a detailed study of the Pz:PFN(9:1) based PSC as a photodetection diode.

This thesis provides a detailed study of perovskite compositional and interface engineering via various optoelectronic measurements. An in-depth device physics is discussed to study the interfacial defects between the charge transport and the perovskite layers. This thesis will be helpful in exploring a new class of perovskite materials and interface modification engineering for fabricating reproducible, stable, and highly efficient hybrid organic-inorganic metal halide perovskite solar cells.

## 8.2 Future Outlooks

### 8.2.1 Tailoring the hole transport layers for different bandgap perovskite solar cells

The energy band alignment between the charge transport layer and the perovskite absorber is essential for efficient photovoltaic performance. Perovskite is a class of material, and different compositions of perovskite can be made with different bandgaps ($E_g$) such as FAPbI$_3$ has $E_g \sim 1.5\ eV$, MAPbI$_3$ has $E_g \sim 1.6\ eV$, CsPbI$_3$ has $E_g \sim 1.7\ eV$, even changing the halides results in a change in the bandgap, especially the valence band, as we have discussed in Chapter 1.[1–3] Therefore, the well matching band alignment of the HTL with perovskite requires various kinds of hole transport layers of different HOMO levels.[4] However, such complications of choosing different HOMO-levelled HTLs can be avoided by choosing a tailored HTL, i.e., mixing Pz:PFN (discussed in Chapter 5). We choose different bandgap perovskites (e.g., 1.49 eV, 1.60 eV, 1.68 eV) and corresponding different mixed Pz:PFN HTLs combinations, and the optimized performance of the PV parameters are shown in **Figure 1** and **Table 1**. Further optimization would result in improved performance. However, for larger bandgap perovskites, e.g., CsPbI$_3$ has a further deeper HOMO level, and in such cases choosing a deeper HOMO level (6.01eV) HTL for instance Br-2PACz (or other HTL) and the aforementioned mixing engineering strategy with PFN-Br would be helpful.[5,6]

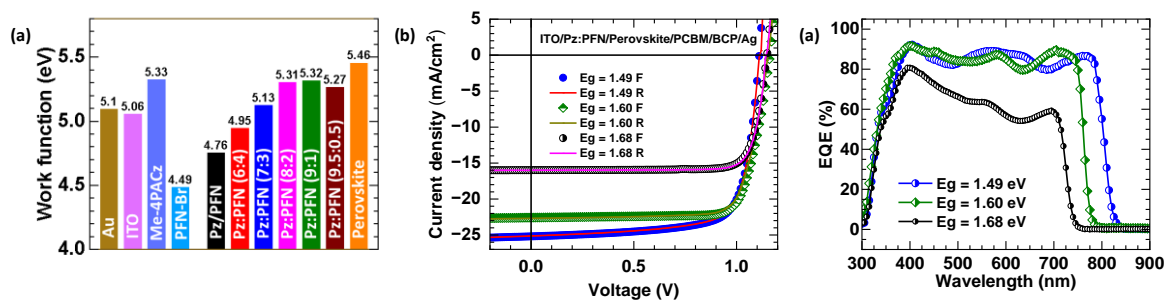

***Figure 1:*** *(**a**) Work function of the mixed Pz:PFN HTLs measured from KPFM study (UPS results discussed in Chapter 5). (**b**) Current density vs. voltage ($J - V$) characteristics of different bandgap perovskite solar cells (PSCs) with different Pz:PFN mixed HTLs. (**c**) The external quantum efficiency of the different bandgap PSCs.*

***Table 1:*** *PV parameters, including integrated J$_{SC}$ measured from the EQE spectrum for different bandgap-based PSCs. Where F and R represent the forward and reverse scan directions respectively.*

| Scan direction | Pz:PFN used | $J_{SC}$ (mA/cm²) | $V_{OC}$ (mV) | FF (%) | PCE (%) | Int. $J_{SC}$ (mA/cm²) |
|---|---|---|---|---|---|---|
| $E_g$ = 1.49 eV - F | Pz:PFN (8:2) | 25.19 | 1.110 | 74.77 | 20.92 | 24.84 |
| $E_g$ = 1.49 eV - R | | 25.03 | 1.117 | 74.32 | 20.77 | |
| $E_g$ = 1.60 eV - F | Pz:PFN (9:1) | 22.58 | 1.161 | 79.04 | 20.72 | 21.53 |
| $E_g$ = 1.60 eV - R | | 22.69 | 1.154 | 77.33 | 20.25 | |
| $E_g$ = 1.68 eV - F | Pz:PFN (9.5:0.5) | 15.99 | 1.149 | 81.97 | 15.06 | 14.20 |
| $E_g$ = 1.68 eV - R | | 16.01 | 1.145 | 82.98 | 15.21 | |

## 8.2.2 Tandem solar cells

In Chapter 7, we have shown the four-terminal tandem solar cell architecture where the individual sub-cells are mechanically stacked on top of each other. However, we used the high bandgap semi-transparent perovskite solar cells as a filter on the top of Si or CdTe solar cells. In this case, the top perovskite solar cell absorbs the high-energy photons, and the low energy unabsorbed photons go to the bottom cells, resulting in photocurrent for the bottom cell. The current produced by the bottom cell depends on the transparency and the bandgap used in the top semi-transparent solar cells. The 4T solar cells can be considered the initial step of 2T tandem solar cells, as optimization of the semi-transparent perovskite solar cells at the individual level is essential in terms of efficiency and transparency. Proper optimization of the top semi-transparent cell will lead to the current matching of the top high bandgap perovskite and bottom Si or CdTe solar cells, which is a prerequisite for the efficient 2T tandem solar cell application,[7,8] which is the future work plan.

# APPENDIX A

# Lowest Dark Current with High Speed Detection in Scalable Perovskite Photodetector

# APPENDIX A

# Lowest Dark Current with High Speed Detection in Scalable Perovskite Photodetector

**Abstract**

A photodetector is an optoelectronic device that transduces an optical signal to an electrical signal. We fabricate *p-i-n* heterostructure perovskite photodetectors (PPDs) with polymeric and self-assembled monolayer (SAM) based hole transport layers (HTLs). The SAM-based PPDs show a lowest dark current $1.48 \times 10^{-10}$ A/cm$^2$ and fast response up to 580 ns for an active area of 10 mm$^2$. Linear dynamic range for this PPD is up to 90 dB using the power based formula. Hence, we scaled up the active area of the SAM-based PPD from 0.100 cm$^2$ to 1.000 cm$^2$. The reverse saturation current density ($J_0$) increases nonlinearly with the increase in the active area of the PPD. Upon increasing the area from 0.100 cm$^2$ to 1.000 cm$^2$, the dark current density at zero bias ($J_d$) increased from $1.48 \times 10^{-10}$ A/cm$^2$ to $3.27 \times 10^{-8}$ A/cm$^2$ and the specific detectivity (D*) reduces from $3.28 \times 10^{13}$ Jones to $2.02 \times 10^{12}$ Jones measured from noise spectral density. Upon increase of the PPD active area, the dark current and the rise time increase but are notably lower compared to the literature reports for the SAM HTL based PPDs.

## A.1 Introduction

In chapter 5, we studied the interface engineering of HTL –perovskite interface in the *p-i-n* architecture based PSC. We observed that the dark current of the device is very low for the Pz:PFN(9:1) HTL based PSC. The low current is immensely useful in the photodetector application. The photodetector figure of merit (FOM) parameters are inversely related to the dark current and hence expect better detectivity (detection of the faint signal) for lower dark current devices. Therefore, we conducted a detailed study of the Pz:PFN(9:1) HTL based PSC as a photodetection diode in this chapter.

The photodetector is an optoelectronic semiconductor device that converts the optical signal to an electrical signal and contributes to many real-life applications e.g., optical communication, image sensing, safety & security, astronomy, etc.[1–3] There are several kinds of semiconducting materials used in photodetectors (PDs) such as Si, InGaAs, GaN, quantum dots, conjugated organic polymers, carbon nanotubes and perovskites etc.[4–8] From the last decade, huge progress is observed with organic-inorganic halide perovskite material in optoelectronic devices such as solar cells,[9] light emitting diodes,[10] transistors[11] and photodetectors[12] etc. Perovskite semiconductor grabbed much attention due to its exciting optoelectronic features such as high absorption coefficient, long carrier diffusion length, low exciton binding energy, a tuneable bandgap, and high defect tolerance, etc.[13–15] Perovskite is a direct bandgap material and has high absorption coefficient ($10^4 - 10^6$ cm$^{-1}$) over Si, Ge, etc., results in less than an μm thin layer is sufficient to absorb the complete 1-Sun solar spectrum light, hence it becomes a potential candidate for photodetector application.[16] Generally, a PD is characterized by some key figure of merit (FOM) parameters such as responsivity (R), linear dynamic range (LDR), noise equivalent power (NEP), specific detectivity (D*), current response bandwidth ($f_{-3dB}$), rise time ($t_r$), etc.[12,17] These FOM parameters of PD depend on the absorber layer, charge transport layer and the active area of the device.

Perovskite is a suitable candidate over commercialized Si detectors in the UV-visible range due to its high responsivity and external quantum efficiency (EQE).[18] Recently, the exploration of achieving high-performance photodetectors in the perovskite field become a core research focus.[19–21] It is to be noted that the dimensions of the photodetectors reported in the literature are small in size (<0.1 cm$^2$).[22–27] However, scaling up the dimension of the photodetector is essential in fields where signal strengths are poor, like in astronomy, environmental monitoring, microscopy, and time–domain spectroscopy.[2,28–30] The scaling up of the photodetectors is always challenging as it requires controlled deposition and uniformity over the large area substrates. As the area of the photodiode increases, the performance goes down (i.e., it is difficult to obtain uniform compact crystalline semiconductor films) hence, it becomes challenging to fabricate large area PDs device stack. Apart from the photodiode dimension, the charge transport layers in the detectors play a crucial role in the FOM parameters, as they are responsible for charge transport and collection.

The perovskite photodetector (PPD) can be either a *n-i-p* (also referred as regular) or *p-i-n* (also referred as inverted) architecture based device in superstrate mode.[31,32] The *p-i-n* structured PPD device is significantly important due to the low-temperature processed charge transport layers. Generally, the different hole transport layers (HTLs) used in the perovskite based devices are such as PEDOT:PSS, NiOx, P3HT, PTAA, Poly-TPD, etc.[32–35] However, the HTLs mentioned above show limited performance and instability in the device. Recently, phosphonic acid-grouped anchored self-assembled monolayer (SAM) HTLs got significant attention in perovskite photovoltaics due to their excellent charge transport, lowest interfacial defects, and high stability properties.[36–39] Nitin Padture *et. al.* showed an improvement in the mechanical reliability of PSC devices by introducing SAM interlayer.[40] Levine *et. al* studied a series of SAMs-based HTLs in the *p-i-n* architecture based photovoltaic (PV) devices and showed [4-(3,6-Dimethyl-9H-carbazol-9-yl)butyl]phosphonic acid (Me-4PACz) has the lowest defect density at the HTL- perovskite interface.[41] This noteworthy feature makes Me-4PACz a suitable candidate for use as HTL in PPD devices.

In this work, we fabricate *p-i-n* architecture based PPDs, where Poly[bis(4-phenyl)(2,4,6-trimethylphenyl)amine (PTAA) and Me-4PACz are used as HTLs. The Me-4PACz SAM HTL-based PPD shows the lowest dark current of $1.48 \times 10^{-10}$ A/cm$^2$ and fast response up to 580 ns for an active area of 10 mm$^2$. The linear dynamic range (LDR) for this PPD is up to 90 dB using the power based formula (in current based formula same LDR would be 180 dB, however using power based formula is standard good practice). Therefore, we scaled up the active area of the SAM-based PPD from 0.100 cm$^2$ to 1.000 cm$^2$. Upon increasing the area of the PPD, the reverse saturation current density ($J_0$) increases nonlinearly. Upon increasing the area from 0.100 cm$^2$ to 1.000 cm$^2$, the dark current density at zero bias ($J_d$) increased from $1.48 \times 10^{-10}$ A/cm$^2$ to $3.27 \times 10^{-8}$ A/cm$^2$ and the specific detectivity (D*) reduces from $3.28 \times 10^{13}$ Jones to $2.02 \times 10^{12}$ Jones. The noise floor and the noise equivalent power (NEP) elevate upon increasing the PPD device area and limit the LDR from 87.46 dB to 68.49 dB. The rise time of the PPD increases linearly with the device area with a slope of $4.53 \ \mu s/cm^2$. The higher rise time limits the speed of photodetection. However, we compared our different active area PPDs with the literature reports, and we observed that, as the PPD device active areas increase, the dark current and the rise time both increase but are notably lower compared to the literature reports on the given active area for Me-4PACz SAM HTL based PPDs.

## A.2 Experimental section

**Materials:** ITO-coated glass substrates (15 $\Omega$/sq) were purchased from Lumtech. Lead iodide (PbI$_2$), formamidinium iodide (FAI), cesium iodide (CsI), lead bromide (PbBr$_2$), and Me–4PACz, all were purchased from TCI chemical and used as received. Methyl ammonium Bromide (MABr) was purchased from greatcell solar. Phenyl-C61-butyric acid methyl ester (PC$_{61}$BM) was purchased from Lumtech and was used as received. The poly(9,9-bis(3'-(N,N-dimethyl)-N-ethylammoinium-propyl-2,7-fluorene)-alt-2,7-(9,9-dioctylfluorene))dibromide (PFN-Br) ordered from Solarmer Material Inc. and used as received. Bathocuproine (BCP) and PTAA were purchased from Sigma Aldrich and used as received.

**Solution preparation:** To make triple cation (FA$_{0.83}$MA$_{0.17}$)$_{0.95}$Cs$_{0.05}$Pb(I$_{0.83}$Br$_{0.17}$)$_3$ perovskite solution, first we added 22.5 mg of MABr, 73.5 mg of PbBr$_2$, 172 mg of FAI, and 507.5 mg of PbI$_2$ in 1 ml of DMF: DMSO (4:1) and stirred at room temperature for 2 hours to make a premixed solution. Separately, we made 1.5 (M) CsI solution in DMSO, i.e., 100 mg of CsI in 257 $\mu$l of DMSO, and stirred for 2 hours. We filtered the premixed solution in a separate vial with a PTFE 45 mm filter. Finally, we added 950 $\mu$l of premixed solution and 50 $\mu$l of CsI solution to get the final triple cation (FA$_{0.83}$MA$_{0.17}$)$_{0.95}$Cs$_{0.05}$Pb(I$_{0.9}$83Br$_{0.17}$)$_3$ perovskite solution and stirred for 1 hour. 1.5 mg/ml PTAA is dissolved in Toluene and stirred for 2 hours before spin coating. To obtain mixed Me–4PACz: PFN-Br solution, first, we prepared 0.4 mg/ml Me-4PACz solution and 0.4 mg /ml PFN-Br solution in anhydrous methanol and stirred overnight. Finally, one hour before spin-coating, we mixed the Me–4PACz and PFN-Br solution in the 9:1 volume ratio.[42] 20 mg/ml PCBM is dissolved in chlorobenzene and stirred overnight. 0.5 mg/ml BCP is dissolved in anhydrous isopropanol and stirred overnight at room temperature and 10 minutes at 70º C just before spin coating.

**Perovskite photodetector device fabrication:** ITO-coated glass (15 $\Omega$/square) substrate was patterned with Zn powder and HCl and then sequentially cleaned with soap solution, deionized (DI) water, acetone, and isopropanol for 10 minutes each. After drying the substrate with a nitrogen gun, we kept the ITO substrates at $80^\circ$C for 10 minutes and then took them inside the oxygen plasma ashing chamber for 20 minutes, and plasma ashing was done at an RF power of 18 watts. After plasma ashing, we immediately take the substrates inside the $N_2$ environment glove box ($O_2$<0.1 ppm, $H_2O$<0.1 ppm) and spin-coat Me-4PACz: PFN-Br (9:1) mixed solution at 4000 rpm for 30 seconds and then annealed at $100^\circ$C for 10 minutes for Me-4PACz based devices. PTAA solution was spin coated at 4000 rpm and annealed at 100 C and after cooling down for 5 minutes, another layer of PFN-Br layer is deposited at 5000 rpm for 30 seconds for PTAA-based devices. After that, the perovskite solution spin-coated on the ITO/ Me-4PACz: PFN-Br and ITO/PTAA/PFN-Br substrates. The CsFAMA perovskite was spin-coated at 5000 rpm for 30 seconds and at the last 7 seconds, we used 150 µl of chlorobenzene as an anti-solvent treatment on a 1.5 cm by 1.5 cm substrate (200 µl of chlorobenzene for 20 cm by 20 cm substrates) and then annealed at $100^\circ$C for 30 minutes. After that, $PC_{61}BM$ was spin-coated at 2000 rpm for 30 seconds and BCP was spin-coated at 5000 rpm for 20 seconds. Finally, 150 nm of Ag was deposited under a vacuum of $2x10^{-6}$ mbar using a metal shadow mask.

**Characterization:** All the photodetector characterizations were carried out at zero bias voltage i.e. self-powered mode under ambient conditions. The current density *vs*. applied voltage (*J-V*) measurement in dark conditions has been carried out using a Keithley 2612B source meter with a scan rate of 50 mV/s. The noise current was measured by taking the fast Fourier transform of dark current *vs.* time at a bandwidth of 24.20 Hz using Keithley 2612B. TPC was measured by using a 638 nm TOPTICA diode laser, THORLABS white lamp S/N M00304198, ArbStudio 1104, and digital oscilloscope Tektronix DPO 4104B. The ArbStudio 1104 function generator is used to modulate the laser from continuous wave (CW) mode to pulse mode. The LDR was measured at different intensities using a 638 nm TOPTICA diode laser in pulse mode at 1 kHz frequency with a duty cycle of 50%. The intensity of the laser was varied using a set of neutral density (ND) filters and the photocurrent was measured using a 2612B source meter. Responsivity and EQE measurements have been carried out to measure the photo response as a function of wavelength using the Bentham quantum efficiency measurement system (Bentham PVE 300). The XRD measurements were carried out in Smartlab, Rigaku diffractometer with Cu K$\alpha$ radiation ($\lambda$=1.54Å). $\Theta$-$2\Theta$ scan has been carried out from $10^\circ$-$50^\circ$ with a step size of $0.001^\circ$. Morphological analysis was done using field emission scanning electron microscopy (FESEM). The contact angle measurement is done in the GBX Digidrop instrument. Optical absorption spectra were carried out using a spectrometer (PerkinElmer LAMBDA 950).

## A.3 Results

Perovskite is a class of material having crystal structure $ABX_3$, where the black ball at the centre of the crystal structure represents the monovalent A cation ($Cs^+$/$FA^+$/$MA^+$), the blue balls at the centre of each octahedral represent the divalent B cations ($Pb^{2+}$) and the red balls at the corner of the octahedral represent X halides ($Br^-$/$I^-$) (see **Figure 1a**).[43] In the perovskite photodetector (PPD) device, we have used the mixed triple monovalent cation perovskite absorber $(FA_{0.83}MA_{0.17})_{0.95}Cs_{0.05}Pb(I_{0.83}Br_{0.17})_3$ abbreviated as CsFAMA, due to its intrinsic lower defect states and the excellent stability.[44–46] To study the PPD performance, we fabricated the inverted (*p-i-n*) heterojunction PPD with a device structure of indium tin oxide (ITO)/HTL /perovskite (CsFAMA) /phenyl-$C_{61}$-butyric acid methyl ester ($PC_{61}BM$)/ bathocuproine(BCP)/ silver (Ag) (see **Figure 1b**).

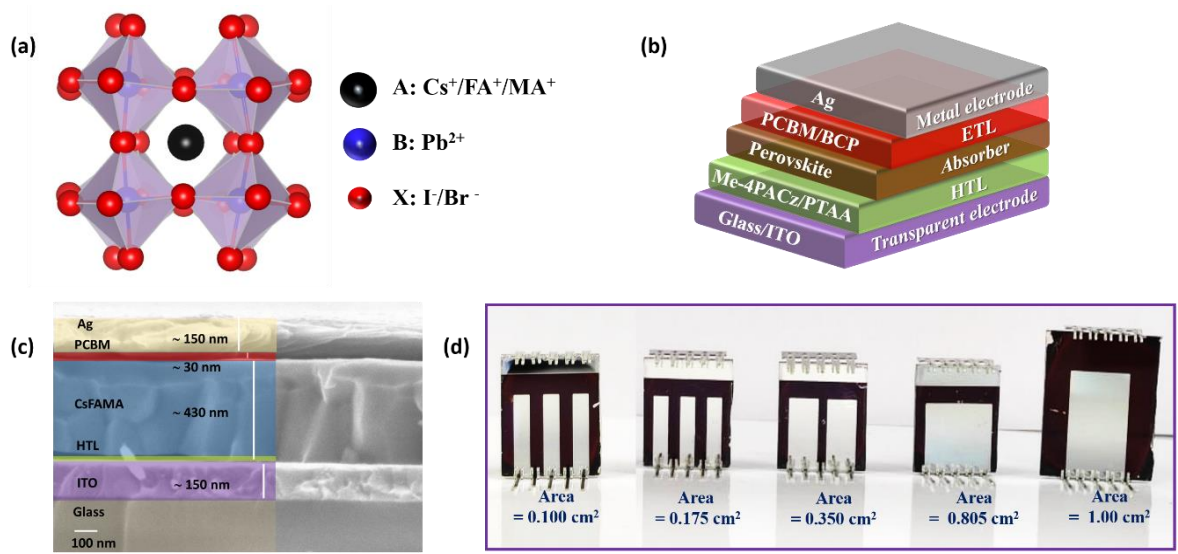

**Figure 1:** (a) Schematic of the $ABX_3$ perovskite crystal structure where the black ball at the centre of the crystal structure represents the monovalent A cation, the blue balls at the centre of each octahedral represent the divalent B cation, and the red balls at the corner of the octahedral represent X halides. (b) Schematic of the inverted (p-i-n) architecture based perovskite photodetector (PPD) device, where HTL represents the hole transport layer i.e. PTAA or Me-4PACz (energy level diagram is shown in Figure 2). (c) The cross-sectional view of the perovskite photodetector (PPD) device indicates the approximate thickness of each layer measured from FESEM imaging.  (d) The scalability of the PPDs: active area 0.100 cm², 0.175 cm², 0.350 cm², 0.805 cm², and 1.000 cm². The details of the active area with dimensions are given in Figure 7.

To fabricate the PPD device, we used two different kinds of hole transport layers (HTLs) such as Poly[bis(4-phenyl)(2,4,6-trimethylphenyl)amine (PTAA) and [4-(3,6-dimethyl-9H-carbazol-9-yl)butyl] phosphonic acid (Me-4PACz). In the device structure (Figure 1b), the ITO acts as a transparent electrode, PTAA or Me-4PACz acts as HTL, perovskite acts as an absorber layer, PCBM and BCP act as an electron transport layer (ETL) and buffer layer, respectively, and finally silver acts as a back metal electrode. The energy level diagram of Figure 1b is shown in **Figure 2**. The PTAA and Me-4PACz HTLs are hydrophobic in nature due to the presence of methyl group (-CH₃) and long alkyl chains resulting in difficulty in forming the uniform perovskite films (see **Figure 3**).[42,47–49] Therefore we used conjugated polyelectrolyte poly(9,9-bis(3′-(N,N-dimethyl)-N-ethylammonium-propyl-2,7-fluorene)-alt-2,7-(9,9-dioctylfluorene)) dibromide (PFN-Br) to get uniform perovskite layer. The details of the HTLs optimization (using PFN-Br) and experimental demonstration are available in our earlier publication.[42,50] The scanning electron microscope (SEM) cross-sectional view of the PPD device structure is shown in **Figure 1c.** The thickness of the perovskite absorber layer is ∼ 430 nm.

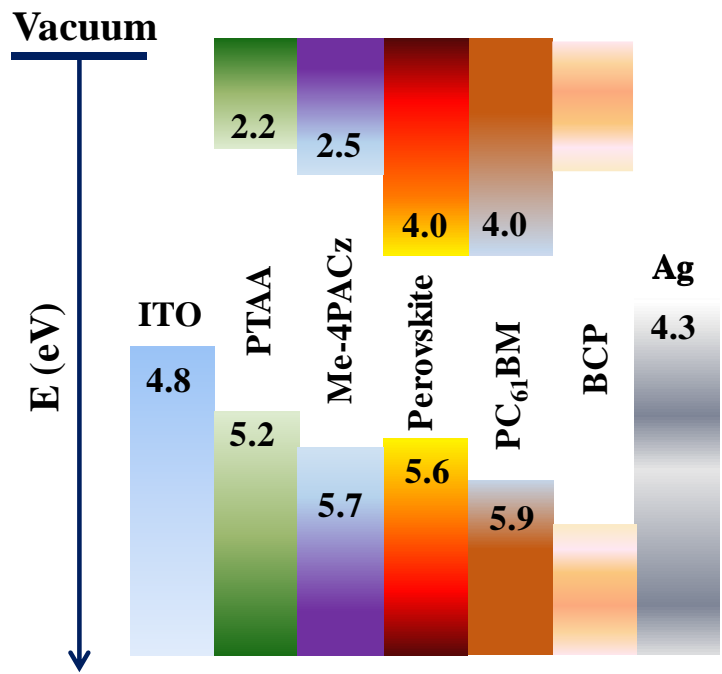

**Figure 2:** *Energy level band diagram of the p-i-n architecture-based perovskite photodetector (PPD). The energy level values are taken from the literature.*[51–53]

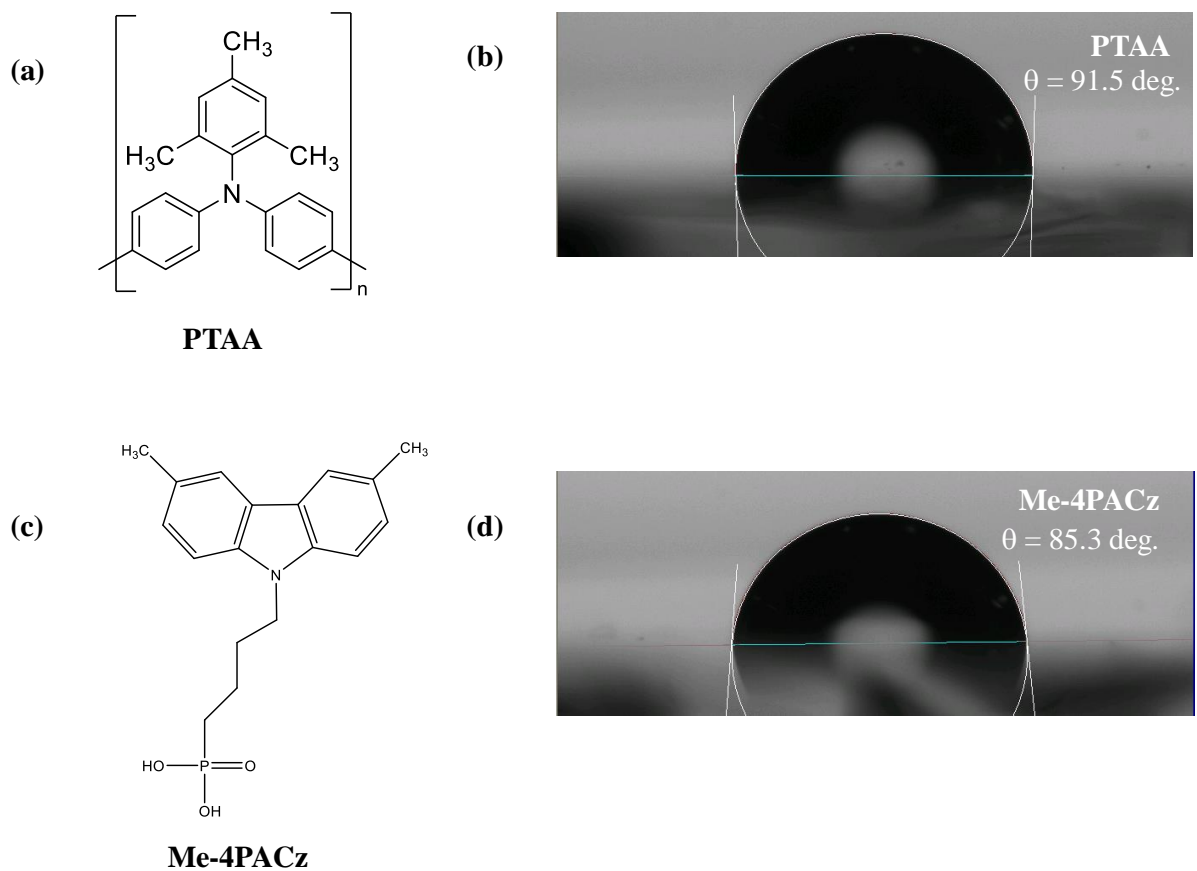

***Figure 3:*** *The water contact angle with the molecular structure of PTAA and Me-4PACz layer deposited on ITO - coated glass substrates.*

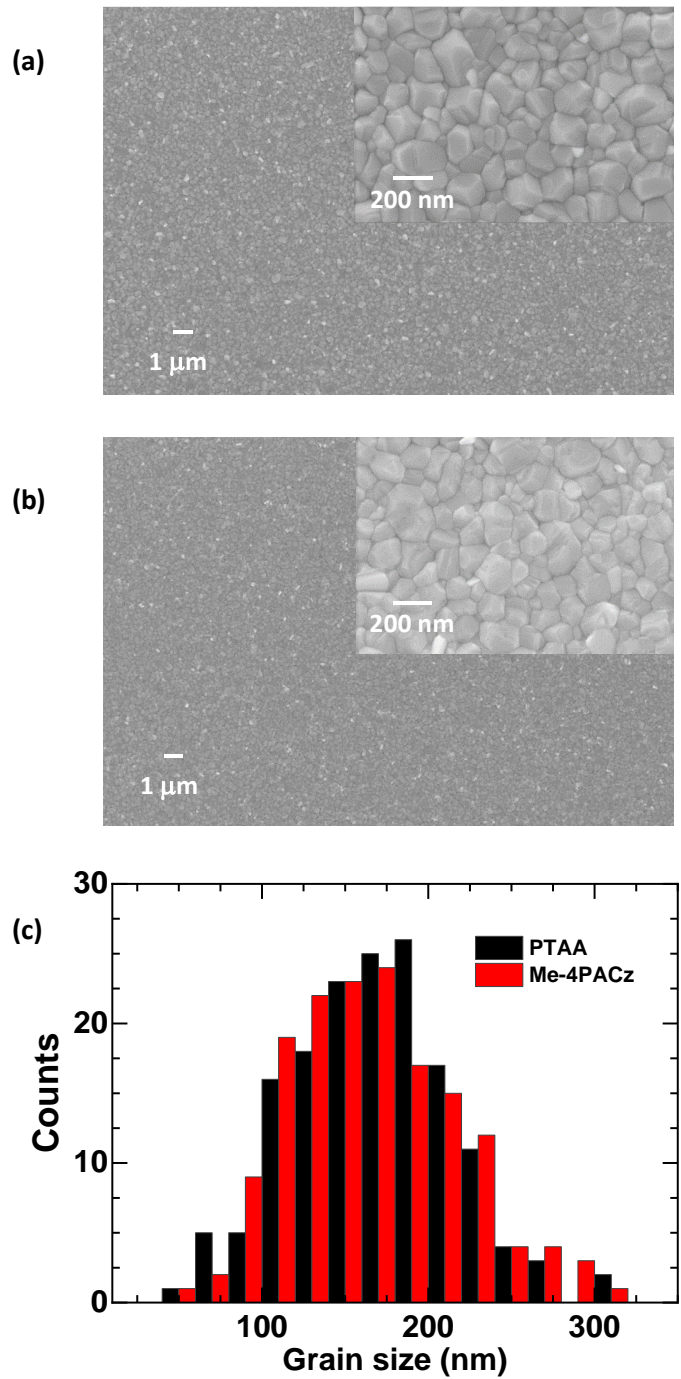

**Figure 4:** *Top surface SEM images of perovskite thin film deposited on* (*a*) *PTAA and* (*b*) *Me-4PACz layer-coated ITO substrates. Both films have an indistinct, uniform, and compact morphology.* (*c*) *The average grain size of both the films are ~ 200 nm.*

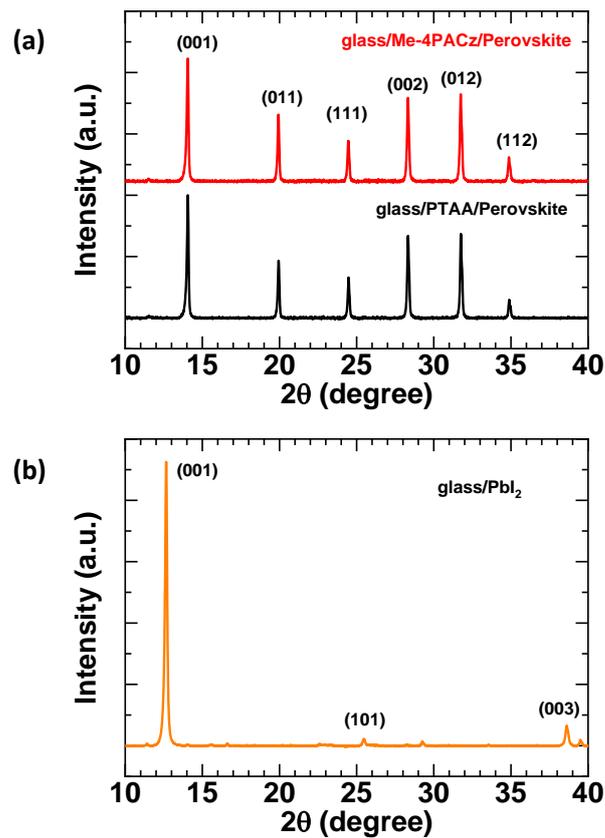

***Figure 5****: (a) The XRD pattern of the perovskite films deposited on PTAA and Me-4PACz coated glass substrates. (b) The XRD pattern of the PbI$_2$ film deposited on the glass substrates.*[54]

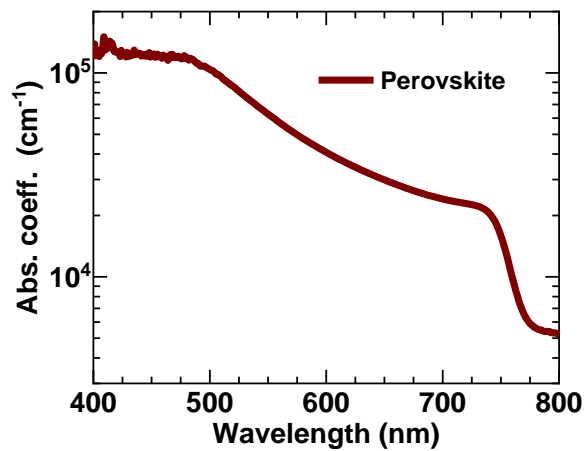

***Figure 6****: The absorption coefficient of the triple cation perovskite (FA$_{0.83}$MA$_{0.17}$)$_{0.95}$Cs$_{0.05}$Pb(I$_{0.83}$Br$_{0.17}$)$_3$ abbreviated as CsFAMA film deposited on a glass substrate.*

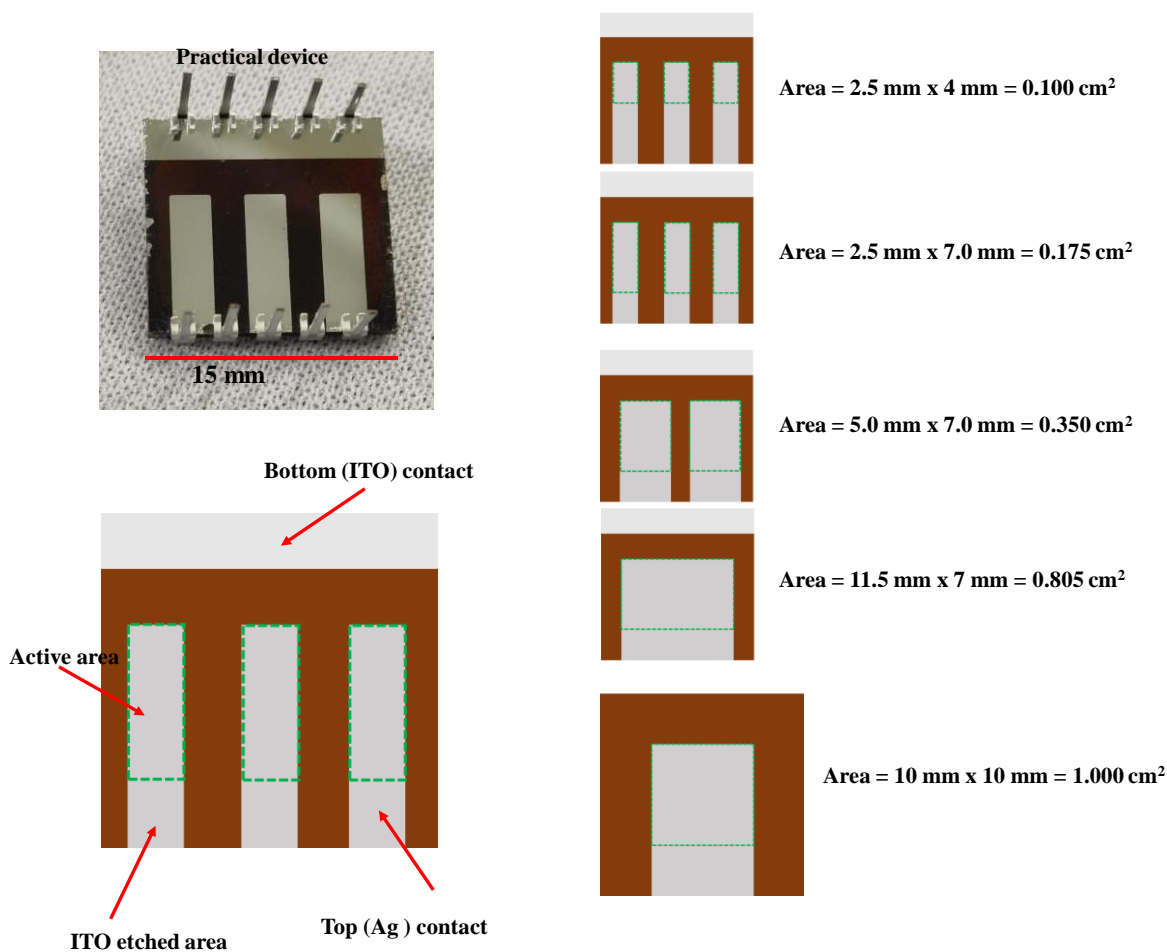

***Figure 7:*** *Schematic of the device active area of the PPD with identifying the active area, etched area, top (Ag) contact, and bottom (ITO) contact. For ease of understanding, the practical device image is shown on the top left panel. The substrate size used was 15 mm x 15 mm for devices of active area 0.100 $cm^2$ to 0.805 $cm^2$, and the size of the substrate was 20 mm x 20 mm for the devices of active area 1.000 $cm^2$.*

The perovskite films deposited on the PTAA and Me-4PCAz layer have indistinct, uniform, compact morphology of average grain size ~ 200 nm (see **Figure 4**). The X-ray diffraction (XRD) patterns show no trace of residual lead iodide peak and no significant difference in the FWHM, i.e., perovskite crystal structure (see **Figure 5**). The perovskite is a direct bandgap semiconducting material and shows a high absorption coefficient ($\sim 10^5$ $cm^{-1}$), hence a suitable candidate for photodetector application (see **Figure 6**).[16] We used different metal masks for back metal electrode (silver) deposition to complete the PPD devices of different active areas (see **Figure 1d**). The active area of the devices used are 0.100 $cm^2$, 0.175 $cm^2$, 0.350 $cm^2$, 0.805 $cm^2$, and 1.00 $cm^2$. The detailed calculation of the active areas is shown in **Figure 7**.

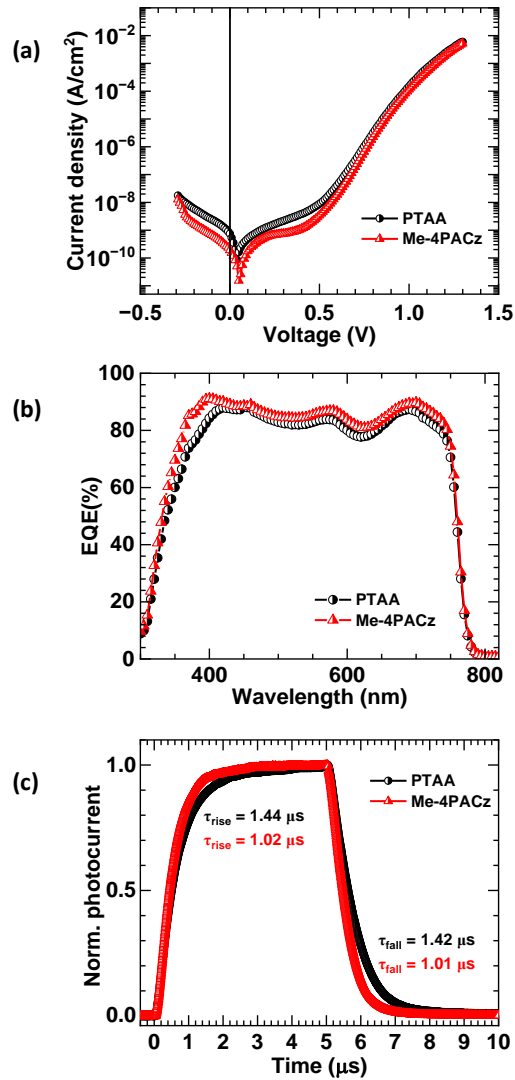

***Figure 8****: **(a)** Current density vs. Voltage (J-V) characteristics of the PTAA and Me-4PACz HTL based perovskite photodetectors (PPDs) under dark conditions. **(b)** External quantum efficiency (EQE) of PTAA and Me-4PACz HTL based PPDs. **(c)** Normalized transient photocurrent (TPC) using a 638 nm modulated laser of pulse duration 5 μs, i.e. repetition rate 10 kHz with a duty cycle of 5%.*

In this chapter, we first compare the role of PTAA and Me-4PACz HTLs in the PPD performance and then analyze the effect of scalability on the PPD performance. We fabricated PTAA and Me-4PACz HTLs-based PPDs of active area of 0.175 cm$^2$. **Figure 8a** represents the current density *vs*. voltage $(J - V)$ characteristics in the semi-log scale under dark conditions. The Keithley 2612B is used to measure the dark *J-V* characteristics which has the minimum current measurement limit of ~10$^{-12}$ A (see **Figure 10**). The dark current density at zero bias for the PTAA HTL-based PPD is $7.49 \times 10^{-10}$ A/cm$^2$ and for the Me-4PACz based device it is $1.91 \times 10^{-10}$ A/cm$^2$ ( see **Figure 9, Table 1**). The lower dark current is observed for the Me-4PACz based PPD, which could be attributed to the better interfacial contact of the HTL with the perovskite absorber (Figure 2).[41] The lower dark current is an important parameter for an efficient photodetector as it is inversely related to the FOM parameters i.e. Specific Detectivity (D*), Linear dynamic range (LDR) etc.[19,55] Lower dark current shows improvement in the photodetection, which will be discussed in more detail in the scalability section later.

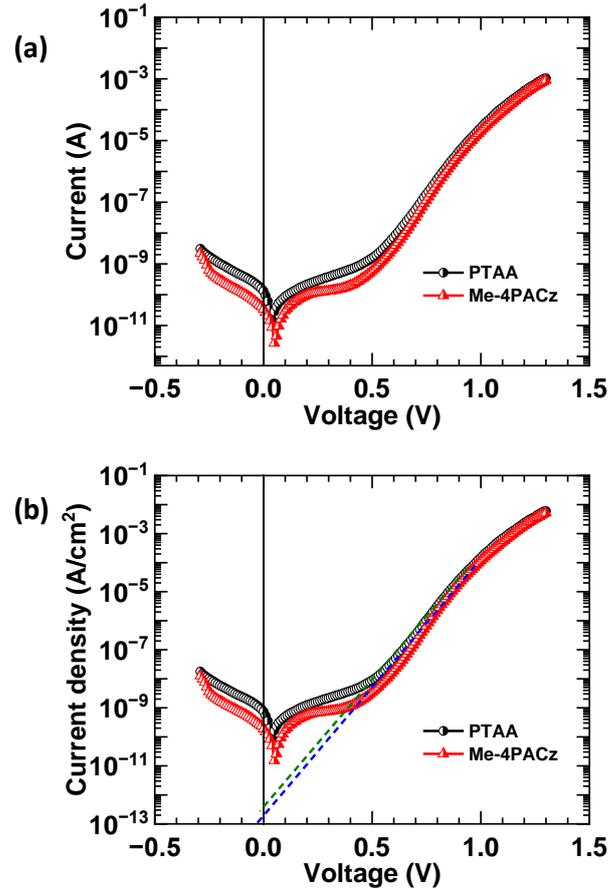

***Figure 9:*** *(**a**) Dark current vs. voltage $(I-V)$ characteristics for PTAA and Me-4PACz HTL-based PPD. (**b**) Dark current density vs. voltage $(J-V)$ characteristics for PTAA and Me-4PACz HTL-based PPD.*

***Table 1:*** *Dark current parameters for the PTAA and Me-4PACz based PPD devices.*

| HTL | Dark current $I_D$ (A) at zero bias | Dark current density $J_D$ (A/cm²) at zero bias | Reverse saturation current density $J_0$ (A/cm²) at zero bias |
|---|---|---|---|
| PTAA | $1.31 \times 10^{-10}$ | $7.49 \times 10^{-10}$ | $3.85 \times 10^{-13}$ |
| Me-4PACz | $3.35 \times 10^{-11}$ | $1.91 \times 10^{-10}$ | $1.92 \times 10^{-13}$ |

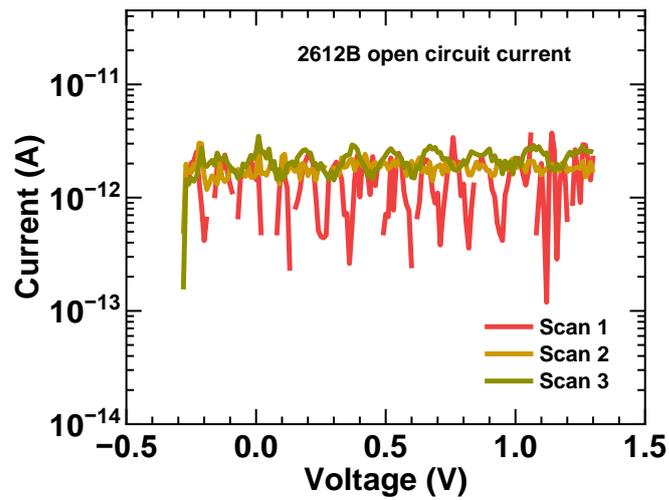

***Figure 10**: Current density vs. voltage  (J − V) characteristics of Keithley 2612B at open circuit condition.*

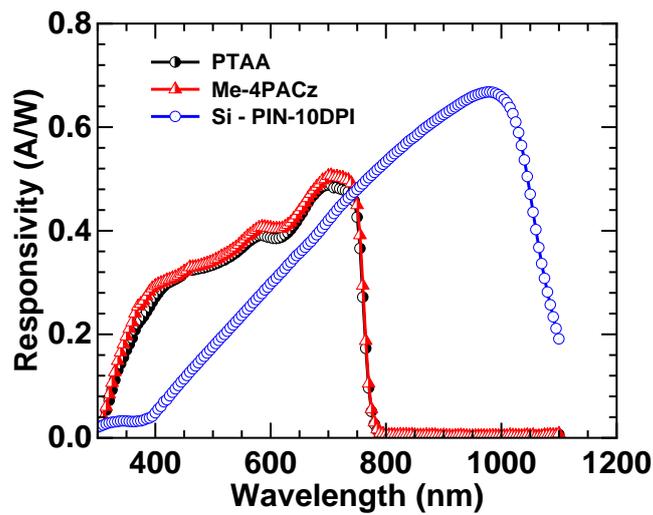

***Figure 11**: Responsivity comparison of the PTAA and Me-4PACz based PPD with respect to Si based PIN-10DPI diode.*

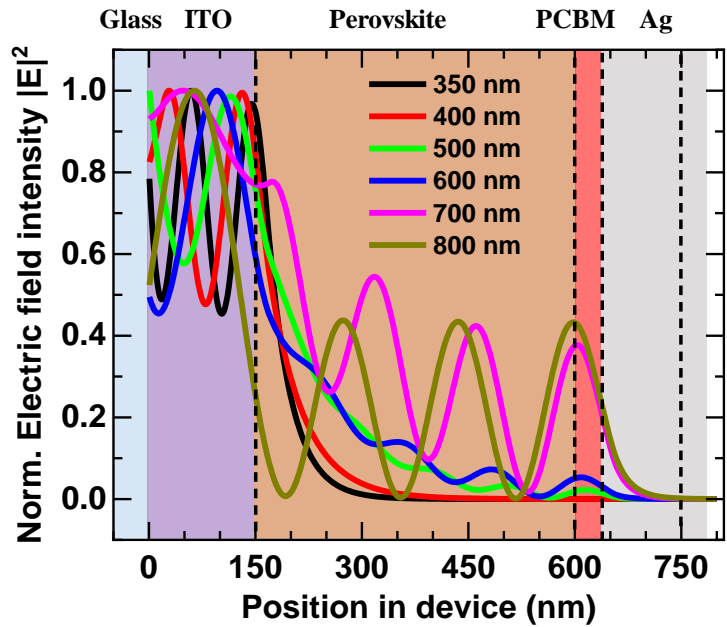

*Figure 12*: *The absorption of the different wavelengths into the device at different positions.*

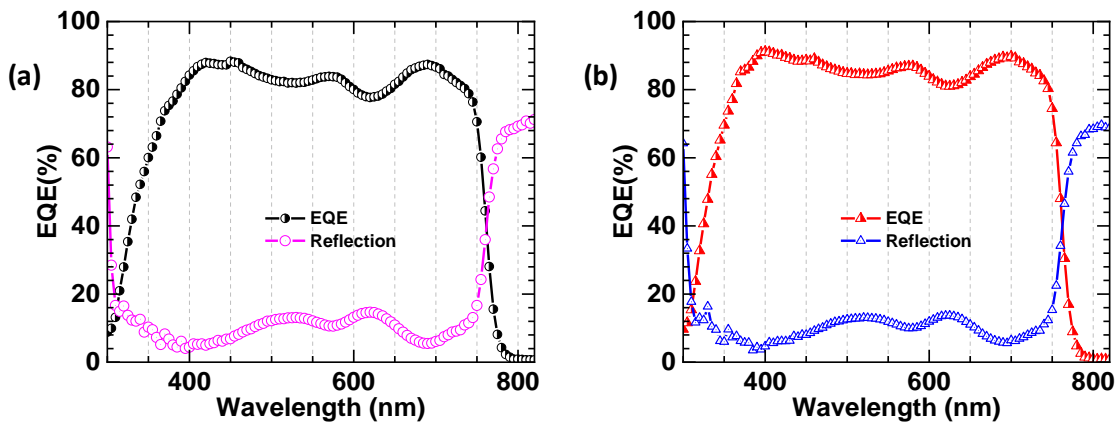

*Figure 13*: *External quantum efficiency (EQE) spectrum including the reflection (R) spectra of (**a**) PTAA and (**b**) Me-4PACz HTL based PPDs.*

Responsivity is a FOM parameter that is defined by the ratio of the amount of current produced (in Ampere) by the photodetector to the input optical power (in watts) at a given wavelength.[20,56] The responsivity of the perovskite photodetector is comparatively higher than the commercialized Si PIN-10DPI photodiode in the UV-visible region (see **Figure 11**). The external quantum efficiency (EQE) of a PPD is defined by the ratio of the number of electrons collected by the external circuit to the number of photons incident at a given wavelength ($\lambda$) and is represented in terms of responsivity by[56],

$$\text{EQE}(\lambda) = \frac{hc}{q\lambda} R(\lambda) \qquad\qquad (1)$$

Where $h$ is the plank constant, $c$ is the speed of light, and $q$ is the elementary charge. The EQE of the Me-4PACz based PPD is slightly higher than the PTAA based PPD and a distinct difference is observed at the lower wavelength (~ 400 nm) (see **Figure 8b).** The lower EQE for the PTAA based PDD near the ultraviolet region may be due to the increased recombination of the charge carriers at the HTL-perovskite interface, as the lower wavelength light absorbed promptly near the absorber surface (see **Figure 12).**[57] The dip in the EQE (as well as in the $R(\lambda)$) spectra correlates well with the hump in the reflection spectra due to the microcavity effect for a multilayer thin film devices with back metal reflecting contact (Ag) **Figure 13**.[50,58] Further, the response speed of a PD is an important FOM parameter that is strongly related to charge transport and collection in the device. **Figure 8c** represents the transient photocurrent (TPC) response of the PPDs, which is measured using a 638 nm laser of pulse duration 5 μs, i.e. repetition rate 10 kHz with a duty cycle of 5%. The rise time ($t_r$) is defined as the time by which the PD device current reaches from 10% to 90%, and the fall time ($t_f$) is vice versa.[59] For the Me-4PACz HTL based PPD, $t_r = 1.02$ μs and $t_f = 1.01$ μs, whereas for the PTAA based PPD $t_r = 1.44$ μs and $t_f = 1.42$ μs. Thus, the Me-4PACz-based PPD is ~1.4 times faster than PTAA-based PPD devices.

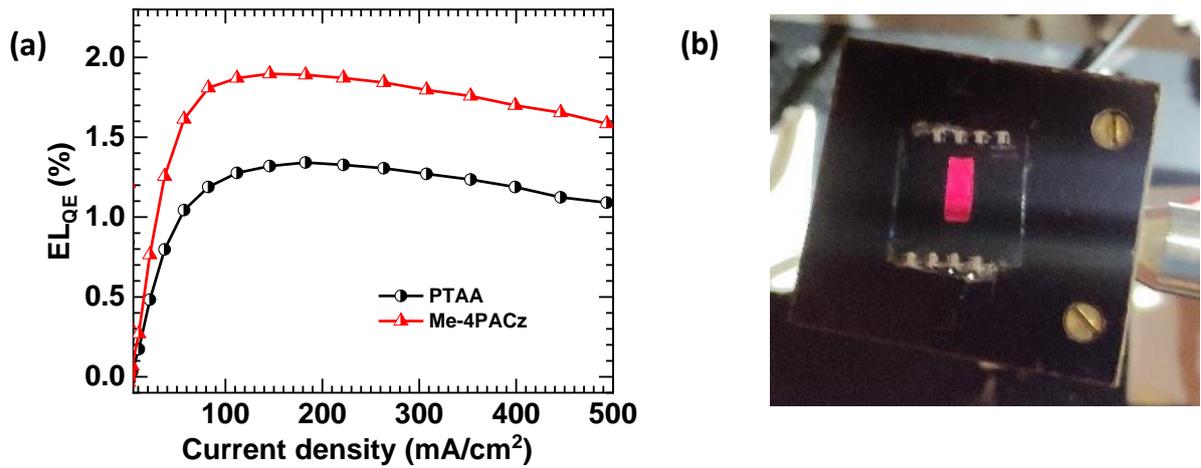

***Figure 14**: The PPD is used as a light-emitting diode (LED). **(a)** Electroluminescence external quantum efficiency of the PPD device as a function of injected current density. **(b)** The photographic image of the device during LED operation.*

The electroluminescence quantum efficiency ($EL_{QE}$) measurements were carried out at different injection current densities (see **Figure 14**). The $EL_{QE}$ of Me-4PACz based PPD is ~1.5 times higher than the PTAA based PPD, indicating improved balanced charge carrier injection. Since only the HTL is different in both the devices, indicating HTL/perovskite interfacial non-radiative traps are significantly reduced in the Me-4PACz HTL based PPD.[60] Thus, choosing an HTL of Me-4PACz based PPD is suitable over the PTAA HTL based PPD.

Using the Me-4PACz HTL, we scale up the active area of the photodetector from 0.100 $cm^2$ to 1.000 $cm^2$ (Figure 1d). **Figure 15a** shows the dark $J - V$ characteristics for different active area-based PPDs. As the area of the photodetector increases, the dark current increases, as shown in **Figure 16** and **Table 2.** It is to be noted that the increase in dark current at zero bias is not linear. As the active area increases, the reverse saturation dark current density ($J_0$) increases exponentially (see **Figure 15b**). The $J_0$ *vs.* Area plot is fitted with $J_0 = \alpha \times e^{\beta A^n}$, where $\alpha$ and $\beta$ are constants, A is the active area of the device, and $n$ is the exponent of the active area. For $n = 1$, the plot fits well at the higher area segment. However, the fitting is well over the whole curve with $n = 2$.

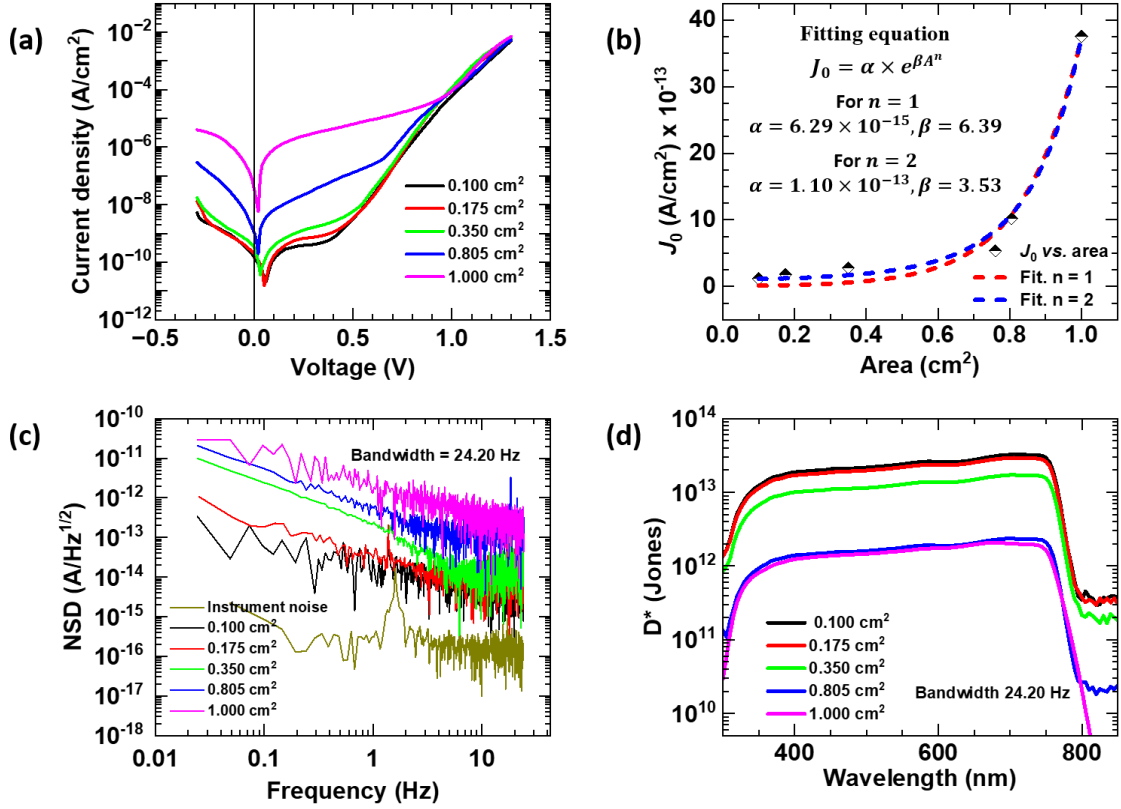

***Figure 15***: *Me-4PACz HTL based perovskite photodetectors (PPDs) performance. Color key- device active area 0.100 cm² is black, 0.175 cm² is red, 0.350 cm² is green, 0.805 cm² is blue, 1.000 cm² is magenta. (**a**) Current density vs. Voltage (J-V) characteristics of the different active area based PPDs under dark conditions. (**b**) Reverse saturation current density ($J_0$) variation with the area (A) of the PPDs. (**c**) Noise spectral density (NSD) for different active area-based PPDs at a bandwidth of 24.20 Hz. (**d**) Specific detectivity estimated using NSD for different active area-based PPDs.*

The FOM parameter specific detectivity ($D^*$) signifies the ability of a PD to detect faint light and can be estimated in two ways, as explained below. When the detectivity is dominated by the dark current, the detectivity is given by[12,19]

$$D^*(\lambda) = \frac{1}{\sqrt{2qJ_d}} R(\lambda) \qquad (4)$$

Where $R(\lambda)$ is the responsivity, $q$ is the elementary charge, and $J_d$ is the dark current density. This formula is widely used in the photodetector community as it is easy to handle and free from the frequency or bandwidth of the measurement. The $R(\lambda)$ and corresponding EQE($\lambda$) for all the areas of PPD are shown in **Figure 17**, and there is no significant difference

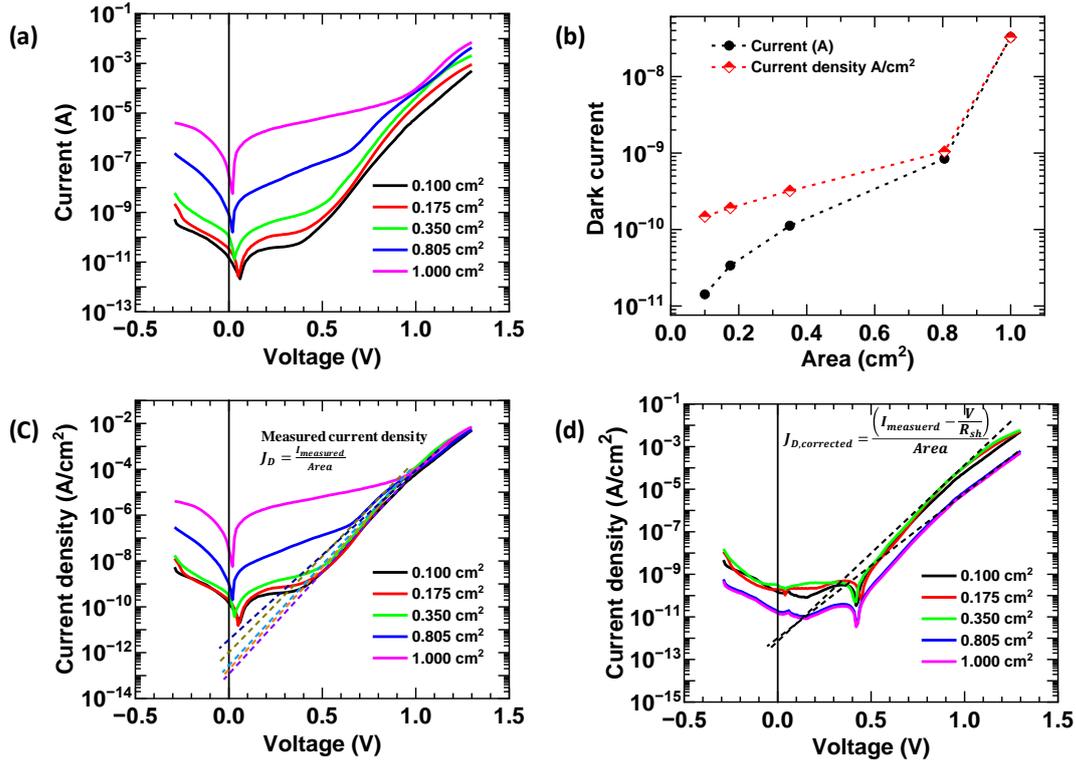

***Figure 16*:** ***(a)*** *Dark current vs. voltage $(I - V)$ characteristics for different active area based PPD.* ***(b)*** *Dark current density vs. voltage $(J - V)$ characteristics for different active area based PPD.* ***(c)*** *dark current (A) and dark current density $(A/cm^2)$ as a function of device area.* ***(d)*** *The dark $J - V$ characteristics after removing the shunt resistance from the dark current.*

***Table 2*:** *Dark current parameters for different active area-based PPDs.*

| Area $(cm^2)$ | Dark current $I_D$ (A) at zero bias | Dark current density $J_D (A/cm^2)$ at zero bias | Reverse saturation current density $J_0 (A/cm^2)$ |
|---|---|---|---|
| 0.100 | $1.48 \times 10^{-11}$ | $1.48 \times 10^{-10}$ | $1.18 \times 10^{-13}$ |
| 0.175 | $3.35 \times 10^{-11}$ | $1.91 \times 10^{-10}$ | $1.82 \times 10^{-13}$ |
| 0.350 | $1.11 \times 10^{-10}$ | $3.18 \times 10^{-10}$ | $2.79 \times 10^{-13}$ |
| 0.805 | $8.39 \times 10^{-10}$ | $1.04 \times 10^{-9}$ | $1.02 \times 10^{-12}$ |
| 1.000 | $3.27 \times 10^{-8}$ | $3.27 \times 10^{-8}$ | $3.76 \times 10^{-12}$ |

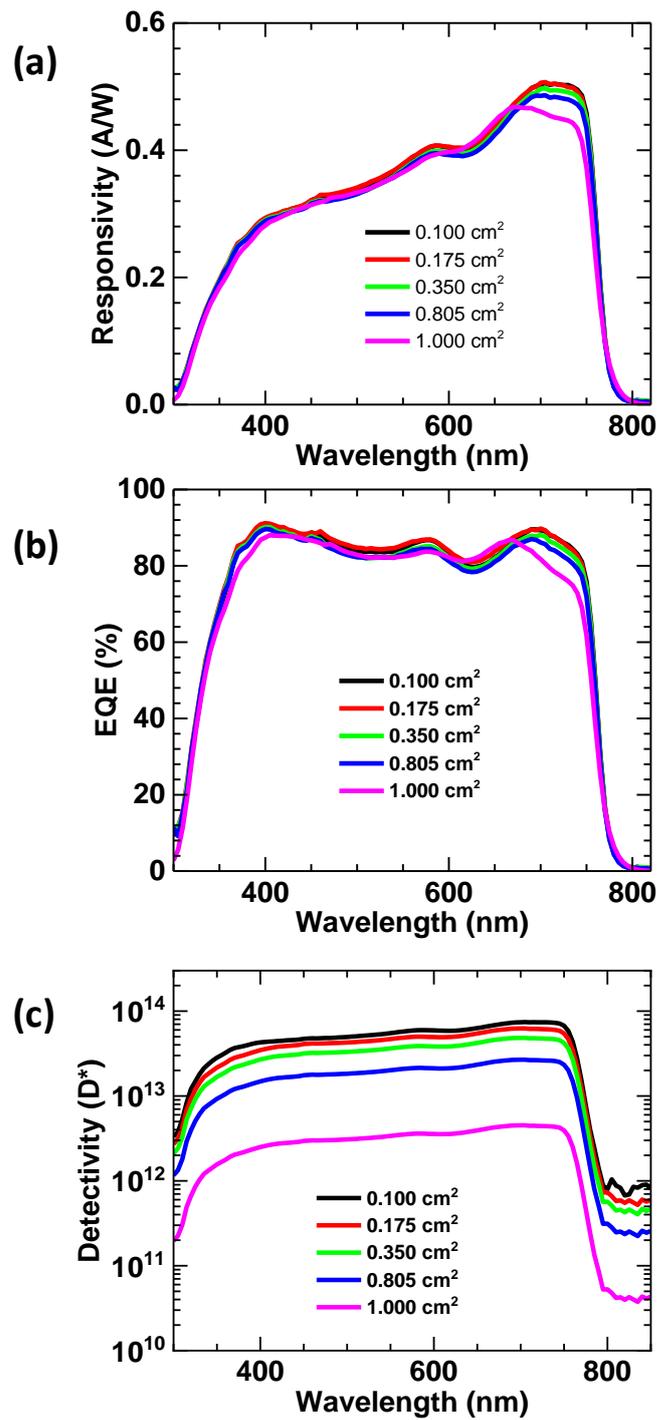

***Figure 17****: (**a**) Responsivity (**b**) External quantum efficiency (EQE) and (**c**) The specific detectivity (D\*) calculated using shot noise current for the different active area-based PPDs.*

***Table 3***: *Responsivity, EQE, and detectivity at wavelength ~ 700 nm for different active area-based PPDs.*

| Area (cm²) | Responsivity | EQE (%) | Detectivity (D*) at zero bias |
|:---:|:---:|:---:|:---:|
| 0.100 | 0.51 | 89.10 | $7.42 \times 10^{13}$ |
| 0.175 | 0.51 | 89.32 | $6.25 \times 10^{13}$ |
| 0.350 | 0.50 | 87.64 | $4.84 \times 10^{13}$ |
| 0.805 | 0.49 | 85.71 | $2.68 \times 10^{13}$ |
| 1.000 | 0.46 | 80.62 | $4.51 \times 10^{12}$ |

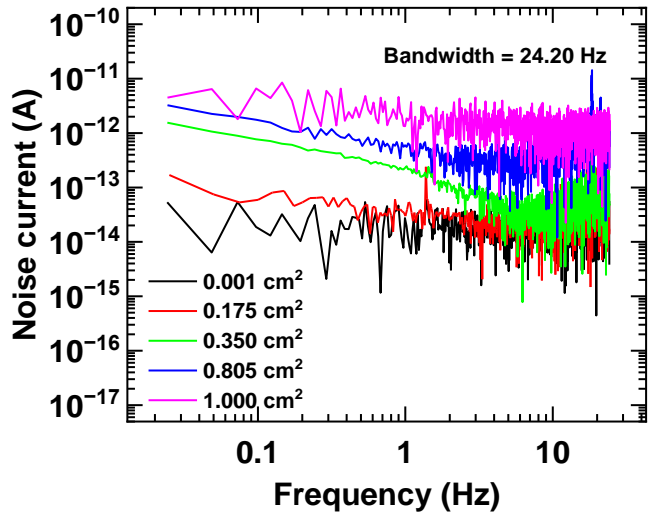

***Figure 18***: *The noise current of the different active area based PPDs is at a bandwidth of 24.20 Hz.*

***Table 4***: *The average saturated noise current of the different active area-based PPD is at a bandwidth of 24.20 Hz.*

| Noise current (A) for 0.100 cm² | Noise current (A) for 0.175 cm² | Noise current (A) for 0.350 cm² | Noise current (A) for 0.805 cm² | Noise current (A) for 1.00 cm² |
|:---:|:---:|:---:|:---:|:---:|
| $2.42 \times 10^{-14}$ | $3.55 \times 10^{-14}$ | $8.40 \times 10^{-14}$ | $9.12 \times 10^{-13}$ | $1.12 \times 10^{-12}$ |

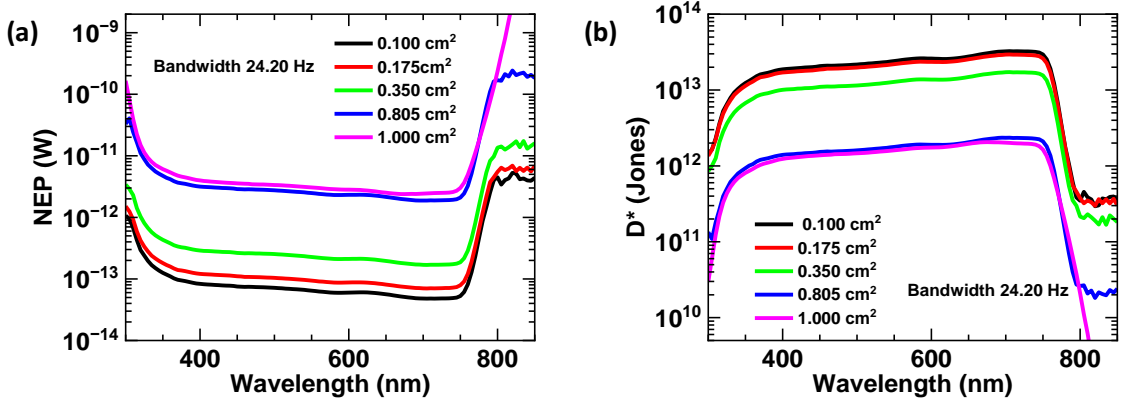

***Figure 19**: (**a**) The noise equivalent power and (**a**) detectivity for the different active area based PPD at a bandwidth of 24.20 Hz.*

***Table 5**: The noise equivalent power and detectivity for the different active area-based PPD at a bandwidth of 24.20 Hz and at wavelength 700 nm.*

| Parameters | Area 0.100 cm$^2$ | Area 0.175 cm$^2$ | Area 0.350 cm$^2$ | Area 0.805 cm$^2$ | Area 1.000 cm$^2$ |
|---|---|---|---|---|---|
| NEP (W) | $4.74 \times 10^{-14}$ | $6.96 \times 10^{-14}$ | $1.68 \times 10^{-13}$ | $1.86 \times 10^{-12}$ | $2.43 \times 10^{-12}$ |
| D* (Jones) | $3.28 \times 10^{13}$ | $2.96 \times 10^{13}$ | $1.73 \times 10^{13}$ | $2.37 \times 10^{12}$ | $2.02 \times 10^{12}$ |

***Table 6**: List of the parameters of $P_{sat}$, NEP, LDR, noise floor current, and dark current at zero bias. The NEP is estimated where the photocurrent deviates from linearity at lower laser power.*

| Area (cm$^2$) | $P_{sat}$ (W) | NEP (W) | LDR (dB) | Noise floor current (A) | Dark current $I_D$ (A) at zero bias |
|---|---|---|---|---|---|
| 0.100 | 0.03 | $5.38 \times 10^{-11}$ | 87.46 | $5.36 \times 10^{-11}$ | $1.48 \times 10^{-11}$ |
| 0.175 | 0.03 | $1.49 \times 10^{-10}$ | 83.04 | $1.44 \times 10^{-10}$ | $3.35 \times 10^{-11}$ |
| 0.350 | 0.03 | $2.73 \times 10^{-10}$ | 80.36 | $4.76 \times 10^{-10}$ | $1.11 \times 10^{-10}$ |
| 0.805 | 0.03 | $2.18 \times 10^{-9}$ | 71.39 | $2.79 \times 10^{-9}$ | $8.39 \times 10^{-10}$ |
| 1.000 | 0.03 | $4.25 \times 10^{-9}$ | 68.49 | $7.88 \times 10^{-9}$ | $3.27 \times 10^{-8}$ |

The dark current density ($J_\text{D}$) at zero bias for 0.175 cm$^2$ PPD is $1.91 \times 10^{-10}$ A/cm$^2$ and using eq. (4), the corresponding specific detectivity (D*) is $6.25 \times 10^{13}$ Jones (cmHz$^{1/2}$W$^{-1}$) at wavelength ~700 nm. The D* for all the areas is shown in **Figure 17c**, and the parameters are listed in **Table 3**. The aforementioned specific detectivity values on the given area of the PPDs are the highest as per literature reports (see **Table 7**). It is to be noted that the above specific detectivity in eq. (4) is overestimated by the shot noise limit, and often, this practice is used in the photodetector community.[61–65] However, the specific detectivity estimation using the total noise current ($i_n$) is as follows,[66]

$$D^*(\lambda) = \frac{\sqrt{A\,\Delta f}}{NEP} \quad \text{……….……… (5)}$$

$$NEP = \frac{i_n}{R(\lambda)} \quad \text{……………. (6)}$$

Where A is the active area of the PD, $\Delta f$ is the detection bandwidth and $NEP$ is the noise equivalent power.

**Figure 15c** represents the noise spectral density (NSD) as a function of frequency for different active area-based PPDs at a bandwidth of 24.20 Hz. The noise current is shown in **Figure 18.** The noise current ($i_n$) or NSD increases with the increase in the active area of the PPD. However, for each active area of the device, the NSD dominates at a lower frequency (<10 Hz), called the flicker noise or $1/f$ noise , and saturates at ~ 10 Hz (Figure 15c).[66,67] The average noise saturation current for the 0.175 cm$^2$ cell is ~$3.55 \times 10^{-14}$A (see **Table 4** for the different active area based devices). The estimated NEP using eq. (6) is $6.96 \times 10^{-14}$ W and the corresponding D* using eq. (5) is $2.96 \times 10^{13}$ Jones at wavelength ~ 700 nm **Figure 15d** (for other areas, see **Figure 19** and **Table 5**). Therefore, as the active area of the device increases, the NEP increases and the D* reduces.

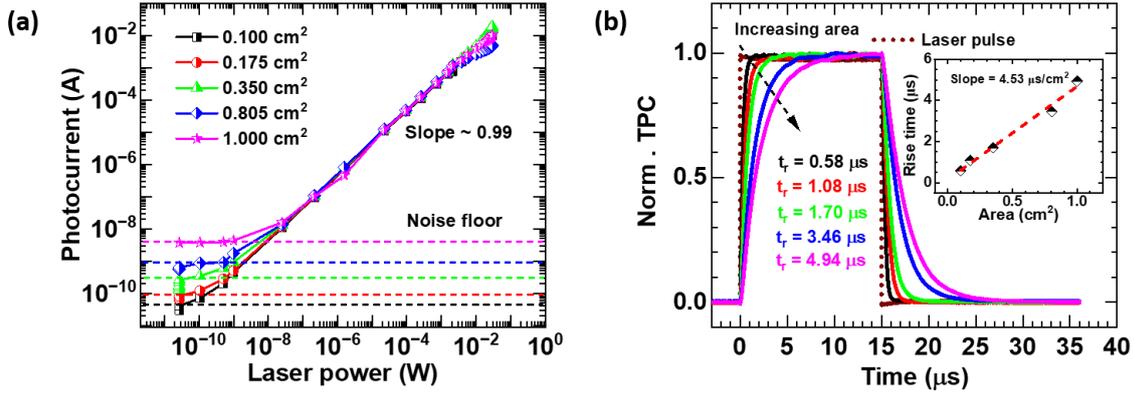

**Figure 20**: *Me-4PACz HTL based perovskite photodetectors (PPDs) performance. Color key-device active area 0.100 cm² is black, 0.175 cm² is red, 0.350 cm² is green, 0.805 cm² is blue, 1.000 cm² is magenta. (**a**) The PPD's linear dynamic range (LDR) is measured with a 638 nm laser and a set of neutral density (ND) filters of different optical densities. (**b**) Normalized transient photocurrent (TPC) is measured using a 638 nm modulated laser of pulse duration 15 μs i.e. repetition rate 10 kHz with a duty cycle 15%.*

A PD is considered to be good if it shows photocurrent response over a large range of intensities linearly, referred to as a linear dynamic range (LDR), an another important FOM parameter and is defined as

$$LDR = 10 \log \left( \frac{P_{sat}}{NEP} \right) \qquad (7)$$

Where $P_{sat}$ is the power at which the photocurrent gets saturated at higher intensity and NEP is the photocurrent saturation power at minimum intensity. **Figure 20a** shows the intensity dependent photocurrent without bias for the different active area PPDs. We used a 638 nm laser to record the photocurrent current using Keithley 2612B. The laser intensity is varied by using a set of neutral density (ND) filters with different optical densities (O.D. of 0.1 to 9.0). As the incident laser power decreases, the photocurrent decreases to the noise equivalent current and saturates, referred to as noise floor.[55,68] The incident laser power at which the photocurrent of the PPD touches the noise floor is called the noise equivalent power (NEP). Interestingly, it is limited by the dark current at zero bias (see **Table 6**). As the area of the PPD device increases, the noise floor elevates to a higher current. The $P_{sat}$ and $NEP$ for the 0.175 cm² PPD is 3.00 × $10^{-2}$ W and 1.49 × $10^{-10}$ W, respectively, and the corresponding LDR is 83.04 dB (see **Table 6** for the other different active areas). Though, we note that $P_{sat}$ is underestimated due to the maximum power available of the laser source used in the measurement. Further, the response time of a PD is an important FOM parameter that is strongly related to charge transport and collection. **Figure 20b** represents the TPC response of the PPDs that is measured using a 638 nm laser of pulse duration 15 μs i.e. repetition rate 10 kHz, and duty cycle of 15%. The TPC signal shows excellent amplitude and temporal coherence over 100 repeated cycles for all active area devices, as shown in **Figures 21** and **22**. The rise time ($t_r$) increases from 0.58 μs to 4.94 μs upon increasing the device area from 0.100 cm$^2$ to 1.000 cm$^2$. The rise time of the PPDs increases linearly with the device area with a slope of 4.53 μs/cm$^2$ (inset Figure 20b).

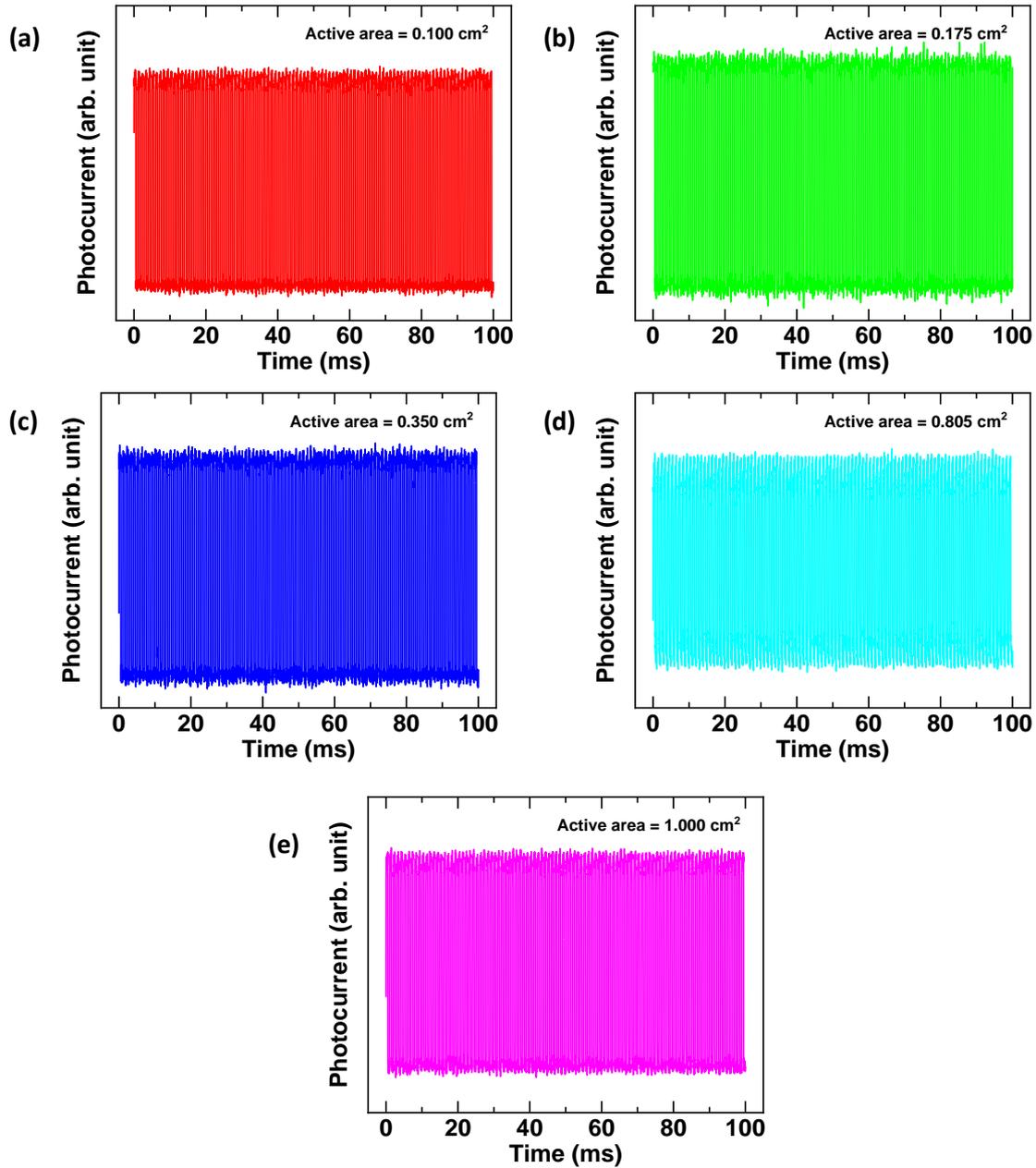

**Figure 21**: *TPC amplitude and temporal coherence over 100 repeated cycles of the perovskite photodetectors having an active area of (**a**) 0.100 cm$^2$, (**b**) 0.175 cm$^2$, (**c**) 0.350 cm$^2$, (**d**) 0.805 cm$^2$, and (**e**) 1.000 cm$^2$, with a repetition rate of 1 kHz and duty cycle of 50% under zero bias condition.*

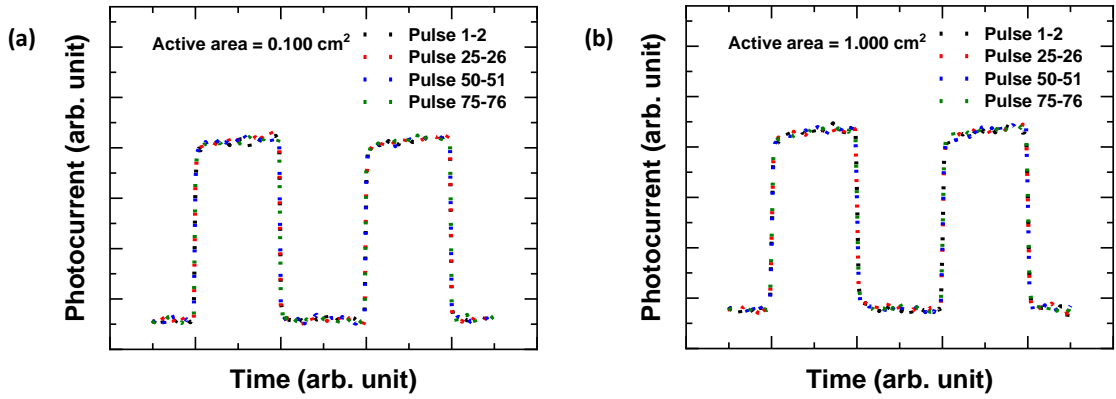

***Figure 22****: TPC amplitude and temporal coherence of the perovskite photodetectors (from figure 21) of pulse 1-2, 25-26, 50-51, and 75-76 having an active area of (**a**) 0.175 cm² and (**b**) 1.000 cm²  with a repetition rate of  1 kHz and duty cycle of 50% under zero bias condition.*

Further, the 638 nm pulsed laser of varying frequency from 200 Hz to 100 kHz is incident to the PPD, and the corresponding photogenerated current is measured using SRS 830 lock-in amplifier (limit 102 kHz) for the 0.175 cm² and 1.000 cm² area based PPD. The frequency at which the photodetector current/voltage drop its amplitude by $\frac{1}{\sqrt{2}}$ times of initial amplitude, called frequency bandwidth $f_{-3dB} \approx \frac{1}{2\pi RC}$.[19,61,69] The $f_{-3dB}$ >100 kHz for 0.175 cm² and ~ 65 kHz for the 1.000 cm² (see **Figure 23**).

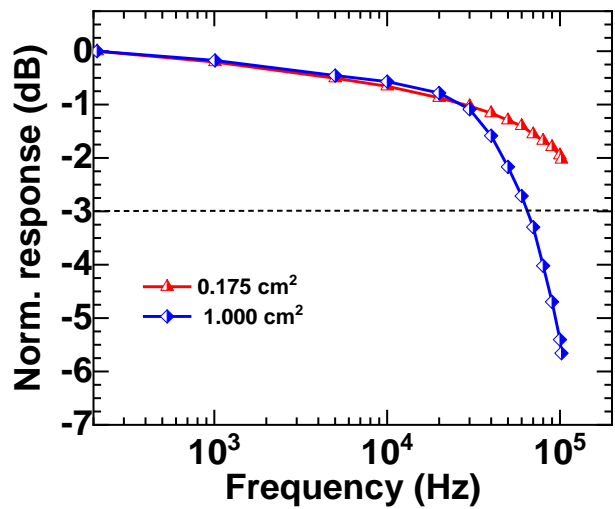

***Figure 23****: The frequency-dependent current response of the PPDs was measured using a modulated 638 nm laser and SR 830 lock-in amplifier for the 0.175 cm²  and 1.000 cm² active area devices. The frequency limit in SRS 830 lock-in amplifier is 102 kHz.*

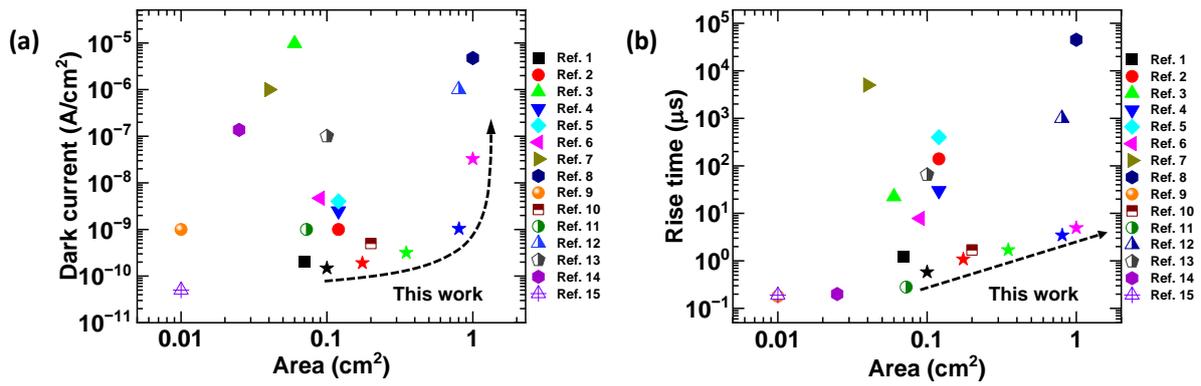

***Figure 24**: **(a)** Dark current density ($J_d$) and **(b)** Rise time ($t_r$) as a function of the active area of our Me-4PACz HTL based perovskite photodetectors compared with the literature reports. The list of reference papers is tabulated in Table 7. The star (\*) symbols indicate this work.*

We compared the dark current at zero bias and the rise time of our Me-4PACz based PPDs on different active areas with the literature reports (see **Figure 24**). The list of reference papers is available in **Table 7**. It is observed that, as the PPD active area increases, the dark current and the rise time increase but are notably lower compared to the literature reports on the given active area of the device. Thus, the scalability of the PPD using the Me-4PACz as HTL is a suitable choice.

***Table 7****: List of the references used in **Figure 24** (of the main text) for the dark current density and rise time comparison for different active area based PPDs.*

| Device structure | Device area (cm$^2$) | Dark current ($J_d$, A/cm$^2$) | rise/fall time (µs) | Detectivity $D^*$(Jones) | Ref. No. in Figure 24 | Ref. |
|---|---|---|---|---|---|---|
| FTO/PCBM/MAPbI$_3$/Spiro/Au | 0.07 | $2 \times 10^{-10}$ | 1.20/3.20 | $4 \times 10^{12}$ | 1 | [31] |
| ITO/SnO$_2$/CsPbBr3/Spiro/Au | 0.12 | $\sim 10^{-9}$ | 140/120 | $4.8 \times 10^{12}$ | 2 | [22] |
| FTO/TiO2/CsI(PbBr$_2$)$_{0.99}$/(AgI)$_{0.01}$/Spiro/Au | 0.06 | $9.71 \times 10^{-6}$ | 22.4/25.7 | $2.46 \times 10^{11}$ | 3 | [70] |
| ITO/SnO2/CsPbBr3/PTAA/Au | 0.12 | $2.5 \times 10^{-9}$ | 30/39 | $7.23 \times 10^{12}$ | 4 | [71] |
| ITO/SnO2/PMMA/CsPbBr3/PTAA/Au | 0.12 | $4 \times 10^{-9}$ | 400/430 | $1 \times 10^{13}$ | 5 | [23] |
| CsPbBr3−Cs4PbBr6/FTO/TiO2/CH3NH3PbI3/Spiro-OMeTAD/Au | 0.09 | $4.7 \times 10^{-10}$ | 7.8/33.6 | $1.2 \times 10^{12}$ | 6 | [72] |
| FTO/TiO2/graphene/MAPbI3/spiro/Au | 0.04 | $\sim 10^{-6}$ | 5000 | $4.5 \times 10^{11}$ | 7 | [24] |
| Glass/PH1000/Perovskite/Spiro/PEDOT:PSS/Cu | 1.00 | $4.76 \times 10^{-6}$ | 45000/46000 | $4.16 \times 10^{12}$ | 8 | [73] |
| ITO/MAPbI$_{3-x}$ Cl$_x$/PCBM/PFN/Al | 0.01 | $\sim 10^{-9}$ | 0.18/0.16 | $4 \times 10^{14}$ | 9 | [25] |
| ITO/PEDOT:PSS/MAPbI3/PC60BM/C60/LiF/Ag | 0.20 | $5 \times 10^{-10}$ | 1.7/1.00 | $10^{12}$ | 10 | [32] |
| ITO/OTPD/CH3NH3PbI3/PCBM/C60/BCP/Al | 0.0725 | $\sim 10^{-9}$ | 0.28 | $7.4 \times 10^{12}$ | 11 | [26] |
| ITO/PEDOT:PSS:/MA0.5FA0.5Pb0.5Sn0.5I3AA/PCBM/Bis-C60/Ag | 0.80 | $\sim 10^{-6}$ | ~ 1000 | $\sim 10^{12}$ | 12 | [74] |
| ITO/PEDOT:PSS/CyPF$_6$/FACsPbI3/PCBM/Ag | 0.10 | $10^{-7}$ | 65/74 | $2 \times 10^{8}$ | 13 | [27] |
| ITO/PEDOT:PSS/FACsPbI3/PCBM/Ag | 0.025 | $1.37 \times 10^{-7}$ | 0.150/0.159 | $2.08 \times 10^{12}$ | 14 | [66] |
| ITO/Poly-TPD/Pb$_{0.5}$Sn$_{0.5}$I3/C60/BCP/Ag | 0.01 | $5 \times 10^{-11}$ | 0.19/0.74 | $2.5 \times 10^{12}$ | 15 | [75] |
| ITO/Me-PACz/CsFAMA/PCBM.BCP/Ag | 0.100 | $1.48 \times 10^{-10}$ | 0.58 | $7.42 \times 10^{13}$ | This work | **Black star** |
| ITO/Me-PACz/CsFAMA/PCBM.BCP/Ag | 0.175 | $1.91 \times 10^{-10}$ | 1.08 | $6.25 \times 10^{13}$ | This work | **Red star** |
| ITO/Me-PACz/CsFAMA/PCBM.BCP/Ag | 0.350 | $3.18 \times 10^{-10}$ | 1.70 | $4.84 \times 10^{13}$ | This work | **Green star** |
| ITO/Me-PACz/CsFAMA/PCBM.BCP/Ag | 0.805 | $1.04 \times 10^{-9}$ | 3.46 | $2.68 \times 10^{13}$ | This work | **Blue star** |
| ITO/Me-PACz/CsFAMA/PCBM.BCP/Ag | 1.000 | $3.27 \times 10^{-8}$ | 4.94 | $4.51 \times 10^{12}$ | This work | **Magenta star** |

## A.4 Discussion

In this work, PTAA and Me-4PACz are used as hole transport layers to fabricate the inverted (*p-i-n*) architecture based PPDs. The PTAA and Me-4PACz HTLs are hydrophobic in nature due to the presence of methyl group (-CH$_3$) and long alkyl chains, resulting in difficulty in forming the uniform perovskite films (see Figure 3).[42,47–49] Therefore, we used conjugated polyelectrolyte polymer (PFN-Br) to get a uniform perovskite layer by using a bi-layer and mixing engineering technique.[42,50] We obtained optimized Me-4PACz HTL by mixing 10% of PFN-Br to it.[42] It is observed that using Me-4PACz HTL, the $J_d$ and the $t_r$ reduced substantially (see **Figure 8a, c**). This smaller leakage current and faster response for the Me-4PACz based PPDs could be attributed to the better interfacial contact of the HTL with the perovskite absorber (see Figure 2). Hence, we scaled up the active area of the Me-4PACz HTL-based PPD from 0.100 cm$^2$ to 1.000 cm$^2$ (Figure 1d). The nonlinear increase of $J_D$ at zero bias of the different active area based PPD is associated with the nonlinear increase in $J_0$ (Figure 15a). The $J_0$ is exponentially dependent on the active area of the PPDs (Figure 15b). Such a nonlinear increase in $J_0$ could be associated with physical or electronic defects in the perovskite thin film.[76,77] The specific reason for the electronic defects is yet to be known.[82] However, the physical defects could be prominent considering solution processable techniques used for perovskite layer deposition and chances of non-uniformity over the larger area substrates.[77–79] However, we note that removing the shunt resistance effects from the dark current results in almost similar $J_0$ for all active area devices (see **Figure 16d**). Such a nonlinear increase of dark current with the active area requires significant attention for scaling up the PD (or PV) devices. However, We did not observe significant differences in the responsivity of the different active area based PPD, unlike Tungsten Telluride (WTe2) based photodetector due to significant non-uniform surface coverage and increased surface defect states.[80,81] The specific detectivity values for the different active areas based on PPDs are calculated using eq. (4), and they are the highest values as per literature reports on the given area (see **Table 7**). It is to be noted that specific detectivity estimation using eq. (4) is overestimated by the shot noise, and this practice is often used in the photodetector community.[61–65] This overestimation impedes progress and results in a lack of clarity in understanding the physical process of the PD field. The eq. (4) is easy to handle as it is free from the bandwidth or frequency limitations due to DC measurements, and only shot noise is considered instead of total noise. It is difficult to measure the absolute noise current that originates from the microscopic voltage fluctuation of a disordered semiconductor.[82,83] However, the total noise current can be measured and have different components such as flicker noise, shot noise, thermal noise, generation-recombination (G-R) noise, etc.[69,83] Hence, it is important to consider all the noise components and measure the true minimum detectable power which produces current equivalent to noise, called as noise equivalent power (NEP). However, for each active area of the device, the flicker noise dominates at a lower frequency (<10 Hz) and saturates at ~ 10 Hz (see **Figure 15c**).[66,67] The average $i_n$ at saturation is used to estimate D* using eq. (5). It is observed that the D* reduces with increase in area due to significant increase of total noise current ($i_n$) in the device (see **Figure 15d**). A PD is considered to be good if shows photocurrent response over a large range of intensities linearly. The photocurrent at NEP is referred the noise floor and is limited by the dark current at zero bias (see **Table 6**). For the large area devices, the noise floor gets elevated to a higher current, i.e., lower NEP, which results in reduced LDR. The measured LDR of different active area devices is limited by the power available of the laser source used in the measurement and, hence, is an underestimated value. The lower rise time (faster response speed) is essential for detecting high frequency AC signals.[69] The rise time increases from 0.58 μs to 4.94 μs upon increasing the device area from 0.100 cm$^2$ to 1.000 cm$^2$. It is observed that the increase of rise time of PPDs with the active area is linear with a slope of 4.53 μs/cm$^2$. The increase in rise time may be qualitatively associated with the increased RC time constant (capacitance: $C = k\varepsilon_0 A/d$) for the larger area (A) of the PPD.[12,69] Which means upon increasing the area, the PD becomes slower responsive. The rise time of 4.94 μs for the 1.000 cm$^2$ active area PPD indicates that any optical AC signal detection at a shorter time < 4.94 μs or higher frequency, the PD lacks to gain the maximum amplitude of the photocurrent (see **Figure 20b**). In some applications, the optical signal is to be collected within a specific bandwidth, e.g., high frequency optical communication, which requires fast extraction of the photogenerated charge carriers without decaying the signal amplitude.[69] Hence, to study the photocurrent response, the frequency dependent photocurrent measurement is carried out and observed that the $f_{-3dB}$ >100 kHz for an active area 0.175 cm$^2$ and ~ 65 kHz for 1.000 cm$^2$ (see **Figure 23**). The reduction in the $f_{-3dB}$ for the large area PPDs may be qualitatively associated with the increase in the RC time constant by increasing the geometric/parasitic capacitance.[12,32,84] Finally, we compare the dark current and the rise time of our different active area based PPDs with the literature reports. We observed that, as the PPD active area increases, the dark current and the rise time increase but are notably lower compared to the literature reports on the given active area for the Me-4PACz HTL based PPDs (see **Figure 24**).[85]

## A.5 Conclusion

In this work, the polymeric molecule PTAA and self-assembled monolayer Me-4PACz are used as HTL in the PPD devices. It is observed that a significant reduction in the dark current and rise time for the Me-4PACz-based PPDs. This could be attributed to the better interfacial contact of the HTL with the perovskite absorber. Thus Me-4PACz based PPD is suitable over the PTAA based PPD. Further, we scaled up the active area of the Me-4PACz based PPD and observed the exponential increase of dark current with the device's active area. However, removing the shunt resistance effects from the dark current results in almost similar $J_0$ for all active area devices. The noise spectral density plot shows that the noise current increases with increasing the device's active area. Therefore, the specific detectivity estimated from the noise current is lower for the larger active area based PPDs. Further, the LDR for the different active area based PPD shows that the noise floor elevates with increased active area due to increased dark current. The increase of rise time or slowing down response speed of the large area PPD may be associated with the increase in the RC time constant (capacitance: $C = k\varepsilon_0 A/d$). Finally, we compared our different active area PPDs with the literature reports and observed that, as the Me-4PACz based PPD active areas increase, the dark current and the rise increase but are notably lower compared to the literature reports. Hence, we conclude that the scalability of the PPD using the Me-4PACz as HTL is a suitable choice.

# APPENDIX B

# List of the variables

# APPENDIX B

## List of the variables

| | Illumination Intensity (mW/cm$^2$) | Carrier generation (cm$^{-3}$s$^{-1}$) | Open-circuit voltage (V) | Carrier density (cm$^{-3}$) |
|---|---|---|---|---|
| Parameters at time t | $I$ | $G$ | $V_{OC}$ | $n$ |
| Steady-state parameters | $I_0$ | $G_0$ | $V_{OC,0}$ | $n_0$ |
| Modulated parameters upon perturbation | $\Delta I = I - I_0$ | $\Delta G = G - G_0$ | $\Delta V_{OC} = V_{OC} - V_{OC,0}$ | $\Delta n = n - n_0$ |

| Parameter | Unit | Definition |
|---|---|---|
| $n, p$ | cm$^{-3}$ | electron, hole density per unit volume. |
| $n_0, p_0$ | cm$^{-3}$ | electron, hole density per unit volume at steady-state conditions. |
| $\Delta n(t) = n - n_0$ | cm$^{-3}$ | Change in electron density at time t. |
| $\Delta n(0)$ | cm$^{-3}$ | Change in perturbed charge carrier density at $t = 0$ (this is also the maximum change in carrier density due to laser pulse illumination) |
| $n_i$ | cm$^{-3}$ | Intrinsic carrier density |
| $n_{BB}$ | cm$^{-3}$ | Carrier density at which radiative recombination dominates trap assisted recombination |
| $n_{Aug}$ | cm$^{-3}$ | Carrier density at which Auger recombination dominates radiative recombination. |
| $T_{ON}$ | s | The laser pulse ON time |
| $\tau$ | s | Carrier lifetime extracted from the transient measurements. |
| $\tau_B$ | s | Trap assisted carrier lifetime in the bulk of the active layer |
| $G_0$ | cm$^{-3}$ s$^{-1}$ | Charge carrier generation per unit volume per unit second upon illumination |
| $\Delta V_{OC,max}$ | V | The maximum change in the open-circuit voltage due to the laser pulse illumination |
| $k_1$ | s$^{-1}$ | Trap assisted recombination coefficient. |
| $k_2$ | cm$^3$ s$^{-1}$ | Radiative recombination coefficient. |
| $k_3$ | cm$^6$ s$^{-1}$ | Auger recombination coefficient. |
| $\alpha$ | NA | Exponent of $\Delta V_{OC}\ vs.\ I_0$ |
| $\beta$ | NA | Exponent of $\tau^{-1}\ vs.\ I_0$ |
| $\gamma$ | NA | Exponent of $J_{SC}\ vs.\ I_0$ |
| $W$ | nm | The thickness of the active layer of the solar cell. |
| $k$ | J/K | Boltzmann constant |
| $T$ | K | Temperature |
| $q$ | C | Electronic charge = $1.6 \times 10^{-19} C$ |
| $kT/q$ | V | Thermal voltage = 25.7 meV |
| $J_{0,dark}$ | mA/cm$^2$ | Reverse saturation current density measured from the $J - V$ scans |
| $J_{0,SV}$ | mA/cm$^2$ | Reverse saturation current density measured from the Suns- $V_{OC}$ measurement. |
| $N_C$ | cm$^{-3}$ | The density of states in the conduction band |
| $N_V$ | cm$^{-3}$ | The density of states in the valence band |
| $N_A$ | cm$^{-3}$ | Doping density |
| $J_{0,SRH}$ | mA/cm$^2$ | Estimated reverse saturation current density due to trap assisted recombination |
| $J_{0,BB}$ | mA/cm$^2$ | Estimated reverse saturation current density due to band to band recombination |
| $S$ | cm/s | Recombination velocity at the interface between the active layer and transport layers |
| $\eta$ | NA | The efficiency of the photovoltaic device |
| $G_{AM\ 1.5G}$ | cm$^3$ s$^{-1}$ | Charge carrier generation per unit volume per unit second upon illumination by AM 1.5G spectrum. |